\documentclass[prb,showkeys,onecolumn,notitlepage,superscriptaddress]{revtex4-2}
\usepackage{amsmath,amsfonts,amssymb,amsthm,epsfig,array}
\usepackage{dsfont}
\usepackage{slashed}
\usepackage{graphics}
\usepackage{float}
\usepackage{verbatim}%\begin{comment} ... \end{comment}
\usepackage{color}
\usepackage[dvipsnames]{xcolor}
\usepackage[hidelinks,colorlinks,linkcolor=blue,
citecolor=blue,urlcolor=blue]{hyperref}
\usepackage[titletoc,title]{appendix}
\usepackage{multirow}
\usepackage{bibentry}
\usepackage{tabularx}
\usepackage[mathscr]{euscript}
\usepackage{mathtools}

\newcommand{\bsl}[1]{\Or{\boldsymbol{\mathrm{#1}}}}

\renewcommand{\mod}{\,\mathrm{mod}\,}

\newcommand{\bra}[1]{\langle #1|}
\newcommand{\ket}[1]{|#1 \rangle}
\newcommand{\braket}[2]{\left\langle #1 | #2  \right\rangle}

\newcommand{\ii}{\mathrm{i}}

\newcommand{\col}{\mathrm{col}}

\newcommand{\dsZ}{\mathbb{Z}}
\newcommand{\dsN}{\mathbb{N}}
\newcommand{\dsR}{\mathbb{R}}

\newcommand{\Tr}{\mathop{\mathrm{Tr}}}

\newcommand{\U}[1]{\mathrm{U}(#1)}

\newcommand{\eqnref}[1]{Eq.\,\eqref{#1}}
\newcommand{\figref}[1]{Fig.\,\ref{#1}}
\newcommand{\tabref}[1]{Tab.\,\ref{#1}}

\newcommand{\appref}[1]{Appendix.\,\ref{#1}}
\newcommand{\refcite}[1]{Ref.\,[\onlinecite{#1}]}

\newcommand{\mat}[1]{\left(\begin{matrix}#1\end{matrix}\right)}

\newcommand{\eq}[1]{\begin{equation} #1 \end{equation}}
\newcommand{\eql}[1]{\begin{widetext}\begin{align}\begin{split} #1 \end{split}\end{align}\end{widetext}}
\newcommand{\eqa}[1]{\begin{align}\begin{split} #1 \end{split}\end{align}}

\newcommand{\Or}[1]{{#1}}

\let\oldAA\AA
\renewcommand{\AA}{\text{\normalfont\oldAA}}
\newcommand{\sgn}[1]{\text{sgn}(#1)}
\newcommand{\ie}{{\Or{i.e.}}}
\newcommand{\eg}{{\Or{e.g.}}}

\newcommand{\s}{\mathrm{s}}
\newcommand{\p}{\mathrm{p}}

\newcommand{\C}{\mathcal{C}}
\newcommand{\G}{\mathcal{G}}
\newcommand{\E}{\mathcal{E}}
\newcommand{\X}{\mathcal{X}}
\renewcommand{\P}{\mathcal{P}}

\newtheorem{definition}{Definition}
\newcommand{\defref}[1]{Def.\,\ref{#1}}

\newcommand{\SN}[1]{\Or{Appendix.\,{#1}}}
\newcommand{\SNs}[1]{\Or{Appendix.\,{#1}}}

\newcommand{\figClassificationDSIFlowchart}{Fig.\,1}%\figref{fig:Classification_DSI_Flowchart}
\newcommand{\figOneDExp}{Fig.\,2}%\figref{fig:1+1D_inver_exp}
\newcommand{\figTwoDFHOTI}{\figref{fig:2+1D_FHOTI}}
\newcommand{\figThreeDFHOTI}{Fig.\,3}

\newcommand{\tabPlaneGSpinless}{Tab.\,1}%\tabref{tab:PlaneG_Spinless}
\newcommand{\tabPlaneGSpinful}{Tab.\,2}%\tabref{tab:PlaneG_Spinful}

\makeatletter
\def\@keys@name{* Correspondence to }
\makeatother

\begin{document}
\title{Dynamical Symmetry Indicators for Floquet Crystals}
\author{Jiabin Yu$^*$}
\affiliation{Condensed Matter Theory Center, Department of Physics,
University of Maryland, College Park, MD 20742, USA}
\author{Rui-Xing Zhang}
\affiliation{Condensed Matter Theory Center, Department of Physics,
University of Maryland, College Park, MD 20742, USA}
\affiliation{Joint Quantum Institute, University of Maryland, College Park, MD 20742, USA}
\author{Zhi-Da Song}
\affiliation{Department of Physics, Princeton University, Princeton, NJ 08544, USA}
\begin{abstract}
Various exotic topological phases of Floquet systems have been shown to arise from crystalline symmetries. 
Yet, a general theory for Floquet topology that is applicable to all crystalline symmetry groups is still in need. 
In this work, we propose such a theory for (effectively) non-interacting Floquet crystals.
We first introduce quotient winding data to classify the dynamics of the Floquet crystals with equivalent symmetry data, and then construct dynamical symmetry indicators (DSIs) to sufficiently indicate the inherently dynamical Floquet crystals.
The DSI and quotient winding data, as well as the symmetry data, are all computationally efficient since they only involve a small number of Bloch momenta.
We demonstrate the high efficiency by computing all elementary DSI sets for all spinless and spinful plane groups using the mathematical theory of monoid, and find a large number of different nontrivial classifications, which contain both first-order and higher-order 2+1D anomalous Floquet topological phases.
Using the framework, we further find a new 3+1D anomalous Floquet second-order topological insulator (AFSOTI) phase with anomalous chiral hinge modes.
\end{abstract}

\keywords{\Or{jiabinyu@umd.edu}}

\maketitle

\section*{Introduction}

Crystalline symmetries are crucial in the study of static band topology~\cite{Hasan2010TI,Qi2010TITSC,Schnyder2008ClassificationTITSC,Kitaev2009TenFoldWayTITSC,Teo2010TenFoldWayTITSC,Ryu2010TenFoldWayTITSC,Chiu2016RMPTopoClas}, because they can protect and can efficiently indicate exotic topological phases~\cite{Fu2011TCI,Shiozeki2014TCITCSC,Chiu2013ReflectionTITSC,
Hughes2011NTRInver,Turner2010Inversion,Liu2014nonsymmorphic,
Fang2015TCI,Zhang2015TMCICorep,Benalcazar2017HOTI,Benalcazar2017HOTI-prb,Schindler2018HOTI,Langbehn2017SOTITSC,
Song2017HOTI,Fang2019RotationAnomaly,Po2018FragileTopo,Cano2018DisEBR,Bradlyn2019Fragile,
Slager2019Wilson,Ahn2019NN,Song2020FragileAffineMonoid,
alexandradinata2020crystallographic,Zhang2020MobiusMnBiTe}.
Powerful theories, topological quantum chemistry~\cite{Bradlyn2017TQC} and symmetry indicators~\cite{Po2017SymIndi,Kruthoff2017TCI}, have been formulated to systematically characterize the crystalline-symmetry-protected (and crystalline-symmetry-indicated) topological phases, which enabled the prediction of thousands of topologically nontrivial materials in a computationally efficient manner~\cite{Bernevig2019TopoMat,Fang2019TopoMat,Wan2019TopoMat,Bernevig2020MTQCMat}.
The power of the two theories relies on the following two features.
First, the two theories can be formally applied to all crystalline symmetry groups.
Second, the topological invariants proposed in the two theories are computationally efficient, as they only involve a small number of high-symmetry momenta~\cite{Turner2012NTRInv,Hughes2011NTRInver,Kruthoff2017TCI,
Song2018TSM,Po2020SI,Bernevig2020MagneticTQC} instead of the entire first Brillouin zone (1BZ).

Beyond the static paradigm, non-interacting Floquet systems~\cite{Sadler2006QuenchecBEC,Oka2009PhotovoltaicHE, Inoue2010FTI, Kitagawa2010TopoFloquet, Lindner2011FTI, Jiang2011MajoranaDriveCAQW, Kitagawa2011PhotoQHI, Dora2012FSHI, Thakurathi2013FloquetMajorana, Wang2013FloquetBloch, Cayssol2013FTI, Rechtsman2013PhotoFTI, Zhou2014FloquetHE, Lababidi2014FloquetQH, Keyserlingk2016FloquetSPT, Else2016FInteractingTP, Zhao2016FloquetSC, Potirniche2017FloquetTPCA, Eckardt2017FloquetReview, Tarnowski2019FloquetCN, Oka2019FloquetReview, Rudner2020FloquetReview, Nakagawa2020WannierRep}---systems with noninteracting time-periodic Hamiltonian---can host anomalous topological phases~\cite{Rudner2013AFTI,Peng2016AFTISound,Maczewsky2017AFTIPhoto,Mukherjee2017AFTIPhoto,Nathan2015TopoSingClas, Fruchart2016FloquetTenFold, Roy2017FloquetTenFold, Yao2017FloquetTenFold,Wintersperger2020ColdAtomAFTI} that have no analogue in any static systems, such as phases with anomalous chiral edge modes in the absence of nonzero Chern numbers~\cite{Rudner2013AFTI,Maczewsky2017AFTIPhoto,Wintersperger2020ColdAtomAFTI}.
Recently, researchers have recognized the important role of crystalline (or space-time) symmetries in protecting driving-induced higher-order topological phases in Floquet systems~\cite{Morimoto2017FTITimeGlide,Xu2018SpaceTime,Franca2018AHOTI, Rodrigues2019FHOTI, Peng2019FHOTI, Ladovrechis2019AFTCI, Seshadri2019FHOTI, Plekhanov2019FHOTI, Nag2019FHOTI, Bomantara2019FHOTI, Chaudhary2019FHOTIPhononInduced, Ghosh2020FHOTI, Hu2020FHOTI, Huang2020FHOTI, Bomantara2020FMCM, Peng2020FHOTIClassSpaceTimeSym, Bomantara2020FHOTSC, Nag2020FHOT, Ghosh2020FHOTSC, Zhang2020FragileFloquet,Zhu2020FHOTI,Zhang2020FHOTI,Chen2020STSym,Zhu2020FloquetMechanical,Nag20213DFloquet}, and predicted exotic physical phenomena like anomalous corner modes.
In particular, \refcite{Zhang2020FHOTI} introduces a systematic theoretical framework of classifying and characterizing 2+1D anomalous Floquet higher-order topological phases protected by point group and chiral symmetries. 

In the field of Floquet topological phases, there are two (among others) open questions that are fundamentally and practically important.
The first one is the topological classification for all crystalline symmetry groups, namely how to efficiently determine whether two generic Floquet crystals with the same crystalline symmetries are topologically equivalent.
The second one is how to efficiently determine whether a generic Floquet crystal is in an anomalous phase that has no analogue in any static systems.
In this work, we refer to such inherently dynamical Floquet crystals as Floquet crystals with obstruction to static limits, in analog to the Wannier obstruction~\cite{Soluyanov2011WannierZ2,Po2017SymIndi,Bradlyn2017TQC,Po2018FragileTopo} for topologically nontrivial phases of static crystals.
Then, the second question can be rephrased as how to efficiently diagnose the obstruction to static limits, which is essential for all the above-mentioned anomalous Floquet topological phenomena.

Unfortunately, there have been few efforts to address the above two open questions in the literature, and the previous related works focused on either specific models or special types of crystalline symmetry groups.
There have been no general theory that is applicable to all crystalline symmetry groups in all spatial dimensions.
Furthermore, the topological invariants proposed in the previous studies have relatively low computational efficiency, since they either do not have an accessible mathematical expression or typically require the information over the entire 1BZ (or a submanifold with nonzero dimensions).
Therefore, a general and computationally efficient theory for Floquet topology that takes crystalline symmetries into account is in need.

\begin{figure*}
    \centering
    \includegraphics[width=0.9\columnwidth]{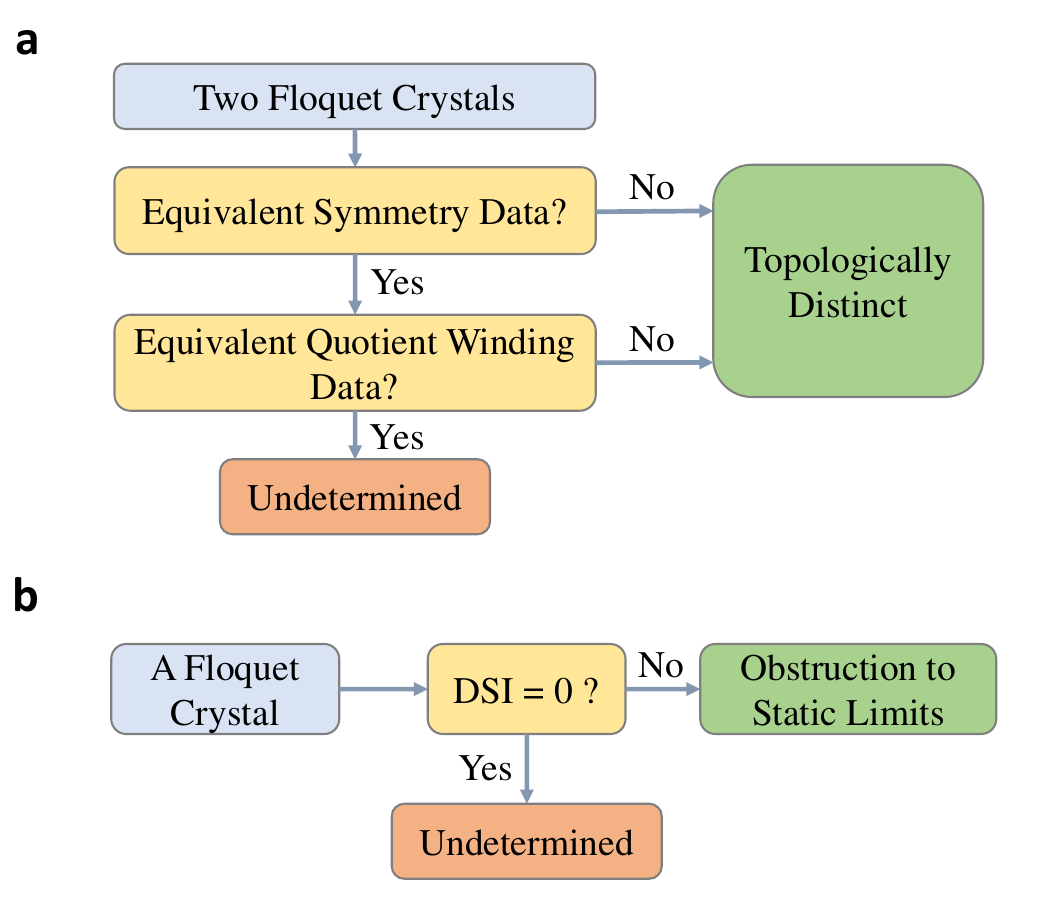}
    \caption{\textbf{Brief summary of main results.} \Or{\textbf{a} is the flowchart for topologically classifying Floquet crystals based on symmetry data and quotient winding data. \textbf{b} is the flowchart for using DSI to indicate obstruction to static limits.}
    }
    \label{fig:Classification_DSI_Flowchart}
\end{figure*}

In this work, we introduce a general theoretical framework to characterize the topological properties of Floquet crystals, which is applicable to all crystalline symmetry groups in all spatial dimensions (up to three).
As a demonstration of our general principle, we focus on non-interacting Floquet crystals in the symmetry class A~\cite{Schnyder2008ClassificationTITSC,Roy2017FloquetTenFold}---in the absence of time-reversal, particle-hole and chiral symmetries---because an applied drive can break the time-reversal symmetry, and exact particle-hole and chiral symmetries hardly appear in normal phases of crystals. 
The brief logic is shown in \figref{fig:Classification_DSI_Flowchart}.
We introduce quotient winding data, which, together with the symmetry data~\cite{Po2017SymIndi,Bradlyn2017TQC} of the quasi-energy bands, provides a topological classification of Floquet crystals (\figref{fig:Classification_DSI_Flowchart}(a)).
In a two-step manner, the symmetry data first provides a coarse classification, which omits the essential information of the dynamics, and the quotient winding data then classifies the dynamics of Floquet crystals with equivalent symmetry data.
Based on our classification scheme, we further introduce the concept of DSIs to indicate the obstruction to static limits (\figref{fig:Classification_DSI_Flowchart}(b)). 
A nonzero DSI is a \Or{sufficient} condition for Floquet crystals to have obstruction to static limits.
\Or{The DSI constructed in this work for Floquet crystals is a dynamical generalization of the symmetry indicator proposed in \refcite{Po2017SymIndi} for static crystals.}

Notably, all indices in our theory---including symmetry data, quotient winding data, and thus the DSI---only involve a small number of Bloch momenta in 1BZ, indicating that the evaluation of them is highly computationally efficient or even analytically feasible.
As a demonstration of the high efficiency, we provide a table of all elementary DSI sets for all spinless and spinful plane groups.
Using specific models, we show that the resultant DSIs can efficiently indicate the nontrivial dynamics in both first-order and higher-order 2+1D anomalous Floquet topological phases.
We further apply our framework to the 3+1D inversion-invariant case (space group P$\overline{1}$ or $\#$2), and find a 3+1D Floquet phase with anomalous chiral hinge modes, which is the first 3+1D AFSOTI phase that is solely protected by static crystalline symmetries.
It is both the generality and high computational efficiency that make our theory remarkably powerful for prediction of new Floquet topological phases.

\section*{Results}

In the following, we will introduce our framework based on symmetry data, quotient winding data, and DSI.
We will use a 1+1D inversion-invariant example to illustrate the main idea.
Then, we will discuss the DSIs for 2+1D systems.
At last, we will introduce the 3+1D AFSOTI phase that we find using DSI.
We also briefly describe the general framework in Methods, and the details can be found in \SNs{B and C}.

The 1+1D inversion-invariant example that we will use is constructed on a 1D lattice with lattice constant being 1, and each lattice site consists of two orbitals at the same position: one spinless $\s$ orbital and one spinless $\p$ orbital.
As we consider the noninteracting cases, we only care about the single-particle Hilbert space, and the symmetry group $\G$ of interest is spanned by the 1D lattice translations and the inversion symmetry.
With bases $\ket{\psi_{k}}=(\ket{\psi_{k,\text{s}}},\ket{\psi_{k,\text{p}}})$, the single-particle Floquet Hamiltonian is represented as $H(k,t)$, where $H(k,t+T)=H(k,t)$ with $T>0$ the time period, and $k$ is the Bloch momentum.
The corresponding unitary time-evolution matrix $U(k,t)$ can be given by Dyson series.
(See \SN{A} for detailed forms of $H$ and $U$ for this 1+1D example.)
Furthermore, the inversion symmetry $\mathcal{P}$ is represented as $\mathcal{P} \ket{\psi_{k}} = \ket{\psi_{-k}} u_{\mathcal{P}}(k)$
with $u_{\P}(k)=\sigma_z$, where $\sigma$'s are the Pauli matrices.
The inversion invariance of the system leads to
\eq{
	\label{eq:1+1D_P_sym_rep}
u_{\P}(k) U(k,t) u_{\P}^\dagger(k) = U(-k,t)\ .
}

The quasi-energy spectrum of $U(k,t)$, derived from diagonalizing $U(k,T)$, is important for our later discussion.
We plot the quasi-energy spectrum for $U(k,t)$ in \figref{fig:1+1D_inver_exp}(a) for one set of parameter values, showing two quasi-energy bands in the phase Brillouin zone (PBZ) $[\Phi_k, \Phi_k+2\pi)$, which are separated by two quasi-energy gaps. (See \SN{A} for details.)
The parameter values used in \figref{fig:1+1D_inver_exp} give us one specific Floquet system; if we change the parameter values, we would get a new time-evolution matrix $U'(k,t)$, featuring another Floquet system.
For this 1+1D example, two Floquet systems are considered to be topologically equivalent if and only if (iff) they are connected by a continuous deformation that preserves the symmetry group $\G$ and both quasi-energy gaps.

In terms of the general terminology discussed in the Methods, we choose both quasi-energy gaps to be the topologically relevant gaps~\cite{Roy2017FloquetTenFold,Yao2017FloquetTenFold} for this 1+1D example, and after choosing the relevant gaps, $U(k,t)$ (or $U'(k,t)$) becomes a \Or{Floquet gapped unitary} (FGU).
In general, the relevant gaps for a generic FGU are chosen based on the physics of interest, and one common choice is to choose all quasi-energy gaps to be relevant, as done in this 1+1D example.
Topological properties of FGUs are the focus of this work.

\subsection*{Symmetry Data}

As the first step of our topological classification, let us describe the symmetry data for the quasi-energy band structure of the 1+1D FGU $U(k,t)$.
Owing to the inversion invariance, the eigenvectors for the quasi-energy bands at an inversion-invariant momentum $k_0=\Gamma/X$ have definite parities $\alpha=\pm$, as shown in \figref{fig:1+1D_inver_exp}(a). 
For each quasi-energy band $\E_{m,k}$ $(m=1,2)$, we can count the number of eigenvectors carrying parity $\alpha$ at each $k_0$, denoted by $n^{m}_{k_0,\alpha}$.
As a result, we have a four-component vector for the $m$-th quasi-energy band as
\eq{
\label{eq:1+1D_P_A_m}
A_m=(n_{\Gamma,+}^m, n_{\Gamma,-}^m,n_{X,+}^m, n_{X,-}^m)^T\ ,
}
of which the values can be read out from \figref{fig:1+1D_inver_exp}(a) as
\eq{
	\label{eq:1+1D_P_sym_data_columns}
	A_1=(1,0,1,0)^T\ ,\ A_2= (0,1,0,1)^T\ .
}
The symmetry data is the matrix $A$ that has $A_1$ and $A_2$ as its two columns
\eq{
	\label{eq:1+1D_P_sym_data}
	A= (A_1\ A_2)\ ,
}
which clearly only involves two momenta in the 1D 1BZ.
The four components of $A_m$ in \eqnref{eq:1+1D_P_A_m} are not independent, as they satisfy the following compatibility relation~\cite{Po2017SymIndi,Bradlyn2017TQC} $n_{\Gamma,+}^m+ n_{\Gamma,-}^m=n_{X,+}^m+ n_{X,-}^m $ or equivalently 
\eq{
\label{eq:comp_rel_1+1D_inversion}
\C A_m = 0
}
with the compatibility matrix $\C$ being
\eq{
\label{eq:1+1D_P_comp_rel_C}
\C=
\mat{
1 & 1 & -1 & -1
}
\ .
}

The above derivation of symmetry data for the quasi-energy band structure is for a given choice of PBZ (as in \figref{fig:1+1D_inver_exp}(a)), which is exactly the same as that for a static crystalline system~\cite{Po2017SymIndi,Bradlyn2017TQC}. 
However, the freedom of choosing PBZ for Floquet crystals leads to an additional subtlety in determining the symmetry data, which is absent in dealing with static crystals.
As shown in \figref{fig:1+1D_inver_exp}(b), we can legitimately shift the PBZ lower bound to $\Phi_k=\pi/4$, which relabels the quasi-energy bands as $1\rightarrow 2$ and $2\rightarrow 1$, resulting in a new symmetry data $\widetilde{A}$
\eq{
\label{eq:1+1D_P_Aprime}
\widetilde{A}= A \mat{ 0 & 1 \\ 1 & 0}\text{ for $\Phi_k=\pi/4$}\ .
}
Therefore, the symmetry data of a Floquet crystal depends on the artificial choice of PBZ, which is in contrast to the static case where the symmetry data of a given static crystal is uniquely determined by the Fermi energy.

We remove this artificial PBZ-dependent ambiguity by defining an equivalence among symmetry data of different FGUs.
We define two FGUs $U'(k,t)$ and $U(k,t)$ to have equivalent symmetry data iff we can find PBZs to make their symmetry data exactly the same. 
In practice, we can first pick a PBZ lower bound $\Phi_k'$ for $U'(k,t)$ and get its symmetry data $A'$.
Then we check whether $A'=A$ (\eqnref{eq:1+1D_P_sym_data}) or $A'=\widetilde{A}$ (\eqnref{eq:1+1D_P_Aprime}); if one of them is true,  $U'(k,t)$ and $U(k,t)$ have equivalent symmetry data, otherwise inequivalent.
Despite the ambiguity of the symmetry data, whether two FGUs have equivalent symmetry data or not is independent of the artificial PBZ choice.
In particular, inequivalent symmetry data infers topological distinction.
Therefore, we can perform a topological classification for FGUs solely based on the symmetry data, similar to what we did for static crystals.

\subsection*{Winding Data}

\begin{figure*}
	\centering
	\includegraphics[width=0.9\columnwidth]{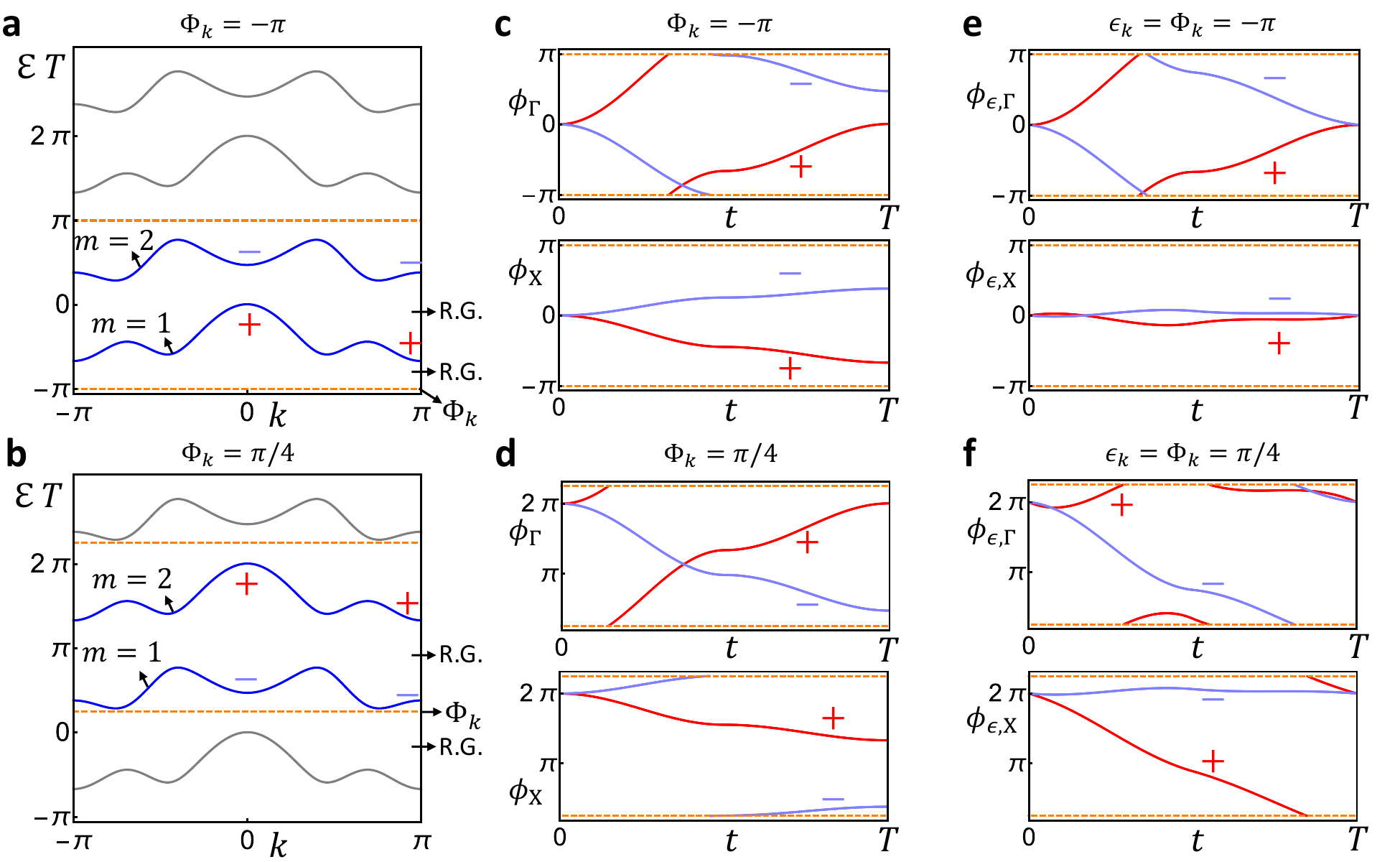}
	\caption{\textbf{Plots of phase and quasi-energy bands for the two-band 1+1D inversion-invariant example.}
		We choose the PBZ lower bound as $\Phi_k=-\pi$ in \Or{\textbf{a,c,e}} and as $\Phi_k=\pi/4$ in \Or{\textbf{b,d,f}}.
		In all plots, the orange dashed lines mark the boundary of the PBZ, and $\pm$ stands for the parity of the eigenvectors at $\Gamma$ $(k=0)$ or $X$ $(k=\pi)$.
		In \Or{\text{a}} and \Or{\text{b}}, we plot the quasi-energy bands (blue lines) given by $U(k,T)$ in the PBZ, while all gray lines are redundant $2\pi$-copies outside the PBZ.
		All quasi-energy gaps are relevant for the topological equivalence, and thereby are relevant gaps (labelled by R.G. in the plots).
		In \Or{\text{c}} and \Or{\text{d}}, we plot the phase bands at $\Gamma$ and $X$ for the time-evolution matrix $U(k,t)$.
		In \Or{\text{e}} and \Or{\text{f}}, we plot the phase bands at $\Gamma$ and $X$ for the return maps $U_{\epsilon=\Phi}(k,t)$.
	}
	\label{fig:1+1D_inver_exp}
\end{figure*}

The above symmetry-data-based classification only involves the time-evolution matrix at $t=T$, missing essential information about the quantum dynamics.
To classify the dynamics of Floquet crystals with equivalent symmetry data, we will construct the winding data, which contains the dynamical information on the entire time period.
A direct visualization of the quantum dynamics for the 1+1D FGU $U(k,t)$ is its phase band spectrum $\phi_{m,k}(t)$ given by directly diagonalizing $U(k,t)$, and we plot the phase bands at two inversion-invariant momenta in \figref{fig:1+1D_inver_exp}(c-d).
However, for the construction of the winding data, it turns out to be inconvenient to directly use $U(k,t)$ or phase bands in \figref{fig:1+1D_inver_exp}(c-d), since they are not time-periodic.

It is much more convenient to use the time-periodic return map~\cite{Nathan2015TopoSingClas,Yao2017FloquetTenFold} $U_\epsilon(k,t)$ defined as
 \eq{
 \label{eq:1+1D_P_RM_def}
 U_\epsilon(k,t)=U(k,t) \left[U(k,T)\right]^{-t/T}_\epsilon\ ,
 }
where $\epsilon_{k}$ is the branch cut for the logrithm used in the return map, and throughout this work, we always set the branch cut to be equal to the PBZ lower bound (\ie, $\epsilon=\Phi$) unless specified otherwise. (See the Methods for more details.)
As we want to make winding data computationally efficient, we only care about the return map at two inversion-invariant momenta $\Gamma$ and $X$.
Since the return map preserves inversion, its eigenvectors at $k_0=\Gamma/X$ have definite parities, as shown in \figref{fig:1+1D_inver_exp}(e) for $\Phi_k=-\pi$.
Then, we can count the total winding (along $t$) of the phase bands of $U_{\epsilon=\Phi}(k_0,t)$ with parity $\alpha$, resulting in an integer-valued winding number $\nu_{k_0,\alpha}$. 
We can read out all four quantized winding numbers from \figref{fig:1+1D_inver_exp}(e) and further group them into a vector
\eq{
\label{eq:1+1D_P_winding_data_V}
V=(\nu_{\Gamma,+},\nu_{\Gamma,-},\nu_{X,+},\nu_{X,-})^T=(1,-1,0,0)^T\ ,
}
which we refer to as the winding data of the FGU $U(k,t)$ for $\Phi_k=-\pi$.
(See more details in the Methods.)
Clearly, the winding data only involves two momenta in the 1D 1BZ.

As exemplified by \eqnref{eq:1+1D_P_winding_data_V}, the four components of the winding data satisfy a compatibility relation
\eq{
\nu_{\Gamma,+}+\nu_{\Gamma,-}=\nu_{X,+}+\nu_{X,-}\ ,
}
since the total winding of all phase bands at each momentum is the same.
As a result, the winding data share the same compatibility relation as that of the symmetry data (\eqnref{eq:comp_rel_1+1D_inversion})
\eq{
	%\label{eq:1+1D_P_comp_rel_V}
	\C V =0\ , 
}
meaning that the winding data takes value in the following set
\eqa{
\label{eq:1+1D_P_winding_data_set}
\{ V \}&= \dsZ^4 \cap \ker \C \\
&= \{ (q_1,q_2,q_3,q_1+q_2-q_3)^T|q_1,q_2,q_3\in\dsZ\} \approx \dsZ^3\ .
}

Shifting the PBZ changes the winding data.
In this 1+1D example, if we shift the PBZ lower bound from $\Phi_k=-\pi$ to $\Phi_k=\pi/4$, the phase bands of return map become \figref{fig:1+1D_inver_exp}(f), and from \figref{fig:1+1D_inver_exp}(f), we know the winding data becomes
\eq{
\label{eq:1+1D_P_winding_data_Vprime}
\widetilde{V}= (0,-1,-1,0)^T = V-A_1\ .
}
Unlike the symmetry data, a $2\pi$-shift of the PBZ $\Phi_k\rightarrow \Phi_k+2\pi$ can also change the winding data
\eq{
\label{eq:1+1D_winding_data_2pi_shift}
V \rightarrow V - (1,1,1,1)^T\ ,
}
suggesting that the 1+1D FGU $U(k,t)$ can have an infinite number of different winding data, which explicitly depend on the artificial choice of PBZ.

The infinite PBZ dependence of the winding data makes it hard to directly generalize the equivalence among symmetry data (which only has a finite number of variants for a single FGU) to define an equivalence among the winding data, since finding a single proper PBZ among an infinite number of possible choices is not straightforward.
Nevertheless, the infinitely many winding data are related by the symmetry data, as shown in \eqnref{eq:1+1D_P_winding_data_Vprime} and \eqnref{eq:1+1D_winding_data_2pi_shift}.
This relation inspires us to define the quotient winding data below, in order to resolve the infinity problem.

\subsection*{Quotient Winding Data}
%\label{sec:1+1D_P_quotient_winding_data}

In this part, we define the quotient winding data to resolve the infinity issue of the winding data.
To have a finite number of different quotient winding data, we define the quotient winding data to be invariant under all PBZ shifts that keep the symmetry data.
For the 1+1D FGU $U(k,t)$, all PBZ shifts that keep the symmetry data are (or equivalent to) the $2\pi n$-shifts of the PBZ, where $n$ labels an arbitrary integer.
Then, we define the quotient winding data $V_Q$ as
\eq{
\label{eq:quotient_winding_data_def}
V_Q = V \mod\ \bar{A}\ ,
}
where
\eq{
\label{eq:1+1D_P_Abar}
\bar{A}=A_1+A_2=(1,1,1,1)^T\ .
}
As $2\pi n$-shifts of the PBZ can only change $V$ by multiples of $\bar{A}$ according to \eqnref{eq:1+1D_winding_data_2pi_shift}, $V_Q$ defined in \eqnref{eq:quotient_winding_data_def} is indeed invariant under $2\pi n$-shifts of the PBZ, just like the symmetry data.
As a result, the 1+1D FGU $U(k,t)$ only has two different quotient winding data derived from the two winding data in \eqnref{eq:1+1D_P_winding_data_V} and \eqnref{eq:1+1D_P_winding_data_Vprime} as
\eqa{
\label{eq:1+1D_P_QWD_example}
&V_Q= V \mod \bar{A}= (0,-2,-1,-1)^T\ \ \text{ for }\Phi_k=-\pi\ ,\\
&\widetilde{V}_Q=\widetilde{V} \mod \bar{A}= (0,-1,-1,0)^T\ \ \text{for }\Phi_k=\pi/4\ ,
}
which are related by
\eq{
\label{eq:1+1D_P_QWD_relation}
\widetilde{V}_Q=V_Q-A_1 \mod \bar{A}\ .
}
(See more details in the Methods.)

To further remove the remaining PBZ-dependent ambiguity, we define an equivalence among quotient winding data of different FGUs.
Recall that the quotient winding data is introduced for a classification of FGUs with equivalent symmetry data, since inequivalent symmetry data already infers topological distinction.
Let us suppose that the two different 1+1D FGUs $U(k,t)$ and $U'(k,t)$ have equivalent symmetry data, meaning that we can always pick PBZ choices $\Phi_k'$ and $\Phi_k$ for $U'(k,t)$ and $U(k,t)$, respectively, such that they have exactly the same symmetry data $A'=A$.
Then, we check whether the quotient winding data of $U'(k,t)$ for $\Phi_k'$ is the same as that of $U(k,t)$ for $\Phi_k$; if so, $U'(k,t)$ and $U(k,t)$ are defined to have equivalent quotient winding data.

Given two FGUs with equivalent symmetry data, the artificial PBZ choice has no influence on whether they have equivalent quotient winding data or not.
In particular, they must have equivalent quotient winding data if they are topologically equivalent, meaning that inequivalent quotient winding data infers topological distinction.
Then, as long as the comparison of $V_Q$ is done for the PBZ choices that yield the same symmetry data, quotient winding data provides a topological classification for all FGUs that have equivalent symmetry data to the 1+1D $U(k,t)$.
Specifically, the classification is given by $\{ V \}$ in \eqnref{eq:1+1D_P_winding_data_set} and $\bar{A}$ in \eqnref{eq:1+1D_P_Abar} as
\eq{
\label{eq:1+1D_P_QWD_set}
\{ V_Q \}=\{ (0,q_2,q_3,q_2-q_3)^T|q_2,q_3\in\dsZ\}\approx\frac{\{ V \}}{\bar{A} \dsZ }  \approx \dsZ^2\ ,
}
where $\bar{A} \dsZ = \{ q \bar{A} = (q,q,q,q)^T | q\in \dsZ\}$. 

Up to now, we have discussed the relative topological classification based on the symmetry data and the quotient winding data, shown in \figref{fig:Classification_DSI_Flowchart}(a).
Nevertheless, the $(A,V_Q)$-based classification fails to tell which FGU has obstruction to static limits, \ie, topologically distinct from all static FGUs with the same symmetries. (See the Methods for general definitions.)
Yet, determining obstruction to static limits is crucial, because it tells whether the Floquet phase of interest has no static analogue or equivalently whether it is allowed to have any anomalous dynamical phenomena.
For this purpose, we construct the DSI below.

\subsection*{DSI}
%\label{sec:1+1D_P_DSI}

In this part, we construct the DSI for the 1+1D example to efficiently indicate its obstruction to static limits.
To determine the obstruction to static limits for the 1+1D FGU $U(k,t)$ with $\G$ spanned by inversion and lattice translation, we only need to consider the $\G$-invariant static FGUs that have symmetry data equivalent to $U(k,t)$, since $U(k,t)$ must be topologically distinct from all other $\G$-invariant static FGUs. 
Then, we can check whether any winding data of $U(k,t)$ is forbidden in all static FGUs with symmetry data equivalent to $U(k,t)$; if so, then $U(k,t)$ must have obstruction to static limits.
Therefore, although we need to use the quotient winding data to give the relative classification, we can directly use winding data to determine the obstruction.
(See \SN{A} for details.)

The DSI is constructed by formalizing the above criterion.
To do so, we first derive a set $\{ V_{SL} \}$ by winding each quasi-energy band of the 1+1D $U(k,t)$ along time, which reads
\eq{
\label{eq:1+1D_P_WD_set_SL}
\{ V_{SL} \}=\{ q_1 A_1 + q_2 A_2=(q_1, q_2, q_1,q_2)^T | q_1,q_2 \in \dsZ \}\ ,
}
where $A_1$ and $A_2$ are two columns of $A$ in \eqnref{eq:1+1D_P_sym_data}.
$\{ V_{SL} \}$ is invariant under the relabelling of the quasi-energy bands ({\ie} $1\leftrightarrow 2$) due to a PBZ shift, and it contains all winding data of all $\G$-invariant static FGUs that have symmetry data equivalent to $U(k,t)$. (See details in \SN{A}.)
Then, according to \eqnref{eq:1+1D_P_winding_data_V}, the wind data $V$ of $U(k,t)$ for $\Phi_k=-\pi$ satisfies $V\notin \{ V_{SL}\}$, since $\nu_{\Gamma,+}- \nu_{X,+}=1$ for $V$ while $\nu_{\Gamma,+}- \nu_{X,+}=0$ for all elements in $\{ V_{SL} \}$.
It means that $V$ cannot exist in any of the static FGUs that have symmetry data equivalent to $U(k,t)$, and thus $U(k,t)$ must have obstruction to static limits.

It turns out that we are allowed to adopt any PBZ choice for $U(k,t)$ to check the above formalized criterion, and we will always get the same result that $U(k,t)$ has the obstruction to static limits, because \eqnref{eq:1+1D_P_winding_data_Vprime} and \eqnref{eq:1+1D_winding_data_2pi_shift} suggests $U(k,t)$ always has $\nu_{\Gamma,+}- \nu_{X,+}=1$ regardless of the PBZ choice.
We refer to the PBZ-independent $(\nu_{\Gamma,+}- \nu_{X,+})$ as the DSI for $U(k,t)$---as well as for all other FGUs that have symmetry data equivalent to $U(k,t)$.
Formally, the DSI is defined to take values from the following set $\X$
\eq{
\label{eq:1+1D_P_DSI}
\X= \frac{\{ V \}}{\{ V_{SL} \}}\approx \{\nu_{\Gamma,+}- \nu_{X,+} \in \dsZ \}\ ,
}
where we have used \eqnref{eq:1+1D_P_winding_data_set} and \eqnref{eq:1+1D_P_WD_set_SL}.
As shown above, the DSI only involves two momenta in the 1BZ, and nonzero DSI means all winding data of $U(k,t)$ are not in $\{ V_{SL} \}$, thus sufficiently indicating the obstruction to static limit.

The idea of using quotient group to mod out the trivial systems (though not exclusively), which is used above, was previously used to construct the static symmetry indicator in \refcite{Po2017SymIndi}.
The difference between \refcite{Po2017SymIndi} and our work is that the quotient is taken for the symmetry contents (like columns of symmetry data) in the construction of the static symmetry indicator in \refcite{Po2017SymIndi} to characterize the static band topology, while the quotient is taken for the winding data in the construction of the DSI to characterize the periodic quantum dynamics here.

Now we discuss the DSIs for all possible 1+1D inversion-invariant FGUs.
To do so, we need to use the Hilbert bases~\cite{Song2020FragileAffineMonoid,Bruns2009Polytopes}, which intuitively speaking, are irreducible bases of the symmetry data. (See Methods and \SN{C} for more details.)
For 1+1D inversion-invariant FGUs, there are four Hilbert bases given by four ways of assigning $\pm$ parities to $\Gamma/X$, which read
\eqa{
\label{eq:1+1D_P_HBs}
a_1=(1,0,1,0)^T\ ,\\
a_2=(0,1,1,0)^T\ ,\\
a_3=(1,0,0,1)^T\ ,\\
a_4=(0,1,0,1)^T\ .
}
Then, any column of any symmetry data of any 1+1D inversion-invariant FGU is the linear combination of the four Hilbert bases with non-negative integer coefficients.

Based on the Hilbert bases, symmetry data of FGUs can be split into two types, irreducible and reducible.
Specifically, we define a symmetry data of a FGU to be irreducible iff all its columns are Hilbert bases; otherwise reducible.
According to the general framework presented in Methods, DSI sets for reducible symmetry data can be constructed from those for irreducible symmetry data.
Therefore, in the following, we focus on the DSI sets for irreducible symmetry data.

For a 1+1D inversion-invariant FGU with irreducible symmetry data, all its symmetry data are spanned by a unique set of the Hilbert bases $\{ a_{j} \}$ with $j$ taking $J\leq 4$ different values in $\{ 1,2,3,4\}$.
According to the general framework discussed in the Methods, we can directly obtain the DSI set for the FGU solely based on the set $\{ a_{j} \}$ and the compatibility matrix $\mathcal{C}$ in \eqnref{eq:1+1D_P_comp_rel_C}.
For the above 1+1D example $U(k,t)$, the two columns of any symmetry data (\eqnref{eq:1+1D_P_sym_data} or \eqnref{eq:1+1D_P_Aprime}) are the Hilbert bases $a_1$ and $a_4$ in \eqnref{eq:1+1D_P_HBs}, and thus are irreducible.
Then, the unique set of the Hilbert bases that span the symmetry data is $\{ a_1, a_4 \}$, and the DSI set \eqnref{eq:1+1D_P_DSI} can be directly derived from $\{ a_1, a_4 \}$ and $\mathcal{C}$ based on the general framework.

In particular, even if two 1+1D inversion-invariant FGUs have inequivalent symmetry data, they have the same $\X$, as long as their irreducible symmetry data are spanned by the same set of Hilbert bases.
This simplification allows us to enumerate all possible DSI sets for irreducible symmetry data by considering all $2^4-1$ nontrivial combinations of Hilbert bases.
As a result, we obtain two nontrivial DSI sets.
One is for the Hilbert basis set $\{a_1,a_4\}$, which is just the above 1+1D example $U(k,t)$, and the DSI is shown in \eqnref{eq:1+1D_P_DSI}.
The other one is for the Hilbert basis set $\{a_2,a_3\}$, and the DSI set reads
\eq{
\label{eq:1+1D_P_DSI_theother}
\X\approx\{ \nu_{\Gamma,+}-\nu_{X,-}\in\dsZ \}\ . 
}

Besides indicating obstruction to static limit, DSI is also a topological invariant---its different values infer topological distinction for FGUs with equivalent symmetry data.
Although the classification given by DSIs is a subset of that given by quotient winding data (like for the 
above 1+1D example), DSIs have the advantage of being PBZ-independent.

\subsection*{2+1D DSI}

The above discussion focused on the 1+1D inversion-invariant case.
We, in this part, discuss the DSIs for 2+1D systems.
Based on the general framework in Methods, we derive the DSI sets for all nontrivial combinations of Hilbert bases for all spinless and spinful 2D plane groups, and list the numbers of nontrivial DSI sets in \tabref{tab:PlaneG_Spinless}-\ref{tab:PlaneG_Spinful}.
These DSI sets are for 2+1D FGUs with irreducible symmetry data, and serve as the building blocks for all 2+1D DSI sets for plane groups. 
The nontrivial DSIs in \tabref{tab:PlaneG_Spinless}-\ref{tab:PlaneG_Spinful} can indicate both first-order and higher-order anomalous Floquet topological phases, as discussed below.

For the first-order phase, we focus on the $\dsZ^3$ set for spinless plane group p2, which is spanned by the 2D lattice translations and the two-fold rotation $C_2$.
We explicitly construct a 2+1D p2-invariant spinless model that has nonzero $\dsZ^3$ DSI (the construction of the model is inspired by the quantum-anomalous-Hall-effect model in \refcite{Qi2010TITSC}); we find that the model has anomalous chiral edge modes in the absence of nonzero Chern numbers, similar as the first-order anomalous Floquet topological phase in \refcite{Rudner2013AFTI}. (See \SN{D} for details.)
Therefore, the $\dsZ^3$ DSI can indicate first-order anomalous Floquet topological phases. 
In particular, all components of the DSI take the same values as the winding number $W$ defined in \refcite{Rudner2013AFTI} in our specific p2-invariant model, but the evaluation of the former is much more efficient than the latter, since the former only involves four $C_2$-invariant momenta while the latter needs the whole 2D 1BZ.
Then, although the winding number $W$ defined in \refcite{Rudner2013AFTI} does not rely on any crystalline symmetries other than lattice translations, our model suggests that in the presence of nontrivial crystalline symmetries, DSIs might efficiently indicate the nontrivial dynamics of the first-order anomalous Floquet topological phases characterized by the $W$ winding number.

For the higher-order phase, we find that the 2+1D anomalous Floquet higher-order topological insulator phase proposed in \refcite{Huang2020FHOTI} can be indicated by the $\dsZ$ DSI of spinful p4mm in \tabref{tab:PlaneG_Spinful}. (See \SN{E} for details.)
In particular, to determine the nontrivial dynamics in the model, the DSI only requires three momenta in the 1BZ, saving us from evaluating the quantized dynamical quadrupole momoent proposed in \refcite{Huang2020FHOTI}, which involves all momenta in the entire 2D 1BZ.

\subsection*{3+1D AFSOTI Phase}

In this part, we apply our framework to the 3+1D inversion-invariant case (P$\overline{1}$ space group), and predict a new 3+1D AFSOTI phase.
We will only present a brief discussion here, and details can be found in \SN{F}.

For P$\overline{1}$, we only need to care about the eight inversion-invariant momenta---$\Gamma(0,0,0)$, $X(\pi,0,0)$, $Y(0,\pi,0)$, $Z(0,0,\pi)$, $V(\pi,\pi,0)$, $U(\pi,0,\pi)$, $T(0,\pi,\pi)$, and $R(\pi,\pi,\pi)$~\cite{Bradlyn2017TQC}---and we have parities $\pm$ at each inversion-invariant momentum.
Then, we choose the winding data to have the form
\eq{
(\nu_{\bsl{K}_1,+}, \nu_{\bsl{K}_1,-},...,\nu_{\bsl{K}_8,+},\nu_{\bsl{K}_8,-})^T
}
with $\bsl{K}_i=(\Gamma,X_,Y,Z,V,U,T,R)_i$, and $\nu$ is the winding number.
Replacing $\nu$ by the number of irreducible representations (irreps) labeled by parities in the above expression gives columns of the symmetry data.

P$\overline{1}$ has 256 Hilbert bases, given by assigning $\pm$ to the eight inversion-invariant momenta in the 3D 1BZ, and thus the number of nontrivial combinations of the Hilbert bases is $2^{256}-1$, which is very large.
For simplicity, we only compute the DSI sets for the 32896 combinations that only include one or two Hilbert bases, resulting in $\dsZ\ (3584)$, $\dsZ^2\ (7168)$, $\dsZ^3\ (8960)$, $\dsZ^4\ (7168)$, $\dsZ^5\ (3584)$, $\dsZ^6\ (1024)$, and, $\dsZ^7\ (128)$, where $\dsZ^n$ labels the DSI set and the number in the bracket labels how many nontrivial combinations of Hilbert bases lead to the DSI set.
For concreteness, we in the following focus on the $\dsZ^7$ DSI set that corresponds to the combination of the following two Hilbert bases
\eqa{
\label{eq:3+1D_P_HB}
& \widetilde{a}_1 = (1,0,0,1,0,1,0,1,0,1,0,1,0,1,1,0)^T \\
& \widetilde{a}_2 = (0,1,1,0,1,0,1,0,1,0,1,0,1,0,0,1)^T\ ,
}
and the DSI is a seven-component vector that reads
\eql{
\label{eq:3+1D_P_DSI}
(\nu_{\Gamma,+}-\nu_{X,-},\nu_{\Gamma,+}-\nu_{Y,-},\nu_{\Gamma,+}-\nu_{Z,-},\nu_{\Gamma,+}-\nu_{V,-},\nu_{\Gamma,+}-\nu_{U,-},\nu_{\Gamma,+}-\nu_{T,-},\nu_{R,-}-\nu_{\Gamma,-})\ .
}

\begin{figure*}
	\centering
	\includegraphics[width=0.9\columnwidth]{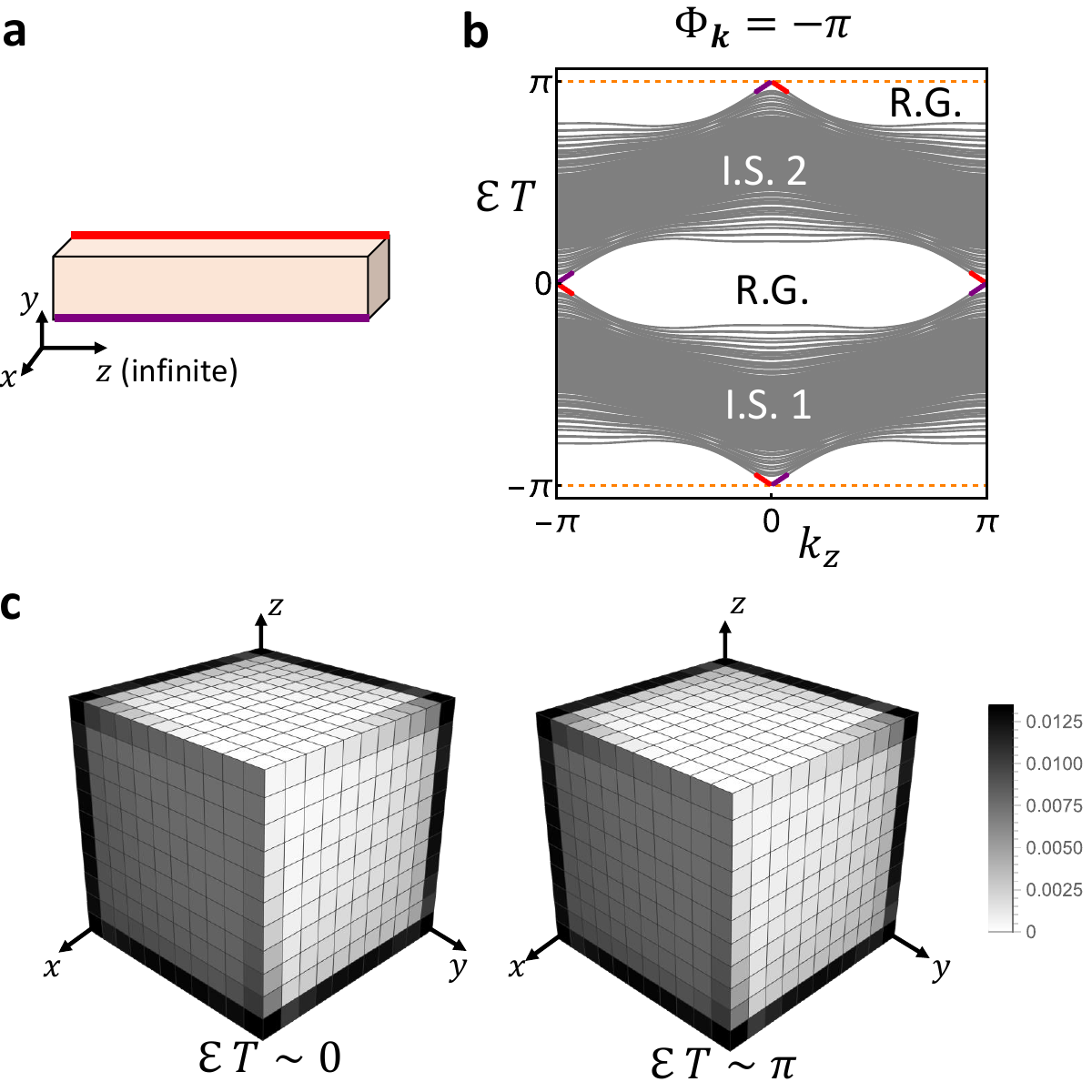}
	\caption{
	\textbf{The 3+1D AFSOTI with anomalous chiral hinge modes.}
	In \Or{\textbf{a}}, we show an inversion-preserving configuration of the 3+1D AFSOTI that is finite along $x,y$ and infinite along $z$. 
	The purple and red lines indicate the hinges that host anomalous chiral hinge modes. 
	In \Or{\textbf{b}}, we plot the quasi-energy band structure of the 3+1D AFSOTI for the configuration in \Or{\textbf{a}}. 
	$\E$ and $T$ label the quasi-energy and time period, respectively. 
	The gray lines are the bulk bands (as well as the dangling surface bands). 
	“I.S.” labels isolated set of bulk quasi-energy bands, and each isolated set contains two bulk quasi-energy bands. 
	“R.G.” stands for the bulk relevant gap, and there are two relevant gaps, one at $\E T=0$ (called 0-gap) and the other at $\E T=\pi$ (called $\pi$-gap). 
	The purple and red lines mark the anomalous chiral hinge modes localized at the purple and red hinges in \Or{\textbf{a}}, respectively. 
	The orange dashed lines mark the boundary of PBZ.  
	In \Or{\textbf{c}}, we consider an inversion-preserving configuration of the 3+1D AFSOTI that is finite along all three spatial directions, and plot total probability density of the two eigenmodes with quasi-energies closest to $0$ or $\pi$ ($\mod 2\pi$).
	Details on the parameter values can be found in \SN{F}.
	}
	\label{fig:3+1D_inver_exp}
\end{figure*}

To demonstrate the dynamical phase indicated by the $\dsZ^7$ DSI, we explicitly construct a 3+1D dynamical tight-binding model with P$\overline{1}$ space group on a cubic lattice with the lattice constant being 1. 
It has four bulk quasi-energy bands, which are split into two isolated sets by two relevant gaps, one 0-gap and one $\pi$-gap. (See \figref{fig:3+1D_inver_exp}(a,b).)
According to the Methods, each isolated set (that consists of two bands) corresponds to one column in the symmetry data, resulting in a two-column symmetry data $A=(A_1\ A_2)$.
Direct calculation shows that $A_1=2 \widetilde{a}_1$ and $A_2 = 2 \widetilde{a}_2$, meaning that the symmetry data is reducible.
Nevertheless, the Hilbert basis set that spans the symmetry data is uniquely $\{\widetilde{a}_1,\widetilde{a}_2\}$, and thus the model is characterized by the DSI in \eqnref{eq:3+1D_P_DSI}, which is evaluated to $(2,2,2,2,2,2,2)$.
As a result of the nontrivial dynamics characterized by the nonzero DSI, the system has chiral hinge modes in each bulk relevant gap, as shown in \figref{fig:3+1D_inver_exp}.
The chiral hinge modes are anomalous, because the static topological invariant, axion angle, for the inversion-protected chiral hinge modes~\cite{Varnava2018AI} is zero for both isolated sets of quasi-energy bands according to the symmetry data~\cite{Turner2010Inversion,Hughes2011NTRInver}.

We emphasize that although hinge modes in 3+1D Floquet insulators were discussed in \refcite{Peng2019FHOTI} and \refcite{Nag20213DFloquet}, our model is fundamentally different from theirs. 
First, the hinge modes originate from the time glide symmetry in \refcite{Peng2019FHOTI} and from the effective spectral symmetry in \refcite{Nag20213DFloquet}, both of which are not static crystalline symmetries, while our model is protected by static inversion symmetry. 
Second, trivial static topology has not been explicitly confirmed for the bulk quasi-energy bands in \refcite{Peng2019FHOTI} and \refcite{Nag20213DFloquet}, while the relevant static topological invariants in our model are confirmed to be trivial.
Therefore, our model, which is constructed based on the DSI, is the first AFSOTI solely protected by the static crystalline symmetries. 

\section*{Discussion}
\label{sec:con_dis}

To summarize, we have established a general and efficient theoretical framework for classifying and characterizing the topological properties of Floquet crystals in the symmetry class A, which is applicable to all crystalline symmetry groups in all spatial dimensions (up to three).

One direct physical implication of the obstruction to static limits is that symmetry breaking or relevant gap closing must appear during any continuous deformation that makes static a Floquet crystal with obstruction.
Here the relevant gap closing refers to the closing of the topologically relevant bulk quasi-energy gaps, according to the definition of the topological equivalence discussed in Methods.
If all relevant symmetries are preserved during the deformation, the gap closing will occur in at least one of the relevant gaps in the bulk quasi-energy spectrum, and is not required to appear in any irrelevant gaps.
Therefore, an experimental test of nonzero DSIs (though not conclusively) would be to observe the quasi-energy gap closing or symmetry breaking as continuously decreasing the driving amplitude to zero while fixing the driving period.
Gap closing in quasi-energy spectrum has been observed in experiments like \refcite{Wintersperger2020ColdAtomAFTI}.

As for more experimental signatures of our theory, it is worth studying the link between nonzero DSIs and nontrivial boundary signature in the future.
A promising direction is p2 plane group, for which the DSI is very likely to contain the information of chiral edge modes.
The intuition is based on the fact that the difference in winding data for different PBZ choices consists of the symmetry contents of quasi-energy bands, which have a mod-$2$ relation to the Chern number~\cite{Turner2010Inversion,Hughes2011NTRInver}.
The 2+1D p2-invariant model presented above also suggests a relation between the DSI and the winding number defined in \refcite{Rudner2013AFTI}, where the latter has a correspondence to the chiral edge modes.
Furthermore, the 3+1D AFSOTI presented above suggests that the 3+1D DSI might indicate the anomalous chiral hinge modes in certain space groups, perhaps related to a dynamical generalization of the static bulk-boundary correspondence between the axion angle and the static chiral hinge modes~\cite{Varnava2018AI}.

As the symmetry-representation theories for static crystals inspired the proposal of fragile topology~\cite{Po2018FragileTopo,Cano2018DisEBR} that is beyond the K-theory classification~\cite{Sato2018KTheorySG}, another interesting direction is to generalize the concept of fragile topology to Floquet crystals~\cite{Yu2021DFragileT}.
Besides, generalizing our theoretical frame work to Floquet crystals with time-reversal, particle-hole, or chiral symmetries is another interesting direction, since it would help identify exotic physical phenomena like anomalous boundary Majorana modes protected by particle-hole symmetries~\cite{Jiang2011MajoranaDriveCAQW,Po2017chiralFloquetKitaev,Peng2020FloquetMajoranaPlanarJosephson,Zhang2020AFCTSC,Vu2021AFTSC}.
As our framework focuses on operators, it is interesting to ask whether it is possible to formalize an equivalent state-based formalism~\cite{Nakagawa2020WannierRep}.
Similar to the symmetry-representation theories~\cite{Po2017SymIndi,Bradlyn2017TQC} for static crystals, our classification is not necessarily complete, since two Floquet crystals with equivalent symmetry and quotient winding data might still be topologically distinct, and the obstruction to static limits might still occur for zero DSI (\figref{fig:Classification_DSI_Flowchart}).
Thereby, the complete topological classification for static and Floquet crystals is a meaningful future direction.

\emph{Note added}: Recently, we noticed \refcite{GongJ2021SymmetryAnalysisAFT}, which proposed to classify Floquet topological phases by using dynamical symmetry inversion points.

\section*{Methods}

\subsection*{Basic Definitions}
In this part, we list the basic definitions used in this work.
Details can be found in \SN{B}.

A Floquet crystal is defined to be a time-evolution operator $\hat{U}(t)$ equipped with a time period $T$, a relevant gap choice, and a crystalline symmetry group $\G$, which is in short denoted by $\hat{U}(t)$.
In the definition of a Floquet crystal, we have implied that $\hat{U}(t)$ is unitary and its matrix representation for any bases is continuous.
A FGU is defined to be a time-evolution matrix $U(\bsl{k},t)$ equipped with a time period $T$, a relevant gap choice, a crystalline symmetry group $\G$, and a symmetry representation $u_g(\bsl{k})$, which is in short denoted by $U(\bsl{k},t)$.
Here $\bsl{k}$ is the momentum.
In the definition of a FGU, we have implied  that $U(\bsl{k},t)$ and $u_g(\bsl{k})$ are unitary, continuous (smooth for $u_g(\bsl{k})$), and invariant under the shift of $\bsl{k}$ by reciprocal lattice vectors.
By choosing bases for a Floquet crystal, we naturally get a FGU with the same time period, relevant gaps and crystalline symmetry group as the Floquet crystal.
FGUs given by the same Floquet crystal with different choices of bases are related by gauge transformations.

Suppose we have two FGUs $U(\bsl{k},t)$ (with $T$, relevant gaps, $\mathcal{G}$, and  $u_g(\bsl{k})$) and $U'(\bsl{k},t)$ (with $T'$, relevant gaps, $\mathcal{G}$, and $u_g'(\bsl{k})$).
The two FGUs $U(\bsl{k},t)$ and $U'(\bsl{k},t)$ are defined to be topologically equivalent under the crystalline symmetry group $\G$ iff there exists a continuous deformation that connects them, preserves $\G$ and preserves all relevant gaps.
As long as the crystalline symmetry group $\G$ for the topological equivalence is specified, we may refer to ``topologically equivalent under $\G$" as ``topologically equivalent" in short.
Suppose we have two Floquet crystals $\hat{U}(t)$ (with $T$, a relevant gap choice, and $\G$) and $\hat{U}'(t)$ (with $T'$, a relevant gap choice, and $\G$).
The two Floquet crystals $\hat{U}(t)$ and $\hat{U}'(t)$ are defined to be topologically equivalent iff there exists a continuous deformation that connects them, preserves $\G$ and preserves all relevant gaps.
If two Floquet crystals are topologically equivalent, they must have topologically equivalent FGUs for any bases choices.
Therefore, the topological distinction among FGUs must infer the topological distinction among the underlying Floquet crystals, and all topological invariants of FGUs can be applied to Floquet crystals.
For this reason, we focus on the FGUs in this work.

The defined topological equivalence for FGUs is similar to the definition in Sec.\,2 of \refcite{Nathan2015TopoSingClas}, except the following two differences.
First, the definition in this work allows the deformation to deviate from the topologically equivalent FGUs by gauge transformations so that the defined topological equivalence is gauge invariant.
Second, the definition in this work allows the symmetry representation and time period to vary along the deformation, and also allows the symmetry representation to depend on momenta.
Furthermore, the topological classification based on the definition in this work may be different from the classification in \refcite{Fruchart2016FloquetTenFold, Roy2017FloquetTenFold, Yao2017FloquetTenFold}.
The topological equivalence defined in this work is immune to any global energy shift, while the same global energy shift may change value of the topological invariant in \refcite{Fruchart2016FloquetTenFold, Roy2017FloquetTenFold, Yao2017FloquetTenFold}.

Last but not least, a static limit is a Floquet crystal with static Hamiltonian; a static FGU is a FGU with static matrix Hamiltonian.
A Floquet crystal (a FGU) with $\G$ is defined to have obstruction to static limits iff it is topologically distinct from all static limits (static FGUs) with $\G$.

\subsection*{Return Map}
The return map for the 1+1D $U(k,t)$ is constructed as follows.
We first expand $U(k,T)$ as
\eq{
\label{eq:1+1D_P_RM_def_prep_1}
U(k,T)=\sum_{m=1}^2 e^{-\ii\E_{m,k} T} P_{k,m}(T)\ ,
}
where $P_{k,m}(T)$ is the projection matrix given by the eigenvector of $U(k,T)$ for $e^{-\ii\E_{m,k} T}$.
With the above expression, the return map reads
\eq{
U_\epsilon(k,t)=U(k,t) \left[U(k,T)\right]^{-t/T}_\epsilon\ ,
}
where 
\eq{
\label{eq:1+1D_P_RM_def_prep_2}
\left[U(k,T)\right]^{-t/T}_\epsilon=\sum_{m=1}^2\exp\left[-\frac{t}{T} \log_{\epsilon_{k}}(e^{-\ii\E_{m,k} T}) \right] P_{k,m}(T).
}
Here $\epsilon_{k}$ serves as the branch cut of the logarithm~\cite{Fruchart2016FloquetTenFold} by requiring $\ii \log_{\epsilon_{k}}(x)\in [\epsilon_{k},\epsilon_{k}+2\pi)$ for all $x\in \U{1}$.
As we always set the branch cut to be equal to the PBZ lower bound (\ie, $\epsilon=\Phi$), we have 
\eq{
\ii \log_{\epsilon_{k}=\Phi_k}(e^{-\ii\E_{m,k} T})= \E_{m,k} T.
}

For the general situation, we just need to generalize the expression of the return map from the $1+1$D two-band case to a $N$-band FGU $U(\bsl{k},t)$ with $T$.
Specifically, we replace $k$ by $\bsl{k}$ and replace $2$ bands by $N$ bands in \eqnref{eq:1+1D_P_RM_def_prep_1}-\eqref{eq:1+1D_P_RM_def_prep_2} to get the return map
\eq{
\label{eq:RM_def_gen}
U_\epsilon(\bsl{k},t)=U(\bsl{k},t) \left[U(\bsl{k},T)\right]^{-t/T}_\epsilon\ ,
}
where 
\eq{
\left[U(\bsl{k},T)\right]^{-t/T}_\epsilon=\sum_{m=1}^N\exp\left[-\frac{t}{T} \log_{\epsilon_{\bsl{k}}}(e^{-\ii\E_{m,\bsl{k}} T}) \right] P_{\bsl{k},m}(T)\ ,
}
and $P_{\bsl{k},m}(T)$ is the projection matrix given by the eigenvector of $U(\bsl{k},T)$ for $e^{-\ii\E_{m,\bsl{k}} T}$.

\subsection*{Winding Data and Modulo Operation}

The winding data of the 1+1D example is mathematically constructed as the follows.
The return map commutes with the inversion symmetry representation at $k_0=\Gamma/X$
\eq{
\label{eq:1+1D_P_RM_k0}
u_{\P}(k_0) U_{\epsilon=\Phi}(k_0,t) u_{\P}^\dagger(k_0)= U_{\epsilon=\Phi}(k_0,t)\ .
}
Combined with the representation of inversion symmetry in \eqnref{eq:1+1D_P_sym_rep}, the return map at $k_0$ has two blocks with opposite parties 
\eq{
U_{\epsilon=\Phi}(k_0,t) = \mat{
U_{\epsilon=\Phi,k_0,+}(t) & \\
 & U_{\epsilon=\Phi,k_0,-}(t)
}\ .
}
Then we can define the following $U(1)$ winding number for each block
\eq{
\label{eq:nu_P}
\nu_{k_0,\alpha}=\frac{\ii}{2\pi}\int_0^T dt \Tr\left[U_{\epsilon=\Phi,k_0,\alpha}^\dagger(t)\partial_t U_{\epsilon=\Phi,k_0,\alpha}(t)\right]\in\mathbb{Z}
}
with $\alpha=\pm$ again labelling the parity.
In particular, the integer-valued nature of $\nu_{k_0,\alpha}$ directly comes from time-periodic nature of the return map. 

From the winding data $V$, a modulo operation $V\mod \bar{A}$ is required to give the quotient winding data, as shown in \eqnref{eq:quotient_winding_data_def}.
In practice, the modulo operation can be taken for the first nonzero component of $\bar{A}$ as discussed in the following.
\eqnref{eq:1+1D_P_Abar} shows that the first nonzero element of $\bar{A}$ is the its first element $\bar{A}_{\Gamma,+}=1$, and then $V_{Q}=V + j \bar{A}$ with integer $j$ satisfying 
\eq{
V_{Q,\Gamma,+}= v_{\Gamma,+} + j \bar{A}_{\Gamma,+} = v_{\Gamma,+} \mod \bar{A}_{\Gamma,+} = 0\ .
}

The generalization from the 1+1D example to a generic FGU is discussed in \SN{C}.

\subsection*{General Framework}

In this part, we briefly introduce the general framework.
Details can be found in \SN{C}.

We consider a generic FGU with a generic crystalline symmetry group $\G$, and discuss its symmetry data first.
The quasi-energy bands of the FGU are separated by relevant gaps into isolated sets of quasi-energy bands, and certain sets may contain more than one bands. 
Then, the symmetry contents, columns of the symmetry data, are defined for the isolated sets of quasi-energy bands.
Each component of a symmetry content is the copy number of the corresponding irrep (like parity in the 1+1D example) of the little group at the corresponding high-symmetry momentum (like the inversion-invariant momentum in the 1+1D example).
Nevertheless, the compatibility relation of all symmetry contents can always be expressed in terms of a compatibility matrix $\mathcal{C}$ as \eqnref{eq:comp_rel_1+1D_inversion}, and all the symmetry contents belong to
\eq{
 \label{eq:BS_set_gen}
 \{BS\}\equiv\mathbb{N}^{K}\cap \ker \mathcal{C}\ ,
}
where $\dsN$ is the set of non-negative integers.
Here $K$ is the number of components of each symmetry content (which is $4$ for the 1+1D example), and both $K$ and the compatibility matrix $\mathcal{C}$ can be determined solely based on $\G$.
The PBZ-dependence of the symmetry data can still be removed by the equivalence among symmetry data defined in Results, and inequivalent symmetry data still infers topological distinction.
Therefore, we can perform a topological classification for FGUs---also for Floquet crystals---solely based on the symmetry data, similar to what we did for static crystals. 

Now we discuss the winding data of the generic FGU.
The winding data is still a vector that consists of the winding numbers resolved by the high-symmetry momenta and irreps. 
We demonstrate that we can always choose the same set of high-symmetry momenta and irreps for the winding data and symmetry data, and the winding data always obeys the same compatibility relation as the symmetry content.
In addition, if an irrep at certain momentum is missing, the corresponding winding number must be zero, which is an extra constraint imposed on the winding data by the symmetry data.
We can always express this extra constraint in terms of a diagonal matrix $\mathcal{D}$ as
\eq{
\label{eq:V_D_constraint}
\mathcal{D} V = 0\ .
}
Then, the winding data takes value from the following group $\{ V \}$
\eq{
\label{eq:V_set_gen}
\{ V \} \equiv \dsZ^K \cap \ker \mathcal{C} \cap \ker \mathcal{D}\ .
}
In the $1+1D$ example (\eqnref{eq:1+1D_P_sym_data}), all inequivalent irreps appear at all high-symmetry momenta, resulting in $\mathcal{D}=0$ and \eqnref{eq:1+1D_P_winding_data_set}.
Similar to the $1+1D$ example, a generic FGU also has an infinite number of winding data, given by varying PBZ.

To solve the infinity issue, we construct the quotient winding data.
For the generic FGU, the quotient winding data is still defined as \eqnref{eq:quotient_winding_data_def}, but $\bar{A}$ needs to be chosen carefully as discussed below.
In the 1+1D example, $\bar{A}$ is the sum of all columns of the symmetry data just because all PBZ shifts that keep the symmetry data are (or equivalent to) the $2\pi n$-shifts of the PBZ.
For the generic FGU with in total $L$ isolated sets of quasi-energy bands, $2\pi$-shift of the PBZ is equivalent to shifting the PBZ lower bound through $L$ isolated sets, which certainly leaves the symmetry data invariant.
Nevertheless, in certain cases, the symmetry data is kept invariant even if we shift the PBZ lower bound through $0<\widetilde{L}<L$ isolated sets, and then we should choose $\bar{A}=\sum_{l=1}^{L_{KSD}} A_l$ for the construction of the quotient winding data, where $L_{KSD}$ is the smallest $\widetilde{L}$ and $A_l$'s are columns of the symmetry data.
After choosing proper $\bar{A}$, the equivalence between quotient winding data defined in Results still holds in the general framework.

Now let us turn to the DSI for the generic FGU.
As mentioned in Results, we need to use Hilbert bases, which will be discussed with more details below.
As shown in \eqnref{eq:BS_set_gen}, the symmetry contents compatible with $\G$ always take value from the set $\{ BS \}$.
Mathematically speaking, $\{ BS \}$ is a monoid rather than a group, since the components of a symmetry content are always non-negative, preventing nonzero elements in $\{ BS\}$ from having inverse.
We call a nonzero element in $\{ BS \}$ irreducible~\cite{Bruns2009Polytopes} if it cannot be expressed as the sum of any two other elements in $\{ BS \}$; otherwise, it is called reducible.
In particular, the irreducible symmetry contents form a unique set of bases of $\{ BS \}$~\cite{Song2020FragileAffineMonoid,Bruns2009Polytopes}, called the Hilbert bases, which we label as $ a_i $ with $i=1,2,...,I$.

As discussed in Results, the symmetry data can be classified as irreducible or reducible.
We first discuss the DSI set for generic FGUs with irreducible symmetry data.
When the symmetry data is irreducible (like the 1+1D example), the symmetry data is spanned by a unique set of Hilbert bases $\{ a_{j} \}$ with $j$ taking $J$ different values in $\{ 1,2,...,I\}$.
In this case, the static winding data set $\{ V_{SL} \} $ for constructing the DSI set $\X$ simply reads
\eq{
\label{eq:V_SL_set_sim}
\{ V_{SL} \} = \{ \sum_{j} a_j q_j | q_j\in \dsZ\}\ .
}
From $\{ a_j\}$, we can also determine the $K$ (as the number of components of $a_j$) and $\mathcal{D}$ (as shown in \SN{C}) in \eqnref{eq:V_set_gen}. 
Then, combined with compatibility matrix $\mathcal{C}$, $\X$ can be directly derived based on \eqnref{eq:V_set_gen} and the first equality in \eqnref{eq:1+1D_P_DSI}, meaning that $\X$ is uniquely determined by $\{ a_j\}$ and $\mathcal{C}$.
In particular, if two FGUs have the same $\G$ and have irreducible symmetry data that involve the same set of Hilbert bases, they have the same $\X$, no matter whether the two FGUs have equivalent symmetry data.
As mentioned in Results, this simplification allows us to enumerate all possible DSI sets for irreducible symmetry data by considering all $2^I-1$ nontrivial combinations of Hilbert bases of a given crystalline symmetry group $\G$.

When the symmetry data is reducible, then it is possible that more than one sets of Hilbert bases can span the symmetry data.
Nevertheless, the DSI set in this case can be constructed from the tensor product of the DSI sets for irreducible symmetry data.
Therefore, the DSI sets for irreducible symmetry data serve as the elementary building blocks for all DSI sets.

\section*{Data Availability}
The datasets generated during and/or analysed during the current study are available from the corresponding author on reasonable request.

\section*{Code Availability}
The Mathematica and SageMath code generated during and/or analyzed for the current study are
available from the corresponding author on reasonable request.

\Or{
\section*{Acknowledgments}
We thank Yu-An Chen, Yang Ge, Biao Huang, Biao Lian, Xiao-Qi Sun, Jian-Xiao Zhang, Junyi Zhang, and in particular Sankar Das Sarma and Zhi-Cheng Yang for helpful discussions. 
J.Y. and R.-X.Z. are supported by the Laboratory for Physical Sciences.
J.Y. acknowledges the UMD Libraries' Open Access Publishing Fund.
R.-X.Z. acknowledges a JQI postdoctoral fellowship.  
Z.-D. S. is supported by the DOE Grant No. DE-SC0016239.

\section*{Author Contributions}

J.Y. conceived the project.
J.Y. performed the theoretical analysis with the input from R.-X.Z. and Z.-D. S.
J.Y., R.-X.Z. and Z.-D. S. wrote the manuscript.

\section*{Competing Interests}
The authors declare no competing interests.
}

\clearpage

\clearpage

\clearpage

\clearpage

\begin{table*}[t]
    \centering
    \begin{tabular}{|c|c|c|}
    \hline
        P.G. & \Or{H.B.N.} & Nontrivial DSI Sets\\
        \hline
        \text{p1} & 1 & None\\
        \hline
        \text{p2} & 16 & $\dsZ\ (2980)$, $\dsZ^2\ (268)$, $\dsZ^3\ (8)$, $\dsZ_2\ (666)$, $\dsZ_2\times \dsZ\ (24)$, $\dsZ_3\ (16)$\\
        \hline
        \text{pm} & 4 & $\dsZ\ (2)$\\
        \hline
        \text{pg} & 1 & None\\
        \hline
        \text{cm} & 2 & None\\
        \hline
        \text{p2mm} & 24 & 
        $\begin{matrix}
 \dsZ\ (1657492)$, $ \dsZ^2\ (372286)$, $ \dsZ^3\ (78060)$, $ \dsZ^4\ (11904)$, $\dsZ^5\ (1200)$, $ \dsZ^6\ (94)$, $\dsZ^7\ (4)$, $\dsZ_2\ (354594)$, $\\
\dsZ_2\times \dsZ\ (63296)$, $\dsZ_2\times \dsZ^2\ (10320)$, $\dsZ_2\times \dsZ^3\ (1264)$, $\dsZ_2\times \dsZ^4\ (65)$, $\dsZ_3\ (10392)$, $\dsZ_3\times\dsZ\ (1024)$, $\\
 \dsZ_3\times \dsZ^2\ (112)$, $\dsZ_3\times \dsZ^3\ (8) $, $\dsZ_4 \ (3424)$, $\dsZ_4\times \dsZ\ (16)$, $\dsZ_5\ (16)$, $\dsZ_6\ (64)
\end{matrix}$
        \\
        \hline
        \text{p2mg} & 6 & $\dsZ\ (15)$, $\dsZ^2\ (3)$\\
        \hline
        \text{p2gg} & 4 & $\dsZ\ (2)$\\
        \hline
        \text{c2mm} & 14 & 
        $\begin{matrix}
        \dsZ\ (3113),\ \dsZ^2\ (686),\ \dsZ^3\ (99),\ \dsZ^4\ (7),\ \dsZ_2\ (476),\ \dsZ_2\times \dsZ\ (168),\ \dsZ_2\times\dsZ^2\ (56),\ \dsZ_2\times\dsZ^3\ (7),\ \dsZ_3\ (12),\ \dsZ_4\ (2)
        \end{matrix}$
        \\
        \hline
        \text{p4} & 32 & 
        $\begin{matrix}
        \dsZ (17587274)$, $\dsZ^2 (491020)$, $\dsZ^3 (20760)$, $\dsZ^4 (336)$, $\dsZ_2 (2175362)$, $\dsZ_2\times \dsZ (56952)$, $\\
        \dsZ_2\times \dsZ^2 (576)$, $\dsZ_3 (27120)$, $\dsZ_3\times \dsZ (384)$, $\dsZ_4 (144)
        \end{matrix}$
\\
        \hline
        \text{p4mm} & 26 & 
        $\begin{matrix} 
        \dsZ \ (6044617)$, $ \dsZ^2 \ (859049)$, $ \dsZ^3 \ (116266)$, $ \dsZ^4 \ (11202)$, $ \dsZ^5 \ (597)$, $
\dsZ^6 \ (14)$, $\dsZ_2 \ (422534)$, $\dsZ_2\times \dsZ \ (81467)$, $\\
\dsZ_2\times \dsZ^2 \ (11010)$, $\dsZ_2\times \dsZ^3 \ (869)$, $\dsZ_2\times \dsZ^4 \ (22)$, $\dsZ_3 \ (3200)$, $\dsZ_3\times \dsZ \ (480)$, $\dsZ_3\times\dsZ^2 \ (56)$, $
\dsZ_3\times \dsZ^3 \ (4)$, $\\
\dsZ_4 \ (2400)$, $\dsZ_4\times \dsZ \ (450)$, $\dsZ_4\times \dsZ^2 \ (42)$, $\dsZ_8 \ (8)$, $\dsZ_8\times \dsZ \ (1)
\end{matrix}$ \\
        \hline
        \text{p4gm} & 11 & $\dsZ\ (615)$, $\dsZ^2\ (99)$, $\dsZ^3\ (7)$, $\dsZ_2\ (1)$\\
        \hline
        \text{p3} & 27 & 
        $
        \begin{matrix}
         \dsZ \ (973458)$, $ \dsZ^2 \ (48762)$, 
$ \dsZ^3 \ (2376)$, $ \dsZ^4 \ (36)$, $\dsZ_2 \ (201690)$, $
\dsZ_2\times \dsZ \ (4968)$, $\dsZ_2\times \dsZ^2 \ (54)$, 
$\dsZ_3 \ (2604)$, $\dsZ_4 \ (324)
\end{matrix}
$
        \\
        \hline
        \text{p3m1} & 12 & $\dsZ\ (378)$, $\dsZ^2\ (27)$, $\dsZ_2\ (360)$, $\dsZ_2\times\dsZ\ (21)$, $\dsZ_4\ (16)$\\
        \hline
        \text{p31m} & 9 & $\dsZ\ (148)$, $\dsZ^2\ (33)$, $\dsZ^3\ (3)$, $\dsZ_2\ (3)$, $\dsZ_2\times\dsZ\ (1)$\\
        \hline
        \text{p6} & 36 & $\dsZ\ (110427458)$, $\dsZ^2\ (2196588)$, $\dsZ^3\ (68760)$, $\dsZ_2\ (16472556)$, $\dsZ_2\times \dsZ\ (254520)$, $\dsZ_3\ (148920)$\\
        \hline
        \text{p6mm} & 20 & 
         $
         \begin{matrix}
         \dsZ\ (189005)$, $\dsZ^2\ (32809)$, $\dsZ^3\ (3301)$, $\dsZ^4\ (168)$, $\dsZ_2\ (6509)$, $
         \dsZ_2\times \dsZ\ (1691)$, $\dsZ_2\times\dsZ^2\ (172)$, $\dsZ_3\ (22)\\
         \end{matrix}
         $\\
         \hline
    \end{tabular}
    \caption{\Or{Numbers of Hilbert bases and nontrivial DSIs for all spinless 2D plane groups.}
    Here we only consider the DSIs for FGUs with irreducible symmetry data. 
    \Or{P.G. means plane group, and H.B.N. means the number of Hilbert bases for each plane group.}
    In the column for nontrivial DSI sets, None means there are no combinations of Hilbert bases that give nontrivial DSIs, and the number in the bracket is the number of Hilbert-bases combinations that give the DSI set in front of the bracket.
    }
    \label{tab:PlaneG_Spinless}
\end{table*}
\clearpage

\begin{table*}
    \centering
    \begin{tabular}{|c|c|c|}\hline
        P.G. & \Or{H.B.N.} & Nontrivial DSI Sets\\
        \hline
        \text{p1} & 1 & None\\
        \hline
        \text{p2} & 16 & $\dsZ\ (2980)$, $\dsZ^2\ (268)$, $\dsZ^3\ (8)$, $\dsZ_2\ (666)$, $\dsZ_2\times \dsZ\ (24)$, $\dsZ_3\ (16)$\\
        \hline
        \text{pm} & 4 & $\dsZ\ (2)$\\
        \hline
        \text{pg} & 1 & None\\
        \hline
        \text{cm} & 2 & None\\
        \hline
        \text{p2mm} & 1 & None
        \\
        \hline
        \text{p2mg} & 6 & $\dsZ\ (15)$, $\dsZ^2\ (3)$\\
        \hline
        \text{p2gg} & 4 & $\dsZ\ (2)$\\
        \hline
        \text{c2mm} & 3 & $\dsZ\ (1),\ \dsZ_2\ (1)$
        \\
        \hline
        \text{p4} & 32 & 
        $\begin{matrix}
        \dsZ (17587274)$, $\dsZ^2 (491020)$, $\dsZ^3 (20760)$, $\dsZ^4 (336)$, $\dsZ_2 (2175362)$, $\dsZ_2\times \dsZ (56952)$, $\\
        \dsZ_2\times \dsZ^2 (576)$, $\dsZ_3 (27120)$, $\dsZ_3\times \dsZ (384)$, $\dsZ_4 (144)
        \end{matrix}$
        \\
        \hline
        \text{p4mm} & 4 &  $\dsZ \ (2)$ \\
        \hline
        \text{p4gm} & 8 & $\dsZ\ (50)$, $\dsZ^2\ (4)$, $\dsZ_2\ (2)$\\
        \hline
        \text{p3} & 27 & 
        $
        \begin{matrix}
         \dsZ \ (973458)$, $ \dsZ^2 \ (48762)$, 
$ \dsZ^3 \ (2376)$, $ \dsZ^4 \ (36)$, $\dsZ_2 \ (201690)$, $
\dsZ_2\times \dsZ \ (4968)$, $\dsZ_2\times \dsZ^2 \ (54)$, 
$\dsZ_3 \ (2604)$, $\dsZ_4 \ (324)
\end{matrix}
$
        \\
        \hline
        \text{p3m1} & 12 & $\dsZ\ (378)$, $\dsZ^2\ (27)$, $\dsZ_2\ (360)$, $\dsZ_2\times\dsZ\ (21)$, $\dsZ_4\ (16)$\\
        \hline
        \text{p31m} & 9 & $\dsZ\ (148)$, $\dsZ^2\ (33)$, $\dsZ^3\ (3)$, $\dsZ_2\ (3)$, $\dsZ_2\times\dsZ\ (1)$\\
        \hline
        \text{p6} & 36 & $\dsZ\ (110427458)$, $\dsZ^2\ (2196588)$, $\dsZ^3\ (68760)$, $\dsZ_2\ (16472556)$, $\dsZ_2\times \dsZ\ (254520)$, $\dsZ_3\ (148920)$\\
        \hline
        \text{p6mm} & 6 & $\dsZ\ (12)$\\
         \hline
    \end{tabular}
    \caption{\Or{Numbers of Hilbert bases and nontrivial DSIs for all spinful 2D plane groups.}
    Here we only consider the DSIs for FGUs with irreducible symmetry data. 
    \Or{P.G. means plane group, and H.B.N. means the number of Hilbert bases for each plane group.}
    In the column for nontrivial DSI sets, None means there are no combinations of Hilbert bases that give nontrivial DSIs, and the number in the bracket is the number of Hilbert-bases combinations that give the DSI set in front of the bracket.
    }
    \label{tab:PlaneG_Spinful}
\end{table*}

\clearpage

\bibliography{bibfile_references.bib}

\appendix

\section{More Details on the 1+1D Inversion-invariant Example}
\label{sec:1+1D_P}

In this part, we show more details on the 1+1D two-band inversion-invariant example. 

\subsection{Model Hamiltonian}

We consider a 1D lattice with lattice constant being $1$, and each lattice site consists of two orbitals at the same position: one spinless $\s$ orbital and one spinless $\p$ orbital.
As we consider the noninteracting cases, we only care about the single-particle Hilbert space, which is spanned by localized states $\ket{R,a}$ with $a=\s,\p$ and $R$ the lattice vector.
The symmetry group $\G$ of interest is spanned by the 1D lattice translations and the inversion symmetry.
Owing to 1D lattice translations, it is convenient to use the Fourier transformation of $\ket{R,a}$ as the bases
\eq{
\label{eq:Bases_1+1D_P}
\ket{\psi_{k,a}}=\frac{1}{\sqrt{\mathcal{N}}}\sum_{R} \ket{R,a} e^{\ii k R}\ ,
}
where $\hbar=1$ is chosen henceforth, $k$ is the momentum, $\mathcal{N}$ is the total number of lattice sites.
Throughout this section, $k\in\text{1BZ}$ is always implied, unless $k\in\dsR$ is explicitly specified.
The bases $\ket{\psi_{k}}=(\ket{\psi_{k,\text{s}}},\ket{\psi_{k,\text{p}}})$ have three key properties: (i) they are orthonormal $\braket{\psi_{k,a}}{\psi_{k',a'}}=\delta_{kk'}\delta_{aa'}$, (ii) $\ket{\psi_{k+G}}=\ket{\psi_{k}}$ for all reciprocal lattice vectors $G$, and (iii) the periodic parts $e^{-\ii \hat{r} k } \ket{\psi_{k,a}}= (1/\sqrt{\mathcal{N}})\sum_{R} \ket{R,a}$ are smooth functions of $k\in \dsR$.
Here we have chosen the localized $\ket{R,a}$ to realize $\hat{r}\ket{R,a}=R\ket{R,a}$.

The bases allow us to express the single-particle Hamiltonian as
\eq{
\hat{H}(t)=\sum_{k} \ket{\psi_{k}} H(k,t) \bra{\psi_{k}}\ ,
}
where $t$ is time.
Since we care about the Floquet crystals, we set
\eq{
H(k,t+T)=H(k,t)
}
with $T>0$ the time period.
Within one period, we choose $H(k,t)$ as the following
\eq{
\label{eq:Floquet_H_1+1D_P}
H(k,t)= \left\{
\begin{array}{ll}
 M_1 (k)\sin( 2 \pi \frac{t}{T}) &,\ 0\leq t < \frac{T}{2} \\
M_2(k)\sin( 2 \pi \frac{t}{T}-\pi) &,\ \frac{T}{2}\leq t < T
\end{array}
\right.
}
where $ M_1(k)=d(k) + t_1 \sin(k) \sigma_x$, $M_2(k) = d(k)/2 + t_2 \sin(k) \sigma_y$, $d(k)=E_1 + B_1 \cos(k) + ( E_2 + B_2 \cos(k)) \sigma_z$, and $\sigma_{x,y,z}$ are the $2\times 2$ Pauli matrices.

Furthermore, the inversion symmetry $\mathcal{P}$ is represented as 
\eq{
\mathcal{P} \ket{\psi_{k}} = \ket{\psi_{-k}} u_{\mathcal{P}}(k)
}
with 
\eq{
%\label{eq:1+1D_P_sym_rep}
u_{\P}(k)=\sigma_z\ .
}
The inversion invariance of the system, $[\hat{H}(t),\mathcal{P}]=0$, is then represented as
\eq{
	\label{eq:1+1D_P_U_sym}
u_{\P}(k) H(k,t) u_{\P}^\dagger(k) = H(-k,t)\ .
}

\subsection{Time-evolution Matrix and Quasi-energy Band}

The corresponding unitary time-evolution operator
\eq{
\label{eq:Floquet_Uhat_1+1D_P}
\hat{U}(t)=\sum_{k} \ket{\psi_{k}} U(k,t) \bra{\psi_{k}}\ ,
}
where $U(k,t)$ is the time-evolution matrix given by Dyson series
\eq{
\label{eq:Floquet_U_1+1D_P}
U(k,t)=\mathcal{T} \exp\left[-\ii \int_0^t d t' H(k,t') \right]\ ,
}
and $\mathcal{T}$ is the time-ordering operator.
Throughout this work, the initial time is set to zero without loss of generality (\appref{app:initial_time}).
Owing to the time-periodic nature of $H(k,t)$, $U(k,t+T)$ is related to $U(k,t)$ via
\eq{
U(k,t+T)=U(k,t)U(k,T)\ ,
}
meaning that all essential information of the dynamics is embedded in one period.
For concreteness, we in the rest of this section choose the following parameter values for $U(k,t)$
\eqa{
\label{eq:1+1D_P_para_val}
& T=2\pi,\ E_1=0.05,\ E_2=0.65,\ B_1=0.2,\\
& B_2=1.2,\ t_1=-0.5,\ t_2=0.6\ .
}

The eigenspectrum of the unitary $U(k,t)$ is important for our later discussion.
Diagonalizing $U(k,t)$ results in two eigenvalues $\exp[-\ii \phi_{m,k}(t)]$ with $m=1,2$,
and the real phase $\phi_{m,k}(t)$ is known as the phase band~\cite{Nathan2015TopoSingClas} of $U(k,t)$, which by definition has a $2\pi$ ambiguity.
Thereby, we can always fix the phase bands in a time-independent $2\pi$ range: $\phi_{m,k}(t)\in [\Phi_k,\Phi_k+2\pi)$, where $[\Phi_k,\Phi_k+2\pi)$ is called the phase Brillouin zone (PBZ) and we call $\Phi_k$ the PBZ lower bound.
In this work, we restrict the PBZ to be time-independent, which is different from the time-dependent PBZ in \refcite{Nathan2015TopoSingClas}.
In particular, the quasi-energy bands $\E_{m,k}$, also known as the Floquet bands, are derived from the phase bands at the end of a driving period 
	\begin{equation}
	\E_{m,k}=\frac{\phi_{m,k}(T)}{T}\ .
	\end{equation}
We plot the quasi-energy spectrum for $U(k,t)$ in {\figOneDExp}(a), where the PBZ lower bound is chosen as $\Phi_k=-\pi$. 
The band index $m$ is assigned to the two quasi-energy bands within the PBZ always following an ascending order: $\E_{1,k}< \E_{2,k}$. 
The quasi-energy bands are separated by two quasi-energy gaps in the PBZ, which are essential for defining topological equivalence for Floquet crystals~\cite{Nathan2015TopoSingClas}.

The parameter values in \eqnref{eq:1+1D_P_para_val} give us one specific Floquet system; if we change the parameter values or even add more symmetry-preserving terms to the two-band Hamiltonian, we would get a different $\G$-invariant Floquet system with a new time-evolution operator $\hat{U}'(t)$ and a new time-evolution matrix $U'(k,t)$.
Throughout this section, two Floquet systems are considered to be topologically equivalent if and only if (iff) they are connected by a continuous deformation that preserves the symmetry group $\G$ and both quasi-energy gaps.

In terms of the terminology adopted in \appref{sec:basic_def}, we choose both quasi-energy gaps to be relevant gaps~\cite{Roy2017FloquetTenFold,Yao2017FloquetTenFold} that must be preserved during any topologically equivalent deformation ({\figOneDExp}(a)).
Then, the time-evolution matrix $U(k,t)$ in \eqnref{eq:Floquet_U_1+1D_P}---equipped with the time period $T$, the relevant gap choice in {\figOneDExp}(a), the symmetry group $\G$, and the symmetry representation of $\G$ like \eqnref{eq:1+1D_P_sym_rep}---is called a \emph{Floquet gapped unitary} (FGU), which is in short denoted by $U(k,t)$.
$U'(k,t)$ stands for another FGU that has the same $\G$ as $U(k,t)$.
On the other hand, the time-evolution operator $\hat{U}(t)$ in \eqnref{eq:Floquet_Uhat_1+1D_P}---equipped with $T$, the relevant gap choice in {\figOneDExp}(a), and $\G$---is called a Floquet crystal, which is in short denoted by $\hat{U}(t)$.
$\hat{U}'(t)$ stands for another $\G$-invariant Floquet crystal.
The above topological equivalence can be defined for both FGUs and Floquet crystals, while the difference is that since a Floquet crystal consists of a FGU and the corresponding bases, topological equivalence among Floquet crystals requires both equivalent FGUs and equivalent bases.
It means that topologically distinction among FGUs must infer topologically distinction among the underlying Floquet crystals, and thereby all topological invariants of FGUs apply to Floquet crystals.
Therefore, to avoid dealing with the deformation of bases, we will focus on the FGUs, unless Floquet crystals are explicitly specified. (See \appref{sec:basic_def} for details.)

\subsection{Symmetry Data of Quasi-energy Band Structure}
\label{sec:1+1D_P_sym_data}

As the first step of our topological classification, let us describe the symmetry data for the quasi-energy band structure of the FGU $U(k,t)$.

First, owing to the inversion invariance, $U(k,t)$ commutes with $u_{\P}(k)$ at an inversion-invariant momentum $k_0$ as
\eq{
\label{eq:1+1D_P_k0}
u_{\P}(k_0) U(k_0,t) u_{\P}^\dagger(k_0)= U(k_0,t),
}
where $k_0$ is $\Gamma$ ($k=0$) or $X$ ($k=\pi$).
Then, the eigenvectors for the quasi-energy bands at $k_0$ (or equivalently the eigenvectors of $U(k_0,T)$) have definite parities $\alpha=\pm$, as shown in {\figOneDExp}(a). 
For each quasi-energy band $\E_{m,k}$, we can count the number of eigenvectors carrying parity $\alpha$ at each $k_0$, denoted by $n^{m}_{k_0,\alpha}$.
As a result, we have a four-component vector for the $m$-th quasi-energy band as
\eq{
%\label{eq:1+1D_P_A_m}
A_m=(n_{\Gamma,+}^m, n_{\Gamma,-}^m,n_{X,+}^m, n_{X,-}^m)^T\ ,
}
of which the values can be read out from {\figOneDExp}(a) as
\eq{
	%\label{eq:1+1D_P_sym_data_columns}
	A_1=(1,0,1,0)^T\ ,\ A_2= (0,1,0,1)^T\ .
}
The symmetry data is the matrix $A$ that has $A_1$ and $A_2$ as its two columns
\eq{
	%\label{eq:1+1D_P_sym_data}
	A= (A_1\ A_2)\ .
}

We emphasize that the four components of $A_m$ in \eqnref{eq:1+1D_P_A_m} are not independent, as they satisfy the following compatibility relation~\cite{Po2017SymIndi,Bradlyn2017TQC}
\eq{
n_{\Gamma,+}^m+ n_{\Gamma,-}^m=n_{X,+}^m+ n_{X,-}^m\ ,
}
or equivalently 
\eq{
%\label{eq:comp_rel_1+1D_inversion}
\C A_m = 0
}
with the compatibility matrix $\C$ as
\eq{
%\label{eq:1+1D_P_comp_rel_C}
\C=
\mat{
1 & 1 & -1 & -1
}
\ .
}

For a given choice of PBZ (as in {\figOneDExp}(a)), the derivation of symmetry data for the quasi-energy band structure is exactly the same as that for a static crystalline system~\cite{Po2017SymIndi,Bradlyn2017TQC}. 
However, the freedom of choosing PBZ for Floquet crystals leads to an additional subtlety in determining the symmetry data, which is absent in dealing with static crystals.
As shown in {\figOneDExp}(b), we can legitimately shift the PBZ lower bound to $\Phi_k=\pi/4$, which relabels the quasi-energy bands as $1\rightarrow 2$ and $2\rightarrow 1$.
As a result, the new $\widetilde{A}_m$ for $\Phi_k=\pi/4$ is related to \eqnref{eq:1+1D_P_sym_data_columns} as $\widetilde{A}_1=A_2$ and $\widetilde{A}_2=A_1$, and the new symmetry data $\widetilde{A}$ is related to \eqnref{eq:1+1D_P_sym_data} by a cyclic permutation 
\eq{
%\label{eq:1+1D_P_Aprime}
\widetilde{A}= A \mat{ 0 & 1 \\ 1 & 0}\text{ for $\Phi_k=\pi/4$}\ .
}
Therefore, the symmetry data of a Floquet crystal depends on the artificial choice of PBZ.
This is in contrast to the static case where the symmetry data of a given static crystal is uniquely determined by the Fermi energy.

We remove this artificial PBZ-dependent ambiguity by defining an equivalence among symmetry data of different FGUs.
Recall that we use $U'(k,t)$ to label another $\G$-invariant two-band FGU.
We define $U'(k,t)$ and $U(k,t)$ to have equivalent symmetry data iff we can find PBZs to make their symmetry data exactly the same. 
In practice, we can first pick a PBZ lower bound $\Phi_k'$ for $U'(k,t)$ and get its symmetry data $A'$.
Then we check whether $A'=A$ (\eqnref{eq:1+1D_P_sym_data}) or $A'=\widetilde{A}$ (\eqnref{eq:1+1D_P_Aprime}); if one of them is true,  $U'(k,t)$ and $U(k,t)$ have equivalent symmetry data, otherwise inequivalent.
Here we use the fact that \eqnref{eq:1+1D_P_sym_data} and \eqnref{eq:1+1D_P_Aprime} are the only two possible symmetry data for $U(k,t)$, since the symmetry data is invariant under $2\pi n$-shift of the PBZ lower bound ($n$ is any integer).

Despite the ambiguity of the symmetry data, whether two FGUs have equivalent symmetry data or not is independent of the artificial PBZ choice.
The equivalence reflects the inherent topological property of FGUs.
If two FGUs $U'(k,t)$ and $U(k,t)$ have inequivalent symmetry data, they must be topologically inequivalent.
Therefore, we can perform a topological classification for FGUs---therefore for Floquet crystals---solely based on the symmetry data, similar to what we did for static crystals. 
However, such symmetry-data-based classification only involves the time-evolution matrix at $t=T$, missing essential information about the quantum dynamics. 
In other words, even if $U'(k,t)$ and $U(k,t)$ have equivalent symmetry data, different quantum dynamics can still make them topologically distinct~\cite{Rudner2013AFTI, Nathan2015TopoSingClas}. 
Thus, we require the dynamical information on the entire time period to classify the dynamics of Floquet crystals with equivalent symmetry data.

\subsection{Winding Data}
\label{sec:1+1D_P_winding_data}

A direct visualization of the quantum dynamics for the given FGU $U(k,t)$ is its phase band spectrum $\phi_{m,k}(t)$ of the time-evolution matrix (\eqnref{eq:Floquet_U_1+1D_P}).
In particular, we focus on the phase bands at two inversion-invariant momenta, which we plot in {\figOneDExp}(c) for $\Phi_k=-\pi$.
Owing to \eqnref{eq:1+1D_P_k0}, the eigenvectors for the phase bands at $\Gamma/X$ can have definite parties. 
We plan to construct a quantized index that can capture the key information of the quantum dynamics at $\Gamma/X$.
For this purpose, it turns out to be inconvenient to directly use $U(k,t)$ in \eqnref{eq:Floquet_U_1+1D_P} or phase bands in {\figOneDExp}(c), which are not time-periodic.
We need a periodized version of them.

The time-periodic return map~\cite{Nathan2015TopoSingClas,Yao2017FloquetTenFold} $U_\epsilon(k,t)$ is what we seek.
To construct it, we first expand $U(k,T)$ as
\eq{
%\label{eq:1+1D_P_RM_def_prep_1}
U(k,T)=\sum_{m=1}^2 e^{-\ii\E_{m,k} T} P_{k,m}(T)\ ,
}
where $P_{k,m}(T)$ is the projection matrix given by the eigenvector of $U(k,T)$ for $e^{-\ii\E_{m,k} T}$.
With the above expression, the return map reads
\eq{
%\label{eq:1+1D_P_RM_def}
U_\epsilon(k,t)=U(k,t) \left[U(k,T)\right]^{-t/T}_\epsilon\ ,
}
where 
\eq{
%\label{eq:1+1D_P_RM_def_prep_2}
\left[U(k,T)\right]^{-t/T}_\epsilon=\sum_{m=1}^2\exp\left[-\frac{t}{T} \log_{\epsilon_{k}}(e^{-\ii\E_{m,k} T}) \right] P_{k,m}(T).
}
Here $\epsilon_{k}$ serves as the branch cut of the logarithm~\cite{Fruchart2016FloquetTenFold} by requiring $\ii \log_{\epsilon_{k}}(x)\in [\epsilon_{k},\epsilon_{k}+2\pi)$ for all $x\in \U{1}$.
Throughout this work, we always set the branch cut to be equal to the PBZ lower bound (\ie, $\epsilon=\Phi$) unless specified otherwise.
Then we have 
\eq{
\ii \log_{\epsilon_{k}=\Phi_k}(e^{-\ii\E_{m,k} T})= \E_{m,k} T.
}
Furthermore, \eqnref{eq:1+1D_P_RM_def} shows that $U_{\epsilon=\Phi}(k,t+T)=U_{\epsilon=\Phi}(k,t)$, $U_{\epsilon=\Phi}(k+G,t)=U_{\epsilon=\Phi}(k,t)$ for all reciprocal lattice vectors $G$, and $U_{\epsilon=\Phi}(k,t)$ is a continuous function of $(k,t)\in \dsR\times \dsR$.

The return map also commutes with the inversion symmetry representation at $k_0=\Gamma/X$
\eq{
%\label{eq:1+1D_P_RM_k0}
u_{\P}(k_0) U_{\epsilon=\Phi}(k_0,t) u_{\P}^\dagger(k_0)= U_{\epsilon=\Phi}(k_0,t)\ .
}
Combined with the representation of inversion symmetry in \eqnref{eq:1+1D_P_sym_rep}, the return map at $k_0$ has two blocks with opposite parties 
\eq{
U_{\epsilon=\Phi}(k_0,t) = \mat{
U_{\epsilon=\Phi,k_0,+}(t) & \\
 & U_{\epsilon=\Phi,k_0,-}(t)
}\ .
}
Then we can define the following $U(1)$ winding number for each block
\eq{
%\label{eq:nu_P}
\nu_{k_0,\alpha}=\frac{\ii}{2\pi}\int_0^T dt \Tr\left[U_{\epsilon=\Phi,k_0,\alpha}^\dagger(t)\partial_t U_{\epsilon=\Phi,k_0,\alpha}(t)\right]\in\mathbb{Z}
}
with $\alpha=\pm$ again labelling the parity.
In particular, the integer-valued nature of $\nu_{k_0,\alpha}$ directly comes from time-periodic nature of the return map. 
Similar to the symmetry data, we can calculate all four quantized winding numbers for our model ($k_0=\Gamma/X$ and $\alpha=\pm$) and further group them into a vector
\eq{
%\label{eq:1+1D_P_winding_data_V}
V=(\nu_{\Gamma,+},\nu_{\Gamma,-},\nu_{X,+},\nu_{X,-})^T=(1,-1,0,0)^T\ .
}
Here we used $\Phi_k=-\pi$ for the second equality.
We call $V$ the winding data of the given FGU $U(k,t)$ for $\Phi_k=-\pi$.

Pictorially, the winding number $\nu_{k_0,\alpha}$ can be understood in the following way.
Similar to the time-evolution unitary, the return map $U_{\epsilon}(k,t)$ is also unitary.
Thereby, its eigenvalues are $U(1)$ numbers $\exp[-\ii \phi_{\epsilon,m,k}(t)]$ with $m=1,2$, and $\phi_{\epsilon,m,k}(t)$ are the phase bands of the return map.
\eqnref{eq:1+1D_P_RM_k0} suggests that the eigenvectors for the phase bands of the return map at $k_0$ also have definite parities, as shown in {\figOneDExp}(e) for $\Phi_k=-\pi$.
Compared with phase bands in {\figOneDExp}(c),
the time-periodic phase bands in {\figOneDExp}(e) can be naively viewed as pushing the quasi-energies in {\figOneDExp}(c) to zero.
The pictorial meaning of $\nu_{k_0,\alpha}$ is simply the total winding (along $t$) of the phase bands of $U_{\epsilon=\Phi}(k_0,t)$ with parity $\alpha$. 
Then the calculated values of $\nu_{k_0,\alpha}$ in \eqnref{eq:1+1D_P_winding_data_V} can be directly read out from {\figOneDExp}(e).

Furthermore, as exemplified by \eqnref{eq:1+1D_P_winding_data_V}, the four winding numbers satisfy a compatibility relation
\eq{
\nu_{\Gamma,+}+\nu_{\Gamma,-}=\nu_{X,+}+\nu_{X,-}\ ,
}
since the total winding of all phase bands at each momentum is the same.
As a result, the winding data share the same compatibility relation as that of the symmetry data (see \eqnref{eq:comp_rel_1+1D_inversion})
\eq{
	%\label{eq:1+1D_P_comp_rel_V}
	\C V =0\ , 
}
indicating that the winding data takes value in the following set
\eqa{
%\label{eq:1+1D_P_winding_data_set}
\{ V \}&= \dsZ^4 \cap \ker \C \\
&= \{ (q_1,q_2,q_3,q_1+q_2-q_3)^T|q_1,q_2,q_3\in\dsZ\} \approx \dsZ^3\ .
}
The same compatibility relation for the winding data and the symmetry data holds for all crystalline symmetry groups in all spatial dimensions (up to three), which is discussed in \appref{sec:gen_frame} and \appref{app:return_map_winding_data}.

Shifting the PBZ changes the winding data.
For example, if we shift the PBZ lower bound from $\Phi_k=-\pi$ to $\Phi_k=\pi/4$, the phase bands of time-evolution unitary and return map become {\figOneDExp}(d-f), and from {\figOneDExp}(f), we know the winding data becomes
\eq{
%\label{eq:1+1D_P_winding_data_Vprime}
\widetilde{V}= (0,-1,-1,0)^T = V-A_1\ .
}
Unlike the symmetry data, a $2\pi$-shift of the PBZ $\Phi_k\rightarrow \Phi_k+2\pi$ can also change the winding data
\eq{
%\label{eq:1+1D_winding_data_2pi_shift}
V \rightarrow V - (1,1,1,1)^T = V - \bar{A}
}
where
\eq{
%\label{eq:1+1D_P_Abar}
\bar{A}=A_1+A_2=(1,1,1,1)^T\ .
}

\eqnref{eq:1+1D_winding_data_2pi_shift} suggests that the given FGU $U(k,t)$ can have an infinite number of different winding data, which explicitly depend on the artificial choice of PBZ. 
This is different from the fact that $U(k,t)$ only has two (which is finite) different symmetry data.
Such difference makes it hard to directly generalize the equivalence among symmetry data to define an equivalence among the winding data, since finding a single proper PBZ among an infinite number of possible choices is not straightforward.
Nevertheless, \eqnref{eq:1+1D_P_winding_data_Vprime} and \eqnref{eq:1+1D_winding_data_2pi_shift} indicate that the infinitely many winding data are related by the symmetry data (which will also be generally demonstrated in \appref{sec:gen_frame}).
This relation inspires us to define the quotient winding data below, in order to resolve the infinity problem.

\subsection{Quotient Winding Data}
\label{sec:1+1D_P_quotient_winding_data}

For the given FGU $U(k,t)$, the number of different symmetry data is finite because the symmetry data is invariant under $2\pi n$-shifts of the PBZ.
Then, in order to have a finite number of different quotient winding data, we can define the quotient winding data to be invariant under all PBZ shifts that keep the symmetry data.
Specifically, we define the quotient winding data $V_Q$ by modding out $\bar{A}$ (\eqnref{eq:1+1D_P_Abar}) from the winding data,
\eq{
%\label{eq:quotient_winding_data_def}
V_Q = V \mod\ \bar{A}
}
In practice, the modulo operation can be taken for the first nonzero component of $\bar{A}$ as discussed in the following.
\eqnref{eq:1+1D_P_Abar} shows that the first nonzero element of $\bar{A}$ is the its first element $\bar{A}_{\Gamma,+}=1$, and then $V_{Q}=V + j \bar{A}$ with integer $j$ satisfying 
\eq{
V_{Q,\Gamma,+}= v_{\Gamma,+} + j \bar{A}_{\Gamma,+} = v_{\Gamma,+} \mod \bar{A}_{\Gamma,+} = 0\ .
}
For the two winding data in \eqnref{eq:1+1D_P_winding_data_V} and \eqnref{eq:1+1D_P_winding_data_Vprime} given by two PBZ lower bounds, we have
\eqa{
%\label{eq:1+1D_P_QWD_example}
&V_Q= V \mod \bar{A}= (0,-2,-1,-1)^T\ \ \text{ for }\Phi_k=-\pi\ ,\\
&\widetilde{V}_Q=\widetilde{V} \mod \bar{A}= (0,-1,-1,0)^T\ \ \text{for }\Phi_k=\pi/4\ .
}
As $2\pi n$-shifts of the PBZ can only change $V$ by multiples of $\bar{A}$ according to \eqnref{eq:1+1D_winding_data_2pi_shift}, $V_Q$ defined in \eqnref{eq:quotient_winding_data_def} is indeed invariant under $2\pi n$-shifts of the PBZ, just like the symmetry data.
As a result, the FGU $U(k,t)$ only has two different quotient winding data in \eqnref{eq:1+1D_P_QWD_example}, which are related by
\eq{
%\label{eq:1+1D_P_QWD_relation}
\widetilde{V}_Q=V_Q-A_1 \mod \bar{A}\ .
}
We emphasize that although $\bar{A}$ used in \eqnref{eq:quotient_winding_data_def} happens to be the sum of all columns of $A$ in this specific $1+1D$ example, $\bar{A}$ in general might only involve a portion of columns of the symmetry data since sometimes PBZ shifts other than $2\pi n$-shifts also leave the symmetry data invariant, as discussed in \appref{sec:gen_frame}.

We have shown that $U(k,t)$ has only two different quotient winding data given by changing the PBZ, and next we show how to remove the remaining PBZ-dependent ambiguity by defining an equivalence among quotient winding data of different FGUs.
Recall that the quotient winding data is introduced for a classification of FGUs with equivalent symmetry data, since inequivalent symmetry data already infers topological distinction.
Then, let us suppose that the two different FGUs $U(k,t)$ and $U'(k,t)$ have equivalent symmetry data.
According to \appref{sec:1+1D_P_sym_data}, we can always pick PBZ choices $\Phi_k'$ and $\Phi_k$ for $U'(k,t)$ and $U(k,t)$, respectively, such that they have exactly the same symmetry data $A'=A$.
Then, we check whether the quotient winding data of $U'(k,t)$ for $\Phi_k'$ is the same as that of $U(k,t)$ for $\Phi_k$; if so (not), we call $U'(k,t)$ and $U(k,t)$ have equivalent (inequivalent) quotient winding data.
The above equivalence among quotient winding data is defined only for FGUs with equivalent symmetry data, and we will not attempt to compare the quotient winding data when the PBZ choices for $U'(k,t)$ and $U(k,t)$ yield different symmetry data, since the quotient winding data can be changed by the PBZ shift that changes symmetry data. 

Given two FGUs with equivalent symmetry data, the artificial PBZ choice has no influence on whether they have equivalent quotient winding data or not.
In particular, they must have equivalent quotient winding data if they are topologically equivalent, meaning that inequivalent quotient winding data provide a topological classification of FGUs (and thereby of Floquet crystals) with equivalent symmetry data.

To illustrate the classification, let us consider all FGUs that have symmetry data  equivalent to the given FGU $U(k,t)$, indicating that the symmetry data of each FGU is either $A$ in \eqnref{eq:1+1D_P_sym_data} or $\widetilde{A}$ in \eqnref{eq:1+1D_P_Aprime} depending on the PBZ choice.
Based on the winding data set $\{ V \}$ in \eqnref{eq:1+1D_P_winding_data_set} and $\bar{A}$ in \eqnref{eq:1+1D_P_Abar}, the quotient winding data of those FGUs take values in the following set
\eq{
%\label{eq:1+1D_P_QWD_set}
\{ V_Q \}=\{ (0,q_2,q_3,q_2-q_3)^T|q_2,q_3\in\dsZ\}\approx\frac{\{ V \}}{\bar{A} \dsZ }  \approx \dsZ^2\ ,
}
where $\bar{A} \dsZ = \{ q \bar{A} = (q,q,q,q)^T | q\in \dsZ\}$. 
To compare the quotient winding data, we always choose the PBZs to yield the same symmetry data for all those FGUs.
With this requirement, we still have two inequivalent types of PBZ choices: (i) the PBZ choices that yield $A$ in \eqnref{eq:1+1D_P_sym_data} for all those FGUs, and (ii) the PBZ choices that yield $\widetilde{A}$ in \eqnref{eq:1+1D_P_Aprime}.
For the type-$A$ PBZ choices, the quotient winding data of each FGU would take a unique value in $\{ V_Q \}$, since the quotient winding data is invariant under the PBZ change that keeps the symmetry data.
In this case, if two FGUs have different quotient winding data, they must be topologically inequivalent according to the above discussion, meaning that $\{ V_Q \}$ serves as a topological classification for those FGUs.
Similarly, for the type-$\widetilde{A}$ PBZ choices, $\{ V_Q \}$ also serves as a topological classification.
Since the quotient winding data for the two types are related according to \eqnref{eq:1+1D_P_QWD_relation}, the two topological classifications for two types are equivalent, \ie, the quotient winding data of two FGUs are the same for the type-$A$ PBZ choices iff they are the same for the type-$\widetilde{A}$ PBZ choices.
As a result, $\{ V_Q \}$ provides a topological classification for all FGUs (and thus for all Floquet crystals) that have equivalent symmetry data to the given FGU $U(k,t)$, as long as the comparison of $V_Q$ is done for the PBZ choices that yield the same symmetry data for all those FGUs.

Up to now, we have shown the scheme shown in {\figClassificationDSIFlowchart}(a), which suggests that the symmetry data and the quotient winding data together provide a classification of FGU and thereby of Floquet crystals.
We emphasize that in general, it is possible that two FGUs with equivalent symmetry and quotient winding data are topologically distinct, indicating that the corresponding classification is not necessarily complete.

\subsection{DSI}
\label{sec:1+1D_P_DSI}

While the $(A,V_Q)$-based classification can tell the relative topological distinction between two FGUs, it fails to tell which FGU is essentially static and which has obstruction to static limits. 
Here static limits are Floquet crystals that have time-independent Hamiltonians, and picking bases for a static limit can give a static FGU. (See more details in \appref{sec:static_limits}.)
The obstruction to static limits means the given Floquet crystal (FGU) with crystalline symmetry group $\G$ is topologically distinct from the all $\G$-invariant static limits (static FGUs).
If picking bases for a $\G$-invariant Floquet crystal gives a FGU that has obstruction to static limits, the Floquet crystal must be topologically distinct from all $\G$-invariant static limits and thereby must have obstruction to static limits.
Thereby, we can focus on the obstruction for FGUs to derive sufficient indices.
In this part, we will define DSI that can sufficiently indicate the obstruction for the FGU $U(k,t)$ (and thereby for the underlying Floquet crystal $\hat{U}(t)$).

To determine the obstruction to static limits for our example, we only need to consider the $\G$-invariant static FGUs that have symmetry data equivalent to $U(k,t)$, since $U(k,t)$ must be topologically distinct from all other $\G$-invariant static FGUs. 
We then check whether $U(k,t)$ has quotient winding data equivalent to any of those static FGUs; if not, $U(k,t)$ must have obstruction to static limits.

To be more specific, recall that we compare quotient winding data by choosing PBZs to yield the same symmetry data.
Let us focus on the PBZ choice $\Phi_k=-\pi$ for $U(k,t)$, which yields symmetry data $A$ in \eqnref{eq:1+1D_P_sym_data}, winding data $V$ in \eqnref{eq:1+1D_P_winding_data_V}, and quotient winding data $V_Q$ in \eqnref{eq:1+1D_P_QWD_example}.
In the following, we will try to find the set $\{ V_{Q,SL}\}$ of all quotient winding data of all static FGUs that have equivalent symmetry data to $U(k,t)$, under the constraint that their PBZ choices yield symmetry data equal to $A$.
Then, we can check whether $V_Q$ is in $\{ V_{Q,SL}\}$ or not; if not, the given FGU $U(k,t)$ must have obstruction to static limits.

To achieve this, let us first consider a subset of those static FGUs, which satisfy $U_{SL}(k,t) = \exp[-\ii h_{SL}(k) t ]$ with 
\eq{
h_{SL}(k)=\sum_{m=1}^2 (\E_{m,k}+q_m \frac{2\pi}{T}) P_{k,m}(T)\ ,
}
where $q_1,q_2\in\dsZ$, and $\E_{m,k}$ and $P_{k,m}(T)$ are shown in \eqnref{eq:1+1D_P_RM_def_prep_1}.
The above equation suggests $U_{SL}(k,T) = U(k,T)$, meaning that $U_{SL}(k,t)$ has the same quasi-energy band structure as $U(k,t)$.
By choosing the PBZ lower bound for $U_{SL}(k,t)$ to be the same as $\Phi_k=-\pi$ for $U(k,t)$, the symmetry data of $U_{SL}(k,t)$ become equal to $A$ in \eqnref{eq:1+1D_P_sym_data}.
The return map of $U_{SL}(k,t)$ with the PBZ lower bound $\Phi_k=-\pi$ reads
\eq{
U_{SL,\epsilon=\Phi}(k,t)=\sum_{m=1}^2 e^{-\ii q_m \frac{2\pi}{T}t} P_{k,m}(T)\ ,
}
As a result, the winding data of static FGUs in the chosen subset with $\Phi_k=-\pi$ must take the form
\eq{
\label{eq:1+1D_P_winding_data_SL}
V_{SL}= q_1 A_1 + q_2 A_2 =(q_1, q_2, q_1,q_2)^T\ ,
}
and the static winding data set reads
\eq{
%\label{eq:1+1D_P_WD_set_SL}
\{ V_{SL} \}=\{ q_1 A_1 + q_2 A_2 | q_1,q_2 \in \dsZ \}\ ,
}
where $A_1$ and $A_2$ are two columns of $A$ in \eqnref{eq:1+1D_P_sym_data}.
We can see that $\{ V_{SL} \}$ only depends on the symmetry data, and thus $\{ V_{SL} \}$ stays invariant even if we include all static FGUs that have equivalent symmetry data to $U(k,t)$, as long as we choose their PBZs to yield symmetry data equal to $A$.
(See a more general and rigorous derivation in \appref{sec:DSI_gen} and \appref{app:DSI}.)

Based on $\{ V_{SL} \}$, we can further derive the desired set of quotient winding data as
\eq{
\label{eq:1+1D_P_QWD_set_SL}
\{ V_{Q,SL} \}=\{(0,q,0,q)^T |q\in \dsZ \}\ .
}
Then, we can check whether the quotient winding data $V_Q$ of $U(k,t)$ for $\Phi_k=-\pi$ (\eqnref{eq:1+1D_P_QWD_example}) is an element of $\{ V_{Q,SL} \}$, and we find that the answer is no, meaning that $U(k,t)$ must have the obstruction to static limits.
\eqnref{eq:1+1D_P_WD_set_SL} suggests that $\{ V_{SL} \}$ is invariant under the relabelling of the quasi-energy bands (i.e. $1\leftrightarrow 2$) due to a shift of $\Phi_k$, indicating that $\{ V_{Q,SL} \}$ does not depend on the PBZ choice $\Phi_k$.
Therefore, we are allowed to adopt any PBZ choice for $U(k,t)$ to check the above criterion, \ie, allowed to use either $V_{Q}$ or $\widetilde{V}_{Q}$ in \eqnref{eq:1+1D_P_QWD_example}, and we will get the same result that $U(k,t)$ has the obstruction to static limits.

The above procedure can be greatly simplified by noting that $V_Q\notin\{V_{Q,SL}\}$ is equivalent to $V\notin \{ V_{SL} \}$.
Here we use $V$ and $V_Q$ to respectively label the winding data and quotient winding data of the 1+1D $U(k,t)$ for a generic PBZ choice $\Phi_k$, and the equivalence can be derived from \eqnref{eq:1+1D_P_winding_data_set}, \eqnref{eq:1+1D_P_QWD_set}, \eqnref{eq:1+1D_P_WD_set_SL} and \eqnref{eq:1+1D_P_QWD_set_SL}.
In fact, the $\Phi_k$-independent nature of $\{ V_{SL} \}$ suggests that $\{ V_{SL} \}$ contains all winding data of all $\G$-invariant static FGUs that have symmetry data equivalent to $U(k,t)$, regardless of the PBZ choices for those static FGUs; then $V\notin \{ V_{SL}\}$ means that $V$ cannot exist in any of those static FGUs, and thus sufficiently indicates that $U(k,t)$ has obstruction to static limits.
To exploit this fact, we define the DSI to take values from the following set $\X$
\eq{
%\label{eq:1+1D_P_DSI}
\X= \frac{\{ V \}}{\{ V_{SL} \}}\approx \{\nu_{\Gamma,+}- \nu_{X,+} \in \dsZ \}\ ,
}
where the last step uses \eqnref{eq:1+1D_P_winding_data_set} and \eqnref{eq:1+1D_P_WD_set_SL}.
Specifically, the DSI for $U(k,t)$---as well as all other FGUs that have symmetry data equivalent to $U(k,t)$---is $(\nu_{\Gamma,+}- \nu_{X,+})$.
Nonzero DSI means $V\notin \{ V_{SL} \}$ and thus infers the obstruction to static limit, which is equivalent to the above procedure of comparing quotient winding data. 
According to \eqnref{eq:1+1D_P_winding_data_V}, \eqnref{eq:1+1D_P_winding_data_Vprime} and \eqnref{eq:1+1D_winding_data_2pi_shift}, $U(k,t)$ has PBZ-independent $\nu_{\Gamma,+}- \nu_{X,+}=1$, coinciding with the above conclusion that $U(k,t)$ has obstruction to static limits.

It turns out that even for a generic FGU, the evaluation of DSI is independent of PBZ as discussed in \appref{sec:DSI_gen}.
We emphasize that a zero DSI does not rule out possible obstruction for a FGU, as shown in {\figClassificationDSIFlowchart}(b), meaning that DSI is a possibly-incomplete topological invariant.
Different DSI values infer topological distinction for FGUs with the same crystalline symmetry group and equivalent symmetry data.
Although the classification given by DSIs is a subset of that given by quotient winding data (like this 1+1D inversion-invariant case), DSIs have the advantage of being PBZ-independent.

At the end of this part, we would like to compare our proposed formalism of DSIs for FGUs (and thus for Floquet crystals) to that of the symmetry indicator~\cite{Po2017SymIndi} for static crystals.
To construct the symmetry indicator, \refcite{Po2017SymIndi} focused on two sets: the set of all possible symmetry contents for a given crystalline symmetry group, and its subset that is given by the atomic limits.
\refcite{Po2017SymIndi} first extended the two sets to two groups by artificially adding negative numbers of bands, and then took the quotient between the two resultant groups to derive the symmetry indicator, which indicates the Wannier obstruction (or equivalently obstruction to atomic limits).
In this work, the quotient in the construction of DSIs is taken between the winding data set \eqnref{eq:1+1D_P_winding_data_set} and its subset given by static limits \eqnref{eq:1+1D_P_WD_set_SL}, in order to indicate the obstruction to static limits.
As the winding number can naturally take negative values, \eqnref{eq:1+1D_P_winding_data_set} and \eqnref{eq:1+1D_P_WD_set_SL} themselves are groups, and thereby we do not need to extend them.
In short, although both \refcite{Po2017SymIndi} and our work used the mathematical concept of quotient group, the quotient is taken for completely different physical quantities and the resultant indicators have completely different physical meanings: the symmetry indicator in \refcite{Po2017SymIndi} is for static band topology while our DSI is for periodic quantum dynamics.  

\subsection{Section Summary}

\begin{figure}
    \centering
    \includegraphics[width=0.8\columnwidth]{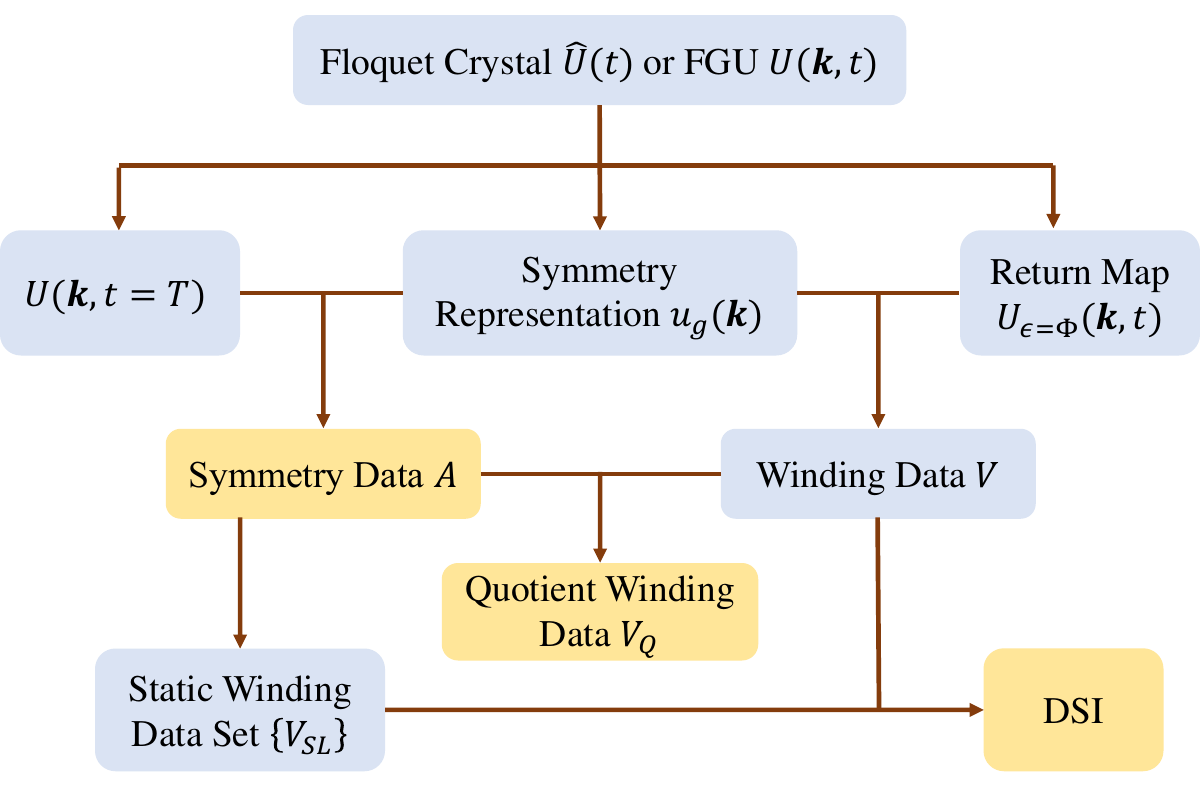}
    \caption{Relations among the key concepts.
    }
    \label{fig:Concepts_Structure}
\end{figure}

The key concepts introduced in this section are summarized in \figref{fig:Concepts_Structure}.
We start by defining a set of bases \eqnref{eq:Bases_1+1D_P} for the time-evolution operator $\hat{U}(t)$, which gives us the time-evolution matrix $U(k,t)$ in \eqnref{eq:Floquet_U_1+1D_P} and the symmetry representation of the crystalline symmetry group $\G$ like \eqnref{eq:1+1D_P_sym_rep}.
We choose both the quasi-energy band gaps to be relevant for the topologically equivalent deformation, resulting in the Floquet crystal $\hat{U}(t)$ and the FGU $U(k,t)$.

On one hand, we combine $U(k,T)$ with the inversion representation \eqnref{eq:1+1D_P_sym_rep} to derive the symmetry data $A$ in \eqnref{eq:1+1D_P_sym_data} for a PBZ lower bound $\Phi_k=-\pi$.
On the other hand, we combine the return map $U_{\epsilon=\Phi}(k,t)$ in \eqnref{eq:1+1D_P_RM_def} with the inversion representation \eqnref{eq:1+1D_P_sym_rep} to obtain the winding data $V$ in \eqnref{eq:1+1D_P_winding_data_V}.
To resolve the infinite ambiguity of the winding data, we construct $\bar{A}$ from $A$ and mod $\bar{A}$ out of the winding data $V$, resulting in the quotient winding data $V_Q$ in \eqnref{eq:1+1D_P_QWD_example}.
We can use the symmetry and quotient winding data to distinguish $U(k,t)$ ($\hat{U}(t)$) from other FGUs (Floquet crystals) with the same crystalline symmetry group according to {\figClassificationDSIFlowchart}(a).

From the symmetry data, we further derive the static winding data set $\{ V_{SL} \}$ in \eqnref{eq:1+1D_P_WD_set_SL}, and we combine $\{ V_{SL} \}$ with the winding data $V$ to obtain the DSI.
The nonzero value of the DSI indicates the obstruction to static limits ({\figClassificationDSIFlowchart}(b)). 
The evaluation of all indices---including symmetry data, quotient winding data, and DSI---is computationally efficient as they only involve two inversion-invariant momenta in 1BZ.

\section{More Details on General Definitions}
\label{sec:basic_def}

In \appref{sec:1+1D_P}, we use a two-band inversion-invariant example in 1+1D to illustrate the main idea of the topological classification and DSI.
In this and next section, we will describe the general framework for the topological classification and DSI, which is applicable to Floquet crystals living in arbitrary spatial dimensions (up to three) with an arbitrary crystalline symmetry group.
We start with the basic definitions in this section.
Although most of the concepts have been introduced in \appref{sec:1+1D_P}, we, in this section, will re-discuss them in a general and detailed manner.

We are interested in noninteracting Floquet crystals described by single-particle Hamiltonians $\hat{H}(t)$ that satisfy
\eq{
\label{eq:Floquet_H}
\hat{H}(t)=\hat{H}(t+T)
}
with the time period $T>0$ (always chosen to be positive throughout the work), and their unitary time-evolution operators have the form
\eq{
\label{eq:U_gen}
\hat{U}(t)=\mathcal{T} \exp\left[-\ii \int_0^t d t' \hat{H}(t') \right]\ ,
}
where this time-ordered form should be replaced by the more general Dyson series when $t<0$.
For convenience, we throughout this work imply that all expressions hold for all values of unspecified parameters, \eg, the above two expressions are implied to hold for all $t\in \dsR$.
Owing to \eqnref{eq:Floquet_H}, $\hat{U}(t+T)$ is related to $\hat{U}(t)$ as 
\eq{
\label{eq:Floquet_U}
\hat{U}(t+T)=\hat{U}(t)\hat{U}(T)\ .
}
Thus, as mentioned in \appref{sec:1+1D_P}, all essential information of the dynamics is included in one time period $t\in[0,T]$.
We set the underlying single-particle Hilbert space, in which the operators $\hat{H}(t)$ and $\hat{U}(t)$ are defined, to be time-independent.

$\hat{H}(t)$ may have various types of symmetries, such as space-time symmetries~\cite{Morimoto2017FTITimeGlide}, crystalline symmetries, and internal symmetries that define the ten-fold way~\cite{Nathan2015TopoSingClas, Fruchart2016FloquetTenFold, Roy2017FloquetTenFold, Yao2017FloquetTenFold}.
In this work, we only consider the time-independent crystalline symmetries of $\hat{H}(t)$, which form a time-independent crystalline symmetry group $\G$, and allow all other symmetries to be freely broken while preserving the particle number and keeping the underlying single-particle Hilbert space well-defined. 
In terms of the ten-fold way~\cite{Nathan2015TopoSingClas, Fruchart2016FloquetTenFold, Roy2017FloquetTenFold, Yao2017FloquetTenFold}, we only consider the symmetry class A.
Then, for any element $g$ in $\G$, $g$ can always be expressed as a combination of a point group operation $R$ and a translation by $\bsl{\tau}$, denoted by $g=\{R|\bsl{\tau}\}$~\cite{Bradley2009MathSSP}.
The time-evolution operator $\hat{U}(t)$ is also invariant under $\G$, \ie,
\eq{
\label{eq:U_g_invariant_gen}
[g, \hat{U}(t)]=0
}
for all $g\in \G$, and again we only care about the symmetries of $\hat{U}(t)$ within $\G$.
$\G$ contains a lattice translation subgroup, and we denote the number of primitive lattice vectors by $d$.
$d$ is no larger than the spatial dimension of system, and together with the extra time dimension, we call the system $d+1$D.
In this work, we require the spatial dimension of the system to be no larger than $3$, thus $d\leq 3$; examples of $\G$ include spatially-three-dimensional space groups, spatially-two-dimensional plane groups, and spatially-one-dimensional line groups.

Owing to the lattice translation symmetry, the Bloch momentum $\bsl{k}\in \text{1BZ}$ is a good quantum number.
Then, we can choose the orthonormal bases of the underlying Hilbert space as $\ket{\psi_{\bsl{k},a}}$ with $a$ taking $N$ different values for all other degrees of freedom like spin, orbital, and so on.
In this work, we require $N$ to be a finite number, and we always imply $\bsl{k}\in \text{1BZ}$ unless $\bsl{k}\in \dsR^d$ is explicitly specified.
As the underlying Hilbert space is time-independent, we always choose $\ket{\psi_{\bsl{k},a}}$ to be independent of time.
We further choose $\ket{\psi_{\bsl{k},a}}=\ket{\psi_{\bsl{k}+\bsl{G},a}}$ to hold for all reciprocal lattice vectors $\bsl{G}$. 
To study the topology, we require the periodic parts of $\ket{\psi_{\bsl{k},a}}$, $\exp[-\ii\bsl{k}\cdot\hat{\bsl{r}}]\ket{\psi_{\bsl{k},a}}$, to be smooth functions of $\bsl{k}\in \dsR^d$.
Such smooth choice always exists in one spatial dimension; in two and three spatial dimensions, the smooth choice exists when the total Chern numbers of all bands are vanishing~\cite{Brouder2007Wannier}.
The above requirements for bases can always be satisfied by a proper Fourier transformation of the real-space bases of any tight-binding model, just like \eqnref{eq:Bases_1+1D_P} in \appref{sec:1+1D_P}.
Nevertheless, our discussion includes the case where the bases cannot be reproduced by physical atomic orbitals or equivalently do not form a band representation~\cite{Bradlyn2017TQC}.
Owing to the smoothness requirement, $\ket{\psi_{\bsl{k},a}}$ may not be the eigenstates of $\hat{H}(t)$ or $\hat{U}(t)$, and thus they are in general called quasi-Bloch states~\cite{Panati2007Bloch, Brouder2007Wannier}.
For convenience, we define a row vector $\ket{\psi_{\bsl{k}}}=(..., \ket{\psi_{\bsl{k},a}}, ...)$.

With $\ket{\psi_{\bsl{k}}}$ as bases, $\hat{U}(t)$ can be represented as
\eq{
\label{eq:U_mat_gen}
\hat{U}(t) = \sum_{\bsl{k}} \ket{\psi_{\bsl{k}}} U(\bsl{k},t) \bra{\psi_{\bsl{k}}}
}
with $ [U(\bsl{k},t)]_{aa'}=\bra{\psi_{\bsl{k},a}} \hat{U}(t) \ket{\psi_{\bsl{k},a'}}$.
We extend the domain of $\bsl{k}$ in $U(\bsl{k},t)$ from 1BZ to $\dsR^d$ by $U(\bsl{k}+\bsl{G},t)=U(\bsl{k},t)$, and the same convention is implied for all other matrix representations furnished by $\ket{\psi_{\bsl{k}}}$ in this work.
We require $U(\bsl{k},t)$ to be a continuous (not necessarily smooth) function of $(\bsl{k},t)\in\dsR^d\times\dsR$, though the matrix representation of the Hamiltonian can be discontinuous along time~\cite{Rudner2013AFTI}.
\eqnref{eq:Floquet_U} suggests
\eq{
\label{eq:time_period_U_gen}
U(\bsl{k},t+T)=U(\bsl{k},t)U(\bsl{k},T)\ .
}
The time-evolution matrix $U(\bsl{k},t)$ for the 1+1D example in \appref{sec:1+1D_P} is shown in \eqnref{eq:Floquet_U_1+1D_P}.

For any $g=\{R|\bsl{\tau}\}\in\G$, $g$ is represented as 
\eq{
\label{eq:sym_rep_g}
g\ket{\psi_{\bsl{k}}}=\ket{\psi_{\bsl{k}_g}} u_g(\bsl{k})\ ,
}
where $\bsl{k}_g=R\bsl{k}$ and $u_g(\bsl{k})$ is unitary.
In the remaining of this work, all symmetry representations (like $u_g(\bsl{k})$ above) are implied to be unitary. 
Owing to the periodicity in reciprocal lattice vectors and the smoothness requirement of the bases, $u_g(\bsl{k}+\bsl{G})=u_g(\bsl{k})$, and $u_g(\bsl{k})$ is a smooth function of $\bsl{k}\in\dsR^d$.
As a representation of $\G$, $u_g(\bsl{k})$ also satisfies
\eq{
\label{eq:sym_rep_g1g2}
u_{g_1 g_2}(\bsl{k})=u_{g_1}(\bsl{k}_{g_2})u_{g_2}(\bsl{k})\ \ \forall g_1,g_2\in\G\ .
}
Furthermore, \eqnref{eq:U_g_invariant_gen} infers  
\eq{
\label{eq:Umat_g_invariant_gen}
u_g(\bsl{k}) U(\bsl{k},t) u_g^\dagger(\bsl{k}) = U(\bsl{k}_g,t)\ .
}
For the 1+1D example in \appref{sec:1+1D_P}, we only show the symmetry representation  for $g=\mathcal{P}$ in \eqnref{eq:1+1D_P_sym_rep}, as the representations of other symmetry operations in $\G$ can be derived from it using \eqnref{eq:sym_rep_g1g2}.

$\ket{\psi_{\bsl{k}}}$ has a $\U{N}$ gauge freedom: 
\eq{
\label{eq:UN_gauge_trans}
\ket{\psi_{\bsl{k}}}\rightarrow \ket{\psi_{\bsl{k}}}  W(\bsl{k})\ ,
}
where the $\U{N}$ gauge transformation matrix $W(\bsl{k})$ is a time-independent $\U{N}$ matrix that satisfies $W(\bsl{k}+\bsl{G})=W(\bsl{k})$ and is a smooth function of $\bsl{k}\in\dsR^d$.
To make sure that $\hat{U}(t)$ and $g$ are invariant under the gauge transformation \eqnref{eq:UN_gauge_trans}, $U(\bsl{k},t)$ and $u_g(\bsl{k})$ should simultaneously transform as
\eqa{
\label{eq:gauge_trans_U_V}
U(\bsl{k},t)\rightarrow W^\dagger(\bsl{k}) U(\bsl{k},t) W(\bsl{k})\\
u_g(\bsl{k})\rightarrow W^\dagger(\bsl{k}_g) u_g(\bsl{k}) W(\bsl{k})\ .
}
Any physical or topological property of the system should be gauge-invariant.

\subsection{Phase Band and Quasi-energy Gap}

\begin{figure}
    \centering
    \includegraphics[width=0.6\columnwidth]{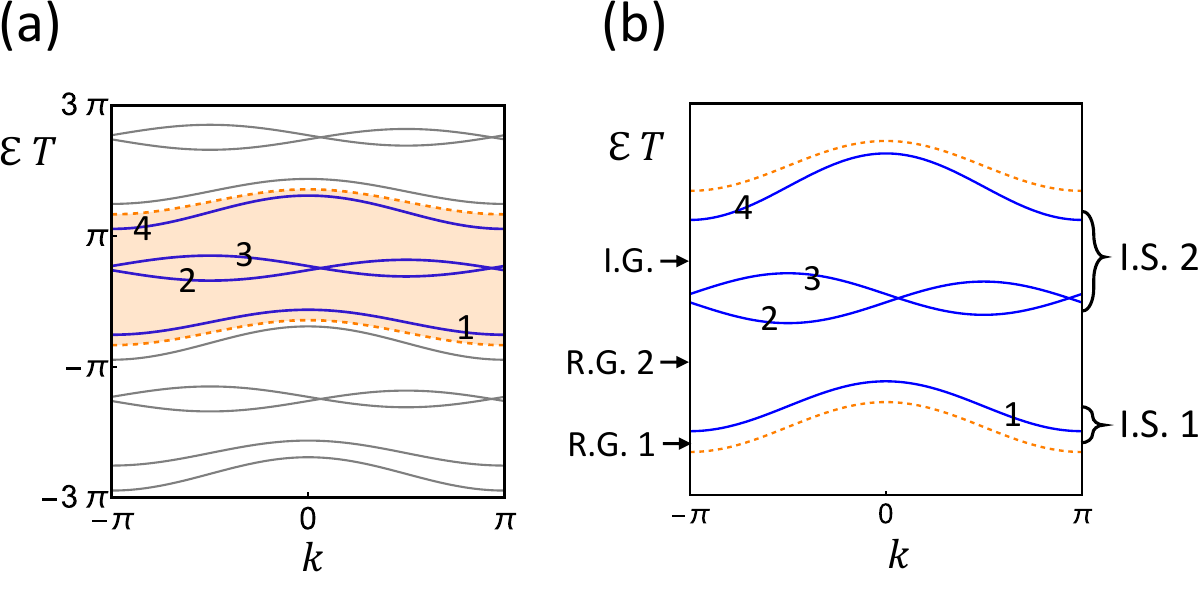}
    \caption{
    Schematic quasi-energy band structures for a 1+1D 4-band model, where the blue solid lines are the quasi-energy bands within the PBZ and the numbers on the blue solid lines stand for the $m$ index.
    In (a), the lower and upper dashed lines are $\Phi_{\bsl{k}}$ and $\Phi_{\bsl{k}}+2\pi$, respectively, and the orange-shaded region is the PBZ.
    The gray solid lines are their redundant copies shifted by multiples of $2\pi$.
    In (b), we only show the quasi-energy bands in the PBZ, and  the dashed lines mark the boundary of the PBZ.
    ``R.G.", ``I.G.", and ``I.S." stand for relevant gap, irrelevant gap, and isolated set, respectively.
    }
    \label{fig:Basic_Def_0}
\end{figure}

We label the eigenvalues of the unitary $U(\bsl{k},t)$ as $e^{-\ii \phi_{m,\bsl{k}}(t)}$ with $m=1,2,...,N$, and the quasi-energy bands are $\E_{m,\bsl{k}}=\phi_{m,\bsl{k}}(T)/T$.
By definition, $e^{-\ii \E_{m,\bsl{k}} T}$ are the eigenvalues of $U(\bsl{k},T)$.
Throughout this work, we only consider $U(\bsl{k},t)$ with at least one quasi-energy gap, \ie, there exists $\Phi_{\bsl{k}}$ such that (i) $\Phi_{\bsl{k}}$ is a real  continuous function of $\bsl{k}\in\dsR^d$, (ii) $\Phi_{\bsl{k}+\bsl{G}}=\Phi_{\bsl{k}}$, (iii) $\Phi_{\bsl{k}_g}=\Phi_{\bsl{k}}$, and (iv) $e^{-\ii \Phi_{\bsl{k}}} \neq e^{-\ii \E_{m,\bsl{k}} T}$ for all $m$ and for all $\bsl{k}$ (or equivalently $\det[e^{-\ii \Phi_{\bsl{k}}}-U(\bsl{k},T)]\neq 0$ for all $\bsl{k}$).
The $2\pi$ redundancy of phase bands, as well as the $2\pi/T$ redundancy of quasi-energy bands, can be removed by requiring $\phi_{m,\bsl{k}}(t)$ to take values only in the PBZ $[\Phi_{\bsl{k}},\Phi_{\bsl{k}}+2\pi)$.
Two $\bsl{k}$-independent examples of $\Phi_{\bsl{k}}$ have been shown in {\figOneDExp}(a-b), and here we show a schematic $\bsl{k}$-dependent $\Phi_{\bsl{k}}$ for a $1+1$D 4-band $U(\bsl{k},t)$ in \figref{fig:Basic_Def_0}(a).

As exemplified by \figref{fig:Basic_Def_0}(a), we can always order the band index $m$ according to the values of $\E_{m,\bsl{k}}$ in the PBZ as $\E_{m+1,\bsl{k}}\geq\E_{m,\bsl{k}}$.
With this convention, we would have $\E_{m,\bsl{k}+\bsl{G}}=\E_{m,\bsl{k}}$, $\E_{m,\bsl{k}_g}=\E_{m,\bsl{k}}$, and $\E_{m,\bsl{k}}$ is continuous in $\dsR^d$.
Furthermore, a quasi-energy gap exists between two quasi-energy bands $\E_{m,\bsl{k}}$ and $\E_{m-1,\bsl{k}}$ iff  $\E_{m,\bsl{k}}> \E_{m-1,\bsl{k}}$ for all $\bsl{k}$, where $\E_{0,\bsl{k}}=\E_{N,\bsl{k}}-2\pi/T$.
In general, $U(\bsl{k},t)$ can have more than one quasi-energy gaps in the PBZ, and $\Phi_{\bsl{k}}$ can be chosen to lie in any of them.
For example, \figref{fig:Basic_Def_0}(a) shows three quasi-energy gaps: one at the PBZ lower bound, one between the bands 1 and 2, and one between the bands 3 and 4.
While the choice of the PBZ should have no influence on any physical and topological properties of the system, a good choice would simplify the derivation, and thus we, in this work, always set the PBZ lower bound in one of the relevant gaps as carefully discussed below.

\subsection{Topological Equivalence}

The topology in Floquet crystals is related to the topology in static crystals~\cite{Hasan2010TI,Qi2010TITSC}, and thereby let us start with a brief review on the latter.
The static crystals are governed by Bloch Hamiltonian, and we care about the symmetry-preserving continuous deformation of the Bloch Hamiltonian.
The deformation may close certain Bloch band gaps, and the key question is whether such a deformation drives a static insulator to a new phase with the same symmetry but different band topology.
The answer lies in a special band gap, which is the gap between the valence (highest occupied) band and conduction (lowest unoccupied) band.
Only this gap is relevant, while all other gaps, either between two occupied bands or between two unoccupied bands, are irrelevant.
As long as the symmetry-preserving continuous deformation of the Bloch Hamiltonian does not close the relevant band gap, the band topology must stay unchanged no matter how many irrelevant gaps are closed~\footnote{The underlying definition of the band topology is the topology of the vector bundle corresponding to the occupied bands.}.

\subsubsection{Floquet Crystal and FGU}

Based on the above brief review, we can see the relevant gaps play a crucial role in the topological equivalence.
So to define the topological equivalence for Floquet crystals, we need to first define the relevant gaps for them.
In Floquet crystals, we care about the deformation of time-evolution operator/matrix instead of the Hamiltonian.
Unlike the static case, it is unintuitive to define the occupied quasi-energy bands for a Floquet crystal that is not in equilibrium.
In this case, we may choose certain number $L$ of the quasi-energy gaps in a PBZ to be the relevant gaps~\cite{Roy2017FloquetTenFold,Yao2017FloquetTenFold}, and the rest of the quasi-energy gaps are irrelevant.
In the schematic example \figref{fig:Basic_Def_0}(b), we choose two of the three quasi-energy gaps to be relevant, resulting in $L=2$.
For the two-band 1+1D example in \appref{sec:1+1D_P}, we choose both quasi-energy gaps in {\figOneDExp}(a,b) to be relevant, also resulting in $L=2$.
If we know the relevant gaps for one PBZ and then change the PBZ choice, the new quasi-energy gaps in the new PBZ must be unique $2\pi n$-shifted copies (for $n\in\dsZ$ with $n=0$ corresponding the unshifted case) of those in the original PBZ, and then a new quasi-energy gap is relevant iff the corresponding original one is relevant.
As a result, the number of relevant gaps is always $L$ for any PBZ choice.

The $L$ relevant gaps in a PBZ separate the quasi-energy bands into $L$ isolated sets, labeled by $l=1,2,...,L$.
Throughout the work, when we talk about a set of bands, we strictly mean a multiset of bands since two degenerate bands are counted as two instead of one.
We emphasize that the quasi-energy bands in each isolated set might not be fully connected due to the possible existence of irrelevant gaps, but quasi-energy bands in different isolated sets must be disconnected owing to the relevant gaps.
As mentioned above, we always set the PBZ lower-bound in one of the relevant gaps in this work.
With this convention, the $L$ isolated sets can be ranked such that the $l+1$th set always has higher quasi-energies than the $l$th set at the same $\bsl{k}$, and the $l$th relevant gap is right beneath the $l$th isolated set.
In the schematic example \figref{fig:Basic_Def_0}(b), the first and second isolated sets contain $m=1$ and $m=2,3,4$ quasi-energy bands, respectively, and the first (second) relevant gap is right beneath the first (second) isolated set of quasi-energy bands.
For the two-band 1+1D example in \appref{sec:1+1D_P}, either of the two isolated sets in {\figOneDExp}(a,b) contain only one quasi-energy band.
By definition, an irrelevant gap can only exist between two quasi-energy bands within the same isolated set.

After picking the relevant gaps, we now are ready to provide explicit definitions for the Floquet crystal and the FGU.
\begin{definition}[Floquet Crystals]
\label{def:FC}
A Floquet crystal is defined to be a time-evolution operator $\hat{U}(t)$ (\eqnref{eq:U_gen}) equipped with a time period $T$ (\eqnref{eq:Floquet_U}), a relevant gap choice, and a crystalline symmetry group $\G$ (\eqnref{eq:U_g_invariant_gen}), which is in short denoted by $\hat{U}(t)$.
\end{definition}
In the definition of a Floquet crystal, we have implied (and will always imply) that $\hat{U}(t)$ is unitary and its matrix representation for any bases is continuous.
\begin{definition}[FGUs]
\label{def:FGU}
A FGU is defined to be a time-evolution matrix $U(\bsl{k},t)$ (\eqnref{eq:U_mat_gen}) equipped with a time period $T$ (\eqnref{eq:time_period_U_gen}), a relevant gap choice, a crystalline symmetry group $\G$, and a symmetry representation $u_g(\bsl{k})$ (\eqnref{eq:sym_rep_g1g2} and \eqnref{eq:Umat_g_invariant_gen}), which is in short denoted by $U(\bsl{k},t)$.
\end{definition}
In the definition of a FGU, we have implied (and will always imply) that $U(\bsl{k},t)$ and $u_g(\bsl{k})$ are unitary, continuous (smooth for $u_g(\bsl{k})$), and invariant under the shift of $\bsl{k}$ by reciprocal lattice vectors.
By choosing bases for a Floquet crystal, we naturally get a FGU with the same time period, relevant gaps and crystalline symmetry group as the Floquet crystal.
When referring to the gauge transformation of FGU, we mean the simultaneous gauge transformation in \eqnref{eq:gauge_trans_U_V}.
So FGUs given by the same Floquet crystal with different choices of bases are related by gauge transformations.

We emphasize that changing the relevant gap choice would give a different Floquet crystal or FGU, even if we keep all other parts (including time-evolution operator/matrix) invariant, since it would dramatically change the topological properties as discussed in \appref{sec:TE_FGUs}.
Moreover, the specified $\G$ does not need to include all crystalline symmetries of a Floquet crystal or a FGU, meaning that the crystalline symmetries outside $\G$ are allowed to be broken for the study of topology.
The choice of $\G$ depends on the physics of interest.

\subsubsection{Topological Equivalence Among FGUs}
\label{sec:TE_FGUs}

\begin{figure}[t]
    \centering
    \includegraphics[width=0.6\columnwidth]{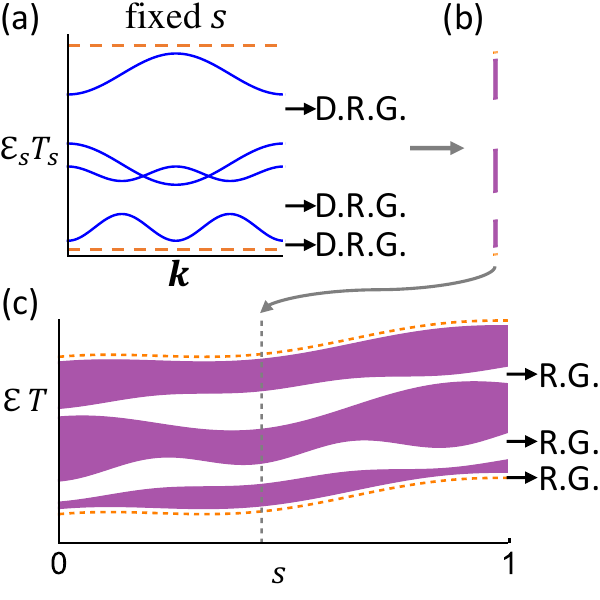}
    \caption{
    A schematic plot of topologically equivalent continuous deformation $U_s(\bsl{k},t)$ for $U(\bsl{k},t)$ and $U'(\bsl{k},t)$ having three relevant gaps (R.G.).
    The orange dashed lines are $\Phi_{\bsl{k},s}$ and $\Phi_{\bsl{k},s}+2\pi$.
    In (a), we schematically plot the quasi-energy bands given by $U_s(\bsl{k},T_s)$ at a fixed $s$ with three deformed relevant gaps (D.R.G.).
    Here we assume that all gaps are indirect at each $s$, while in general direct gaps are enough.
    In (b), we show the quasi-energy range of the quasi-energy bands in (a) by the purple region.
    The nonzero width of the purple region is given by the dispersion of the quasi-energy bands with respect to $\bsl{k}$, while the white parts indicate the deformed relevant gaps.
    We group the (b)-type plots for all values of $s$ to get (c).
    In (c), the three purple regions show how three isolated sets of quasi-energy bands evolve along $s$, and the white regions stand for three deformed relevant gaps which are not closed during the entire deformation.
    The indirect gaps allow us to always make $\Phi_{\bsl{k},s}$ independent of $\bsl{k}$ (thus of zero width in the quasi-energy).
    }
    \label{fig:continuous_deformation}
\end{figure}

With the definition of relevant gaps and FGUs, we next discuss the topological equivalence.
Before addressing the topological equivalence among Floquet crystals, let us first focus on the topological equivalence among FGUs.
Suppose we have two FGUs $U(\bsl{k},t)$ (with $T$, relevant gaps, $\mathcal{G}$, and  $u_g(\bsl{k})$) and $U'(\bsl{k},t)$ (with $T'$, relevant gaps, $\mathcal{G}$, and $u_g'(\bsl{k})$).
Note that the two FGUs are invariant under the same crystalline symmetry group $\G$.
In analogy to the static case, we can operationally define the topological equivalence for FGUs as the following.
\begin{definition}[Topological Equivalence for FGUs]
\label{def:topo_equi}
The two FGUs $U(\bsl{k},t)$ and $U'(\bsl{k},t)$ are defined to be topologically equivalent under the crystalline symmetry group $\G$ iff there exists a continuous deformation that connects them, preserves $\G$ and preserves all relevant gaps.
\end{definition}
As long as the crystalline symmetry group $\G$ for the topological equivalence is specified, we may refer to ``topologically equivalent under $\G$" as ``topologically equivalent" in short.
The topological equivalence defined in \defref{def:topo_equi} is an equivalence relation.
Specifically, a FGU is always topologically equivalent to itself; if a FGU $U(\bsl{k},t)$ is equivalent to another FGU $U'(\bsl{k},t)$, then $U'(\bsl{k},t)$ is equivalent to $U(\bsl{k},t)$; if $U(\bsl{k},t)$ is equivalent to $U'(\bsl{k},t)$ and $U'(\bsl{k},t)$ is equivalent to $U''(\bsl{k},t)$, then $U(\bsl{k},t)$ is equivalent to $U''(\bsl{k},t)$.
The relation between \defref{def:topo_equi} and the related previous literature~\cite{Nathan2015TopoSingClas,Fruchart2016FloquetTenFold, Roy2017FloquetTenFold, Yao2017FloquetTenFold} will be addressed in \appref{sec:BasicDef_Topo_Comp}.
In the rest of this part, we elaborate on each part of \defref{def:topo_equi}.

A continuous deformation between $U(\bsl{k},t)$ and $U'(\bsl{k},t)$ is a unitary matrix function $U_{s}(\bsl{k},t)$ with $s\in[0,1]$ such that (i) $U_{s}(\bsl{k}+\bsl{G},t)=U_{s}(\bsl{k},t)$ and $U_{s}(\bsl{k},t+T_s)=U_{s}(\bsl{k},t) U_{s}(\bsl{k},T_s)$, (ii) $U_{s}(\bsl{k},t)$ is a continuous function of $(\bsl{k},t,s)\in \dsR^d\times\dsR\times [0,1]$ and $T_s>0$ is continuous in $[0,1]$, and (iii) $T_{s=0}=T$, $T_{s=1}=T'$, and there exist $U(N)$ gauge transformation matrices $W_{0,1}(\bsl{k})$ such that 
\eqa{
\label{eq:continuous_def_U}
& U_{s=0}(\bsl{k},t)=W_{0}^\dagger(\bsl{k})U(\bsl{k},t)W_{0}(\bsl{k}) \\
& U_{s=1}(\bsl{k},t)=W_{1}^\dagger(\bsl{k})U'(\bsl{k},t)W_{1}(\bsl{k})\ .
}
The existence of $U_s$ infers that $U(\bsl{k},t)$ and $U'(\bsl{k},t)$ must have the same matrix dimension.

Preserving $\G$ means that there exist unitary $u_{s,g}(\bsl{k})$ such that (i) $u_{s,g}(\bsl{k})$ is a continuous function of $(\bsl{k},s)\in \dsR^d\times[0,1]$  and $u_{s,g}(\bsl{k}+\bsl{G})=u_{s,g}(\bsl{k})$, (ii) $u_{s,g}(\bsl{k})$ satisfies \eqnref{eq:sym_rep_g1g2} for each value of $s$, (iii) 
\eq{
u_{s,g}(\bsl{k}) U_s(\bsl{k},t) u_{s,g}^\dagger(\bsl{k}) = U_s(\bsl{k}_g,t)\ ,
}
and (iv) 
\eqa{
\label{eq:continuous_def_u}
& u_{s=0,g}(\bsl{k})= W_0^\dagger(\bsl{k}_g) u_g(\bsl{k}) W_0(\bsl{k})\\
& u_{s=1,g}(\bsl{k})= W_1^\dagger(\bsl{k}_g) u_g'(\bsl{k}) W_1(\bsl{k})\ .
}
Owing to \eqnref{eq:continuous_def_U} and \eqnref{eq:continuous_def_u}, the topological equivalence between FGUs is $U(N)$ gauge invariant.

Preserving all relevant gaps first requires that there exist a proper $\Phi_{\bsl{k},s}$ that allows us to plot the quasi-energy bands given by $U_s(\bsl{k},T_s)$ within $(\Phi_{\bsl{k},s},\Phi_{\bsl{k},s}+2\pi)$.
Then, preserving all relevant gaps further requires that if we track the relevant gaps of $U(\bsl{k},t)$ as varying $s$ from $0$ to $1$, (i) none of the relevant gaps close and (ii) the deformed relevant gaps of $U(\bsl{k},t)$ would exactly coincide with the relevant gaps of $U'(\bsl{k},t)$ as $s$ reaches $1$.
To be more specific, a proper $\Phi_{\bsl{k},s}$ is required to satisfy that $\Phi_{\bsl{k}+\bsl{G},s}=\Phi_{\bsl{k},s}$, it is a real continuous function of $(\bsl{k},s)\in \dsR^d \times [0,1]$, $\Phi_{\bsl{k}_g,s}=\Phi_{\bsl{k},s}$, $\det(e^{-\ii \Phi_{\bsl{k},s}}-U_s(\bsl{k},T_s))\neq 0$, and $\Phi_{\bsl{k},s=0}$ is a PBZ lower bound of $U(\bsl{k},t)$.
If the relevant gaps are preserved, $\Phi_{\bsl{k},s=1}$ must lie in a relevant gap of $U'(\bsl{k},t)$.
Owing to this requirement, two topologically equivalent FGUs must have the same number of relevant gaps.
\figref{fig:continuous_deformation} schematically shows an example of the topologically continuous deformation for the case with three indirect relevant gaps, though in general direct gaps are enough.
In particular, the three white regions in \figref{fig:continuous_deformation}(c) show that the three relevant gaps of $U(\bsl{k},t)$ keep open as $s$ continuously increases and eventually become the three relevant gaps of $U'(\bsl{k},t)$.
A more mathematical but equivalent way to express this requirement is that $U(\bsl{k},t)$ and $U'(\bsl{k},t)$ have $L$ relevant gaps, and for any PBZ lower bound $\Phi_{\bsl{k}}$ of $U(\bsl{k},t)$, there exists $\Phi_{l,\bsl{k},s}$ with $l=1,2,...,L$ such that (i) $\Phi_{l,\bsl{k},s}$ is a continuous function of $(\bsl{k},s)\in\dsR^d\times[0,1]$ and satisfies $\Phi_{l,\bsl{k}+\bsl{G},s}=\Phi_{l,\bsl{k},s}$ and $\Phi_{l,\bsl{k}_g,s}=\Phi_{l,\bsl{k},s}$, (ii) $\Phi_{l,\bsl{k},s=0}$ lies in the $l$th relevant gap of $U(\bsl{k},t)$ and $\Phi_{1,\bsl{k},s=0}=\Phi_{\bsl{k}}$, (iii) $\Phi_{l,\bsl{k},s=1}$ ($l=1,...,L$) respectively lie in all $L$ relevant gaps of $U'(\bsl{k},t)$, and (iv) $\det\left(e^{-\ii \Phi_{l,\bsl{k},s}}-U_s(\bsl{k},T_s)\right)\neq 0$.
Another equivalent statement can be obtained by replacing ``for any PBZ lower bound $\Phi_{\bsl{k}}$ of $U(\bsl{k},t)$" by ``for at least one PBZ lower bound $\Phi_{\bsl{k}}$ of $U(\bsl{k},t)$" in the above requirement.

We emphasize that the choice of relevant gaps is crucial for determining whether two FGUs are topologically equivalent according to \defref{def:topo_equi}.
Even if two FGUs have exactly the same time-evolution matrix $U(\bsl{k},t)=U'(\bsl{k},t)$, different choices of relevant gaps can make them topologically distinct.
As mentioned above, if we choose different numbers of relevant gaps for $U(\bsl{k},t)$ and $U'(\bsl{k},t)$, they must be topologically distinct since no continuous deformation can change the number of relevant gaps without closing any of them.

Even if we choose the same number of relevant gaps for $U(\bsl{k},t)=U'(\bsl{k},t)$, it is still possible to make them topologically distinct by choosing different quasi-energies for the relevant gaps.
Let us consider two $0+1D$ two-band FGUs with trivial $\G$, and suppose they have the same time-evolution matrix $U(t)=U'(t)$ (the Bloch momentum is not needed) as schematically shown in \figref{fig:Basic_Def_1}(a).
Suppose we only pick one of the two quasi-energy gaps to be relevant.
If we choose different relevant gaps for the two FGUs, it is impossible to establish the topological equivalence between them according to \defref{def:topo_equi}, since it is impossible to continuously deform the relevant gap of $U$ into the relevant gap of $U'$ without closing it.
In reality, choosing the relevant gaps normally requires careful consideration based on the physics of interest.
One common choice is to treat all quasi-energy gaps as relevant, just like the 1+1D example in \appref{sec:1+1D_P}. 
In the remaining of this work, we will not address the issue of choosing the relevant gaps, and we always discuss FGUs with relevant gaps already specified, unless specified otherwise.

\subsubsection{Topological Equivalence Among Floquet Crystals}

Now let us turn to the topological equivalence among Floquet crystals.
Suppose we have two Floquet crystals $\hat{U}(t)$ (with $T$, a relevant gap choice, and $\G$) and $\hat{U}'(t)$ (with $T'$, a relevant gap choice, and $\G$).
Similar to \defref{def:topo_equi}, we have the following definition for Floquet crystals.
\begin{definition}[Topological Equivalence for Floquet Crystals]
\label{def:topo_equi_FC}
The two Floquet crystals $\hat{U}(t)$ and $\hat{U}'(t)$ are defined to be topologically equivalent iff there exists a continuous deformation that connects them, preserves $\G$ and preserves all relevant gaps.
\end{definition}

Specifically, the deformation that connects $\hat{U}(t)$ and $\hat{U}'(t)$ is a unitary operator $\hat{U}_s(t)$ depending on $s\in[0,1]$ such that (i) $\hat{U}_{s=0}(t)=\hat{U}(t)$ and $\hat{U}_{s=1}(t)=\hat{U}'(t)$, (ii) $\hat{U}_{s}(t+T_s)=\hat{U}_s(t)\hat{U}_s(T_s)$ with $T_s>0$ satisfying $T_{s=0}=T$ and $T_{s=1}=T'$.
The deformation being continuous means that there exist $\ket{\psi_{\bsl{k},s}}$ serving as bases of $\hat{U}_s(t)$ at each value of $s$ (thus satisfying all requirements for bases at each value of $s$) such that (i) 1BZ is independent of $s$ and the periodic part of the bases $e^{-\ii \bsl{k}\cdot\hat{\bsl{r}}}\ket{\psi_{\bsl{k},s}}$ is a continuous function of $(\bsl{k},s)\in\dsR^d\times[0,1]$, and (ii) the matrix representation of $\hat{U}_{s}(t)$, denoted by  $U_s(\bsl{k},t)$, is a continuous function of $(\bsl{k},t,s)\in \dsR^d\times\dsR\times [0,1]$, and (iii) $T_s$ is continuous in $[0,1]$.
The deformation preserving symmetry means that $[\hat{U}_{s}(t),g]=0$ and $g\ket{\psi_{\bsl{k},s}}=\ket{\psi_{\bsl{k}_g,s}} u_{s,g}(\bsl{k})$.
The deformation preserving the relevant gaps means that after choosing the relevant gaps of $U_{s=0}(\bsl{k},t)$ ($U_{s=1}(\bsl{k},t)$) to be the same as $\hat{U}(t)$ ($\hat{U}'(t)$), the relevant gaps of $U_{s=0}(\bsl{k},t)$ are kept open as $s$ increases from $0$ and eventually becomes the relevant gaps of $U_{s=1}(\bsl{k},t)$ as $s$ reaches $1$.

The defined topological equivalence between two Floquet crystals is a equivalence relation, \ie, (i) a Floquet crystal is always equivalent to itself, (ii) $\hat{U}(t)$ being equivalent to $\hat{U}'(t)$ infers that $\hat{U}'(t)$ being equivalent to $\hat{U}(t)$, and (iii) $\hat{U}(t)$ being equivalent to $\hat{U}'(t)$ and $\hat{U}'(t)$ being equivalent to $\hat{U}''(t)$ infer that $\hat{U}(t)$ being equivalent to $\hat{U}''(t)$.

As discussed in \appref{sec:TE_FGUs}, we can naturally define a FGU for any given Floquet crystal upon choosing bases.
If two Floquet crystals are topologically equivalent, they must have topologically equivalent FGUs for any bases choices, where the equivalence between the FGUs is established by $U_s(\bsl{k},t)$ (together with $T_s$) and $u_{s,g}(\bsl{k})$ furnished by $\ket{\psi_{\bsl{k},s}}$ in the above discussion.
Therefore, the topological distinction among FGUs must infer the topological distinction among the underlying Floquet crystals, and all topological invariants of FGUs can be applied to Floquet crystals.
As we do not require the completeness of the topological invariants, we in this work focus on the topological equivalence among FGUs unless specified otherwise.

\subsubsection{Comparison to Previous Literature}
\label{sec:BasicDef_Topo_Comp}

\begin{figure}
    \centering
    \includegraphics[width=0.6\columnwidth]{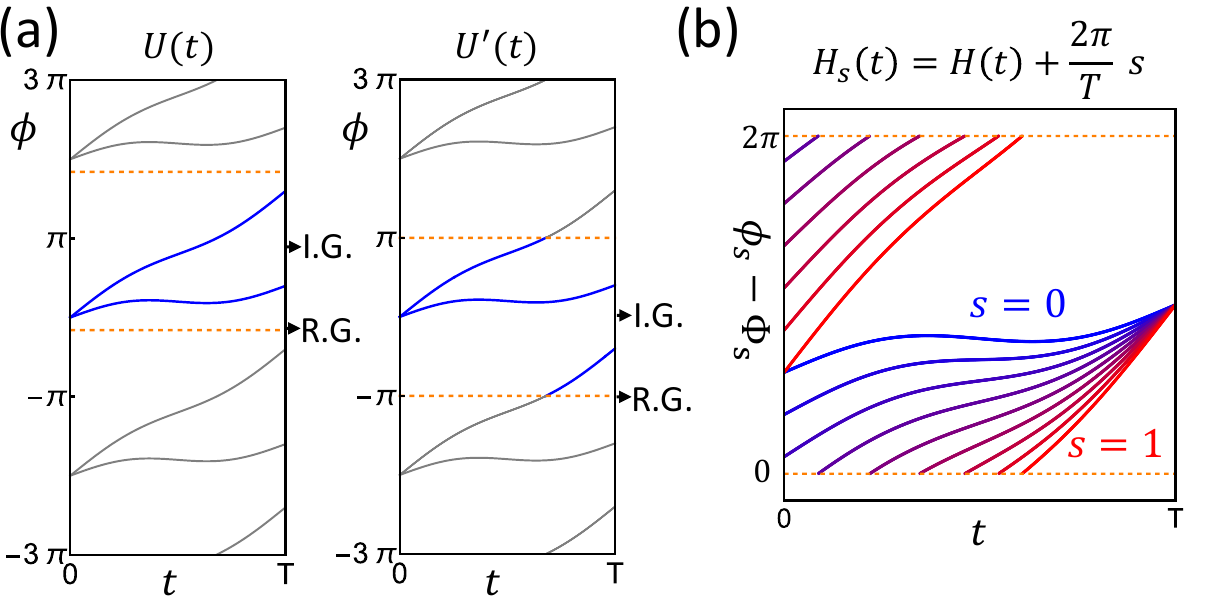}
    \caption{
    Schematic plots of the phase bands for 0+1D class-A models.
    ``R.G." and ``I.G." stand for relevant gap and irrelevant gap, respectively.
    In (a), we consider two $0+1$D class-A 2-band FGUs with the same time-evolution matrix $U(t)=U'(t)$.
    Due to the different choices of the relevant gaps, the two FGUs are topologically distinct according to \defref{def:topo_equi}.
    In (b), we schematically show the continuous deformation described in \eqnref{eq:deformation_enshift_gen_U} for a $0+1$D class-A case: $U_s(t)= e^{-\ii \phi_s(t)}$ with $\phi_s(t)=\phi(t)+ 2\pi s t/T$.
    From blue to red, $s$ varies from $0$ to $1$.
    The deformed PBZ $[\Phi_s, \Phi_s+2\pi)$ (bounded by the dashed orange lines) is given by $\Phi_s= \phi_s(T) - \pi$, which always lies in the only quasi-energy gap.
    The quasi-energy gap is kept open along the deformation as $\phi_s(T)-\Phi_s$ cannot pass through 0.
    }
    \label{fig:Basic_Def_1}
\end{figure}

\defref{def:topo_equi} for FGUs is similar to the definition in Sec.\,2 of \refcite{Nathan2015TopoSingClas}, except the following two key differences.
First, \defref{def:topo_equi} allows the deformation to deviate from the topologically equivalent FGUs by $\U{N}$ gauge transformations (\eqnref{eq:continuous_def_U} and \eqref{eq:continuous_def_u}) so that the defined topological equivalence is gauge invariant.
Second, \defref{def:topo_equi} allows the symmetry representation and time period to vary along the deformation, and also allows the symmetry representation to depend on momenta.
Next, we discuss the possible difference between the topological classification based on \defref{def:topo_equi} and the classification in \refcite{Fruchart2016FloquetTenFold, Roy2017FloquetTenFold, Yao2017FloquetTenFold}.

For the topological equivalence defined in \defref{def:topo_equi}, the PBZ is allowed to continuously evolve along with the deformation $U_{s}(\bsl{k},t)$ (\eg, \figref{fig:continuous_deformation}), or in other words the quasi-energy bands (times $T_s$) given by $U_{s}(\bsl{k},T_s)$ do not need to be confined in a $s$-independent $2\pi$ range (like $[-\pi,\pi)$).
The reason for us to adopt this definition is demonstrated by the following deformation.

Let us consider a Floquet Hamiltonian in class A parametrized by $s\in[0,1]$ as
\eq{
\label{eq:deformation_enshift_gen_H}
\hat{H}_s(t)=\hat{H}(t)+\frac{2\pi}{T}s\ ,
}
where $\hat{H}(t+T)=\hat{H}(t)$ and different values of $s$ just correspond to different calibrations of the energy (or different global energy shifts).
We emphasize that even in the Fock space for many-body Hamiltonians, $2\pi s/T$ should still be proportional to the identity operator instead of the particle-number operator, and thus it does not change the particle number.
Therefore, varying $s$ in $\hat{H}_s(t)$ should not change any physical property (like the crystalline symmetry group $\G$) or topological property (like topological distinction).

We can choose a set of $s$-independent bases, and then the corresponding time-evolution matrix reads
\eq{
\label{eq:deformation_enshift_gen_U}
U_{s}(\bsl{k},t)=U_0(\bsl{k},t) \exp\left(-\ii \frac{2\pi s }{T} t\right)\ ,
}
and the representation of $\G$ reads $u_g(\bsl{k})$.
Let us focus on $U_{0}(\bsl{k},t)$ and $U_{1}(\bsl{k},t)$. 
Since $U_{0}(\bsl{k},T)=U_{1}(\bsl{k},T)$, we can choose the same relevant gaps for $U_{0}(\bsl{k},t)$ and $U_{1}(\bsl{k},t)$.
Then, we have two FGUs $U_{0}(\bsl{k},t)$ and $U_{1}(\bsl{k},t)$ with the same $T$, same relevant gap choice, same $\G$, and same $u_g(\bsl{k})$, provided that all other requirements are satisfied.
$U_{s}(\bsl{k},t)$ in \eqnref{eq:deformation_enshift_gen_U}, together with $T_s=T$ and $u_{s,g}(\bsl{k})=u_{g}(\bsl{k})$, establishes the topological equivalence between $U_{0}(\bsl{k},t)$ and $U_{1}(\bsl{k},t)$, since the relevant gaps are preserved and all other conditions are satisfied.
Specifically for the relevant gaps, suppose $\Phi_{\bsl{k}}$ is a PBZ lower bound of $U_{0}(\bsl{k},T)$, and we can choose the deformed PBZ lower-bound during the deformation to be $\Phi_{\bsl{k},s}=\Phi_{\bsl{k}}+2\pi s$, resulting in the deformed quasi-energy bands $\E_{m,\bsl{k}}^s=\E_{m,\bsl{k}}^0+2 \pi s/T$ within $[\Phi_{\bsl{k},s},\Phi_{\bsl{k},s}+2\pi)/T$.
Then, varying $s$ can only shift all quasi-energy bands simultaneously by the same amount and thus cannot close any of the quasi-energy gaps.
Moreover, $\E_{m,\bsl{k}}^{s=1}=\E_{m,\bsl{k}}^0+2 \pi/T$ means that the quasi-energy gaps at $s=1$ are nothing but $2\pi$-shifts of those at $s=0$.
As a result, the relevant gaps of $U_{0}(\bsl{k},t)$ are kept open and eventually coincide with the relevant gaps of $U_{1}(\bsl{k},t)$ as $s$ reaches $1$, since $U_0(\bsl{k},t)$ and $U_1(\bsl{k},t)$ have the same relevant gaps.
The topological equivalence between $U_0(\bsl{k},t)$ and $U_{1}(\bsl{k},t)$ according to \defref{def:topo_equi} coincides with above statement that the global energy shift should not change any physical or topological property in class A. 
As all quasi-energy bands are continuously shifted by $2\pi/T$ as $s$ changes from $0$ to $1$, no bands (times $T_s$) can be confined in a $s$-independent $2\pi$ range during this deformation.

Owing to \defref{def:topo_equi}, the topological classification that we obtain might differ from the previous classification~\cite{Fruchart2016FloquetTenFold, Roy2017FloquetTenFold, Yao2017FloquetTenFold}.
One example would be $0+1$D one-band class-A case without any crystalline symmetries.
In this case, the Bloch momentum is not needed, and we consider a FGU with a $1\times 1$ time-evolution matrix $U_{0}(t)=e^{-\ii \phi(t)}$, where $\phi(t)$ is real and $U_{0}(t+T)=U_{0}(t)U_{0}(T)$.
For the PBZ lower bound $\Phi=\phi(T)-\pi$, we have only one quasi-energy $\phi(T)$ and only one quasi-energy gap---the one between $\phi(T)$ and $\phi(T)-2\pi$---which we have chosen to be relevant.
Let us again shift the global energy to give a deformation  $U_{s}(t)=e^{-\ii \phi(t)-\ii 2\pi s t /T }$ with $s\in [0,1]$, which is a special case of \eqnref{eq:deformation_enshift_gen_U}.
As schematically shown in \figref{fig:Basic_Def_1}(b), the deformation indeed does not close the quasi-energy gap, and thereby $U_{0}(t)$ is topologically equivalent to $U_{1}(t)$ as long as we also choose the only quasi-energy gap of $U_{1}(t)$ to be relevant, according to \defref{def:topo_equi} and the above discussion.
In contrast, according to the classification in \refcite{Fruchart2016FloquetTenFold, Roy2017FloquetTenFold, Yao2017FloquetTenFold}, $U_{0}(t)$ and $U_{1}(t)$ are topologically distinct since they have different winding numbers if we impose the same branch cut for their return maps.
Such a difference arises from the different definition of topological equivalence in \refcite{Fruchart2016FloquetTenFold, Roy2017FloquetTenFold, Yao2017FloquetTenFold}.

\subsection{Return Map}
\label{sec:return_map}

The definition of the return map has been discussed in \appref{sec:1+1D_P_winding_data}.
Here we just need to generalize it from the $1+1$D two-band case to a $N$-band FGU $U(\bsl{k},t)$ with $T$, a relevant gap choice, a generic $\G$, and $u_g(\bsl{k})$.
Specifically, we replace $k$ by $\bsl{k}$ and replace $2$ bands by $N$ bands in \eqnref{eq:1+1D_P_RM_def_prep_1}-\eqref{eq:1+1D_P_RM_def_prep_2} to get the return map
\eq{
%\label{eq:RM_def_gen}
U_\epsilon(\bsl{k},t)=U(\bsl{k},t) \left[U(\bsl{k},T)\right]^{-t/T}_\epsilon\ ,
}
where 
\eq{
\left[U(\bsl{k},T)\right]^{-t/T}_\epsilon=\sum_{m=1}^N\exp\left[-\frac{t}{T} \log_{\epsilon_{\bsl{k}}}(e^{-\ii\E_{m,\bsl{k}} T}) \right] P_{\bsl{k},m}(T)\ ,
}
and $P_{\bsl{k},m}(T)$ is the projection matrix given by the eigenvector of $U(\bsl{k},T)$ for $e^{-\ii\E_{m,\bsl{k}} T}$.
Under the gauge transformation \eqnref{eq:gauge_trans_U_V}, $U_\epsilon(\bsl{k},t)$ transforms as
\eq{
\label{eq:gauge_trans_Uepsilon}
U_\epsilon(\bsl{k},t)\rightarrow W^\dagger(\bsl{k}) U_\epsilon(\bsl{k},t) W(\bsl{k})\ .
}
Recall that we always choose the PBZ lower bound $\Phi_{\bsl{k}}$ as the branch cut $\epsilon_{\bsl{k}}=\Phi_{\bsl{k}}$ unless specified otherwise.
Then, $U_{\epsilon=\Phi}(\bsl{k},t+T)=U_{\epsilon=\Phi}(\bsl{k},t)$, $U_{\epsilon=\Phi}(\bsl{k},t)$ is a continuous function of $(\bsl{k},t)\in\dsR^d\times\dsR$, $U_{\epsilon=\Phi}(\bsl{k}+\bsl{G},t)=U_{\epsilon=\Phi}(\bsl{k},t)$, and $U_{\epsilon=\Phi}(\bsl{k},t)$ is $\G$-invariant
\eq{
\label{eq:sym_Uepsilon}
u_g(\bsl{k}) U_{\epsilon=\Phi}(\bsl{k},t) u_g^\dagger(\bsl{k}) = U_{\epsilon=\Phi}(\bsl{k}_g,t)\ .
}
(See more details in \appref{app:return_map_winding_data}.)
Since the PBZ lower bound $\Phi_{\bsl{k}}$ is required to lie in a relevant gap, $U_{\epsilon=\Phi}(\bsl{k},t)$ is continuously deformed under any topologically equivalent continuous deformation of $U(\bsl{k},t)$, as long as we continuously deform the PBZ lower bound along with the deformed relevant gap.

\subsection{Obstruction to Static Limits}
\label{sec:static_limits}

\defref{def:topo_equi} only defines the relative topological equivalence, but it does not tell us which side is topologically nontrivial.
In static crystals, the obstruction to the atomic limits is used to define the topologically nontrivial systems~\cite{Po2017SymIndi,Bradlyn2017TQC}.
As mentioned in \appref{sec:1+1D_P_DSI}, here we are interested in the obstruction to static limits for the characterization of Floquet dynamics.
Specifically, a Floquet crystal $\hat{U}(t)$ has obstruction to static limits iff given any continuous deformation that starts from $\hat{U}(t)$ and ends as the time-evolution operator of a static Hamiltonian, the deformation must break certain symmetries or close certain RGs of $\hat{U}(t)$.
It turns out that for later derivations, it is more convenient to use an equivalent formal definition of obstruction to static limits based on a formal definition of static limits, which are discussed below.

The explicit definition that we adopt for static limits and static FGUs in this work is the following.
\begin{definition}[Static Limits and Static FGUs]
\label{def:SL_def_gen}
A static limit is a Floquet crystal with static Hamiltonian; a static FGU is a FGU with static matrix Hamiltonian.
\end{definition}

As a static limit (static FGU) satisfies the definition of Floquet crystal (FGU), we can study its topological properties according to proposed definition of topological equivalence.

Now we discuss how to construct static limits given a static Hamiltonian $\hat{H}_{SL}$.
The time-evolution operator $\hat{U}_{SL}(t)=\exp(-\ii \hat{H}_{SL} t)$ and crystalline symmetry group $\G$ can be naturally determined from $\hat{H}_{SL}$.
However, to make it a static limit that satisfies the definition of a Floquet crystal, we need to have the time period and the relevant gaps.
This is where a subtlety appears.
When we refer to the period $T$ of a Floquet crystal, we actually mean the fundamental period---the smallest positive $T$ that satisfies $\hat{H}(t+T)=\hat{H}(t)$.
Static Hamiltonians do not have a fundamental period since they are invariant under any time shift.
To resolve this issue, we can assign a period $T_{SL}>0$ to a given $\hat{U}_{SL}(t)$, and determine the quasi-energy bands and pick the relevant gaps according to $\hat{U}_{SL}(T_{SL})$.
Another way to fix this issue is to add an infinitesimal drive with period $T_{SL}$ to the static Hamiltonian and define the Floquet crystal based on the resultant driven Floquet system.
Both ways are equivalent, and we, in this work, stick to assigning $T_{SL}$ to $\hat{U}_{SL}(t)$.
The resultant static limit is just the time-evolution operator $\hat{U}_{SL}(t)= e^{-\ii \hat{H}_{SL} t}$ with the assigned $T_{SL}$, the relevant gaps chosen according to $\hat{U}_{SL}(T_{SL})$, and the crystalline symmetry group $\G$ of $\hat{H}_{SL}$.
The same procedure can be applied to a static matrix Hamiltonian $H_{SL}(\bsl{k})$ with a crystalline symmetry group $\G$ and the representation $u_g(\bsl{k})$, resulting in a static FGU as $U_{SL}(\bsl{k},t)= e^{-\ii H_{SL}(\bsl{k}) t}$ with the assigned $T_{SL}$, the relevant gaps chosen according to $U_{SL}(\bsl{k},T_{SL})$, $\G$, and $u_g(\bsl{k})$.
A static FGU can be naturally given by picking bases for a static limit.
We emphasize that different assigned $T_{SL}$'s or different relevant gap choices by definition give different static limits (or static FGUs).

\begin{definition}[Obstruction to Static Limits]
A Floquet crystal (a FGU) with $\G$ is defined to have obstruction to static limits iff it is topologically distinct from all static limits (static FGUs) with $\G$.
\end{definition}
According to the definition, it seems that we need take into account all possible choices of $T_{SL}$ to determine the obstruction.
It turns out that given a FGU with time period $T$ and crystalline symmetry group $\G$, we can neglect static FGUs with $T_{SL}\neq T$ in order to determine the obstruction for the FGU.
It is because for any static FGU $U_{SL}(\bsl{k},t)= e^{-\ii H_{SL}(\bsl{k}) t}$ with $T_{SL}$, relevant gap choice, $\G$, and $u_g(\bsl{k})$, we always have another static FGU $U_{SL}'(\bsl{k},t)= e^{-\ii H_{SL}(\bsl{k}) \frac{T_{SL}}{T} t }$ with $T$, relevant gap choice same as $U_{SL}$, $\G$, and $u_g(\bsl{k})$, such that $U_{SL}(\bsl{k},t)$ and $U_{SL}'(\bsl{k},t)$ are topologically equivalent.
The same relevant gap choice is allowed by $U_{SL}(\bsl{k},T_{SL})=U_{SL}'(\bsl{k},T)$, and the topological equivalence is established by $U_s(\bsl{k},t)=e^{-\ii H_s(\bsl{k}) t}$ with $s\in [0,1]$, $T_s=(1-s) T_{SL}+s T$, $H_{s}(\bsl{k})=H_{SL}(\bsl{k}) T_{SL}/T_s$, and $u_{s,g}(\bsl{k})=u_{g}(\bsl{k})$.
In other words, to have obstruction to static limits, a $\G$-invariant FGU with period $T$ must and only need to be topologically distinct from all $\G$-invariant static FGUs with $T_{SL}=T$.
The same conclusion can also be drawn for the Floquet crystals.
Furthermore, if a FGU of a $\G$-invariant Floquet crystal for certain bases has obstruction to static limits, the Floquet crystal then must be topologically distinct from all $\G$-invariant static limits, and thus has obstruction to static limits.
Therefore, in the remaining of this work, we will focus on the obstruction of FGUs, and when determining the obstruction for a FGU with period $T$, we always assign $T_{SL}=T$ to all static FGUs unless specified otherwise.

In contrast to \defref{def:SL_def_gen} adopted in this work, the static limit was sometimes implied as the $T\rightarrow 0$ limit in previous literature~\cite{Titum2015DisorderFTI}.
According to \defref{def:SL_def_gen}, $T\rightarrow 0$ is just one way to make a Floquet crystal (or FGU) static, while there are infinite many other ways, including continuously decreasing the driving amplitude to zero while fixing $T$.
In this work, if a Floquet crystal (or FGU) has the obstruction to static limits, all continuous deformations that make it static are forbidden (or equivalently must break certain symmetries or close certain relevant gaps).

\section{Details on General Framework}
\label{sec:gen_frame}

In this section, we follow \figref{fig:Concepts_Structure} to introduce the symmetry data, the quotient winding data, and the DSI for class-A $d+1D$ FGUs (and thus for Floquet crystals) with a generic crystalline symmetry group $\G$ and $d\leq 3$.
Henceforth, when we discuss different FGUs, we always imply that they have the same crystalline symmetry group $\G$.

\subsection{Symmetry Data of Quasi-energy Band Structure}
\label{sec:sym_data}

We start with introducing the symmetry data of the quasi-energy bands.
We first follow \refcite{Bradley2009MathSSP,Po2017SymIndi,Bradlyn2017TQC}, and then discuss the subtlety that is absent in static crystals.

Let us first consider a generic FGU $U(\bsl{k},t)$ with time period $T$, a relevant gap choice, a generic crystalline symmetry group $\G$ and a symmetry representation $u_{g}(\bsl{k})$.
According to \eqnref{eq:sym_rep_g}, an element $g=\{ R |\bsl{\tau} \}$ of the crystalline symmetry group $\G$ can change the Bloch momentum $\bsl{k}$ to $\bsl{k}_g=R \bsl{k}$.
Iff there exists a reciprocal lattice vector $\bsl{G}$ such that $\bsl{k}_g=\bsl{k}+\bsl{G}$, we say $g$ leaves $\bsl{k}$ invariant.
For any $\bsl{k}\in$1BZ, all elements of $\G$ that leave $\bsl{k}$ invariant form a group, which is called the little group~\cite{Bradley2009MathSSP} of $\bsl{k}$ and denoted by $\G_{\bsl{k}}$.
$\G_{\bsl{k}}$ must contain all lattice translations in $\G$; if $\G_{\bsl{k}}$ contains more than lattice translations, such as the little groups for $\Gamma$ and $X$ discussed in \appref{sec:1+1D_P_sym_data}, we call $\bsl{k}$ a high-symmetry momentum~\cite{Bradley2009MathSSP}.

Now we focus on $\G_{\bsl{k}}$.
When restricting $g\in\G_{\bsl{k}}$, the representation $u_{g}(\bsl{k})$ satisfies a simpler version of \eqnref{eq:sym_rep_g1g2}, which reads
\eq{
\label{eq:small_sym_rep}
u_{g_1 g_2}(\bsl{k})=u_{g_1}(\bsl{k})u_{g_2}(\bsl{k})\ \ \forall g_1,g_2\in\G_{\bsl{k}}\ ,
}
where the Bloch momenta other than $\bsl{k}$ are not involved.
\eqnref{eq:small_sym_rep} suggests $u_{g}(\bsl{k})$ with fixed $\bsl{k}$ is a representation of $\G_{\bsl{k}}$, which is called a small representation of $\G_{\bsl{k}}$.
In particular, the small representation $u_{g}(\bsl{k})$ commutes with the time-evolution matrix $U(\bsl{k},t)$:
\eq{
\label{eq:little_group_commutes_U_gen}
u_{g}(\bsl{k}) U(\bsl{k},t) u_{g}^\dagger(\bsl{k}) = U(\bsl{k},t)\ \forall g\in \G_{\bsl{k}}\ .
}
\eqnref{eq:little_group_commutes_U_gen} suggests that each eigenvector of $U(\bsl{k},T)$ participates in furnishing a definite small irreducible representation (irrep) of $\G_{\bsl{k}}$, and thereby we can associate each quasi-energy band in a given PBZ with a small irrep of $\G_{\bsl{k}}$.  

Let us now pick a generic PBZ lower bound $\Phi_{\bsl{k}}$ for the given FGU $U(\bsl{k},t)$.
Recall that the quasi-energy bands in the $\Phi_{\bsl{k}}$-PBZ are separated into isolated sets by the relevant gaps. 
For each small irrep $\alpha$ of $\G_{\bsl{k}}$, we can count the number of quasi-energy bands in the $l$th isolated set that are associated with it, labelled as $\tilde{n}^{l}_{\bsl{k},\alpha}$, where $\alpha$ ranges over all inequivalent small irreps of $\G_{\bsl{k}}$.
For convenience, we do not directly use $\tilde{n}^{l}_{\bsl{k},\alpha}$ but use the number of copies of irreps, which is $n^{l}_{\bsl{k},\alpha}=\tilde{n}^{l}_{\bsl{k},\alpha}/d_{\alpha}$ with $d_{\alpha}$ the dimension of the small irrep $\alpha$ of $\G_{\bsl{k}}$.
In the $1+1$D example discussed in \appref{sec:1+1D_P_sym_data}, the small irreps at high-symmetry momenta are labelled by the parity and have dimension $1$, resulting in $n^{l}_{\bsl{k},\alpha}=\tilde{n}^{l}_{\bsl{k},\alpha}$.
$n^{l}_{\bsl{k},\alpha}$ is invariant under the gauge transformation \eqnref{eq:gauge_trans_U_V}, since it is derived from the trace of symmetry representations.

For the given crystalline symmetry group $\G$, we do not need to include all momenta in 1BZ for the study of $n^{l}_{\bsl{k},\alpha}$.
To see this, we classify the momenta in 1BZ into a finite number of types based on the following definition.
Two momenta $\bsl{k}$ and $\bsl{k}'$ in 1BZ are defined to be of the same type iff there exists a symmetry $g\in\G$, a reciprocal lattice vector $\bsl{G}$, and a continuous path $\bsl{k}_s$ with $s\in[0,1]$ such that (i) $\bsl{k}_{s=0}=\bsl{k}_g+\bsl{G}$ and $\bsl{k}_{s=1}=\bsl{k}'$, and (ii) $\G_{\bsl{k}_{s=0}}=\G_{\bsl{k}_{s=1}}\subset \G_{\bsl{k}_s}$. 
There are two elementary cases: (i) $\bsl{G}=0$ and $g$ is identity, meaning that $\bsl{k}_{s=0}=\bsl{k}$ and $\bsl{k}_{s=1}=\bsl{k}'$, and (ii) $\bsl{k}_s=\bsl{k}'$ is constant in $s$, meaning that $\bsl{k}_g+\bsl{G}=\bsl{k}'$.
Note that the path $\bsl{k}_s$ is allowed to take values outside of 1BZ if needed.
Moreover, we allow $\G_{\bsl{k}_s}$ to be larger than $\G_{\bsl{k}_{s=0}}$ and $\G_{\bsl{k}_{s=1}}$ (\eg, even if $\G_{\bsl{k}_{s=0}}$ and $\G_{\bsl{k}_{s=1}}$ only contain lattice translations, the path is allowed to pass a mirror plane), though we typically do not need a larger $\G_{\bsl{k}_s}$.
According to the definition, being in the same type is an equivalence relation, and thereby the types of momenta are just the corresponding equivalence classes.
It turns out $n^{l}_{\bsl{k},\alpha}=n^{l}_{\bsl{k}',\alpha}$ as long as $\bsl{k}$ and $\bsl{k}'$ are of the same type, and thus we only need to consider one representative in each type of momenta.

Now turn to the symmetry data of the given FGU $U(\bsl{k},t)$ for the given PBZ choice $\Phi_{\bsl{k}}$.
The symmetry content for the $l$th isolated set of quasi-energy bands is the vector 
\eq{
\label{eq:sym_content_gen}
A_l = (..., n^{l}_{\alpha,\bsl{k}}, ...)^T\ ,
}
where $\bsl{k}$ and $\alpha$ respectively range over all types of Bloch momenta and all inequivalent small irreps of $\G_{\bsl{k}}$.
All components of $A_l$ are non-negative integers.
As exemplified by \eqnref{eq:1+1D_P_comp_rel_C}, not all components of $A_l$ are independent, as they satisfy the compatibility relation $\mathcal{C}$ 
\eq{
\label{eq:comp_rel_gen}
\mathcal{C} A_l = 0\ .
}
The compatibility relations of all crystalline symmetry groups (spatial dimensions up to three) can be found on the website of \emph{Bilbao Crystallographic Server}~\cite{Bradlyn2017TQC}.
Owing to the compatibility relation, we are allowed to omit certain types of momenta (especially those whose little groups are not maximal subgroups of $\G$) without affecting the results.
As a result, only a small number of high-symmetry momenta are included in general, like the $1+1$D example discussed in \appref{sec:1+1D_P_sym_data}; if $\G$ has no high-symmetry momenta, we only need to pick one generic momentum.
After picking the momentum types, the number of components of $A_l$ is fixed for the given $\G$, which we label as $K$.
Then, combined with \eqnref{eq:comp_rel_gen}, we have 
\eq{
%\label{eq:BS_set_gen}
A_l \in \{BS\}\equiv\mathbb{N}^{K}\cap \ker \mathcal{C}\ .
}
The symmetry data $A$ of $U(\bsl{k},t)$ for $\Phi_{\bsl{k}}$ is the $K\times L$ matrix with $A_l$ as its columns
\eq{
A= \mat{A_1 & A_2 & ... & A_L}\ ,
}
where $L$ is the total number of isolated sets in any PBZ.

The above discussion of symmetry data is for a fixed PBZ choice, which is the same as the discussion for static crystals~\cite{Bradlyn2017TQC,Po2017SymIndi}.
As discussed in \appref{sec:1+1D_P_sym_data}, the freedom of choosing PBZ for FGUs leads to a subtlety that is absent in static crystals.
Specifically, changing the artificial PBZ choice might change the symmetry data by a cyclic permutation.
Nevertheless, a given FGU can only have a finite number of different symmetry data given by varying PBZ choices, as discussed below.

Suppose $\widetilde{\Phi}_{\bsl{k}}$ is another possible PBZ lower bound of the given FGU $U(\bsl{k},t)$, which yields symmetry data $\widetilde{A}$. 
To relate $\widetilde{A}$ to the symmetry data $A$ given by $\Phi_{\bsl{k}}$, we can consider a continuous deformation $(1-s)\Phi_{\bsl{k}}+s\widetilde{\Phi}_{\bsl{k}}$ which connects $\Phi_{\bsl{k}}$ to $\widetilde{\Phi}_{\bsl{k}}$ as $s$ continuously evolves from $0$ to $1$.
We define $\widetilde{L}$ as the number of isolated sets of quasi-energy bands, as well as their $2\pi n$-copies (with $n$ integer), swept through by the deformation as $s$ continuously increases from $0$ to $1$.
When $\widetilde{L}\neq 0$, $\sgn{\widetilde{L}}=\sgn{\widetilde{\Phi}_{\bsl{k}}-\Phi_{\bsl{k}}}$.
For examples, $\widetilde{L}=0$ iff $\widetilde{\Phi}_{\bsl{k}}$ lies in the same relevant gap as $\Phi_{\bsl{k}}$, $\widetilde{L}=n L$ if $\widetilde{\Phi}_{\bsl{k}}=\Phi_{\bsl{k}}+n 2\pi$, and $\widetilde{L}=l-1$ ($l=2,...,L$) iff $\widetilde{\Phi}_{\bsl{k}}$ lies in the $l$th relevant gap in the PBZ defined by $\Phi_{\bsl{k}}$.
With this convention, we say $\widetilde{\Phi}_{\bsl{k}}$ is given by a $\widetilde{L}$-shift of $\Phi_{\bsl{k}}$, and then $2\pi n$-shifts are equivalent to $n L$-shifts.
For example, the PBZ lower bound in {\figOneDExp}(b) is given by a $1$-shift of that in {\figOneDExp}(a).
In general, a $\widetilde{L}_1$-shift followed by a $\widetilde{L}_2$-shift is always equivalent to a $(\widetilde{L}_1+\widetilde{L}_2)$-shift.

Suppose $\widetilde{\Phi}_{\bsl{k}}$ is given by a $\widetilde{L}$-shift of $\Phi_{\bsl{k}}$ with $0<\widetilde{L}<L$.
Then, $\widetilde{\Phi}_{\bsl{k}}$ lies in the $(\widetilde{L}+1)$th relevant gap of the $\Phi_{\bsl{k}}$-PBZ, and the first isolated set of quasi-energy bands in the $\widetilde{\Phi}_{\bsl{k}}$-PBZ would be the $(\widetilde{L}+1)$th isolated set of quasi-energy bands in the $\Phi_{\bsl{k}}$-PBZ.
As a result, the symmetry data $\widetilde{A}$ for $\widetilde{\Phi}_{\bsl{k}}$ should have $A_{\widetilde{L}+1}$ as its first column and reads 
\eq{
\widetilde{A}=\mat{A_{\widetilde{L}+1} & ... & A_{L} & A_{1} & ... & A_{\widetilde{L}} }\ .
}
Furthermore, adding $n L$ to $\widetilde{L}$ is equivalent to further shifting $\widetilde{\Phi}_{\bsl{k}}$ by $2\pi n$, which leaves the symmetry data invariant. 
Then, for general $\widetilde{L}\in \dsZ$, we have 
\eq{
\widetilde{A}=A P_{\widetilde{L}}\ ,
}
where $P_{\widetilde{L}}$ is a $L\times L$ cyclic permutation matrix taking the form
\eq{
\label{eq:cyclic_permutation_P_sym_data}
\left[ P_{\widetilde{L}} \right]_{ij} = \delta_{i, j+(\widetilde{L}\mod L)}+\delta_{i, j-L+(\widetilde{L}\mod L)}\ .
}
As shown by \eqnref{eq:cyclic_permutation_P_sym_data}, the symmetry data for $\widetilde{\Phi}_{\bsl{k}}$ is determined by $\widetilde{L}$ and the symmetry data for $\Phi_{\bsl{k}}$, without caring about the detailed forms of $\Phi_{\bsl{k}}$ and $\widetilde{\Phi}_{\bsl{k}}$.
It is because the symmetry data is invariant under any deformation of PBZ lower bound within one relevant gap.
\eqnref{eq:cyclic_permutation_P_sym_data} also suggests that 
\eq{
\label{eq:CP_commutation}
P_{\widetilde{L}_1}  P_{\widetilde{L}_2}= P_{\widetilde{L}_1+\widetilde{L}_2} =P_{\widetilde{L}_2} P_{\widetilde{L}_1}  \ ,
}
which coincides with the additive nature of PBZ-shifts.

\begin{figure}[t]
    \centering
    \includegraphics[width=0.6\columnwidth]{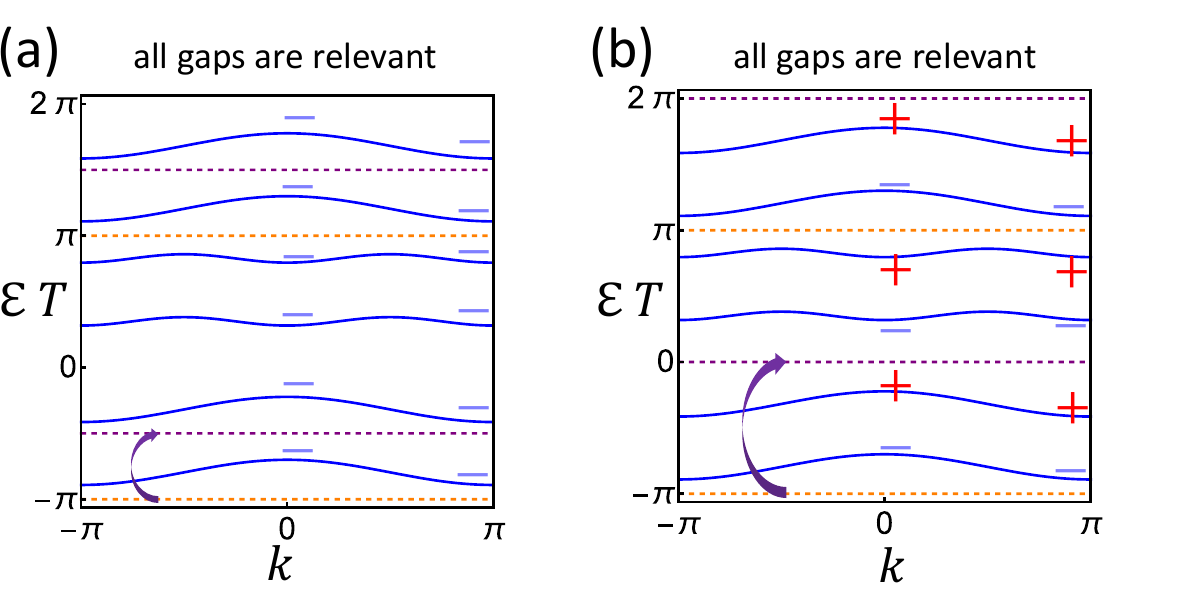}
    \caption{Two schematic quasi-energy band structures for $1+1$D inversion-invariant FGUs.
    In both (a) and (b), all quasi-energy gaps are relevant, and $\pm$ indicate the parities.
    The orange and purple dashed lines indicate two different choices of PBZ that yield the same symmetry data, and there are four isolated sets of quasi-energy bands (each set has one band) for each choice of PBZ.
    In (a), moving a PBZ lower-bound through $L_{KSD}=1$ isolated set (purple arrow) leaves the symmetry data invariant.
    In (b), moving a PBZ lower-bound through 2 isolated sets (purple arrow) leaves the symmetry data invariant while moving through 1 isolated set fails, resulting in $L_{KSD}=2$.
    }
    \label{fig:L_KSD}
\end{figure}

For the given FGU $U(\bsl{k},t)$, we focus on the smallest positive $\widetilde{L}$ that satisfies $A=A P_{\widetilde{L}}$, which we label as $L_{KSD}$ with ``KSD" short for keeping-symmetry-data.
An $\widetilde{L}$-shift of $\Phi_{\bsl{k}}$ leaves $A$ invariant iff $\widetilde{L}\mod L_{KSD} =0$, because if $A P_{(\widetilde{L} \mod L_{KSD})}=A$ holds for $0< \widetilde{L}\mod L_{KSD} < L_{KSD}$, $L_{KSD}$ cannot be smallest.
Thus, the number of different symmetry data given by changing PBZ is just $L_{KSD}$.
For examples, $L_{KSD}=1$ for \figref{fig:L_KSD}(a), $L_{KSD}=2$ for \figref{fig:L_KSD}(b), and $L_{KSD}=2$ for the 1+1D example discussed in \appref{sec:1+1D_P_sym_data}.
Although we derive $L_{KSD}$ from the symmetry data $A$ given by the PBZ lower bound $\Phi_{\bsl{k}}$, $L_{KSD}$ is in fact independent of PBZ choices, owing to the commutation relation of the cyclic permutations in \eqnref{eq:CP_commutation}.
It coincides with the fact that the number of different symmetry data possessed by a FGU should not rely on specific PBZ choices.

The finite $L_{KSD}$ allows us to define the equivalent symmetry data to remove the artificial ambiguity of the symmetry data brought by changing PBZ, as discussed in \appref{sec:1+1D_P_sym_data}.
We define two FGUs $U(\bsl{k},t)$ and $U'(\bsl{k},t)$ to have equivalent symmetry data iff there exist PBZ choices that yield exactly the same symmetry data for them.
Alternatively, we can define $[ A ]$ to be the set of all symmetry data of $U(\bsl{k},t)$ given by varying PBZ, similarly $[ A' ]$ for $U'(\bsl{k},t)$.
Then, having equivalent symmetry data is equivalent to $[ A ]=[ A' ]$.
Based on this equivalent statement, the equivalence among symmetry data of FGUs does not depend on specific PBZ choices, and is an equivalence relation.

As shown in \figref{fig:continuous_deformation}, given two topologically equivalent FGUs $U(\bsl{k},t)$ and $U'(\bsl{k},t)$, we can always pick a PBZ lower bound $\Phi_{\bsl{k},0}$ for $U(\bsl{k},t)$, and continuously deform $\Phi_{\bsl{k},0}$ into a PBZ lower bound $\Phi_{\bsl{k},1}$ for $U'(\bsl{k},t)$ without touching the deformed quasi-energy bands.
Since no relevant gaps are closed during the deformation, we know the symmetry data are exactly the same for $U(\bsl{k},t)$ with $\Phi_{\bsl{k},0}$ and $U'(\bsl{k},t)$ with $\Phi_{\bsl{k},1}$, meaning that topologically equivalent FGUs must have equivalent symmetry data.
As the contrapositive, inequivalent symmetry data infers topological distinction among FGUs (thus among Floquet crystals) and provides a topological classification that only involves the time-evolution operators at $t=T$.
For two FGUs with equivalent symmetry data, they must have the same number of bands and the same number of relevant gaps, but the dynamics can still make them topologically distinct.
Next, we will introduce the quotient winding data that can help classify the dynamics of FGUs (thus of Floquet crystals) with equivalent symmetry data.

\subsection{Winding Data}
\label{sec:winding_data}
To introduce quotient winding data, we first discuss the winding data for a generic FGU $U(\bsl{k},t)$ with time period $T$, a relevant gap choice, a generic crystalline symmetry group $\G$ and a symmetry representation $u_{g}(\bsl{k})$.
By picking a generic PBZ lower bound $\Phi_{\bsl{k}}$, \appref{sec:sym_data} suggests that we can derive the symmetry data $A$ of $U(\bsl{k},t)$.
As exemplified in \appref{sec:1+1D_P_winding_data}, the winding data is derived from the return map $U_{\epsilon=\Phi}(\bsl{k},t)$.
Since the return map is invariant under $\G$ as discussed in \appref{sec:return_map}, we have 
\eq{
\label{eq:little_group_commutes_RM_gen}
u_{g}(\bsl{k}) U_{\epsilon=\Phi}(\bsl{k},t) u_{g}^\dagger(\bsl{k}) = U_{\epsilon=\Phi}(\bsl{k},t)\ \forall g\in \G_{\bsl{k}}\ .
}
It enables us to simultaneously block diagonalize $U_{\epsilon=\Phi}(\bsl{k},t)$ and $u_{g}(\bsl{k})$ (for all $g\in \G_{\bsl{k}}$) according to inequivalent small irreps of $\G_{\bsl{k}}$:
\eqa{
\label{eq:RM_block_gen}
& W_{\G_{\bsl{k}}}^\dagger U_{\epsilon=\Phi}(\bsl{k},t) W_{\G_{\bsl{k}}} = 
\mat{
\ddots &  & \\
 &U_{{\epsilon=\Phi},\bsl{k},\alpha}(t)  & \\
 &  & \ddots 
}\\
& W_{\G_{\bsl{k}}}^\dagger u_g(\bsl{k}) W_{\G_{\bsl{k}}} = 
\mat{
\ddots &  & \\
 &\tilde{u}_g^\alpha(\bsl{k})  & \\
 &  & \ddots 
}\ ,
}
where $W_{\G_{\bsl{k}}}$ is a unitary matrix, and $U_{{\epsilon=\Phi},\bsl{k},\alpha}(t)$ and $\tilde{u}_g^\alpha(\bsl{k})$ are the blocks of the return map and the symmetry representation that correspond to the small irrep $\alpha$  of $\G_{\bsl{k}}$, respectively.
Specifically, $\tilde{u}_g^\alpha(\bsl{k})$ is a small representation of $\G_{\bsl{k}}$ that can be unitarily transformed to $\mathds{1}_{n_{\bsl{k},\alpha}}\otimes u^{\alpha}_{g}(\bsl{k})$ in a $g$-independent way, where 
$u^{\alpha}_{g}(\bsl{k})$ is the small irrep $\alpha$ of $\G_{\bsl{k}}$, $n_{\bsl{k},\alpha}=\sum_{l=1}^L n_{\bsl{k},\alpha}^l$ is the total number of copies of small irrep $\alpha$ that occur in $u_{g}(\bsl{k})$, and $L$ is the total number of isolated sets in any PBZ.
For the 1+1D example in \appref{sec:1+1D_P_winding_data}, $W_{\G_{\bsl{k}}}$ happens to be an identity matrix.
Based on \eqnref{eq:RM_block_gen}, we can define the following $U(1)$ winding number
\eq{
\label{eq:winding_number_gen}
\nu_{\bsl{k},\alpha}=\frac{\ii}{2\pi} \frac{1}{ d_{\alpha} }\int^T_0 dt \Tr[U_{{\epsilon=\Phi},\bsl{k},\alpha}^\dagger(t) \partial_t U_{{\epsilon=\Phi},\bsl{k},\alpha}(t) ]\ ,
}
where $d_{\alpha}$ was defined as the dimension of small irrep $\alpha$ in \appref{sec:sym_data}.
We emphasize that this expression of $\nu_{\bsl{k},\alpha}$ requires $U_{{\epsilon=\Phi},\bsl{k},\alpha}(t)$ to be a piece-wise differentiable function of $t$, while a more general definition of $\nu_{\bsl{k},\alpha}$ is the winding number of the continuous phase angle of $\det[U_{{\epsilon=\Phi},\bsl{k},\alpha}(t)]$ divided by $d_{\alpha}$.
The integer-valued $\nu_{\bsl{k},\alpha}$ is invariant under the gauge transformation \eqnref{eq:gauge_trans_U_V} and \eqnref{eq:gauge_trans_Uepsilon}, as discussed in \appref{app:return_map_winding_data}.

Interestingly, $\nu_{\bsl{k},\alpha}=\nu_{\bsl{k}',\alpha}$ if $\bsl{k}$ and $\bsl{k}'$ are in the same type, and $\nu_{\bsl{k},\alpha}$ has the same compatibility relation as the symmetry content $A_l$ in \eqnref{eq:comp_rel_gen}. 
(See more details in \appref{app:return_map_winding_data}.)
Therefore, we can pick the same momentum types as the symmetry data, and get the winding data of the given FGU $U(\bsl{k},t)$ for the given PBZ choice $\Phi_{\bsl{k}}$ as
\eq{
V= ( ..., \nu_{\bsl{k},\alpha} ,...)^T\ ,
}
where $\bsl{k}$ and $\alpha$ respectively range over all chosen types of Bloch momenta and all inequivalent small irreps of $\G_{\bsl{k}}$.
As a result, $V$ has the same number $K$ of components as the symmetry content $A_l$, and the compatibility relation reads
\eq{
\label{eq:comp_rel_V_gen}
\mathcal{C} V = 0\ .
}
However, unlike $A_l$, the components of $V$ are allowed to take negative values.

Besides the compatibility relation, there is a possible extra constraint on $V$ imposed by the symmetry data $A$, which is
\eq{
n_{\bsl{k},\alpha}=0 \Rightarrow \nu_{\bsl{k},\alpha}=0\ .
} 
Specifically, $n_{\bsl{k},\alpha}=0$ means that the block-diagonalized $u_g(\bsl{k})$ in \eqnref{eq:RM_block_gen} has no blocks for the small irrep $\alpha$ of $\G_{\bsl{k}}$, and thereby $U_{\epsilon=\Phi,\bsl{k},\alpha}(t)$ has zero dimension, resulting in $\nu_{\bsl{k},\alpha}=0$.
This extra constraint can be expressed in terms of a diagonal matrix $\mathcal{D}$ 
\eq{
%\label{eq:V_D_constraint}
\mathcal{D} V = 0\ ,
}
where a diagonal element of $\mathcal{D}$ is $0$ ($1$) if the corresponding $n_{\bsl{k},\alpha}$ is nonzero (zero).
Combining \eqnref{eq:comp_rel_V_gen} with \eqnref{eq:V_D_constraint}, the winding data takes value from the following group $\{ V \}$
\eq{
%\label{eq:V_set_gen}
\{ V \} \equiv \dsZ^K \cap \ker \mathcal{C} \cap \ker \mathcal{D}\ .
}
As $K$, $\mathcal{C}$, and $\mathcal{D}$ are independent of PBZ choices, so is $\{ V \}$.
Therefore, all winding data of all FGUs with equivalent symmetry data should belong to the same $\{ V \}$.

In general, for a given crystalline symmetry group $\G$, the largest winding data set $\{ V \}$ occurs when the FGUs of interest contain all inequivalent small irreps of $\G_{\bsl{k}}$ for all chosen momenta $\bsl{k}$.
In this case, the constraint \eqnref{eq:V_D_constraint} disappears, and the winding data set $\{ V \}$ becomes $\overline{\{ V \}}$ as
\eq{
\label{eq:Vbar_set_gen}
\overline{\{ V \}} \equiv \dsZ^K \cap \ker \mathcal{C}\ .
}
This is what happens for the 1+1D example {\figOneDExp}(a) discussed in \appref{sec:1+1D_P_winding_data}, where we do not need to consider the $\mathcal{D}$ constraint.
We further list $\overline{\{ V \}}$ for all spinless and spinful 2D plane groups in \tabref{tab:vbar_PG}, which can be reproduced by artificially allowing the negative numbers of band in the $\{ BS \}$ set~\cite{Kruthoff2017TCI,Po2017SymIndi} (\eqnref{eq:BS_set_gen}).

\begin{table*}[t]
    \centering
    \begin{tabular}{|c|ccccccccccccccccc|}
    \hline
         P.G. & p1 & p2 & pm & pg & cm & p2mm & p2mg & p2gg & c2mm & p4 & p4mm & p4gm & p3 & p3m1 & p31m& p6 & p6mm \\ 
         \hline
         $\overline{\{ V \}}$ for spinless P.G. & $\dsZ$ & $\dsZ^5$ & $\dsZ^3$ & $\dsZ$ & $\dsZ^2$ & $\dsZ^9$ & $\dsZ^4$ & $\dsZ^3$ & $\dsZ^6$ & $\dsZ^8$ & $\dsZ^9$ & $\dsZ^6$ & $\dsZ^7$ & $\dsZ^5$ & $\dsZ^5$ & $\dsZ^9$ & $\dsZ^8$ \\
         \hline
         $\overline{\{ V \}}$ for spinful P.G. & $\dsZ$ & $\dsZ^5$ & $\dsZ^3$ & $\dsZ$ & $\dsZ^2$ & $\dsZ$ & $\dsZ^4$ & $\dsZ^3$ & $\dsZ^2$ & $\dsZ^8$ & $\dsZ^3$ & $\dsZ^4$ & $\dsZ^7$ & $\dsZ^5$ & $\dsZ^5$ & $\dsZ^9$ & $\dsZ^4$ \\
         \hline
    \end{tabular}
    \caption{The largest winding data sets $\overline{\{ V \}}$ (\eqnref{eq:Vbar_set_gen}) for all spinless and spinful plane groups.
    ``P.G." means plane group.
    Note that $\overline{\{ V \}}$ does not provide a topological classification, while the quotient group $\overline{\{ V \}}/\bar{A}\dsZ$ does as shown in \eqnref{eq:Vbar_classification_gen}.
    }
    \label{tab:vbar_PG}
\end{table*}

Nevertheless, as mentioned in \appref{sec:1+1D_P_winding_data}, $\{ V \}$ does not give topological classifications since different the PBZ choices can result in different winding data.
Specifically for the given FGU $U(\bsl{k},t)$, a $\widetilde{L}$-shift of the PBZ lower bound $\Phi_{\bsl{k}}$ leads to
\eq{
\label{eq:V_PBZ_shift}
V \rightarrow V - \frac{\widetilde{L}-(\widetilde{L}\mod L)}{L}\sum_{l=1}^L A_{l}- \sum_{l=1}^{\widetilde{L}\mod L} A_{l}\ ,
}
where $A_l$ are labelled according to the original $\Phi_{\bsl{k}}$.
(See more details in \appref{app:return_map_winding_data}.)
Therefore, a FGU has infinitely many different winding data given by varying the PBZ, causing difficulty for comparing winding data to determine topological distinction.
Next, we introduce the quotient winding data to resolve this issue.

\subsection{Quotient Winding Data}
\label{sec:QWD_gen}

The quotient winding data is defined as the following.
Let us consider a FGU $U(\bsl{k},t)$, and by choosing a PBZ lower bound $\Phi_{\bsl{k}}$, we can derive symmetry data $A$ and winding data $V$ of $U(\bsl{k},t)$. 
As discussed in \appref{sec:sym_data}, the symmetry data is invariant and only invariant under $n L_{KSD}$-shifts of $\Phi_{\bsl{k}}$ (with $n$ integer).
Then, similar to the discussion in \appref{sec:1+1D_P_quotient_winding_data}, we want to make the quotient winding data $V_Q$ also invariant under all the $n L_{KSD}$-shifts.
According to \eqnref{eq:V_PBZ_shift}, we can achieve the invariance for $V_Q$ by defining
\eq{
\label{eq:QWD}
V_Q = V \mod \bar{A}\ ,
}
where
\eq{
\bar{A}=\sum_{l=1}^{L_{KSD}}A_l\ .
}
By exploiting \eqnref{eq:cyclic_permutation_P_sym_data}, one can show that $\bar{A}$ is independent of PBZ choices, and in fact $\bar{A}$ is the same for all FGUs with equivalent symmetry data.
The modulo operation in \eqnref{eq:QWD} can be taken for the first nonzero component of $\bar{A}$ as specified in \appref{sec:1+1D_P_quotient_winding_data}.
In contrast to the winding data, the given FGU $U(\bsl{k},t)$ only has $L_{KSD}$ different quotient winding data upon changing the PBZ, just like the symmetry data.
As discussed in \appref{sec:1+1D_P_quotient_winding_data}, we can then define an equivalence among quotient winding data of FGUs with equivalent symmetry data as the following.

For two FGUs with equivalent symmetry data, we define them to have equivalent quotient winding data iff they have the same quotient winding data for all PBZ choices that yield the same symmetry data.
We do not compare the quotient winding data when the PBZ choices yield different symmetry data, since the quotient winding data can be changed by any artificial PBZ shift that changes symmetry data.
In the following, we provide two other equivalent definitions for equivalent quotient winding data.
Given a FGU $U(\bsl{k},t)$, we can pick a PBZ lower bound $\Phi_{\bsl{k}}$ to get the symmetry and quotient winding data $(A,V_Q)$; similarly, for another FGU $U'(\bsl{k},t)$, we have $(A',V_Q')$ for a $\Phi_{\bsl{k}}'$.
Then, $U(\bsl{k},t)$ and $U'(\bsl{k},t)$ have equivalent symmetry and quotient winding data iff there exist $\Phi_{\bsl{k}}$ and $\Phi_{\bsl{k}}'$ such that $(A,V_Q)=(A',V_Q')$.
Moreover, by varying the PBZ lower bound $\Phi_{\bsl{k}}$ of $U(\bsl{k},t)$, we can get a set $[(A,V_Q)]$ of all symmetry and quotient winding data of $U(\bsl{k},t)$; similarly $[(A',V_Q')]$ for $U'(\bsl{k},t)$.
Then, $U(\bsl{k},t)$ and $U'(\bsl{k},t)$ have equivalent symmetry and quotient winding data iff $[(A,V_Q)]=[(A',V_Q')]$.
The equivalence among the three definitions relies on the correlated changes of the symmetry data \eqnref{eq:cyclic_permutation_P_sym_data} and winding data \eqnref{eq:V_PBZ_shift} given by the PBZ shifts.

The first two definitions provide an efficient way to determine equivalent quotient winding data.
Provided that $U(\bsl{k},t)$ and $U'(\bsl{k},t)$ have equivalent symmetry data and we have picked $\Phi_{\bsl{k}}$ and $\Phi_{\bsl{k}}'$ to yield $A=A'$, then the first definition suggests that $V_Q\neq V_Q'$ infers inequivalent quotient winding data, and the second definition suggests that $V_Q= V_Q'$ infers equivalent quotient winding data.
The third definition shows that having equivalent symmetry and quotient winding data is independent of the specific PBZ choices and is an equivalence relation.

As discussed in \appref{sec:1+1D_P_quotient_winding_data}, the equivalence of the quotient winding data should be related to the topological equivalence.
Suppose the above-mentioned $U(\bsl{k},t)$ and $U'(\bsl{k},t)$ are topologically equivalent.
Then, according to \figref{fig:continuous_deformation}, we have a continuously evolving in-gap $\Phi_{\bsl{k},s}$ with $\Phi_{\bsl{k},s=0}=\Phi_{\bsl{k}}$, and we can always pick $\Phi_{\bsl{k},s=1}$ as $\Phi_{\bsl{k}}'$.
With this choice, we would have $A=A'$ and the same winding data $V=V'$, resulting in $\bar{A}=\bar{A}'$ and $V_Q=V_Q'$.
Therefore, two topologically equivalent FGUs have equivalent symmetry and quotient winding data.
The contrapositive suggests if two FGUs have equivalent symmetry data but have inequivalent quotient winding data, they must be topologically distinct.

The quotient winding data does not lose any essential information compared to the winding data, because if two FGUs have equivalent symmetry and quotient winding data, there must exist PBZ choices for them to have the same symmetry and winding data.
To be more specific, when PBZ choices give the same symmetry and quotient winding data for two FGUs, the two FGUs must have the same $\bar{A}$, $L$ and $L_{KSD}$, always allowing us to compensate the difference in winding data by performing a $n L_{KSD}$-shift on the PBZ lower bound of one FGU without changing the symmetry data.
Nevertheless, the quotient winding data has the advantage of directly providing a topological classification for FGUs (and thus for Floquet crystals) with equivalent symmetry data, as discussed in the following.

Let us consider all FGUs that have symmetry data equivalent to a given FGU $U(\bsl{k},t)$, and we can always choose PBZs for them such that they have the same symmetry data.
As mentioned in \appref{sec:1+1D_P_quotient_winding_data} and \appref{sec:sym_data}, we still have $L_{KSD}$ different types of the PBZ choices, which respectively yield the $L_{KSD}$ different symmetry data of $U(\bsl{k},t)$ for all those FGUs.
For each type of PBZ choices, the quotient winding data of each FGU takes a unique value in the following set 
\eqa{
\label{eq:QWD_classification}
& \{ V_Q \}=\{ V\mod \bar{A} | V\in \{V\}\}\\
&\approx\frac{\{ V \} }{\bar{A}\dsZ}= \frac{\dsZ^K\cap \ker \C \cap \ker \mathcal{D}}{\bar{A}\dsZ}\ ,
}
where $\bar{A}\dsZ=\{q \bar{A}| q\in \dsZ\}$.
As different $V_Q$ in this case infers topological distinction, $\{ V_Q \}$ serves as a topological classification of those FGUs for each type of PBZ choices.
Since the winding data given by different PBZs are related (\eqnref{eq:V_PBZ_shift}), the quotient winding data for different types of PBZs are also related, suggesting that the $\{ V_Q \}$-based classifications for different types of PBZ choices are equivalent.
Specifically, for any two types of PBZ choices, two of those FGUs have the same quotient winding data for one type iff they have the same quotient winding data for the other type.
Therefore, $\{ V_Q \}$ provides a topological classification for FGUs with equivalent symmetry data, as long as the comparison of $V_Q$ is done for the PBZ choices that yield the same symmetry data.

The classification \eqnref{eq:QWD_classification} given by $\{ V_Q \}$ is fully determined by the symmetry group $\G$ and the PBZ-independent $\bar{A}$ of the FGUs with equivalent symmetry data.
To see the reason, recall that $\mathcal{D}$ is determined by whether the copy number $n_{\bsl{k},\alpha}$ of each small irrep $\alpha$ at each $\bsl{k}$ is zero, and then $ n_{\bsl{k},\alpha}=(L/L_{KSD}) \bar{A}_{\bsl{k},\alpha}$
suggests that $\mathcal{D}$ can be fully determined by $\bar{A}$.
\appref{sec:sym_data} further suggests that $K$ and $\mathcal{C}$ are determined by $\G$, resulting in the above statement.
It is worth mentioning that if the FGUs contain all inequivalent small irreps at all chosen momenta like {\figOneDExp}(a), we have $\mathcal{D}=0$, and the classification becomes
\eq{
\label{eq:Vbar_classification_gen}
\frac{\overline{\{ V \}} }{\bar{A}\dsZ}=\frac{\dsZ^K\cap \ker \C }{\bar{A}\dsZ}\ .
}

As discussed in \appref{sec:1+1D_P_quotient_winding_data}, the symmetry data and quotient winding data together provide a topological classification of FGUs (and thus of Floquet crystals).
However, the classification is not necessarily complete, \ie, if two FGUs have equivalent symmetry and quotient winding data, they can still be topologically distinct.
We do not resolve this completeness issue in this work as it is in general highly nontrivial.
On the other hand, \eqnref{eq:QWD_classification} only gives a relative classification without telling us which side is nontrivial.
Next, we will resolve this issue by constructing the DSI.

\subsection{DSI}
\label{sec:DSI_gen}

In this part, we will construct the DSI to sufficiently indicate the obstruction to static limits for a given FGU $U(\bsl{k},t)$ with $\G$ its crystalline symmetry group.

In order to determine the obstruction to static limits, we only need to consider $\G$-invariant static FGUs with symmetry data equivalent to $U(\bsl{k},t)$ since inequivalent symmetry data must infer topological distinction.
Then, based on the classification in \appref{sec:QWD_gen}, if all those $\G$-invariant static FGUs have quotient winding data inequivalent to $U(\bsl{k},t)$, then $U(\bsl{k},t)$ must have obstruction to static limits.
Specifically, we can pick a PBZ lower bound $\Phi_{\bsl{k}}$ for the FGU $U(\bsl{k},t)$ to get its symmetry data $A$, winding data $V$, and quotient winding data $V_Q$.
We further enumerate all winding data $V_{SL}$ and quotient winding data $V_{Q,SL}$ of all those static FGUs for all PBZ choices that yield symmetry data $A_{SL}=A$, resulting in a static winding data set $\{ V_{SL}\}$ and a static quotient winding data set $\{ V_{Q,SL} \}$.
Then, if $V_Q\notin \{ V_{Q,SL} \}$, we know $U(\bsl{k},t)$ has obstruction to static limits.

It turns out for the obstruction to static limits, we can use the winding data instead of the quotient winding data owing to
\eq{
\label{eq:V_Q_V_check}
V_Q\notin \{ V_{Q,SL} \} \Leftrightarrow V\notin \{ V_{SL} \}\ ,
}
which saves us from an extra modulo operation.
The reasoning is the following.
Since the static FGUs have symmetry data equivalent to $U(\bsl{k},t)$, the static FGUs have the same $\bar{A}$ as $U(\bsl{k},t)$.
If $V=V_{SL}$, we have $V_{Q}=V\mod \bar{A}=V_{SL}\mod \bar{A}=V_{Q,SL}$; if $V_Q=V_{Q,SL}$, the difference in the winding data can always be compensated by a PBZ shift for the static FGU without changing the symmetry data, as discussed in \appref{sec:QWD_gen}.
Therefore, a sufficient condition for $U(\bsl{k},t)$ to have the obstruction to static limits is $V\notin \{ V_{SL} \}$, which is the underlying idea for constructing DSI.
As discussed in \appref{sec:1+1D_P_DSI}, besides indicating obstruction to static limit, DSI is also a topological invariant---its different values infer topological distinction for FGUs (and thus for Floquet crystals) with equivalent symmetry data---though the resultant classification is a subset of that given by quotient winding data.
Although the idea of constructing DSI is the same as \appref{sec:1+1D_P_DSI}, there are some subtleties in the construction of $\{ V_{SL} \}$ and the derivation of DSI, which will be discussed below.

\subsubsection{Positive Affine Monoid and Hilbert Bases}

To derive $\{ V_{SL} \}$ for the given FGU $U(\bsl{k},t)$ with $\Phi_{\bsl{k}}$, let us first discuss several properties of the symmetry contents $A_l$ of isolated sets of quasi-energy bands.
As shown in \eqnref{eq:BS_set_gen}, the symmetry content compatible with the given crystalline symmetry group $\G$ always takes value from the set $\{ BS \}$.
We call a nonzero element in $\{ BS \}$ irreducible~\cite{Bruns2009Polytopes} if it cannot be expressed as the sum of any two other elements in $\{ BS \}$; otherwise, it is called reducible.
If an isolated set of quasi-energy bands has an irreducible symmetry content $A_l$, the quasi-energy bands in the isolated set must be connected, since if there is a gap that splits the isolated set into two isolated subsets of bands, the symmetry contents of the two subsets, labeled as $A_{l,1}$ and $A_{l,2}$, would satisfy $A_l=A_{l,1}+A_{l,2}$ and $A_{l,1}\neq A_l$ and $A_{l,2}\neq A_l$, violating $A_l$ being irreducible.
We further define the symmetry data $A$ of $U(\bsl{k},t)$ for $\Phi_{\bsl{k}}$ to be irreducible if all its columns are irreducible symmetry contents; otherwise, $A$ is reducible.

For the given $\G$, the irreducible symmetry contents form a unique set of bases of $\{ BS \}$~\cite{Song2020FragileAffineMonoid,Bruns2009Polytopes}.
Mathematically speaking, $\{ BS \}$ is a monoid because $\{ BS \}$ has an identity for the addition and the addition is closed and associative in $\{ BS \}$, while all symmetry contents have non-negative integer components and thus typically have no inverse.
More specifically, $\{ BS \}$ is a positive affine monoid~\cite{Song2020FragileAffineMonoid,Bruns2009Polytopes}, whose irreducible elements form a unique minimal set of bases of the monoid.
Here ``affine" means the monoid is a finitely generated submonoid of $\dsZ^K$, ``positive" means that only the zero element in the affine monoid has inverse, and ``bases" means that any element of the positive affine monoid can be expressed as the linear combination of the bases with non-negative integer coefficients.
The irreducible bases are called the Hilbert bases, and all irreducible symmetry contents are just the Hilbert bases of the $\{ BS \}$, labeled as $ a_i $ with $i=1,2,...,I$.
Here $I$ is the total number of distinct Hilbert bases.
For the 1+1D inversion-invariant case in \appref{sec:1+1D_P_sym_data}, there are four Hilbert bases given by four ways of assigning $\pm$ parities to $\Gamma/X$, which read
\eqa{
%\label{eq:1+1D_P_HBs}
a_1=(1,0,1,0)^T\ ,\\
a_2=(0,1,1,0)^T\ ,\\
a_3=(1,0,0,1)^T\ ,\\
a_4=(0,1,0,1)^T\ .
}
In general, we can obtain the Hilbert bases using the \emph{SageMath} software~\cite{SageMath}, together with the \emph{4ti2} package~\cite{4ti2} and the data~\cite{Bradlyn2017TQC} of atomic limits from \emph{Bilbao Crystallographic Server}.
We have obtained the Hilbert bases for all spinless and spinful 2D plane groups as mentioned in {\tabPlaneGSpinless}-2 and listed in \appref{app:HB}.
Next we will discuss how we use the Hilbert bases to construct the $\{V_{SL}\}$ and DSI.

\subsubsection{DSI for Irreducible Symmetry Data}
\label{sec:irreducible_sym_data}

We start with the case where the symmetry data $A$ of the given FGU $U(\bsl{k},t)$ for the PBZ choice $\Phi_{\bsl{k}}$ is irreducible, such as {\figOneDExp}(a).
Owing to the irreducible symmetry data, each isolated set of the quasi-energy bands is connected; thereby, there are no irrelevant gaps and all quasi-energy gaps are relevant.
In this case, the static winding data set $\{ V_{SL} \}$ in \eqnref{eq:V_Q_V_check} has the following expression
\eq{
\label{eq:V_SL_set}
\{ V_{SL} \} = \{ \sum_{l=1}^L A_l q_l | q_l\in \dsZ\}\ ,
}
where $A_l$ is the $l$th column of $A$ and $L$ is the number of isolated sets (or equivalently the number of relevant gaps) in any PBZ. (See \appref{app:DSI} for details.)

\eqnref{eq:V_SL_set} can be simplified.
For $A_l=A_{l'}$, $q_{l} A_{l} + q_{l'} A_{l'}=(q_{l}  + q_{l'})A_{l}$ means that only the sum $(q_{l}  + q_{l'})$ contributes to the winding data, allowing us to only include the distinct columns of $A$ for $\{ V_{SL} \} $.
Then, we can list all different $A_l$, relabeled as $a_{j}$ with $j$ taking $J$ different values in $\{ 1,2,...,I\}$, which are the Hilbert bases involved in the irreducible symmetry data of $U(\bsl{k},t)$.
Here $I$ is the total number of distinct Hilbert bases for $\G$.
As a result, \eqnref{eq:V_SL_set} is simplifed to
\eq{
%\label{eq:V_SL_set_sim}
\{ V_{SL} \} = \{ \sum_{j} a_j q_j | q_j\in \dsZ\}\ .
}
Combined with the winding data set $\{V\}$ in \eqnref{eq:V_set_gen}, the DSI takes values in the following quotient group
\eq{
\label{eq:DIS_IR_gen}
\X=\frac{\{V\} } { \{ V_{SL} \} }\ .
}
Strictly speaking, each element of $\X$ is a set of winding data; for the given winding data $V$ of $U(\bsl{k},t)$ for $\Phi_{\bsl{k}}$, we can find the $x$ in $\X$ such that $V\in x$, and then $x$ is the DSI of $U(\bsl{k},t)$.
We label $x$ in $\X$ as zero iff $x$ contains $0$.
Then, the zero DSI for $U(\bsl{k},t)$ means $V-0\in \{ V_{SL} \}$; nonzero DSI infers $V\notin \{ V_{SL} \}$ and thus infers the obstruction to static limits for the FGU (and thus for the underlying Floquet crystal).
Normally, we can use certain simple index to label $x$, just like the expression used for the 1+1D example in \eqnref{eq:1+1D_P_DSI}.

Recall that the symmetry data $A$ is derived after a PBZ lower bound $\Phi_{\bsl{k}}$ is picked for $U(\bsl{k},t)$.
Therefore, the above discussion is for a particular PBZ choice $\Phi_{\bsl{k}}$ for $U(\bsl{k},t)$.
If we change $\Phi_{\bsl{k}}$, $\{ V_{SL} \}$ stays invariant since any cyclic permutation of columns of $A$ leaves \eqnref{eq:V_SL_set} invariant.
In other words, all winding data of all $\G$-invariant static FGUs with symmetry data equivalent to the given FGU $U(\bsl{k},t)$ belong to the same $\{ V_{SL} \}$, even if the PBZ choices for static FGUs yield symmetry data $A_{SL}\neq A$.
Combined with the fact that $\{ V \}$ is PBZ-independent (\appref{sec:winding_data}), we know $\X$ is PBZ-independent.
The change of the winding data brought by changing $\Phi_{\bsl{k}}$ is a linear combination of symmetry contents (\eqnref{eq:V_PBZ_shift}), which is contained in $\{ V_{SL} \}$.
Therefore, the evaluation of DSI is independent of PBZ choice $\Phi_{\bsl{k}}$ for $U(\bsl{k},t)$.

\eqnref{eq:V_SL_set_sim} suggests that $\{ V_{SL} \}$ only depends on the set of Hilbert bases  $\{ a_{j} \}$ involved in the irreducible symmetry data.
On the other hand, as the vanishing components of $\sum_{l} A_l$ are the same as the vanishing components of $\sum_{j} a_{j}$, the $\mathcal{D}$ constraint in \eqnref{eq:V_set_gen} is also determined by the set $\{ a_{j} \}$.
Specifically, a diagonal element of $\mathcal{D}$ is $0$ ($1$) if the corresponding component of $\sum_{j} a_{j}$ is nonzero (zero).
Therefore, if two FGUs have the same $\G$ and have irreducible symmetry data that involve the same set of Hilbert bases, they have the same $\{ V_{SL} \}$, $\{ V \}$, and $\X$, no matter whether the two FGUs have equivalent symmetry data.
This simplification allows us to enumerate all possible DSI sets for irreducible symmetry data by considering all possible combinations of Hilbert bases of a given crystalline symmetry group $\G$.
All possible combinations of Hilbert bases can be enumerated by considering the presence and absence of each Hilbert basis, resulting in $2^I-1$ nontrivial combinations, where the only trivial one corresponds to the absence of all bases.
We emphasize that not all the combinations of Hilbert bases can be reproduced by FGUs since certain symmetry contents are forbidden for isolated sets of bands~\cite{Turner2010Inversion}.
Nevertheless, the above derivation can guarantee none of the physical combinations of Hilbert bases are missed.

We perform this derivation for the $1+1$D inversion-invariant class-A FGUs with irreducible symmetry data, and obtain only two nontrivial DSI sets.
One is for the case where the irreducible symmetry data is spanned by $a_1$ and $a_4$ in \eqnref{eq:1+1D_P_HBs}, which is just {\figOneDExp}(a) and the DSI is shown in \eqnref{eq:1+1D_P_DSI}.
The other one is for the irreducible symmetry data spanned by $a_2$ and $a_3$ in \eqnref{eq:1+1D_P_HBs}, and the DSI set reads
\eq{
%\label{eq:1+1D_P_DSI_theother}
\X\approx\{ \nu_{\Gamma,+}-\nu_{X,-}\in\dsZ \}\ . 
}
We further derive the DSI sets for all nontrivial combinations of Hilbert bases for all spinless and spinful 2D plane groups, and list the numbers of nontrivial DSI sets in {\tabPlaneGSpinless}-2.
We do not list the exact forms of DSIs since the paper would be too long otherwise, but we present in \appref{app:DSI} the detailed method that we adopt to obtain {\tabPlaneGSpinless}-2.
Rigorously speaking, the final results given by the method are not exactly equal to the DSI sets, but there are one-to-one correspondences (bijections) between them.

\subsubsection{DSI for Reducible Symmetry Data}

\label{sec:reducible_SD_DSI}

In this part, we discuss the DSI for the case where the symmetry data $A$ of the given FGU $U(\bsl{k},t)$ for $\Phi_{\bsl{k}}$ is reducible.
As we can see below, the DSI sets for irreducible symmetry data actually serve as elementary building blocks for the construction of DSIs for reducible symmetry data.

\begin{figure}[t]
    \centering
    \includegraphics[width=0.6\columnwidth]{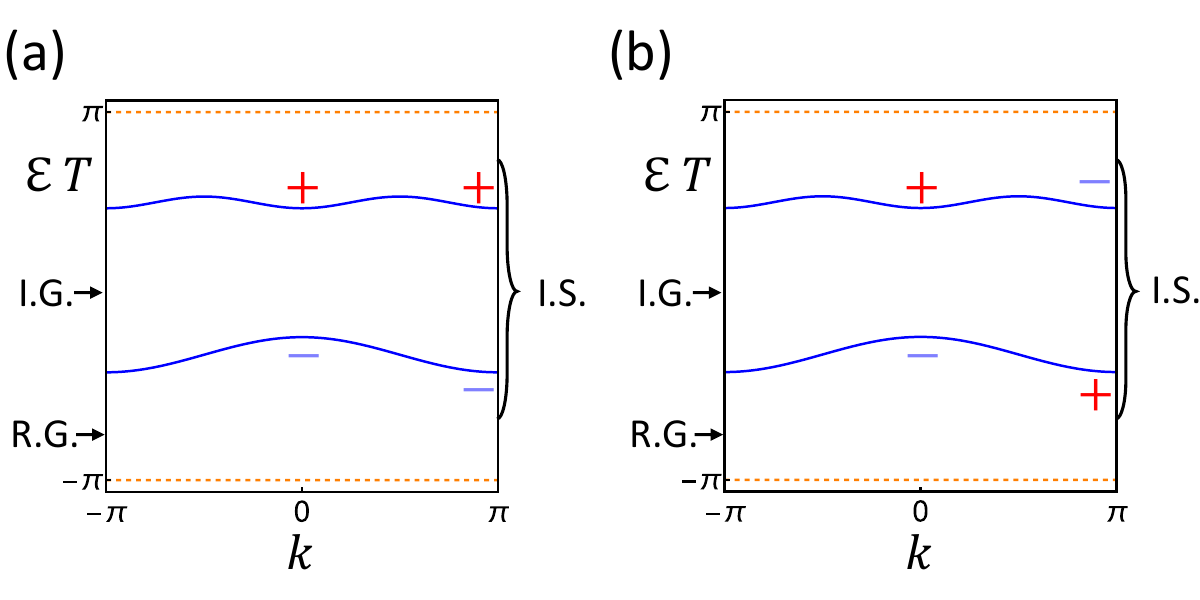}
    \caption{The schematic quasi-energy band structure (blue) for two $1+1$D inversion-invariant FGUs with the same reducible symmetry data.
    ``R.G.", ``I.G.", and ``I.S." stand for relevant gap, irrelevant gap, and isolated set, respectively.
    The dashed lines mark the boundary of the PBZ, and $\pm$ mark the parity.
    In either (a) or  (b), there is one isolated set of quasi-energy bands with symmetry content $A_1=(1,1,1,1)^T$, which is separated into two connected subsets by one irrelevant gap.
    The symmetry contents for the connected subsets are $A_{1,1}=(0,1,0,1)^T$ and $A_{1,2}=(1,0,1,0)^T$ for (a), and are $A_{1,1}=(0,1,1,0)^T$ and $A_{1,2}=(1,0,0,1)^T$ for (b).
    }
    \label{fig:1+1D_P_reducible_sym_data}
\end{figure}

Two examples of reducible symmetry data for 1+1D inversion-invariant case are shown in \figref{fig:1+1D_P_reducible_sym_data}.
Suppose the $l$th isolated set of quasi-energy bands of $U(\bsl{k},t)$ has reducible symmetry content $A_l$ and contains irrelevant gaps.
The irrelevant gaps separate the $l$th isolated set into isolated connected subsets, and we label the symmetry content of the $r_l$th connected subset as $A_{l,r_l}$, which satisfies $A_l=\sum_{r_l} A_{l,r_l}$.
As a result, the relevant gaps and irrelevant gaps together reduce the symmetry data $A$ into a finer matrix $(... A_{l,r_l}...)$, and we call $(... A_{l,r_l}...)$ a reduction of $A$.
We emphasize that the definition of a reduction $(... A_{l,r_l}...)$ of $A$ requires and only requires that $\sum_{r_l} A_{l,r_l} = A_l $ and $A_{l,r_l}\in \{ BS \}$ is nonzero, while we do not require that $(... A_{l,r_l}...)$ can be reproduced by a FGU.
$A$ is also a reduction of $A$, since $r_l$ is allowed to take only one value.
\figref{fig:1+1D_P_reducible_sym_data} provides two different reductions of the same reducible symmetry data.
Owing to the existence of the reduction of $A$, the winding data $V_{SL}$ of any $\G$-invariant static FGU with any PBZ yielding symmetry data $A$ takes value in the following set 
\eq{
\label{eq:V_SL_bar}
\overline{\{ V_{SL} \}} = \bigcup_{(... A_{l,r_l}...)\text{ reduces $A$}} \{ \sum_{l,r_l} A_{l,r_l} q_{l,r_l} | q_{l,r_l} \in \dsZ \}\ ,
}
and we have 
\eq{
\{ V_{SL}\} \subset \overline{\{ V_{SL} \}}\ .
}
(See \appref{app:DSI} for more details.)

Unlike \eqnref{eq:V_SL_set} where each $A_l$ is reproducible by an isolated set of bands since it appears in the given FGU, it is possible that not all reduction of $A$ can be reproduced by FGUs, and thus it is possible that $\overline{\{ V_{SL} \}}$ is strictly larger than $\{ V_{SL}\}$.
Nevertheless, for the winding data $V$ of the FGU $U(\bsl{k},t)$, $V\notin \overline{\{ V_{SL}\}}$ infers $V\notin \{ V_{SL}\}$ and thus sufficiently indicates the obstruction to static limits.
Since we only require DSIs to be sufficient indices for obstruction to static limits, we in this part use $\overline{\{ V_{SL} \}}$ instead of $\{ V_{SL} \}$.

$\overline{\{ V_{SL} \}}$ can be simplified by the irreducible inductions of $A$, \ie, each column of those reductions is irreducible.
For each irreducible induction of $A$, the story becomes similar to \appref{sec:irreducible_sym_data}.
We can use $\{ a_j \}$ to label the set of all distinct columns of an irreducible reduction $(... A_{l,r_l}...)$ of $A$, and then 
\eq{
\{ \sum_{l,r_l} A_{l,r_l} q_{l,r_l} | q_{l,r_l} \in \dsZ \} = \{ \sum_j q_j a_j | q_j \in \dsZ\}
}
similar to \eqnref{eq:V_SL_set_sim}.
The $\{ a_j\}$ is a set of Hilbert bases, and we say $\{ a_j \}$ spans $A$ since $\{ a_j\}$ consists of all distinct columns of an irreducible reduction $(... A_{l,r_l}...)$ of $A$.
Suppose $A_{i}=A_{j}+A_{j'}$, then $A_{i} \dsZ \subset \{q_1 A_{j} + q_2 A_{j'} | q_1,q_2\in \dsZ \}$, indicating that $\{ \sum_{l,r_l} A_{l,r_l} q_{l,r_l} | q_{l,r_l} \in \dsZ \}$ constructed from a reducible reduction must be a subset of that constructed from certain irreducible reduction.
Then, $\overline{\{ V_{SL} \}}$ can be simplified to
\eq{
\label{eq:V_SL_set_union_HB_gen}
\overline{\{ V_{SL} \}} = \bigcup_{ \{ a_{j} \}\text{ spans A}}  \{\sum_{j} q_{j} a_{j}| q_{j}\in \dsZ \}\ .
}
Similar to the discussion in \appref{sec:irreducible_sym_data}, $\overline{\{ V_{SL} \}}$ is actually independent of the PBZ choice $\Phi_{\bsl{k}}$ for $U(\bsl{k},t)$, and contains all winding data of all $\G$-invariant static FGUs with symmetry data equivalent to the given FGU $U(\bsl{k},t)$, regardless of the PBZ choices for static FGUs.

Let us take \figref{fig:1+1D_P_reducible_sym_data} as an example.
The reducible symmetry data shown in \figref{fig:1+1D_P_reducible_sym_data} has four irreducible reductions $(a_1\ a_4)$, $(a_4\ a_1),$ $(a_2\ a_3)$, and $(a_3\ a_2)$, where $a_i$ are shown in \eqnref{eq:1+1D_P_HBs}.
As a result, only two sets of Hilbert bases---$\{a_1,a_4\}$ and $\{a_2,a_3\}$---span the symmetry data, and $\overline{\{ V_{SL} \}}$ for \figref{fig:1+1D_P_reducible_sym_data} would just be
\eq{
\overline{\{ V_{SL} \}}= \overline{\{ V_{SL} \}}' \cup  \overline{\{ V_{SL} \}}''\ ,
}
where
\eqa{
\overline{\{ V_{SL} \}}'=\{q_1 a_1+q_2 a_4 | q_1,q_2\in \dsZ \}\ \\
\overline{\{ V_{SL} \}}''=\{q_1 a_2+q_2 a_3 | q_1,q_2\in \dsZ \}\ .
}
(See \appref{app:DSI} for a general method of determining Hilbert bases sets that span symmetry data.)

If all irreducible reductions of the reducible $A$ correspond to the same set of Hilbert bases (or equivalently only one set of Hilbert bases that spans $A$), then $\overline{\{ V_{SL} \}}$ is still a group for addition and we can calculate the DSI according to \eqnref{eq:DIS_IR_gen}.
If more than one sets of Hilbert bases are involved, it is very likely that $\overline{\{ V_{SL} \}}$ is not a group anymore.
In this case, we can define the DSI set for each set of Hilbert bases $\{ a_{j} \}$ that spans $A$ as
\eq{
\label{eq:DSI_irre_reduction}
\X[\{ a_{j} \}] = \frac{\{ V \}}{ \{\sum_{j} q_{j} a_{j}| q_{j}\in \dsZ \} }\ .
}
Since the DSI for one set of Hilbert bases has been addressed in \appref{sec:irreducible_sym_data}, we only need to find out all distinct sets that span $A$ to specify $\{ a_{j} \}$ for \eqnref{eq:DSI_irre_reduction}.
For example, we know $\{a_1,a_4\}$ and $\{a_2,a_3\}$ are the two sets of Hilbert bases that span the symmetry data for \figref{fig:1+1D_P_reducible_sym_data}, and the corresponding DSI sets $\X[\{a_1,a_4\}]$ and $\X[\{a_2,a_3\}]$ are given in \eqnref{eq:1+1D_P_DSI} and \eqnref{eq:1+1D_P_DSI_theother}, respectively.

We focus on the direct product of the DSI sets for all sets of Hilbert bases that span $A$, which reads
\eqa{
\label{eq:X_bar}
& \overline{\X}=\prod_{\{ a_{j} \} \text{ spans } A} \X[\{ a_{j} \}] \\
& = \X[\{ a_{j} \}]  \times \X[\{ a_{j'} \}] \times \X[\{ a_{j''}\}]\times ...\ .
}
It means that element $\bar{x}$ in $\overline{\X}$ is a vector, and each component of $\bar{x}$ is an element of \eqnref{eq:DSI_irre_reduction}.
Similar to the irreducible case \appref{sec:irreducible_sym_data}, $\overline{\X}$ and the evaluation of $\bar{x}$ are independent of the PBZ choice $\Phi_{\bsl{k}}$ for the given $U(\bsl{k},t)$.
Moreover, $\overline{\X}$ provides a topological classification for FGUs with equivalent symmetry data.

The given FGU (and thus its underlying Floquet crystal) must have obstruction to static limits if all components of its $\bar{x}$ are nonzero.
Then, we can define the DSI for the given FGU as the product of all components of its $\bar{x}$, meaning that nonzero DSI infers obstruction to static limits.
(If certain components of $\bar{x}$ are vectors, we can treat both numbers and vectors as matrices, and use the more general Kronecker product to define DSI.)
For \figref{fig:1+1D_P_reducible_sym_data}, the $\overline{\X}$ set is 
\eq{
\overline{\X}\approx\{  \nu_{ \Gamma,+}- \nu_{ X,+}\in \dsZ \}\times \{  \nu_{\Gamma,+}-\nu_{X,-} \in \dsZ \}\ ,
}
and then the DSI is $(\nu_{ \Gamma,+}- \nu_{ X,+})(\nu_{\Gamma,+}-\nu_{X,-} )$.
Then, a 1+1D inversion-invariant FGU with the symmetry data shown in \figref{fig:1+1D_P_reducible_sym_data} must have obstruction to static limits if $(\nu_{ \Gamma,+}- \nu_{ X,+})(\nu_{\Gamma,+}-\nu_{X,-}) \neq 0$.

%%%%%%%%%%%%%%%%%%%%%%%%%%%%%%%%%%%%%%%%%%%%%%%%%%%%%%%%%%%%
%%%%%%%%%%%%%%%%%%%%%%%%%%%%%%%%%%%%%%%%%%%%%%%%%%%%%%%%%%%%

\section{Details on DSI For A 2+1D Anomalous Floquet First-order Topological Insulator}

In this section, we construct a 2+1D model with plane group p2, which has chiral edge modes in the absence of nonzero Chern numbers.
Plane group p2 is spanned by a two-fold rotation $C_2$ and the 2D lattice translations.
Since spinless p2 is equivalent to spinful p2, we will focus on the spinless p2 in the following, \ie, $\G=$ spinless p2.
We will use the DSI to indicate its obstruction to static limits.

We consider a 2D square lattice with lattice constant being $1$, and each lattice site consists of one spinless $\s$ orbital and one spinless $\p$ orbital at the same position.
We use $\ket{\bsl{R},a}$ to label the Wannier bases, where $a=\s,\p$ and $\bsl{R}$ the 2D lattice vector, and then the Bloch bases are 
\eq{
\label{eq:Bases_2+1D_C2}
\ket{\psi_{\bsl{k},a}}=\frac{1}{\sqrt{\mathcal{N}}}\sum_{\bsl{R}} \ket{\bsl{R},a} e^{\ii \bsl{k}\cdot\bsl{R}}\ .
}
With $\ket{\psi_{\bsl{k}}}=(\ket{\psi_{\bsl{k},s}},\ket{\psi_{\bsl{k},p}})$, $C_2$ is represented as
\eq{
\label{eq:2+1D_C2_sym_rep}
u_{C_2}(\bsl{k})=\sigma_z\ ,
}
according the \eqnref{eq:sym_rep_g}.

The Floquet Hamiltonian that we choose has the form 
\eq{
\hat{H}(t)=\sum_{\bsl{k}} \ket{\psi_{\bsl{k}}} H(\bsl{k},t) \bra{\psi_{\bsl{k}}}
}
with
\eq{
\label{eq:FFOTI_h}
H(\bsl{k},t)= \left\{
\begin{array}{ll}
 \frac{1}{4}[1-2 \cos k_x - 2\cos k_y - 2 \cos(k_x+k_y)]\sigma_z &,\ 0\leq t < \frac{T}{2} \\
\frac{1}{2}[\sin(k_x) \sigma_x + \sin(k_y)\sigma_y] &,\ \frac{T}{2}\leq t < T
\end{array}
\right.\ .
}
The construction of the above model is inspired by the quantum-anomalous-Hall-effect model in \refcite{Qi2010TITSC}.
The time-evolution matrix $U(\bsl{k},t)$ can be derived from \eqnref{eq:FFOTI_h} based on \eqnref{eq:U_gen}.
For concreteness, we choose $T=2\pi$ in the following.

\begin{figure}[t]
    \centering
    \includegraphics[width=0.8\columnwidth]{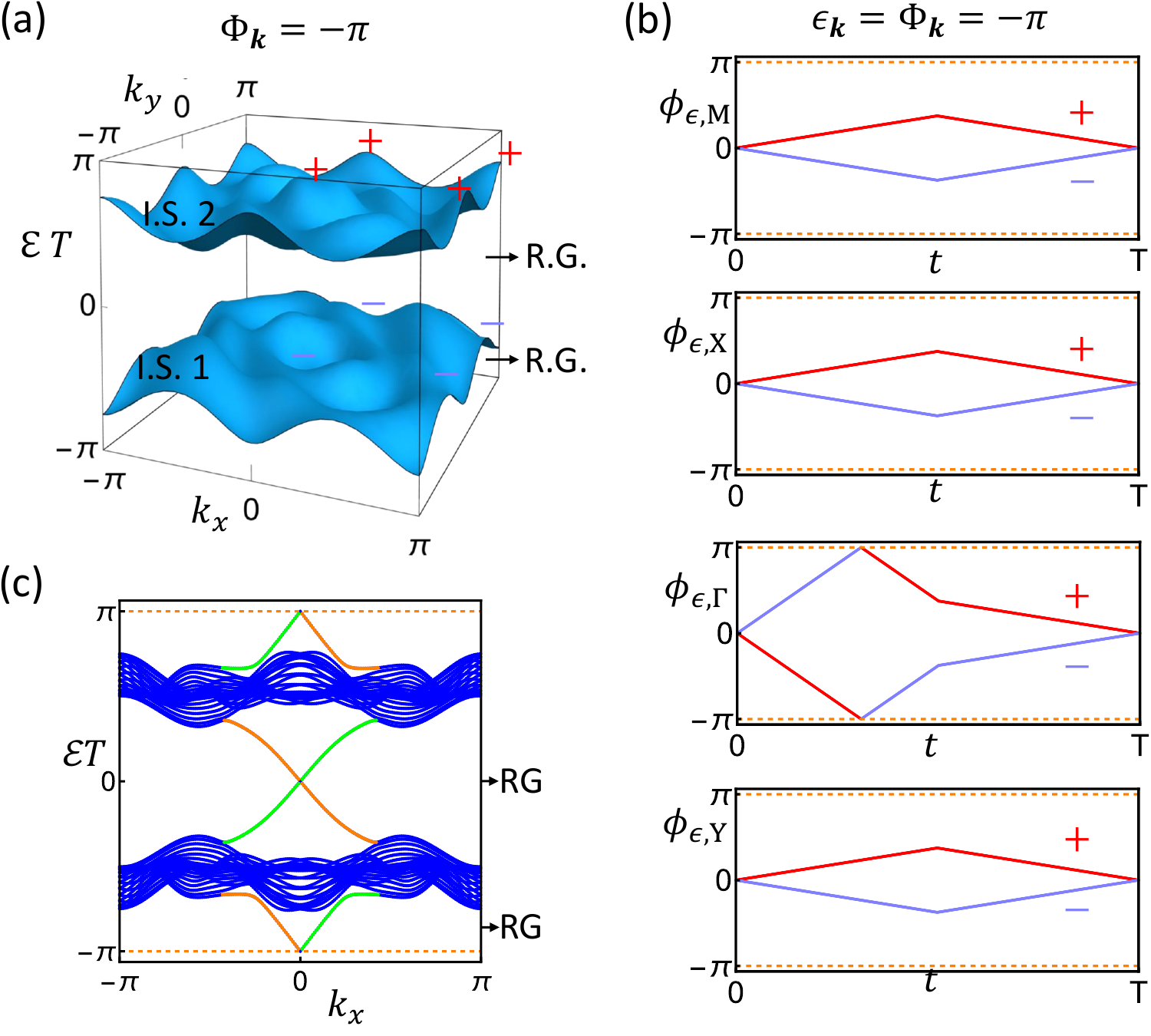}
    \caption{The symmetry data, winding data and boundary modes for the $2+1$D anomalous Floquet first-order topological insulator (\eqnref{eq:FFOTI_h}).
    ``R.G." and ``I.S." stand for relevant gap and isolated set, respectively.
    In (a), we plot the two quasi-energy bands in $[-\pi,\pi)$.
    Both quasi-energy gaps are chosen as relevant gaps, resulting in two isolated sets of quasi-energy bands, and $\Phi_{\bsl{k}}=-\pi$ is the PBZ lower bound.
    The $C_2$-parities for each band at $\Gamma$, $X$, $Y$ and $M$ are marked.
    In (b), we plot the phase bands of the return map at $\Gamma$, $X$, $Y$ and $M$ for $\epsilon_{\bsl{k}}=\Phi_{\bsl{k}}=-\pi$.
    The dashed lines label the boundary of the PBZ.
    In (c), we plot quasi-energy bands for open boundary condition along $y$ with $N_y=20$ layers along $y$.
    The orange (green) lines mark the chiral modes at $y=20$ ($y=1$).
    The dashed lines label the boundary of the PBZ.
    }
    \label{fig:2+1D_FFOTI}
\end{figure}

As shown in \figref{fig:2+1D_FFOTI}(a), the system has two quasi-energy bands, and we choose both quasi-energy gaps to be relevant.
Then, the two quasi-energy bands are separated into two isolated sets, of which each contains one band.
According to \emph{Bilbao Crystallographic Server}~\cite{Bradlyn2017TQC}, we only need to consider four $C_2$-invariant momenta for p2 in the study of the symmetry data, namely $\Gamma (0,0)$,  $X (\pi,0)$,  $Y (0,\pi)$, and $M (\pi,\pi)$.
Then, the symmetry content of each isolated set should have the form
\eq{
A_l=(n^l_{M,+},n^l_{M,-},n^l_{X,+},n^l_{X,-},n^l_{\Gamma,+},n^l_{\Gamma,-},n^l_{Y,+},n^l_{Y,-})^T\ ,
}
where $l=1,2$ labels the isolated sets of quasi-energy bands, and $n^l_{\bsl{k},\alpha}$ represents the number of parity-$\alpha$ states at $\bsl{k}$ in the $l$th set of quasi-energy bands.
According to \figref{fig:2+1D_FFOTI}(a), we have the symmetry data $A$ of $U(\bsl{k},t)$ for $\Phi_{\bsl{k}}=-\pi$ as
\eq{
\label{eq:2+1D_FFOTI_SD}
A=\mat{A_1 & A_2}\ ,
}
where 
\eqa{
& A_1=(0,1,0,1,0,1,0,1)^T \\
& A_2=(1,0,1,0,1,0,1,0)^T\ .
}
Then, according to the list of HBs in \appref{app:HB} for plane group p2, the symmetry data is irreducible. (When comparing to \appref{app:HB}, we should perform $X\rightarrow B$, $M\rightarrow A$, $+\rightarrow 1$ and $-\rightarrow 2$ on our convention to match the convention in \appref{app:HB}.)

As discussed in \appref{sec:winding_data}, the momenta and irreps for winding data $V$ are the same as those for the symmetry data, and we derive the winding data from the return map at those momenta.
Based on \eqnref{eq:winding_number_gen}, we have the winding data of $U(\bsl{k},t)$
\eq{
V=(\nu_{M,+},\nu_{M,-},\nu_{X,+},\nu_{X,-},\nu_{\Gamma,+},\nu_{\Gamma,-},\nu_{Y,+},\nu_{Y,-})^T\ ,
}
which can be intuitively read out from the winding of the phase bands of the return map for each irrep. 
Then, \figref{fig:2+1D_FFOTI}(b) suggests
\eq{
\label{eq:2+1D_FFOTI_WD}
V=(0,0,0,0,-1,1,0,0)^T
}
for $\Phi_{\bsl{k}}=-\pi$.
Indeed, direct calculation based on \eqnref{eq:winding_number_gen} also yields \eqnref{eq:2+1D_FFOTI_WD}.
With this preparation, we next derive the DSI.

According to \emph{Bilbao Crystallographic Server}, the compatibility relation for spinless p2 reads
\eq{
n^l_{\Gamma,+}+n^l_{\Gamma,-}=n^l_{X,+}+n^l_{X,-}=n^l_{Y,+}+n^l_{Y,-}=n^l_{M,+}+n^l_{M,-}\ ,
}
or equivalently the compatibility relation matrix $\mathcal{C}$ reads
\eq{
\mathcal{C}=\mat{
1 & 1 & -1 & -1 & 0 & 0 & 0 & 0\\
1 & 1 & 0 & 0 & -1 & -1 & 0 & 0\\
1 & 1 & 0 & 0 & 0 & 0 & -1 & -1 
}\ .
}
According to \eqnref{eq:2+1D_FFOTI_SD}, $U(\bsl{k},t)$ contains all inequivalent small irreps, and thereby the $\mathcal{D}$ matrix in \eqnref{eq:V_set_gen} is zero.
Then, all winding data of all $\G$-invariant FUGs with symmetry data equivalent to $U(\bsl{k},t)$ belong to the following set
\eql{
\{ V \}=\dsZ^8\cap \ker{\mathcal{C}} = \left\{\left.\mat{ A_2 & A_1 & \text{Vec}_3 & \text{Vec}_4 & \text{Vec}_5 }\mat{ \nu_{\Gamma,+} \\ \nu_{\Gamma,-} \\ \nu_{M,+}-\nu_{\Gamma,+} \\ \nu_{X,+}-\nu_{\Gamma,+} \\ \nu_{Y,+}-\nu_{\Gamma,+}  }\right| \nu_{\Gamma,+},\nu_{\Gamma,-},\nu_{X,+},\nu_{Y,+},\nu_{M,+}\in \dsZ\right\} \ ,
}
where $\text{Vec}_3=(1,-1,0,0,0,0,0,0)^T$, $\text{Vec}_4=(0,0,1,-1,0,0,0,0)^T$, and $\text{Vec}_5=(0,0,0,0,0,0,1,-1)^T$.
Since the symmetry data is irreducible, according to \eqnref{eq:V_SL_set}, all winding data of all $\G$-invariant static FGUs with symmetry data equivalent to $U(\bsl{k},t)$ belong to
\eq{
\{ V_{SL}\}=\{q_1 A_1 + q_2 A_2 | q_1,q_2\in \dsZ\} \ .
}
As a result, the DSIs for all $\G$-invariant FUGs with symmetry data equivalent to $U(\bsl{k},t)$ take values in 
\eq{
\X=\frac{\{ V \}}{\{ V_{SL}\}}\approx \{ (\nu_{M,+}-\nu_{\Gamma,+},\nu_{X,+}-\nu_{\Gamma,+} , \nu_{Y,+}-\nu_{\Gamma,+} )\in \dsZ^3\}\ ,
}
meaning that the DSI is $(\nu_{M,+}-\nu_{\Gamma,+}, \nu_{X,+}-\nu_{\Gamma,+} , \nu_{Y,+}-\nu_{\Gamma,+} )$.
In fact, this is one example for the $\dsZ^3$ DSI set of spinless p2 in {\tabPlaneGSpinless}. 
Then, according to \eqnref{eq:2+1D_FFOTI_WD}, we know the DSI of the FGU $U(\bsl{k},t)$ is $(1,1,1)\neq 0$, indicating the obstruction to static limits.

One signature of the obstruction to static limits is the anomalous edge modes shown in \figref{fig:2+1D_FFOTI}(c).
The edge modes are anomalous because both bulk bands have zero Chern numbers~\cite{TKNN}.
The $\pi_3$ winding number defined in \refcite{Rudner2013AFTI} is evaluated as $W=1$ for the model, where 
\eq{
\label{eq:W_pi_3}
W=\frac{1}{24\pi^2}\int_0^T dt  \int d k_x d k_y \epsilon^{i_1 i_2 i_3}\Tr[U_{\epsilon}^\dagger \partial_{i_1} U_{\epsilon} U_{\epsilon}^\dagger \partial_{i_2} U_{\epsilon} U_{\epsilon}^\dagger \partial_{i_3} U_{\epsilon}]
}
with $i_1,i_2,i_3\in\{ 0,1,2 \}$, $(\partial_0,\partial_1,\partial_2)=(\partial_t,\partial_{k_x},\partial_{k_y})$ and $\epsilon=\Phi=-\pi$, verifying the anomalous nature of the chiral modes.
Furthermore, we can see in this specific model, all components of the DSI take the same value of the $\pi_3$ winding number $W$, implying a relation between the DSI and $W$. 
Nevertheless, the evaluation of DSI is more efficient than that of $W$, since the former only cares about four $C_2$-invariant momenta while the latter needs the entire 1BZ.

%%%%%%%%%%%%%%%%%%%%%%%%%%%%%%%%%%%%%%%%%%%%%%%%%%%%%%%%%%%%
%%%%%%%%%%%%%%%%%%%%%%%%%%%%%%%%%%%%%%%%%%%%%%%%%%%%%%%%%%%%

\section{Details on DSI For A 2+1D Anomalous Floquet Higher-order Topological Insulator}
\label{sec:DSI_2+1D_FHOTI}

In this section, we derive the DSI for the 2+1D model proposed in \refcite{Huang2020FHOTI}, which has an anomalous Floquet higher-order topological insulator phase.
We will show that DSI is indeed nonzero in the anomalous phase, even if we only consider the crystalline symmetries in the model and neglect the internal symmetries like chiral symmetry.

The model in \refcite{Huang2020FHOTI} is a dynamical version of the static quadruple insulator model proposed in \refcite{Benalcazar2017HOTI}, which is constructed on a square lattice with four sublattices at each lattice site.
As a result, we have a $4\times 4$ matrix Hamiltonian~\cite{Huang2020FHOTI}
\eql{
\label{eq:FHOTI_h}
h(\bsl{k},t)=\left\{
\begin{array}{cc}
 \gamma (\tau_x\sigma_0-\tau_{y}\sigma_y)     & ,\ t\in [0, \frac{T}{4}]\\
    \lambda  \left[ \cos(k_x)\tau_x \sigma_0-\sin(k_x)\tau_y\sigma_z-\cos(k_y)\tau_y\sigma_y-\sin(k_y)\tau_y\sigma_x \right]& ,\ t\in (\frac{T}{4}, \frac{3 T}{4}] \\
     \gamma (\tau_x\sigma_0-\tau_{y}\sigma_y)     & ,\ t\in (\frac{3T}{4}, T) 
\end{array}
\right.\ ,
}
where $h(\bsl{k},t)=h(\bsl{k},t+T)$, $\tau$'s are also Pauli matrices, and the lattice constant is set to be 1.
The time-evolution matrix $U(\bsl{k},t)$ can be derived from \eqnref{eq:FHOTI_h} based on \eqnref{eq:U_gen}.

The model effectively has the spinful p4mm plane group as the crystalline symmetry group $\G$, which is spanned by a four-fold rotation $C_4$ along $z$, a mirror $m_y$ perpendicular to $y$, and lattice translations.
Specifically, $C_4$ and $m_y$ are represented as
\eqa {
\label{eq:FHOTI_sym_rep}
& u_{C_4}(\bsl{k})= \mat{ & \ii \sigma_y \\ \sigma_0 &} \\
& u_{m_y}(\bsl{k})= -\ii \tau_x\sigma_x \ .
}
The model also has other symmetries like the chiral symmetry, but we choose to omit them, meaning that we allow the continuous deformation of $U(\bsl{k},t)$ to break chiral symmetry, as well as other symmetries that are not in $\G$.
In this case, the model can be treated as a class-A system with a time-independent crystalline symmetry group $\G$.

\begin{figure}[t]
    \centering
    \includegraphics[width=0.8\columnwidth]{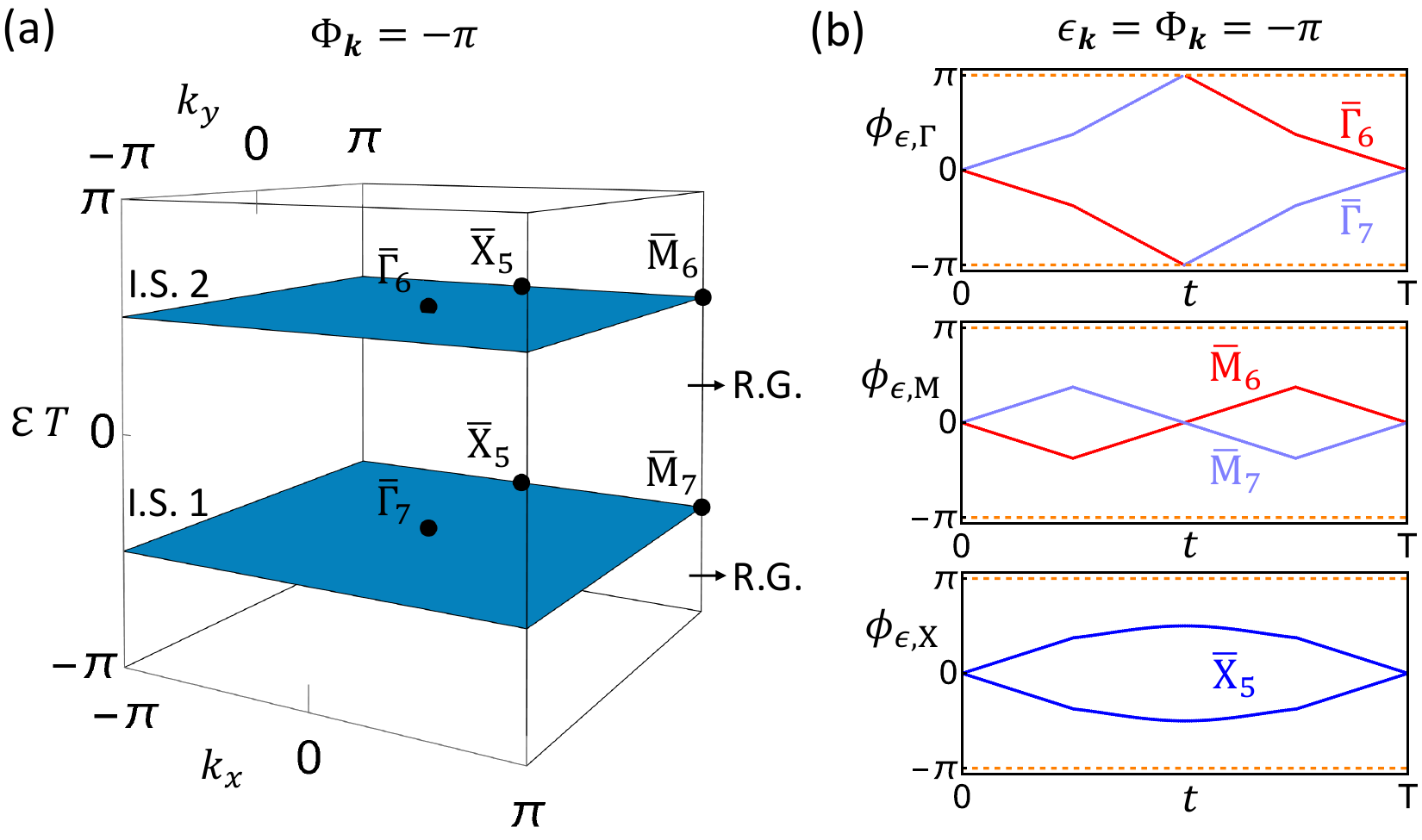}
    \caption{The symmetry data and winding data for the $2+1$D anomalous Floquet higher-order topological insulator (\eqnref{eq:FHOTI_h} with \eqnref{eq:2+1D_FHOTI_paraval}).
    ``R.G." and ``I.S." stand for relevant gap, and isolated set, respectively.
    In (a), we plot the two doubly-degenerate quasi-energy bands in $[-\pi,\pi)$.
    Both quasi-energy gaps are chosen as relevant gaps, resulting in two isolated sets of quasi-energy bands, and $\Phi_{\bsl{k}}=-\pi$ is the PBZ lower bound.
    The black dots label $\Gamma$, $M$, and $X$, and the irreps at the three momenta for each isolated set are marked.
    In (b), we plot the phase bands of the return map at $\Gamma$, $M$, and $X$ for $\epsilon_{\bsl{k}}=\Phi_{\bsl{k}}=-\pi$.
    Each phase band is doubly degenerate, and the corresponding irrep is marked.
    The dashed lines label the boundary of the PBZ.
    }
    \label{fig:2+1D_FHOTI}
\end{figure}

For concreteness, we choose 
\eq{
\label{eq:2+1D_FHOTI_paraval}
T=2,\ \sqrt{2}\gamma=\frac{\pi}{2},\ \sqrt{2}\lambda=\pi\ ,
}
for which the model is in the anomalous Floquet higher-order topological insulator phase according to \refcite{Huang2020FHOTI}.
We emphasize that the topological properties of $U(\bsl{k},t)$ determined with \eqnref{eq:2+1D_FHOTI_paraval} should hold for the entire phase, since other parameter values in the same phase should be topological equivalent to \eqnref{eq:2+1D_FHOTI_paraval}.
With \eqnref{eq:2+1D_FHOTI_paraval}, $U(\bsl{k},T)$ can be analytically diagonalized, and we get two doubly degenerate eigenvalues $\pm \ii$ at each $\bsl{k}$.
In {\figTwoDFHOTI}(a), we plot the two doubly degenerate flat quasi-energy bands of $U(\bsl{k},t)$ in $[-\pi,\pi)$, showing two quasi-energy gaps.
According to \refcite{Huang2020FHOTI}, both quasi-energy gaps are relevant, and then combined with time period $T$, $\G=$ spinful p4mm and the symmetry representations like \eqnref{eq:FHOTI_sym_rep}, we have a FGU $U(\bsl{k},t)$.
Furthermore, $\Phi_{\bsl{k}}=-\pi$ is a legitimate PBZ lower bound for $U(\bsl{k},t)$ since it lies in a relevant gap.
As shown in {\figTwoDFHOTI}(a), we have two isolated sets of quasi-energy bands, and each set consists of one doubly degenerate band.

According to \emph{Bilbao Crystallographic Server}~\cite{Bradlyn2017TQC}, we only need to consider three momenta for spinful p4mm in the study of the symmetry data, namely $\Gamma (0,0)$, $M (\pi,\pi)$ and $X (0,\pi)$, which are shown as black dots in {\figTwoDFHOTI}(a). 
Here picking $X$ as $(0,\pi)$ for spinful p4mm  is the convention used in \emph{Bilbao Crystallographic Server}, since $(0,\pi)$ and $(\pi,0)$ are equivalent owing to $C_4$.
At each of the three momenta, the little group only has two-dimensional small irreps.
Specifically, we have two small irreps $\overline{\Gamma}_6$ and $\overline{\Gamma}_7$ for $\G_{\Gamma}=\G$, two small irreps $\overline{M}_6$ and $\overline{M}_7$ for $\G_{M}=\G$, and one small irrep $\overline{X}_5$ for $\G_{X}=\ $spinful p2mm.
Moreover, trace of the representation of $C_4$ distinguishes  $\overline{i}_6$ ($\Tr(C_4)=-\sqrt{2}$) from $\overline{i}_7$ ($\Tr(C_4)=\sqrt{2}$), where $i=\Gamma, M$.
As a result, the symmetry content of each isolated set should be 
\eq{
A_l=(n^l_{\overline{\Gamma}_6},n^l_{\overline{\Gamma}_7},n^l_{\overline{M}_6},n^l_{\overline{M}_7}, n^l_{\overline{X}_5})^T\ ,
}
where $l=1,2$ labels the two isolated sets, recall that $n^l_{\bsl{k},\alpha}$ labels the copy number of the small irrep $\alpha$ at $\bsl{k}$ in the $l$th isolated set, and we do not need to separately label the momentum for each component of $A_l$ since the name of each irrep contains the label of the momentum.
According to {\figTwoDFHOTI}(a), we have the symmetry data $A$ of $U(\bsl{k},t)$ for $\Phi_{\bsl{k}}=-\pi$ as
\eq{
\label{eq:2+1D_FHOTI_SD}
A=\mat{A_1 & A_2}\ ,
}
where 
\eqa{
& A_1=(0,1,0,1,1)^T \\
& A_2=(1,0,1,0,1)^T\ .
}

As discussed in \appref{sec:winding_data}, the momenta for winding data $V$ are the same as those for the symmetry data, and we derive the winding data from the return map at those momenta.
Based on \eqnref{eq:winding_number_gen}, we have the winding data of $U(\bsl{k},t)$
\eq{
V=(\nu_{\overline{\Gamma}_6},\nu_{\overline{\Gamma}_7},\nu_{\overline{M}_6},\nu_{\overline{M}_7}, \nu_{\overline{X}_5})^T\ ,
}
which can be intuitively read out from the winding of the phase bands of the return map for each irrep. 
Then, {\figTwoDFHOTI}(b) suggests
\eq{
\label{eq:2+1D_FHOTI_WD}
V=(-1,1,0,0,0)^T
}
for $\Phi_{\bsl{k}}=-\pi$.
Indeed, direct calculation based on \eqnref{eq:winding_number_gen} also yields \eqnref{eq:2+1D_FHOTI_WD}.
With this preparation, we next derive the DSI.

According to \emph{Bilbao Crystallographic Server}, the compatibility relation for spinful p4mm reads
\eq{
n^l_{\overline{\Gamma}_6}+n^l_{\overline{\Gamma}_7}=n^l_{\overline{M}_6}+n^l_{\overline{M}_7}=n^l_{\overline{X}_5}\ ,
}
or equivalently the compatibility relation matrix $\mathcal{C}$ reads
\eq{
\mathcal{C}=\mat{
1 & 1 & -1 & -1 & 0\\
1 & 1 & 0 & 0 & -1
}\ .
}
According to \eqnref{eq:2+1D_FHOTI_SD}, $U(\bsl{k},t)$ contains all inequivalent small irreps, and thereby the $\mathcal{D}$ matrix in \eqnref{eq:V_set_gen} is zero.
Then, all winding data of all $\G$-invariant FUGs with symmetry data equivalent to $U(\bsl{k},t)$ belong to the following set
\eql{
\{ V \}=\dsZ^5\cap \ker{\mathcal{C}} = \{\nu_{\overline{\Gamma}_6} A_2 + \nu_{\overline{\Gamma}_7} A_1 + (\nu_{\overline{\Gamma}_6}-\nu_{\overline{M}_6}) (0,0,-1,1,0)^T| \nu_{\overline{\Gamma}_6},\nu_{\overline{\Gamma}_7},\nu_{\overline{\Gamma}_6}-\nu_{\overline{M}_6}\in \dsZ\} \ .
}
On the other hand, both columns of $A$ in \eqnref{eq:2+1D_FHOTI_SD} are Hilbert bases according to \appref{app:HB}, and thereby $A$ is irreducible.
Then, according to \eqnref{eq:V_SL_set}, all winding data of all $\G$-invariant static FGUs with symmetry data equivalent to $U(\bsl{k},t)$ belong to
\eq{
\{ V_{SL}\}=\{q_1 A_1 + q_2 A_2 | q_1,q_2\in \dsZ\} \ .
}
As a result, the DSIs for all $\G$-invariant FUGs with symmetry data equivalent to $U(\bsl{k},t)$ take values in 
\eq{
\X=\frac{\{ V \}}{\{ V_{SL}\}}\approx \{ \nu_{\overline{\Gamma}_6}-\nu_{\overline{M}_6} \in \dsZ\}\ ,
}
meaning that the DSI is $\nu_{\overline{\Gamma}_6}-\nu_{\overline{M}_6}$.
In fact, this is one example for the $\dsZ$ DSI set of spinful p4mm in {\tabPlaneGSpinful}. 
Then, according to \eqnref{eq:2+1D_FHOTI_WD}, we know the DSI of the FGU $U(\bsl{k},t)$ in the anomalous Floquet higher-order topological phase is $\nu_{\overline{\Gamma}_6}-\nu_{\overline{M}_6}=-1\neq 0$, indicating the obstruction to static limits.

The above analysis shows that the anomalous Floquet higher-order topological insulator phase in \refcite{Huang2020FHOTI} has obstruction to static limits as long as the spinful p4mm is preserved, regardless of the chiral symmetry.
In other words, although the chiral symmetry is needed to pin the corner modes in the quasi-energy spectrum, it is not essential for the ``inherently dynamical" nature of the phase.
Furthermore, to determine the obstruction, the DSI only requires three momenta in the 1BZ, saving us from evaluating the quantized dynamical quadrupole momoent proposed in \refcite{Huang2020FHOTI}, which involves all momenta in the entire 1BZ.

%%%%%%%%%%%%%%%%%%%%%%%%%%%%%%%%%%%%%%%%%%%%%%%%%%%%%%%%%%%%
%%%%%%%%%%%%%%%%%%%%%%%%%%%%%%%%%%%%%%%%%%%%%%%%%%%%%%%%%%%%

\section{Details on The 3+1D AFSOTI}

In this section, we construct a 3+1D model with space group P$\overline{1}$, which has chiral hinge modes in the absence of nonzero axion angles.
P$\overline{1}$ is space group $\#2$, and is spanned by the inversion $\P$ and the 3D lattice translations.
Since spinless P$\overline{1}$ is the same as spinful P$\overline{1}$, we will focus on the spinless P$\overline{1}$ in the following, \ie, $\G=$ spinless P$\overline{1}$.
We will use the DSI to indicate its obstruction to static limits.

We consider a 3D cubic lattice with lattice constant being $1$, and each lattice site consists of two spinless $\s$ orbitals and two spinless $\p$ orbitals at the same position.
We use $\ket{\bsl{R},a}$ to label the Wannier bases, where $a=\s_1,\s_2,\p_1,\p_2$, $\s_{1,2}$ labels the two $\s$ orbitals, $\p_{1,2}$ labels the two $\p$ orbitals, and $\bsl{R}$ is the 3D lattice vector.
Then, the Bloch bases are 
\eq{
\label{eq:Bases_3+1D_P}
\ket{\psi_{\bsl{k},a}}=\frac{1}{\sqrt{\mathcal{N}}}\sum_{\bsl{R}} \ket{\bsl{R},a} e^{\ii \bsl{k}\cdot\bsl{R}}\ .
}
With $\ket{\psi_{\bsl{k}}}=(\ket{\psi_{\bsl{k},\s_1}},\ket{\psi_{\bsl{k},\s_2}},\ket{\psi_{\bsl{k},\p_1}},\ket{\psi_{\bsl{k},\p_2}})$, the inversion $\P$ is represented as
\eq{
\label{eq:3+1D_P_sym_rep}
u_{\P}(\bsl{k})=\tau_z\sigma_0\ ,
}
based on the convention \eqnref{eq:sym_rep_g}.
Here $\tau's$ are also the Pauli matrices.

The Floquet Hamiltonian that we choose has the form 
\eq{
\hat{H}(t)=\sum_{\bsl{k}} \ket{\psi_{\bsl{k}}} H(\bsl{k},t) \bra{\psi_{\bsl{k}}}
}
with
\eq{
\label{eq:3+1D_FSOTI_h}
H(\bsl{k},t)= \left\{
\begin{array}{ll}
 \frac{1}{4}[2+\cos k_x + \cos k_y +  \cos(k_z)]\tau_z\sigma_0 +0.02(\tau_0\sigma_x+\tau_0\sigma_y+\tau_0\sigma_z) &,\ 0\leq t < \frac{T}{2} \\
\frac{1}{4}[\sin(k_x) \tau_y\sigma_x + \sin(k_y)\tau_y\sigma_y+\sin(k_z)\tau_y\sigma_z] +0.02(\tau_0\sigma_x+\tau_0\sigma_y+\tau_0\sigma_z) &,\ \frac{T}{2}\leq t < T
\end{array}
\right.\ .
}
The time-evolution matrix $U(\bsl{k},t)$ can be derived from \eqnref{eq:3+1D_FSOTI_h} based on \eqnref{eq:U_gen}.
For concreteness, we choose $T=2\pi$ in the following.

\begin{figure}[t]
    \centering
    \includegraphics[width=0.4\columnwidth]{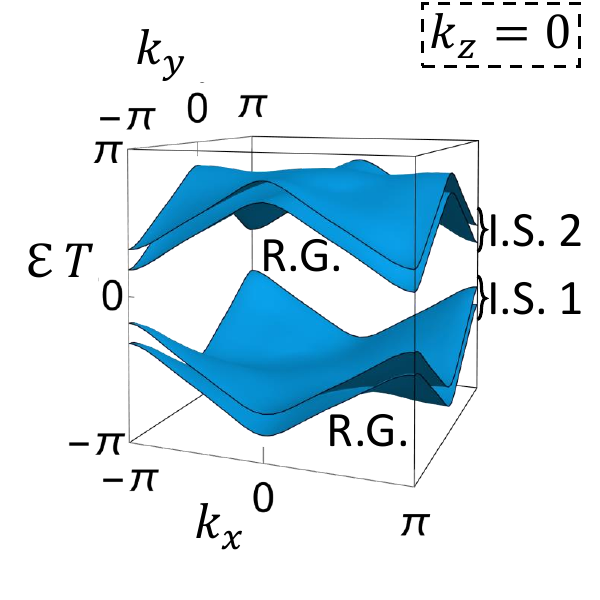}
    \caption{The bulk quasi-energy bands for the $3+1$D inversion-invariant  AFSOTI (\eqnref{eq:3+1D_FSOTI_h}) at $k_z=0$ in $[-\pi,\pi)$.
    ``R.G.", and ``I.S." stand for relevant gap and isolated set, respectively.
    Both quasi-energy gaps are chosen as relevant gaps, resulting in two isolated sets of quasi-energy bands; each isolated set consists of two quasi-energy bands.
    $\Phi_{\bsl{k}}=-\pi$ is the PBZ lower bound.
    Although we only plot the quasi-energy bands at $k_z=0$, the relevant gap choice, the separation of isolated sets, and the PBZ lower bound hold for the entire 1BZ.
    }
    \label{fig:3+1D_P}
\end{figure}

As shown in \figref{fig:3+1D_P}, the system has four quasi-energy bands, and two quasi-energy gaps at $0$ and $\pi$.
We choose both quasi-energy gaps to be relevant, and then, the four quasi-energy bands are separated into two isolated sets, of which each consists of two bands.
According to \emph{Bilbao Crystallographic Server}~\cite{Bradlyn2017TQC}, we only need to consider eight inversion-invariant momenta for P$\overline{1}$ in the study of the symmetry data, namely $\Gamma(0,0,0)$, $X(\pi,0,0)$, $Y(0,\pi,0)$, $Z(0,0,\pi)$, $V(\pi,\pi,0)$, $U(\pi,0,\pi)$, $T(0,\pi,\pi)$, and $R(\pi,\pi,\pi)$.
Then, the symmetry content of each isolated set should have the form
\eq{
A_l=(n^l_{\Gamma,+},n^l_{\Gamma,-},n^l_{X,+},n^l_{X,-},n^l_{Y,+},n^l_{Y,-},n^l_{Z,+},n^l_{Z,-},n^l_{V,+},n^l_{V,-},n^l_{U,+},n^l_{U,-},n^l_{T,+},n^l_{T,-},n^l_{R,+},n^l_{R,-})^T\ ,
}
where $l=1,2$ labels the isolated sets of quasi-energy bands, and $n^l_{\bsl{k},\alpha}$ represents the number of parity-$\alpha$ states at $\bsl{k}$ in the $l$th set of quasi-energy bands.
Direct calculation yields the symmetry data $A$ of $U(\bsl{k},t)$ for $\Phi_{\bsl{k}}=-\pi$ as
\eq{
\label{eq:3+1D_P_SD}
A=\mat{A_1 & A_2}\ ,
}
where 
\eqa{
& A_1=(2,0,0,2,0,2,0,2,0,2,0,2,0,2,2,0)^T \\
& A_2=(0,2,2,0,2,0,2,0,2,0,2,0,2,0,0,2)^T\ .
}

As discussed in \appref{sec:winding_data}, the momenta and irrep labels for winding data $V$ are the same as those for the symmetry data, and we derive the winding data from the return map at those momenta.
Based on \eqnref{eq:winding_number_gen}, we have the winding data of $U(\bsl{k},t)$
\eq{
\label{eq:3+1D_P_WD}
V=(2,-2,0,0,0,0,0,0,0,0,0,0,0,0,0,0)^T
}
for $\Phi_{\bsl{k}}=-\pi$.
With this preparation, we next derive the DSI.

The compatibility matrix reads
\eq{
\mathcal{C}
=
\left(
\begin{array}{cccccccccccccccc}
 1 & 1 & -1 & -1 & 0 & 0 & 0 & 0 & 0 & 0 & 0 & 0 & 0 & 0 & 0 & 0 \\
 1 & 1 & 0 & 0 & -1 & -1 & 0 & 0 & 0 & 0 & 0 & 0 & 0 & 0 & 0 & 0 \\
 1 & 1 & 0 & 0 & 0 & 0 & -1 & -1 & 0 & 0 & 0 & 0 & 0 & 0 & 0 & 0 \\
 1 & 1 & 0 & 0 & 0 & 0 & 0 & 0 & -1 & -1 & 0 & 0 & 0 & 0 & 0 & 0 \\
 1 & 1 & 0 & 0 & 0 & 0 & 0 & 0 & 0 & 0 & -1 & -1 & 0 & 0 & 0 & 0 \\
 1 & 1 & 0 & 0 & 0 & 0 & 0 & 0 & 0 & 0 & 0 & 0 & -1 & -1 & 0 & 0 \\
 1 & 1 & 0 & 0 & 0 & 0 & 0 & 0 & 0 & 0 & 0 & 0 & 0 & 0 & -1 & -1 \\
\end{array}
\right)
\ .
}
According to \eqnref{eq:3+1D_P_SD}, $U(\bsl{k},t)$ contains all inequivalent small irreps, and thereby the $\mathcal{D}$ matrix in \eqnref{eq:V_set_gen} is zero.
Then, all winding data of all $\G$-invariant FUGs with symmetry data equivalent to $U(\bsl{k},t)$ belong to the following set
\eql{
\{ V \}=\dsZ^{16}\cap \ker{\mathcal{C}} = \left\{\left. M \mat{ \nu_{\Gamma,+} \\ \nu_{\Gamma,-}  \\ \nu_{\Gamma,+}-\nu_{X,-} \\ \nu_{\Gamma,+}-\nu_{Y,-} \\ \nu_{\Gamma,+}-\nu_{Z,-}  \\ \nu_{\Gamma,+}-\nu_{V,-} \\ \nu_{\Gamma,+}-\nu_{U,-} \\ \nu_{\Gamma,+}-\nu_{T,-}  \\ \nu_{R,-}-\nu_{\Gamma,-} } \right| \nu_{\Gamma,+},\nu_{\Gamma,-},\nu_{X,-},\nu_{Y,-},\nu_{Z,-},\nu_{V,-},\nu_{U,-},\nu_{T,-},\nu_{R,-}\in \dsZ\right\} \ ,
}
where 
\eq{
M = \mat{ a_1 & a_2 & \text{Vec}_3 & \text{Vec}_4 & \text{Vec}_5 & \text{Vec}_6 & \text{Vec}_7 & \text{Vec}_8 & \text{Vec}_9 }
}
with 
\eqa{
&\text{Vec}_3=(0,0,1,-1,0,0,0,0,0,0,0,0,0,0,0,0)^T\\
&\text{Vec}_4=(0,0,0,0,1,-1,0,0,0,0,0,0,0,0,0,0)^T\\
&\text{Vec}_5=(0,0,0,0,0,0,1,-1,0,0,0,0,0,0,0,0)^T\\
&\text{Vec}_6=(0,0,0,0,0,0,0,0,1,-1,0,0,0,0,0,0)^T\\
&\text{Vec}_7=(0,0,0,0,0,0,0,0,0,0,1,-1,0,0,0,0)^T\\
&\text{Vec}_8=(0,0,0,0,0,0,0,0,0,0,0,0,1,-1,0,0)^T\\
&\text{Vec}_9=(0,0,0,0,0,0,0,0,0,0,0,0,0,0,-1,1)^T
\ .}

According to the compatibility matrix, P$\overline{1}$ has 256 Hilbert bases, which are given by the 256 permutations of the following vector
\eq{
(1,0,1,0,1,0,1,0,1,0,1,0,1,0,1,0) \ ,
}
where the 256 permutations are given by permuting any two elements with the same momentum.
Then, the symmetry data \eqnref{eq:3+1D_P_SD} is reducible, but it is can be spanned by only one set of Hilbert basis $\{ a_1, a_2 \}$, where
\eqa{
& a_1=(1,0,0,1,0,1,0,1,0,1,0,1,0,1,1,0)^T \\
& a_2=(0,1,1,0,1,0,1,0,1,0,1,0,1,0,0,1)^T\ .
}
According to \eqnref{eq:V_SL_set_union_HB_gen}, all winding data of all $\G$-invariant static FGUs with symmetry data equivalent to $U(\bsl{k},t)$ belong to
\eq{
\overline{\{ V_{SL}\}}=\{q_1 a_1 + q_2 a_2 | q_1,q_2\in \dsZ\} \ ,
}
where we have used the fact that the symmetry data is spanned by only one Hilbert basis set.
Combined with \eqnref{eq:X_bar}, we have  
\eqa{
\overline{\X}&=\frac{\{ V \}}{\{q_1 a_1 + q_2 a_2 | q_1,q_2\in \dsZ\}}\\
&\approx \{ (\nu_{\Gamma,+}-\nu_{X,-} , \nu_{\Gamma,+}-\nu_{Y,-} , \nu_{\Gamma,+}-\nu_{Z,-}  , \nu_{\Gamma,+}-\nu_{V,-} , \nu_{\Gamma,+}-\nu_{U,-} , \nu_{\Gamma,+}-\nu_{T,-}  , \nu_{R,-}-\nu_{\Gamma,-})\in \dsZ^7\}\ ,
}
meaning that the DSI is $(\nu_{\Gamma,+}-\nu_{X,-} , \nu_{\Gamma,+}-\nu_{Y,-} , \nu_{\Gamma,+}-\nu_{Z,-}  , \nu_{\Gamma,+}-\nu_{V,-} , \nu_{\Gamma,+}-\nu_{U,-} , \nu_{\Gamma,+}-\nu_{T,-}  , \nu_{R,-}-\nu_{\Gamma,-})$ since there is only one Hilbert basis set that spans $A$.
In fact, this is one example for the $\dsZ^7$ DSI set of P$\overline{1}$. 
Then, according to \eqnref{eq:3+1D_P_WD}, we know the DSI of the FGU $U(\bsl{k},t)$ is $(2,2,2,2,2,2,2)\neq 0$, indicating the obstruction to static limits.

One signature of the obstruction to static limits is the anomalous chiral hinge modes shown in \figThreeDFHOTI.
The hinge modes are anomalous because both bulk bands have zero axion angle $\theta\mod\ 2\pi$.
The axion angle for each isolated set can be derived from the symmetry data \eqnref{eq:3+1D_P_SD} according to the following expression~\cite{Qi2008TFT,Turner2010Inversion,Varnava2018AI}
\eq{
%\label{eq:nu_P}
\frac{\theta_l}{\pi}\mod 2 =\sum_{\bsl{K}} \frac{n_{\bsl{K},+}^l-n_{\bsl{K},-}^l}{4}\mod 2 = 0\ ,
}
where $\bsl{K}$ ranges over all eight inversion-invariant momenta, and $\theta^l$ is the axion angle of the $l$th isolated set.

At last of this section, we specify the parameters that we use to plot {\figThreeDFHOTI} of the main text.
In \figThreeDFHOTI(a-b), we choose $N_x=11$ and $N_y=11$, where $N_i$ is the number of lattice sites along $i$ direction with $i\in\{x,y,z\}$.
In \figThreeDFHOTI(a-b), the red color is marked when the mode has total probability at $(x,y)=(1,N_y),(1,N_y-1),(2,N_y),(2,N_y-1)$ larger than 1/2, and the purple color is marked when the mode has total probability at $(x,y)=(N_x,1),(N_x,2),(N_x-1,1),(N_x-1,2)$ larger than 1/2.
In \figThreeDFHOTI(c), we choose $N_x=N_y=N_z=11$.

%%%%%%%%%%%%%%%%%%%%%%%%%%%%%%%%%%%%%%%%%%%%%%%%%%%%%%%%%%%%
%%%%%%%%%%%%%%%%%%%%%%%%%%%%%%%%%%%%%%%%%%%%%%%%%%%%%%%%%%%%

\section{Nonzero Initial Time}
\label{app:initial_time}

In this section, we will show that setting the initial time to zero does not lose any generality for the study of topology.
We will focus on the FGUs, since a similar argument can be applied to Floquet crystals.

Consider a time-evolution matrix $U(\bsl{k},t)$ with zero initial time, time period $T$, a crystalline symmetry group $\G$, and a symmetry representation $u_g(\bsl{k})$.
It is not a FGU yet since we have not picked the relevant gaps.
Let us now shift the initial time to $t_0$, and the time-evolution matrix then reads
\eq{
\label{eq:U_initial_time}
U(\bsl{k},t+t_0,t_0)=\mathcal{T}e^{-\ii \int_{t_0}^{t_0+t} dt' H(\bsl{k},t') }\ ,
}
where $H(\bsl{k},t)$ is the underlying matrix Hamiltonian.
By defining $H_{t_0}(\bsl{k},t)=H(\bsl{k},t+t_0)$, we have an equivalent expression of \eqnref{eq:U_initial_time} as
\eq{
\label{eq:U_initial_time_alt}
U(\bsl{k},t+t_0,t_0)=U_{t_0}(\bsl{k},t)\equiv\mathcal{T}e^{-\ii \int_{0}^{t} dt' H_{t_0}(\bsl{k},t') }\ ,
}
and $U(\bsl{k},t)=U_{t_0=0}(\bsl{k},t)$.
Based on \eqnref{eq:U_initial_time_alt}, we can view $U_{t_0}(\bsl{k},t)$ as the time-evolution matrix of a new matrix Hamiltonian $H_{t_0}(\bsl{k},t)$ for zero initial time.
$U_{t_0}(\bsl{k},t)$ still has time period $T$ as $U_{t_0}(\bsl{k},t+T)=U_{t_0}(\bsl{k},t)U_{t_0}(\bsl{k},T)$, and has crystalline symmetry group $\G$ and symmetry representation $u_g(\bsl{k})$ owing to $u_g(\bsl{k}) U_{t_0}(\bsl{k},t) u_g^\dagger(\bsl{k})=U_{t_0}(\bsl{k}_g,t)$.

The quasi-energy bands given by $U_{t_0}(\bsl{k},T)$ are the same as those given by $U(\bsl{k},T)$.
To see this, first note that
\eq{
U(\bsl{k},T+t_0,t_0)= U(\bsl{k},t_0+T,T)U(\bsl{k},T,0)U(\bsl{k},0,t_0)\ .
}
Combined with $H(\bsl{k},t+T)=H(\bsl{k},t)$ and $U^\dagger(\bsl{k},t_0,t+t_0)=U(\bsl{k},t+t_0,t_0)$, we have 
\eq{
U(\bsl{k},T+t_0,t_0)= U(\bsl{k},t_0,0)U(\bsl{k},T,0)U^\dagger(\bsl{k},t_0,0)\ ,
}
resulting in
\eq{
\label{eq:U_initial_time_T}
U_{t_0}(\bsl{k},T)=U(\bsl{k},t_0) U(\bsl{k},T) [U(\bsl{k},t_0)]^\dagger\ .
}
Owing to the same quasi-energy bands, we can always choose the same relevant gaps for $U_{t_0}(\bsl{k},t)$ and $U(\bsl{k},t)$.
Therefore, we have two FGUs---one is $U(\bsl{k},t)$ (with T, a relevant gap choice, $\G$, $u_{g}(\bsl{k})$) and the other one is $U_{t_0}(\bsl{k},t)$ (with $T$, the relevant gap choice same as $U(\bsl{k},t)$, $\G$, $u_g(\bsl{k})$)---which are related by a shift of the initial time.

It turns out $U_{t_0}(\bsl{k},t)$ is topologically equivalent to $U(\bsl{k},t)$.
The deformation that establishes the topological equivalence is $U_{s}(\bsl{k},t)=U(\bsl{k},t+s t_0,s t_0)$, $T_s=T$, and $u_{s,g}(\bsl{k})=u_{g}(\bsl{k})$ with $s\in[0,1]$.
Since $s$ is just changing the initial time, $U_{s}(\bsl{k},t)$ is a continuous function of $(\bsl{k},t,s)\in\dsR^d\times \dsR\times [0,1]$, and the quasi-energy bands given by $U_{s}(\bsl{k},T_s)$ are the same as those of $U_{s=0}(\bsl{k},t)=U(\bsl{k},t)$ for all $s\in[0,1]$.
It means that all relevant gaps of $U(\bsl{k},t)$ will be kept open as $s$ continuously increases and eventually become the relevant gaps of $U_{s=1}(\bsl{k},t)=U_{t_0}(\bsl{k},t)$.
All other requirements of the continuous deformation for topological equivalence in \defref{def:topo_equi} can be straightforwardly checked.
Therefore, shifting the initial time of a FGU while keeping the relevant gap choice always results in an topologically equivalent FGU.
A similar argument can show that the same conclusion holds for Floquet crystals.
Then, for the study of topology of FGUs and Floquet crystals, we only need to consider $t_0=0$.

\section{Details on Return Map and Winding Data}
\label{app:return_map_winding_data}

In this section, we present more details on the return map and winding data.
Within this section, we allow the return map to have branch cut $\epsilon_{\bsl{k}}$ different from the PBZ lower bound.
We still require the continuous real $\epsilon_{\bsl{k}}$ (i) to lie either in a relevant gap in the PBZ or in a redundant $2\pi n$-copy of a relevant gap, (ii) to satisfy $\epsilon_{\bsl{k}+\bsl{G}}=\epsilon_{\bsl{k}}$ for all reciprocal lattice vectors $\bsl{G}$, and (iii) to satisfy $\epsilon_{\bsl{k}_g}=\epsilon_{\bsl{k}}$ for all $g\in \G$.
In other words, $\epsilon_{\bsl{k}}$ is required to satisfy the requirement for PBZ lower bounds.

\subsection{Return Map: Symmetry Properties and Change of Branch Cut}

Let us start with the return map of a given FGU $U(\bsl{k},t)$ with time period $T$, a relevant gap choice, a crystalline symmetry group $\G$, and a symmetry representation $u_g(\bsl{k})$.
After picking the PBZ lower bound $\Phi_{\bsl{k}}$, we can label the quasi-energy bands and their projection matrices as discussed in \appref{sec:basic_def}.
For the convenience of latter discussion, we relabel the quasi-energy bands and their corresponding projection matrices as $\E_{\bsl{k},l,m_l}$ and $P_{\bsl{k},l,m_l}(T)$, respectively, where $l=1,2,...,L$ labels the isolated sets of quasi-energy bands, $m_l= 1,2,...,n_l$ labels the quasi-energy bands in the $l$th isolated set, and $n_l$ is the total number of quasi-energy bands in the $l$th isolated set.
The relabelling is required to make sure
\eqa{
& \E_{\bsl{k},l+1,m_{l+1}}> \E_{\bsl{k},l,m_l}\\
& \E_{\bsl{k},l,m_{l}+1} \geq \E_{\bsl{k},l,m_l}\ .
}
As mentioned in \appref{sec:basic_def}, each quasi-energy band is a continuous function of $\bsl{k}\in\dsR^d$, is $\bsl{G}$-periodic ($\E_{\bsl{k}+\bsl{G},l,m_l}=\E_{\bsl{k},l,m_l}$), and is $\G$-symmetric ($\E_{\bsl{k}_g,l,m_l}=\E_{\bsl{k},l,m_l}$).

With the relabelling, the definition of $[U(\bsl{k},T)]_{\epsilon}^{-t/T}$ in \eqnref{eq:RM_def_gen} is re-expressed as
\eqa{
& [U(\bsl{k},T)]_{\epsilon}^{-t/T}\\
& =\sum_{l=1}^L\sum_{m_l=1}^{n_l}\exp\left[-\frac{t}{T} \log_{\epsilon_{\bsl{k}}}(e^{-\ii\E_{\bsl{k},l,m_l} T}) \right] P_{\bsl{k},l,m_l}(T)\ ,
}
where 
\eq{
\ii\log_{\epsilon_{\bsl{k}}}(e^{-\ii\E_{\bsl{k},l,m_l} T})=\E_{\bsl{k},l,m_l} T + 2\pi j_l \in [\epsilon_{\bsl{k}},\epsilon_{\bsl{k}}+2\pi)
}
and $j_l\in\dsZ$.
$j_l$ does not depend on $m_l$ or $\bsl{k}$ since $\epsilon_{\bsl{k}}$ lies in a relevant gap (or one of its redundant copies) and thus $\epsilon_{\bsl{k}}$ and $\E_{\bsl{k},l,m_l}$ are continuous.
Then, the return map defined in \eqnref{eq:RM_def_gen} becomes
\eq{
\label{eq:RM_def_gen_app}
U_\epsilon(\bsl{k},t)=U(\bsl{k},t) \sum_{l=1}^L\sum_{m_l=1}^{n_l} e^{\ii \E_{\bsl{k},l,m_l} t +\ii 2\pi j_l t/T} P_{\bsl{k},l,m_l}(T)\ .
}
Since $e^{\ii \E_{\bsl{k},l,m_l} t +\ii 2\pi j_l t/T}$ has the same degeneracy property as $e^{-\ii\E_{\bsl{k},l,m_l} T}$, $[U(\bsl{k},T)]_{\epsilon}^{-t/T}$ should have the same symmetry and continuity properties as $U(\bsl{k},T)$.
Therefore, $[U(\bsl{k}_g,T)]_{\epsilon}^{-t/T}$ is continuous in $\dsR^d\times\dsR$, is $\bsl{G}$-periodic, and satisfies
\eq{
u_g(\bsl{k}) [U(\bsl{k},T)]_{\epsilon}^{-t/T}u_g^\dagger(\bsl{k}) = [U(\bsl{k}_g,T)]_{\epsilon}^{-t/T}\ \forall g\in\G\ .
}
As a result, combined with \eqnref{eq:Umat_g_invariant_gen}, we know $U_\epsilon(\bsl{k},t)$ is continuous, is $\bsl{G}$-periodic, and satisfies
\eq{
\label{eq:RM_sym_app}
u_g(\bsl{k}) U_\epsilon(\bsl{k},t) u_g^\dagger(\bsl{k}) = U_\epsilon(\bsl{k}_g ,t)\ \forall g\in\G\ ,
}
which further yields \eqnref{eq:little_group_commutes_RM_gen} after choosing $\epsilon=\Phi$.

According to \eqnref{eq:RM_def_gen_app}, changing the branch cut can only change $j_l$.
Specifically, when the branch cut lies in the $l$th relevant gap in the PBZ, denoted by $\epsilon_l$, $j_{l'}=0$ for $l'\geq l$ and $j_{l'}=1$ for $l'<l$.
If shifting the branch cut by $\epsilon\rightarrow \epsilon- 2\pi q $ with $q$ integer, then $j_l\rightarrow j_l - q$ for all $l$.
As a result, we have
\eq{
\label{eq:RM_epsilon_shift}
U_{\epsilon_l-2\pi q}(\bsl{k},t) = U_{\epsilon=\Phi}(\bsl{k},t) e^{-\ii q 2\pi \frac{t}{T}}  \sum_{l'=1}^L e^{\ii 2\pi \theta(l-l') \frac{t}{T}} P_{\bsl{k},l'}(T)\ ,
}
where $\theta(x)=0$ for $x\leq 0$, $\theta(x)=1$ for $x> 0$, 
\eq{
 P_{\bsl{k},l}(T) = \sum_{m_l=1}^{n_l} P_{\bsl{k},l,m_l}(T)\ ,
}
and we use $U_{\epsilon_1}(\bsl{k},t)=U_{\epsilon=\Phi}(\bsl{k},t)$ since the PBZ lower bound $\Phi_{\bsl{k}}$ lies in the first relevant gap.

\subsection{Winding Data: Gauge Invariance and Change of Branch Cut}

In order to show the effect of changing the branch cut, we will focus on the $\epsilon$-dependent winding vector in the following.
First, similar to \eqnref{eq:RM_block_gen}, \eqnref{eq:RM_sym_app} suggests that we can block diagonalize $U_{\epsilon}(\bsl{k},t)$ and $u_g(\bsl{k})$ simultaneously by a unitary $W_{\G_{\bsl{k}}}$ as 
\eqa{
\label{eq:RM_block_gen_app}
& W_{\G_{\bsl{k}}}^\dagger U_{\epsilon}(\bsl{k},t) W_{\G_{\bsl{k}}} = 
\mat{
\ddots &  & \\
 &U_{{\epsilon},\bsl{k},\alpha}(t)  & \\
 &  & \ddots 
}\\
& W_{\G_{\bsl{k}}}^\dagger u_g(\bsl{k}) W_{\G_{\bsl{k}}} = 
\mat{
\ddots &  & \\
 &\tilde{u}_g^\alpha(\bsl{k})  & \\
 &  & \ddots 
}\ ,
}
where $U_{{\epsilon},\bsl{k},\alpha}(t)$ and $\tilde{u}_g^\alpha(\bsl{k})$ are the blocks of the return map and the symmetry representation that correspond to the small irrep $\alpha$  of $\G_{\bsl{k}}$, respectively.
Recall that $\tilde{u}_g^\alpha(\bsl{k})$ is a small representation of $\G_{\bsl{k}}$ that can be unitarily transformed to $\mathds{1}_{n_{\bsl{k},\alpha}}\otimes u^{\alpha}_{g}(\bsl{k})$, where 
$u^{\alpha}_{g}(\bsl{k})$ is the small irrep $\alpha$ of $\G_{\bsl{k}}$, and $n_{\bsl{k},\alpha}=\sum_{l=1}^L n_{\bsl{k},\alpha}^l$ is the total number of copies of small irrep $\alpha$ that occur in $u_{g}(\bsl{k})$.
Then, we define the following $\epsilon$-dependent $U(1)$ winding number
\eq{
\label{eq:winding_app}
\nu_{\epsilon,\bsl{k},\alpha}=\frac{\ii}{2\pi} \frac{1}{ d_{\alpha} }\int^T_0 dt \Tr[U_{{\epsilon},\bsl{k},\alpha}^\dagger(t) \partial_t U_{{\epsilon},\bsl{k},\alpha}(t) ]\ ,
}
and the $\epsilon$-dependent winding vector
\eq{
V_\epsilon= ( ..., \nu_{\epsilon,\bsl{k},\alpha} ,...)^T
}
with $\bsl{k}$ and $\alpha$ respectively ranging over all chosen types of momenta and all inequivalent small irreps of $\G_{\bsl{k}}$.
The choice of momenta for the winding vector is based on the fact that $\nu_{\epsilon,\bsl{k},\alpha}$ obeys all compatibility relations for symmetry contents, which will be elaborated in the last part of this section.
$V_\epsilon$ becomes the winding data if the PBZ lower bound $\Phi$ is chosen as the branch cut $\epsilon=\Phi$.

$V_\epsilon$ is gauge invariant.
The $U(N)$ gauge transformation of $U_{\epsilon}(\bsl{k},t)$ is $U_{\epsilon}(\bsl{k},t)\rightarrow W(\bsl{k})^\dagger U_{\epsilon}(\bsl{k},t) W(\bsl{k})$, which can be canceled by simultaneously performing $W(\bsl{k})\rightarrow W_{\G_{\bsl{k}}}^\dagger W(\bsl{k})$ according to \eqnref{eq:RM_block_gen_app}.
However, the $U(N)$ gauge freedom or the choice of bases is not the only redundancy that we need to consider. 
After fixing the $U(N)$ gauge or the choice of bases, the choice of $W_{\G_{\bsl{k}}}$ in \eqnref{eq:RM_block_gen_app} allows a new gauge freedom
\eq{
W_{\G_{\bsl{k}}}\rightarrow W_{\G_{\bsl{k}}}
\mat{
\ddots &  & \\
 & W_{\G_{\bsl{k}},\alpha}  & \\
 &  & \ddots 
}\ ,
}
where $W_{\G_{\bsl{k}},\alpha}$ is a unitary matrix.
Under this gauge transformation, we have 
\eq{
U_{{\epsilon},\bsl{k},\alpha}(t) \rightarrow   W_{\G_{\bsl{k}},\alpha}^\dagger U_{{\epsilon},\bsl{k},\alpha}(t) W_{\G_{\bsl{k}},\alpha}\ ,
}
which leaves $V_\epsilon$ invariant according to \eqnref{eq:winding_app}.
Therefore, $V_\epsilon$ is gauge invariant, and so does the winding data $V_{\epsilon=\Phi}$.

Now we show the components of $V_{\epsilon}$ must be integers.
Owing to the gauge invariance of $V_{\epsilon}$, we can always choose $W_{\G_{\bsl{k}}}$ such that $\tilde{u}_g^\alpha(\bsl{k})=\mathds{1}_{n_{\bsl{k},\alpha}}\otimes u_g^\alpha(\bsl{k})$.
Then according to Schur's Lemma~\cite{Tung1985GroupTheory}, $U_{{\epsilon},\bsl{k},\alpha}(t)$ in \eqnref{eq:RM_block_gen_app} has the form
\eq{
U_{{\epsilon},\bsl{k},\alpha}(t) = \widetilde{U}_{{\epsilon},\bsl{k},\alpha}(t)\otimes \mathds{1}_{d_{\alpha}}\ ,
}
where $\widetilde{U}_{{\epsilon},\bsl{k},\alpha}(t)$ is a $n_{\bsl{k},\alpha}\times n_{\bsl{k},\alpha}$ matrix, and $d_\alpha$ is the dimension of $u_g^\alpha(\bsl{k})$.
Substituting the above equation into \eqnref{eq:winding_app}, we arrive at 
\eq{
\label{eq:winding_app_sim}
\nu_{\epsilon,\bsl{k},\alpha}=\frac{\ii}{2\pi} \int^T_0 dt \Tr[\widetilde{U}_{{\epsilon},\bsl{k},\alpha}^\dagger(t) \partial_t \widetilde{U}_{{\epsilon},\bsl{k},\alpha}(t) ]\ ,
}
which must be an integer since it represents the winding number of the continuous phase angle of $\det[\widetilde{U}_{{\epsilon},\bsl{k},\alpha}(t)]$ over one time period.
Therefore, the components of $V_\epsilon$, as well as the winding data $V_{\epsilon=\Phi}$, must be integers.

At the end of this part, we show how $V_{\epsilon}$ changes upon changing the branch cut $\epsilon$.
Combining \eqnref{eq:RM_epsilon_shift} with \eqnref{eq:RM_block_gen_app}, we have
\eq{
 U_{\epsilon_l-2\pi q,\bsl{k},\alpha}(t)= U_{\epsilon=\Phi,\bsl{k},\alpha}(t) e^{-\ii q 2\pi t / T } U_{l,\alpha,\bsl{k}}(t)\ ,
}
where $U_{l,\alpha,\bsl{k}}(t)$ is given by
\eq{
\label{eq:P_alpha_block_app}
W_{\G_{\bsl{k}}}^\dagger \sum_{l'=1}^L e^{\ii 2\pi \theta(l-l') \frac{t}{T}} P_{\bsl{k},l'}(T) W_{\G_{\bsl{k}}} = 
\mat{
\ddots &  & \\
 & U_{l,\alpha,\bsl{k}}(t)  & \\
 &  & \ddots 
}\ .
}
Combined with \eqnref{eq:winding_app}, we get
\eqa{
\label{eq:n_epsilon_shift_intermediate_app}
& \nu_{\epsilon_l-2\pi q,\bsl{k},\alpha}=\nu_{\bsl{k},\alpha}+ q\ n_{\bsl{k},\alpha} \\
& + \frac{\ii}{2\pi} \frac{1}{ d_{\alpha} }\int^T_0 dt \Tr[U_{l,\alpha,\bsl{k}}^\dagger(t) \partial_t U_{l,\alpha,\bsl{k}}(t) ]\ .
}
As $\nu_{\epsilon_l-2\pi q,\bsl{k},\alpha}$ is gauge invariant, we can choose $W_{\G_{\bsl{k}}}$ such that each of its columns not only corresponds to certain small irrep of $\G_{\bsl{k}}$ but also belongs to a definite isolated set of quasi-energy bands at $\bsl{k}$, labeled as $Y_{l,m_{l,\alpha}}^\alpha$ with $m_{l,\alpha}=1,...,n^{l}_{\bsl{k},\alpha} d_{\alpha}$.
Then, 
\eq{
P_{\bsl{k},l'}^\alpha(T) Y_{l,m_{l,\alpha}}^\alpha=\delta_{l'l} Y_{l,m_{l,\alpha}}^\alpha\ .
}
We can collect all columns of $W_{\G_{\bsl{k}}}$ belonging to $\alpha$ irrep to form a matrix
\eq{
W_{\G_{\bsl{k}}}^\alpha=\mat{... & Y_{l,m_{l,\alpha}}^\alpha &...}\ ,
}
where $...$ ranges over $l,m_{l,\alpha}$.
As a result, we have
\eqa{
\label{eq:n_epsilon_shift_intermediate_2_app}
& U_{l,\alpha,\bsl{k}}(t)=[W_{\G_{\bsl{k}}}^\alpha]^\dagger \sum_{l'=1}^L e^{\ii 2\pi \theta(l-l') \frac{t}{T}} P_{\bsl{k},l'}(T) W_{\G_{\bsl{k}}}^\alpha \\
& =
\mat{
\ddots &  & \\
 & e^{\ii 2\pi t \theta(l-l')/T } \mathds{1}_{n^{l}_{\bsl{k},\alpha} d_{\alpha}} & \\
 &  & \ddots_{l'} 
}\ .
}
Combined with \eqnref{eq:n_epsilon_shift_intermediate_app}, we arrive at 
\eq{
\nu_{\epsilon_l-2\pi q,\bsl{k},\alpha}=\nu_{\bsl{k},\alpha} + q\ n_{\bsl{k},\alpha} - \sum_{l'=1}^{l-1} n^{l'}_{\bsl{k},\alpha}\ ,
}
and thereby
\eq{
V_{\epsilon_l-2\pi q}= V +q \sum_{l'=1}^{L} A_{l'} -  \sum_{l'=1}^{l-1} A_{l'}\ .
}
The above expression suggests that we do not need to choose branch cut for the winding data different from the PBZ lower bound since they are related by the symmetry data.

If we choose the PBZ lower bound $\Phi_{\bsl{k}}$ as the branch cut, a new PBZ lower bound $\widetilde{\Phi}_{\bsl{k}}$ given by a $\widetilde{L}$-shift of $\Phi_{\bsl{k}}$ would be equivalent to $\epsilon_{\tilde{l}+1}+2\pi q$ with $\widetilde{l}=\widetilde{L}\mod L$ and $q= (\widetilde{L}- \widetilde{l})/L$.
(Recall that $L$ is the number of isolated sets of quasi-energy bands in one PBZ.)
Then, the new winding data would be $V_{\epsilon_{\tilde{l}+1}+2\pi q}$, resulting in \eqnref{eq:V_PBZ_shift}.

\subsection{Compatibility Relation of Winding Numbers}

At the end of this section, we demonstrate that $\nu_{\epsilon,\bsl{k},\alpha}$ obeys all compatibility relations for symmetry contents.
Again, we demonstrate it for tunable branch cut $\epsilon$.
In the above discussion, when we talk about the small irreps of $\G_{\bsl{k}}$, we always imply those small irreps are furnished by bases at $\bsl{k}$.
However, in the remaining of this section, we sometimes need to consider the small irreps of $\G_{\bsl{k}}$ furnished by bases at another $\bsl{k}'$.
Then, we need complicate our notation to emphasize the bases: we use $\alpha_{\bsl{k}}$ instead of $\alpha$ to label inequivalent small irreps of $\G_{\bsl{k}}$ furnished by bases at $\bsl{k}$, unless specified otherwise.

\subsubsection{Same Winding Numbers for Momenta of Same Type}

We start with showing that the winding number $\nu_{\epsilon,\bsl{k},\alpha_{\bsl{k}}}$ is the same for two momenta of the same type.
Recall the definition of two momenta being in the same type discussed in \appref{sec:sym_data}:  two momenta $\bsl{k}$ and $\bsl{k}'$ in 1BZ are defined to be of the same type iff there exists a symmetry $h\in\G$, a reciprocal lattice vector $\bsl{G}$, and a continuous path $\bsl{k}_s$ with $s\in[0,1]$ such that (i) $\bsl{k}_{s=0}=\bsl{k}_h+\bsl{G}$ and $\bsl{k}_{s=1}=\bsl{k}'$, and (ii) $\G_{\bsl{k}_{s=0}}=\G_{\bsl{k}_{s=1}}\subset \G_{\bsl{k}_s}$ for all $s\in[0,1]$. 
Note that we do not need to confine $\bsl{k}_s$ in 1BZ.
Based on the definition, we split the derivation into two steps below.

First, we show the winding number is the same for $\bsl{k}_{s=0}$ and $\bsl{k}_{s=1}$ (in short denoted by $\bsl{k}_0$ and $\bsl{k}_1$ below, respectively) in the definition.
Since $\G_{\bsl{k}_0}\subset \G_{\bsl{k}_s}$, $\bsl{k}_s$ is invariant under $\G_{\bsl{k}_0}$, and thus $u_{g}(\bsl{k}_s)$ satisfies
\eq{
u_{g_1}(\bsl{k}_s)u_{g_2}(\bsl{k}_s)=u_{g_1 g_2}(\bsl{k}_s)\ \forall\ g_1,g_2\in\G_{\bsl{k}_0}\ . 
}
Therefore, $u_{g}(\bsl{k}_s)$ is a small representation of $\G_{\bsl{k}_0}$ furnished by bases at $\bsl{k}_s$ instead of $\bsl{k}_0$.
Recall that we use $\alpha_{\bsl{k}_0}$ to label the small irreps of $\G_{\bsl{k}_0}$ at $\bsl{k}_0$.
Owing to the continuous path, we are allowed to use the $\alpha_{\bsl{k}_0}$ to label the small irreps of $\G_{\bsl{k}_0}$ at $\bsl{k}_s$~\cite{Bradley2009MathSSP}.
Such a correspondence is enabled by tracking the small irreps continuously along the path (or more mathematically based on the underlying projective representations of $\G_{\bsl{k}_0}/\mathds{T}$ with $\mathds{T}$ the lattice translation group).
In this case, we can use a unitary $W_{\G_{\bsl{k}_0}}(\bsl{k}_s)$ to block diagonalize the return map and symmetry representation at $\bsl{k}_s$ according to the inequivalent small irreps of $\G_{\bsl{k}_0}$ at $\bsl{k}_s$ as 
\eqa{
\label{eq:RM_block_kpath_app}
& \left[W_{\G_{\bsl{k}_0}}(\bsl{k}_s)\right]^\dagger U_{\epsilon}(\bsl{k}_s,t) W_{\G_{\bsl{k}_0}}(\bsl{k}_s) = 
\mat{
\ddots &  & \\
 &U_{{\epsilon},\bsl{k}_s,\alpha_{\bsl{k}_0}}(t)  & \\
 &  & \ddots 
}\\
& \left[W_{\G_{\bsl{k}_0}}(\bsl{k}_s)\right]^\dagger u_{g}(\bsl{k}_s) W_{\G_{\bsl{k}_0}}(\bsl{k}_s) = 
\mat{
\ddots &  & \\
 &\widetilde{u}_{g}^{\alpha_{\bsl{k}_0}}(\bsl{k}_s)  & \\
 &  & \ddots 
}
\ ,
}
where $g\in\G_{\bsl{k}_0}$, $\widetilde{u}_{g}(\bsl{k}_s)$ can be unitarily transformed to $\mathds{1}_{n_{\bsl{k}_0,\alpha_{\bsl{k}_0}} }\otimes u^{\alpha_{\bsl{k}_0}}_g(\bsl{k}_s)$ in a $g$-independent way, and $u^{\alpha_{\bsl{k}_0}}_g(\bsl{k}_s)$ is the small irrep $\alpha_{\bsl{k}_0}$ of $\G_{\bsl{k}_0}$ at $\bsl{k}_s$.
In the above equation, we used the fact that the number of $u^{\alpha_{\bsl{k}_0}}_g(\bsl{k}_s)$ in $u_{g}(\bsl{k}_s)$ is equal to $n_{\bsl{k}_0,\alpha_{\bsl{k}_0}}$ that is the number of $u^{\alpha_{\bsl{k}_0}}_g(\bsl{k}_0)$ in $u_{g}(\bsl{k}_0)$ since the symmetry contents respect the momentum type.
We emphasize that $W_{\G_{\bsl{k}_0}}(\bsl{k}_s)$ is not $W_{\G_{\bsl{k}_s}}$ suggested in \eqnref{eq:RM_block_gen_app} since $\G_{\bsl{k}_s}$ may not equal to $\G_{\bsl{k}_0}$.
We can always choose $W_{\G_{\bsl{k}_0}}(\bsl{k}_s)$ to be a continuous of $s$ since the columns of $W_{\G_{\bsl{k}_0}}(\bsl{k}_s)$ that correspond to the same small irrep of $\G_{\bsl{k}_0}$ are sections of a vector bundle with 1D base space, resulting that  $U_{{\epsilon},\bsl{k}_s,\alpha_{\bsl{k}_0}}(t)$ is a continuous function of $(s,t)$.
Based on \eqnref{eq:RM_block_kpath_app}, we can further define 
\eq{
\label{eq:winding_s_app}
\nu_{\epsilon,\bsl{k}_s,\alpha_{\bsl{k}_0}}=\frac{\ii}{2\pi d_{\alpha_{\bsl{k}_0}}}\int_0^T dt\Tr[U_{{\epsilon},\bsl{k}_s,\alpha_{\bsl{k}_0}}(t) \partial_t U_{{\epsilon},\bsl{k}_s,\alpha_{\bsl{k}_0}}(t)]\ ,
}
where we use the fact that the dimension of $u^{\alpha_{\bsl{k}_0}}_g(\bsl{k}_s)$ is equal to $d_{\alpha_{\bsl{k}_0}}$ that is the dimension of the $\alpha_{\bsl{k}_0}$ small irrep of $\G_{\bsl{k}_0}$ at $\bsl{k}_0$.
Since $\nu_{\epsilon,\bsl{k}_s,\alpha_{\bsl{k}_0}}$ is a continuous function of $s$ and is quantized to integers, we have $\nu_{\epsilon,\bsl{k}_0,\alpha_{\bsl{k}_0}}=\nu_{\epsilon,\bsl{k}_1,\alpha_{\bsl{k}_0}}$.
Combined with the fact that $\G_{\bsl{k}_0}=\G_{\bsl{k}_1}$ and thus $\alpha_{\bsl{k}_0}$ can enumerates all small irreps of $\G_{\bsl{k}_1}$ at $\bsl{k}_1$, we arrive at
\eq{
\nu_{\epsilon,\bsl{k}_0,\alpha_{\bsl{k}_0}}=\nu_{\epsilon,\bsl{k}_1,\alpha_{\bsl{k}_0}}\ ,
}
where the same label for small irreps of $\G_{\bsl{k}_0}$ and $\G_{\bsl{k}_1}$ is given by the continuous path as discussed above.

Second, we show the winding number is the same for $\bsl{k}_{0}$ and $\bsl{k}$.
Owing to $\bsl{k}_0=\bsl{k}_h+\bsl{G}$ with $h\in \G$, $\G_{\bsl{k}}=h^{-1}\G_{\bsl{k}_0} h$ and thereby $\G_{\bsl{k}}$ and $\G_{\bsl{k}_0}$ are isomorphic.
Then, we know the small irreps of $\G_{\bsl{k}}$ at $\bsl{k}$ are one-to-one corresponding to those of $\G_{\bsl{k}_0}$ at $\bsl{k}_0$, which can both be labeled as $\alpha$.
Specifically, we can choose $u_{g_0}^{\alpha}(\bsl{k}_0)=u_{h^{-1} g_0 h}^{\alpha}(\bsl{k})$ for all $g_0\in\G_{\bsl{k}_0}$ and all inequivalent $\alpha$.
Suppose we choose unitary $W_{\G_{\bsl{k}}}$ to give 
\eqa{
& W_{\G_{\bsl{k}}}^\dagger u_{g}(\bsl{k}) W_{\G_{\bsl{k}}} = 
\mat{
\ddots &  & \\
 & \mathds{1}_{n_{{\bsl{k}},\alpha}}\otimes u_{g}^{\alpha}(\bsl{k})  & \\
 &  & \ddots 
}\\
& W_{\G_{\bsl{k}}}^\dagger U_{\epsilon}(\bsl{k},t) W_{\G_{\bsl{k}}} = 
\mat{
\ddots &  & \\
 & U_{{\epsilon},\bsl{k},\alpha}(t)  & \\
 &  & \ddots 
}
}
for $g\in \G_{\bsl{k}}$.
Then, owing to $u_h(\bsl{k}) u_{ h^{-1} g_0 h}(\bsl{k}) u_h^\dagger(\bsl{k}) = u_{g_0}(\bsl{k}_0)$ that holds for all $g_0\in \G_{\bsl{k}_0}$, we can choose \eq{
 W_{\G_{\bsl{k}_0}}=u_h(\bsl{k}) W_{\G_{\bsl{k}}}
}
such that 
\eq{
W_{\G_{\bsl{k}_0}}^\dagger u_{g_0}(\bsl{k}_0) W_{\G_{\bsl{k}_0}} = 
\mat{
\ddots &  & \\
 & \mathds{1}_{n_{{\bsl{k}},\alpha}}\otimes u_{h^{-1} g_0 h}^{\alpha}(\bsl{k})  & \\
 &  & \ddots 
}
}
for all $g_0\in \G_{k_0}$, and 
\eq{
W_{\G_{\bsl{k}_0}}^\dagger U_{\epsilon}(\bsl{k}_0,t) W_{\G_{\bsl{k}_0}} = 
\mat{
\ddots &  & \\
 &U_{{\epsilon},\bsl{k},\alpha}(t)  & \\
 &  & \ddots 
}\ .
}
Owing to $u_{g_0}^{\alpha}(\bsl{k}_0)=u_{h^{-1} g_0 h}^{\alpha}(\bsl{k})$ and $n_{{\bsl{k}},\alpha}=n_{{\bsl{k}_0},\alpha}$, we know the blocks of return map for $\bsl{k}_0$ and $\bsl{k}$ are equal $U_{{\epsilon},\bsl{k}_0,\alpha}(t) =U_{{\epsilon},\bsl{k},\alpha}(t)$.
Combined with \eqnref{eq:winding_app}, we arrive at 
\eq{
\nu_{\epsilon,\bsl{k},\alpha}=\nu_{\epsilon,\bsl{k}_0,\alpha}\ ,
}
where the same label for small irreps of $\G_{\bsl{k}}$ and $\G_{\bsl{k}_0}$ is given by the symmetry $h\in\G$ as discussed above.

Combining two steps, we have
\eq{
\nu_{\epsilon,\bsl{k},\alpha}=\nu_{\epsilon,\bsl{k}_0,\alpha}=\nu_{\epsilon,\bsl{k}_1,\alpha}=\nu_{\epsilon,\bsl{k}',\alpha}
}
for all $\alpha$.
Therefore, the winding numbers are the same for two momenta of the same type, and thus we only need to consider one momentum for each type.

\subsubsection{Winding Numbers Obey All Compatibility Relations for Symmetry Contents}

Now, we show that $\nu_{\epsilon,\bsl{k},\alpha_{\bsl{k}}}$ obeys all compatibility relations for symmetry contents.
For symmetry contents, there are two types of compatibility relations~\cite{Bradlyn2017TQC,Po2017SymIndi}.
The first type comes from two momenta $\bsl{k}_0$ and $\bsl{k}_1$ (in 1BZ) that are connected by a continuous path $\bsl{k}_{s}$ with $s\in [0,1]$ and satisfy $\G_{\bsl{k}_0}\subsetneq\G_{\bsl{k}_1}$ and $\G_{\bsl{k}_0}\subset \G_{\bsl{k}_s}$ for all $s$.
Here we require $\G_{\bsl{k}_1}$ to be strictly larger than $\G_{\bsl{k}_0}$, since otherwise $\bsl{k}_0$ and $\bsl{k}_1$ become in the same type.
In practice, we can try to make $\bsl{k}_0$ and $\bsl{k}_1$ infinitesimally close to each other~\cite{Bradlyn2018TQCGraph}.

Suppose that $u_{g_1}^{\alpha_{\bsl{k}_1}}(\bsl{k}_1)$ is the $\alpha_{\bsl{k}_1}$ small irrep of $\G_{\bsl{k}_1}$ at $\bsl{k}_1$ for $g_1\in \G_{\bsl{k}_1}$.
Owing to $\G_{\bsl{k}_0}\subsetneq\G_{\bsl{k}_1}$, $u_{g_0}^{\alpha_{\bsl{k}_1}}(\bsl{k}_1)$ is also a small representation of $\G_{\bsl{k}_0}$ at $\bsl{k}_1$ with $g_0\in\G_{\bsl{k}_0}$. 
$u_{g_0}^{\alpha_{\bsl{k}_1}}(\bsl{k}_1)$ might not be irreducible for $\G_{\bsl{k}_0}$, and then we can express $u_{g_0}^{\alpha_{\bsl{k}_1}}(\bsl{k}_1)$ as the direct sum of small irreps of $\G_{\bsl{k}_0}$ at $\bsl{k}_1$
\eq{
\label{eq:u_group_subgroup_rel}
u_{g_0}^{\alpha_{\bsl{k}_1}}(\bsl{k}_1)=
\mat{
\ddots &  & \\
 & \mathds{1}_{w_{\alpha_{\bsl{k}_1},\alpha_{\bsl{k}_0}}}\otimes u_{g_0}^{ \alpha_{\bsl{k}_0}}(\bsl{k}_1)  & \\
 &  & \ddots_{\alpha_{\bsl{k}_0}} 
}
}
for all $g_0\in\G_{\bsl{k}_0}$, where the diagonal blocks range over $\alpha_{\bsl{k}_0}$, and a proper gauge is chosen.
It is the continuous path that allows us to use $\alpha_{\bsl{k}_0}$, which is originally the label for the small irreps of $\G_{\bsl{k}_0}$ at $\bsl{k}_0$, to label the small irreps of $\G_{\bsl{k}_0}$ at $\bsl{k}_1$ as $u_{g_0}^{ \alpha_{\bsl{k}_0}}(\bsl{k}_1)$.
In particular, $w_{\alpha_{\bsl{k}_1},\alpha_{\bsl{k}_0}}$ is the number of $u_{g_0}^{ \alpha_{\bsl{k}_0}}(\bsl{k}_1)$ in $u_{g_0}^{\alpha_{\bsl{k}_1}}(\bsl{k}_1)$, which is determined by $\alpha_{\bsl{k}_1}$, $\alpha_{\bsl{k}_0}$, $\G_{\bsl{k}_1}$, and $\G_{\bsl{k}_0}$~\cite{Tung1985GroupTheory,Bradley2009MathSSP}.
\refcite{Po2017SymIndi,Bradlyn2017TQC} suggests that the symmetry data satisfies 
\eq{
\label{eq:n_comprel_1}
n_{\bsl{k}_0,\alpha_{\bsl{k}_0}} ^l= \sum_{\alpha_{\bsl{k}_1}} w_{\alpha_{\bsl{k}_1},\alpha_{\bsl{k}_0}} n_{\bsl{k}_1,\alpha_{\bsl{k}_1}}^l\ ,
}
where $l$ labels the isolated set of quasi-energy bands, and $n_{\bsl{k},\alpha_{\bsl{k}}}^l$ was defined in \appref{sec:sym_data}.

We want to demonstrate that the relation \eqnref{eq:n_comprel_1} holds between $\nu_{\epsilon,\bsl{k}_0,\alpha_{\bsl{k}_0}}$ and $\nu_{\epsilon,\bsl{k}_1,\alpha_{\bsl{k}_1}}$, where $\nu_{\epsilon,\bsl{k},\alpha_{\bsl{k}}}$ was defined in \eqnref{eq:winding_app}.
According to \eqnref{eq:winding_s_app}, we can construct $\nu_{\epsilon,\bsl{k}_1,\alpha_{\bsl{k}_0}}$, which is the winding number of the return map block for the $\alpha_{\bsl{k}_0}$ small irrep of $\G_{\bsl{k}_0}$ at $\bsl{k}_1$, and we have 
\eq{
\label{eq:winding_k0_k1_comprel_app}
\nu_{\epsilon,\bsl{k}_0,\alpha_{\bsl{k}_0}}=\nu_{\epsilon,\bsl{k}_1,\alpha_{\bsl{k}_0}} 
}
with $\alpha_{\bsl{k}_0}$ ranging over all inequivalent small irreps of $\G_{\bsl{k}_0}$.
However, since $\G_{\bsl{k}_1}$ is strictly larger than $\G_{\bsl{k}_0}$, $\alpha_{\bsl{k}_0}$ cannot be used to label all small irreps of $\G_{\bsl{k}_1}$ at $\bsl{k}_1$.
Thus, we need to connect $\nu_{\epsilon,\bsl{k}_1,\alpha_{\bsl{k}_0}}$ to $\nu_{\epsilon,\bsl{k}_1,\alpha_{\bsl{k}_1}}$.

To do so, we can use a special unitary $W_{\G_{\bsl{k}_1}}$ to give \eqnref{eq:u_group_subgroup_rel} as well as 
\eqa{
& W_{\G_{\bsl{k}_1}}^\dagger U_{\epsilon}(\bsl{k}_1,t) W_{\G_{\bsl{k}_1}} = 
\mat{
\ddots &  & \\
 &\widetilde{U}_{{\epsilon},\bsl{k}_1,\alpha_{\bsl{k}_1}}(t)\otimes \mathds{1}_{d_{\alpha_{\bsl{k}_1}}}  & \\
 &  & \ddots 
}\\
&W_{\G_{\bsl{k}_1}}^\dagger u_{g_1}(\bsl{k}_1) W_{\G_{\bsl{k}_1}} = 
\mat{
\ddots &  & \\
 &\mathds{1}_{n_{\bsl{k}_1,\alpha_{\bsl{k}_1}}}\otimes u_{g_1}^{\alpha_{\bsl{k}_1}}(\bsl{k}_1)  & \\
 &  & \ddots 
}\ ,
}
where $g_1\in \G_{\bsl{k}_1}$.
$\nu_{\epsilon,\bsl{k}_1,\alpha_{\bsl{k}_1}}$ is given by $\widetilde{U}_{{\epsilon},\bsl{k}_1,\alpha_{\bsl{k}_1}}(t)$ according to \eqnref{eq:winding_app_sim}.
The $n_{\bsl{k}_1,\alpha_{\bsl{k}_1}} d_{\alpha_{\bsl{k}_1}}$ columns in $W_{\G_{\bsl{k}_1}}$ that furnish the copies of $\alpha_{\bsl{k}_1}$ small irrep of $\G_{\bsl{k}_1}$ can be labeled as $Y^{\alpha_{\bsl{k}_1}}_{\bsl{k}_1,j_{\bsl{k}_1,\alpha_{\bsl{k}_1}},i_{\alpha_{\bsl{k}_1}}}$ with $j_{\bsl{k}_1,\alpha_{\bsl{k}_1}}=1,...,n_{\bsl{k}_1,\alpha_{\bsl{k}_1}}$ labels the copies of the small irrep and $i_{\alpha_{\bsl{k}_1}}=1,...,d_{\alpha_{\bsl{k}_1}}$ labels the components for each copy.
Owing to the \eqnref{eq:u_group_subgroup_rel}, the $i_{\alpha_{\bsl{k}_1}}$ index can be relabeled as $(\alpha_{\bsl{k}_0},j_{\alpha_{\bsl{k}_1},\alpha_{\bsl{k}_0}},i_{\alpha_{\bsl{k}_0}})$ with $j_{\alpha_{\bsl{k}_1},\alpha_{\bsl{k}_0}}=1,...,w_{\alpha_{\bsl{k}_1},\alpha_{\bsl{k}_0}}$ and $i_{\alpha_{\bsl{k}_0}}=1,...,d_{\alpha_{\bsl{k}_0}}$.
Then, we have $Y^{\alpha_{\bsl{k}_1},\alpha_{\bsl{k}_0}}_{\bsl{k}_1,j_{\bsl{k}_1,\alpha_{\bsl{k}_1}},j_{\alpha_{\bsl{k}_1},\alpha_{\bsl{k}_0}},i_{\alpha_{\bsl{k}_0}}}$ as columns of $W_{\G_{\bsl{k}_1}}$ satisfying
\begin{widetext}
\eqa{
&U_{\epsilon}(\bsl{k}_1,t)Y^{\alpha_{\bsl{k}_1},\alpha_{\bsl{k}_0}}_{\bsl{k}_1,j_{\bsl{k}_1,\alpha_{\bsl{k}_1}},j_{\alpha_{\bsl{k}_1},\alpha_{\bsl{k}_0}},i_{\alpha_{\bsl{k}_0}}}= \sum_{j_{\bsl{k}_1,\alpha_{\bsl{k}_1}}'}Y^{\alpha_{\bsl{k}_1},\alpha_{\bsl{k}_0}}_{\bsl{k}_1,j_{\bsl{k}_1,\alpha_{\bsl{k}_1}}',j_{\alpha_{\bsl{k}_1},\alpha_{\bsl{k}_0}},i_{\alpha_{\bsl{k}_0}}} [\widetilde{U}_{{\epsilon},\bsl{k}_1,\alpha_{\bsl{k}_1}}(t)]_{j_{\bsl{k}_1,\alpha_{\bsl{k}_1}}'j_{\bsl{k}_1,\alpha_{\bsl{k}_1}}}\\
& u_{g_0}(\bsl{k}_1)Y^{\alpha_{\bsl{k}_1},\alpha_{\bsl{k}_0}}_{\bsl{k}_1,j_{\bsl{k}_1,\alpha_{\bsl{k}_1}},j_{\alpha_{\bsl{k}_1},\alpha_{\bsl{k}_0}},i_{\alpha_{\bsl{k}_0}}}= \sum_{i_{\alpha_{\bsl{k}_0}}'}Y^{\alpha_{\bsl{k}_1},\alpha_{\bsl{k}_0}}_{\bsl{k}_1,j_{\bsl{k}_1,\alpha_{\bsl{k}_1}},j_{\alpha_{\bsl{k}_1},\alpha_{\bsl{k}_0}},i_{\alpha_{\bsl{k}_0}}'} [u_{g_0}^{\alpha_{\bsl{k}_0}}(\bsl{k}_1)]_{i_{\alpha_{\bsl{k}_0}}'i_{\alpha_{\bsl{k}_0}}}\ \forall g_0\in\G_{\bsl{k}_0}\ .
}
We can then regroup $Y^{\alpha_{\bsl{k}_1},\alpha_{\bsl{k}_0}}_{\bsl{k}_1,j_{\bsl{k}_1,\alpha_{\bsl{k}_1}},j_{\alpha_{\bsl{k}_1},\alpha_{\bsl{k}_0}},i_{\alpha_{\bsl{k}_0}}}$ with the same $\alpha_{\bsl{k}_0}$ together and give a unitary $W_{\G_{\bsl{k}_0}}(\bsl{k}_1)$ that satisfies
\eqa{
& W_{\G_{\bsl{k}_0}}^\dagger(\bsl{k}_1) u_{g_0}(\bsl{k}_1) W_{\G_{\bsl{k}_0}}(\bsl{k}_1) = \mat{
\ddots &  & \\
 & \mathds{1}_{n_{{\bsl{k}_1},\alpha_{\bsl{k}_0}}}\otimes u_{g_0}^{\alpha_{\bsl{k}_0}}(\bsl{k}_1)  & \\
 &  & \ddots 
}\ \forall g_0\in\G_{\bsl{k}_0} \\
& W_{\G_{\bsl{k}_0}}^\dagger(\bsl{k}_1) U_{\epsilon}(\bsl{k}_1,t) W_{\G_{\bsl{k}_0}}(\bsl{k}_1) = \mat{
\ddots &  & \\
 & U_{\epsilon,\bsl{k}_1,\alpha_{\bsl{k}_0}}(t)  & \\
 &  & \ddots 
}\ ,
}
where $n_{{\bsl{k}_1},\alpha_{\bsl{k}_0}}=\sum_{\alpha_{\bsl{k}_1}} w_{\alpha_{\bsl{k}_1},\alpha_{\bsl{k}_0}} n_{\bsl{k}_1,\alpha_{\bsl{k}_1}}$ is the number of $u_{g_0}^{ \alpha_{\bsl{k}_0}}(\bsl{k}_1)$ in $u_{g_0}(\bsl{k}_1)$, and 
\eq{
U_{{\epsilon},\bsl{k}_1,\alpha_{\bsl{k}_0}}(t) = \mat{
\ddots &  & \\
 & \widetilde{U}_{{\epsilon},\bsl{k}_1,\alpha_{\bsl{k}_1}}(t)\otimes \mathds{1}_{w_{\alpha_{\bsl{k}_1},\alpha_{\bsl{k}_0}}  d_{\alpha_{\bsl{k}_0}}}  & \\
 &  & \ddots_{\alpha_{\bsl{k}_1}}
}\ .
}
We then have 
\eqa{
& \nu_{\epsilon,\bsl{k}_1,\alpha_{\bsl{k}_0}} \\
&=\frac{\ii}{2\pi} \frac{1}{ d_{\alpha_{\bsl{k}_0}} }\int^T_0 dt \Tr[U_{{\epsilon},\bsl{k}_0,\alpha_{\bsl{k}_0}}^\dagger(t) \partial_t U_{{\epsilon},\bsl{k}_0,\alpha_{\bsl{k}_0}}(t)] \\
& =\sum_{\alpha_{\bsl{k}_1}} w_{\alpha_{\bsl{k}_1},\alpha_{\bsl{k}_0}} \frac{\ii}{2\pi} \int^T_0 dt \Tr[\widetilde{U}_{{\epsilon},\bsl{k}_1,\alpha_{\bsl{k}_1}}^\dagger(t) \partial_t \widetilde{U}_{{\epsilon},\bsl{k}_1,\alpha_{\bsl{k}_1}}(t)]\\
& = \sum_{\alpha_{\bsl{k}_1}} w_{\alpha_{\bsl{k}_1},\alpha_{\bsl{k}_0}} \nu_{\epsilon,\bsl{k}_1,\alpha_{\bsl{k}_1}}\ .
}
Combined with \eqnref{eq:winding_k0_k1_comprel_app}, we arrive at
\eq{
\nu_{\epsilon,\bsl{k}_0,\alpha_{\bsl{k}_0}} = \nu_{\epsilon,\bsl{k}_1,\alpha_{\bsl{k}_0}} = \sum_{\alpha_{\bsl{k}_1}} w_{\alpha_{\bsl{k}_1},\alpha_{\bsl{k}_0}} \nu_{\epsilon,\bsl{k}_1,\alpha_{\bsl{k}_1}}\ ,
}
which is the same as \eqnref{eq:n_comprel_1} for symmetry contents.
\end{widetext}

The first type of compatibility relation is enough for all symmorphic crystalline groups.
For non-symmorphic crystalline groups, we need to include the second type.
To introduce the second type, first note that $\G_{\bsl{k}}=\G_{\bsl{k}+\bsl{G}}$.
The compatibility relation arises when $\bsl{k}$ and $\bsl{k}+\bsl{G}$ can be connected by a continuous path $\bsl{k}_s$ with $s\in [0,1]$ such that $\bsl{k}_0=\bsl{k}$, $\bsl{k}_1=\bsl{k}+\bsl{G}$, and $\G_{\bsl{k}}\subset \G_{\bsl{k}_s}$ for all $s$.
Then, according to the first part of the definition of the momentum type, the small irreps of $\G_{\bsl{k}+\bsl{G}}$ at $\bsl{k}+\bsl{G}$ can be labeled by $\alpha_{\bsl{k}}$ (originally for the small irreps of $\G_{\bsl{k}}$ at $\bsl{k}$) based on the continuous path, and we know
\eqa{
\label{eq:comprel_2_intermediate_app}
\nu_{\epsilon,\bsl{k},\alpha_{\bsl{k}}}=\nu_{\epsilon,\bsl{k}+\bsl{G},\alpha_{\bsl{k}}}\ .
}
With this convention, for certain momentum $\bsl{k}$ whose little group $\G_{\bsl{k}}$ contains non-symmorphic symmetries, the $\alpha_{\bsl{k}}$ small irrep of $\G_{\bsl{k}+\bsl{G}}$ at $\bsl{k}+\bsl{G}$, labeled as $u_{g}^{\alpha_{\bsl{k}}}(\bsl{k}+\bsl{G})$, may not equal to the $\alpha_{\bsl{k}}$ small irrep of $\G_{\bsl{k}}$ at $\bsl{k}$, labeled as $u_{g}^{\alpha_{\bsl{k}}}(\bsl{k})$  for $g\in \G_{\bsl{k}}=\G_{\bsl{k}+\bsl{G}}$; instead they satisfy (up to a $g$-independent unitary transformation)
\eq{
u_{g}^{\alpha_{\bsl{k}}}(\bsl{k}+\bsl{G})=u_{g}^{p_{\bsl{G}}(\alpha_{\bsl{k}})}(\bsl{k})
}
for all $g\in\G_{\bsl{k}}$, where $p_{\bsl{G}}$ labels a permutation of the small irreps.
As a result, we have
\eqa{
\nu_{\epsilon,\bsl{k}+\bsl{G},\alpha_{\bsl{k}}}=\nu_{\epsilon,\bsl{k},p_{\bsl{G}}(\alpha_{\bsl{k}})}\ ,
}
where the $\bsl{G}$-periodic nature of $u_g(\bsl{k})$ and $U_\epsilon(\bsl{k},t)$ is used.
Combining this equation with \eqnref{eq:comprel_2_intermediate_app}, we arrive at 
\eqa{
\label{eq:comprel_2_app}
\nu_{\epsilon,\bsl{k},\alpha_{\bsl{k}}}=\nu_{\epsilon,\bsl{k},p_{\bsl{G}}(\alpha_{\bsl{k}})}\ .
}
On the other hand the symmetry contents also obey $ n_{\bsl{k},\alpha_{\bsl{k}}}^l=n_{\bsl{k},p_{\bsl{G}}(\alpha_{\bsl{k}})}^l$, showing that the winding numbers possess the second type of compatibility relation of the symmetry contents.
In short, there are two ways of labelling small irreps at $\bsl{k}+\bsl{G}$: one is based on the continuous path, and the other is to make small irreps $\bsl{G}$-periodic.
The second type of compatibility relation is nothing but the result of compromising these two ways.

Since the winding numbers obey all compatibility relations for the symmetry contents, we can choose the same types of momenta for the symmetry data and winding data.

\section{Details on Static Winding Data Set and DSI}
\label{app:DSI}

In this section, we present more details on the static winding data set $\{ V_{SL} \}$ for a given FGU $U(\bsl{k},t)$ with time period $T$, a relevant gap choice, a crystalline symmetry group $\G$, and a symmetry representation $u_g(\bsl{k})$ of  $\G$.
Then, we elaborate the core method for the calculation of the DSI set given the Hilbert bases.
At last, we discuss how to determine the Hilbert bases sets that span a given symmetry data.

\subsection{$\{ V_{SL} \}$}

In this part, we show how to construct $\{ V_{SL} \}$ for the given FGU $U(\bsl{k},t)$.
Let us pick a PBZ choice for $U(\bsl{k},t)$ that yields symmetry data $A$.
As discussed in \appref{sec:static_limits} and \appref{sec:DSI_gen}, we only need to consider the $\G$-invariant static FGUs with time period $T_{SL}=T$ and symmetry data equivalent to $U(\bsl{k},t)$, labelled as $U_{SL}(\bsl{k},t)=e^{-\ii H_{SL}(\bsl{k}) t}$ with the corresponding relevant gap choice and symmetry representation.

$H_{SL}(\bsl{k})$ can always be expanded by the projection matrices as
\eq{
H_{SL}(\bsl{k}) = \sum_{r=1}^R\sum_{m_r=1}^{M_r} E_{\bsl{k}, r, m_r} P_{\bsl{k}, r, m_r}\ ,
}
where $P_{\bsl{k}, r, m_r}$ is the time-independent projection matrix onto the subspace corresponding to the band $E_{\bsl{k}, r, m_r}$.
Here we use $r$ to label the isolated connected set of bands and use $m_r$ to label the bands in the $r$th isolated connected set.
Being connected means the for any $m_r<M_r$, there exist $\bsl{k}_0$ such that $E_{\bsl{k}_0,r,m_{r}+1}=E_{\bsl{k}_0,r,m_r}$.
We can always choose $E_{\bsl{k},r,m_r}$ to be continuous in $\dsR^d$, $\bsl{G}$-periodic, and $\G$-symmetric for all $r,m_r$, and we also choose $E_{\bsl{k},r+1,m_{r+1}}>E_{\bsl{k},r,m_r}$ and $E_{\bsl{k},r,m_{r}+1}\geq E_{\bsl{k},r,m_r}$.
Then, the time-evolution matrix reads 
\eq{
U_{SL}(\bsl{k},t)=\sum_{r,m_r}e^{-\ii E_{\bsl{k}, r, m_r} t} P_{\bsl{k}, r, m_r}\ .
}

The relevant gaps of the static FGU are picked based on the quasi-energy band structure given by $U_{SL}(\bsl{k},T)$.
We further choose a PBZ lower bound $\Phi_{SL,\bsl{k}}$ for the static FGU.
As a result, we can have the quasi-energy bands as 
\eq{
\E_{\bsl{k},r,m_r}=\frac{\ii}{T}\log_{\epsilon_{\bsl{k}}=\Phi_{SL,\bsl{k}}} e^{-\ii E_{\bsl{k}, r, m_r} T}= E_{\bsl{k}, r, m_r} + \frac{2\pi}{T}q_{r}\ ,
}
where $q_{r}\in \dsZ$ realizes $\E_{\bsl{k},r,m_r} T\in [\Phi_{SL,\bsl{k}},\Phi_{SL,\bsl{k}}+2\pi)$.
Here $q_r$ is independent of $m_r$ and $\bsl{k}$ because (i) $\Phi_{SL,\bsl{k}}$ lies in a gap of $U_{SL}(\bsl{k},T)$, (ii) $\Phi_{SL,\bsl{k}}$ and $ E_{\bsl{k}, r, m_r}$ are continuous functions of $\bsl{k}$, and (iii) $ E_{\bsl{k}, r, m_r}$ $(m_r=1,...,M_r)$ is a connected set for each $r$.
Although $\E_{\bsl{k},r,m_r}$ and $\E_{\bsl{k},r',m_{r'}'}$ have no definite relations for $r\neq r'$ before determining $q_r$, $\E_{\bsl{k},r,m_r}$ with $m_r=1,...,M_r$, denoted by $\E_{\bsl{k},r}$, must always be a connected set.
Then, each connected set $\E_{\bsl{k},r}$ must lie in a unique isolated set of quasi-energy bands of $U_{SL}(\bsl{k},t)$, and thereby we can relabel the index $r$ as $(l,r_l)$, where $l$ labels the isolated set of quasi-energy bands in which $\E_{\bsl{k},r}$ lies, and $r_l$ is the index of $\E_{\bsl{k},r}$ in the $l$th isolated set.
With this notation, the bands of $H_{SL}(\bsl{k})$ are now labeled as $E_{\bsl{k},l,r_l,m_{l,r_l}}$ with $(l,r_l)$ still labelling the isolated connected set of bands of $H_{SL}(\bsl{k})$, and we have 
\eqa{
\label{eq:H_U_E_SL}
& H_{SL}(\bsl{k}) = \sum_{l=1}^L\sum_{r_l, m_{l,r_l}} E_{\bsl{k}, l, r_l, m_{l,r_l}} P_{\bsl{k}, l, r_l, m_{l,r_l}} \\
& U_{SL}(\bsl{k},t)=\sum_{l=1}^L\sum_{r_l, m_{l,r_l}} e^{-\ii t E_{\bsl{k}, l, r_l, m_{l,r_l}} } P_{\bsl{k}, l, r_l, m_{l,r_l}} \\
& \E_{\bsl{k},l, r_l,m_{l,r_l}}= E_{\bsl{k}, l, r_l, m_{l,r_l}} + \frac{2\pi}{T}q_{l,r_l}\ .
}

To derive the corresponding $V_{SL}$, we need to make sure the relevant gap choice and the PBZ choice give $A_{SL}=A$.
Since $(l,r_l)$ labels the isolated connected set of bands of $H_{SL}(\bsl{k})$, $P_{\bsl{k}, l, r_l}=\sum_{m_{l,r_l}}P_{\bsl{k}, l, r_l, m_{l,r_l}}$ provides a nonzero symmetry content $A_{l,r_l}\in \{ BS \}$, which is also the symmetry content of $\E_{\bsl{k},l,r_l}$.
Owing to $A_{SL}=A$, we have $\sum_{r_l} A_{l,r_l}=A_l$, and therefore 
$(...A_{l,r_l} ...)$ is a reduction of $A$.
(See the definition of reduction in \appref{sec:reducible_SD_DSI}.)
On the other hand, $V_{SL}$ is directly derived from the return map, which is given by \eqnref{eq:RM_def_gen} as 
\eq{
U_{SL,\epsilon=\Phi_{SL}}(\bsl{k},t)=\sum_{l, r_l}  e^{\ii q_{l,r_l} \frac{2\pi}{T} t} P_{\bsl{k}, l, r_l}\ .
}
Based on a derivation similar to \eqnref{eq:n_epsilon_shift_intermediate_2_app}, we have 
\eq{
V_{SL}= -\sum_{l,r_l} q_{l,r_l} A_{l,r_l}\ .
}
Since $(...A_{l,r_l} ...)$ is a reduction of $A$ and $-q_{l,r_l}\in\dsZ$, we arrive at 
\eq{
\{ V_{SL}\} \subset \overline{\{ V_{SL} \}}
}
with $\overline{\{ V_{SL} \}}$ defined in \eqnref{eq:V_SL_bar}.

The above derivation does not specify whether $A$ is irreducible or not.
(Recall that we define the symmetry data $A$ of a FGU to be irreducible if all its columns are irreducible symmetry contents; otherwise, $A$ is reducible.)
If $A$ is reducible, it is possible that $\overline{\{ V_{SL} \}}$ is strictly larger than $\{ V_{SL}\}$ since certain reduction of $A$ might be not reproducible by isolated sets of bands. 
If $A$ is irreducible, then we only have one reduction of $A$, which is $A$ itself, and this reduction can be reproduced by isolated sets of bands since $U(\bsl{k},t)$ has it, resulting in \eqnref{eq:V_SL_set}.

\subsection{DSI Set for Irreducible Symmetry Data}
In this part, we will derive \eqnref{eq:DIS_IR_gen} from  \eqnref{eq:V_set_gen} and  \eqnref{eq:V_SL_set_sim} given the set of Hilbert bases $\{ a_j \}$ with $J$ elements.
The derivation will show how to construct the DSI set for FGUs with irreducible symmetry data.

The winding data set in \eqnref{eq:V_set_gen} can be rewritten as
\eq{
\label{eq:V_set_app}
\{ V \} = \dsZ^K\cap \ker \mat{ \mathcal{C} \\ \mathcal{D}}\ ,
}
and \eqnref{eq:V_SL_set_sim} gives $\{ V_{SL} \}$.
Recall that the diagonal $\mathcal{D}$ is determined as: a diagonal element of $\mathcal{D}$ is $0$ ($1$) if the corresponding component of $\sum_{j} a_{j}$ is nonzero (zero).
To derive \eqnref{eq:DIS_IR_gen}, let us first define a matrix with $a_j$ as its columns:
\eq{
M_a = \mat{ ... a_j ...}\ .
}
Since $M_a$ is a $K\times J$ matrix with integer elements, it always has the so-called Smith normal form (SNF)~\cite{Storjohann1996SNF,Song2020FragileAffineMonoid}, \ie, there exists a unimodular $K\times K$ matrix $U_{L}$ and a unimodular $J\times J$ matrix $U_{R}$ such that
\eq{
\label{eq:SNF_Ma}
M_a   = 
U_{L} \Lambda U_R\ ,
}
where the $K\times J$ matrix $\Lambda$ satisfies
\eq{
\Lambda_{ij}=\left\{ 
\begin{array}{ll}
\lambda_i\ ,     &  i=j\in \{ 1,2,...,r\}\\
0\ ,    & \text{otherwise}
\end{array}
\right.\ ,
}
$\lambda_{1,..,r}$ are positive integers, $r$ is the matrix rank of $M_a$, and $\lambda_{i+1}/\lambda_{i}$ is a positive integer for all $i=1,...,r-1$.
Here being unimodular means that (i) the square matrix is invertible and (ii) itself and its inverse are all matrices with integer elements.
Then, the DSI set reads
\eq{
\label{eq:DSI_set_IrreSD_app}
\X=\frac{\{ V \}}{\{ V_{SL} \}}\approx \dsZ^{K-\widetilde{r}-r}\times \dsZ_{\lambda_1\times\lambda_2 \times ... \times \lambda_r}\ ,
}
where $\widetilde{r}$ is the rank of $\mat{ \mathcal{C} \\ \mathcal{D}}$.
In the following, we derive \eqnref{eq:DSI_set_IrreSD_app} explicitly.

Let us focus on the first $r$ columns of $U_{L}$, denoted by $U_{L,1}, ... , U_{L,r}$, which form a matrix $B$
\eq{
B=\mat{U_{L,1} ... U_{L,r}}\ .
}
Combining this definition with \eqnref{eq:SNF_Ma}, we have
\eq{
M_a q_J  = 
B \Lambda_r q_r\ ,
}
where 
\eq{
\Lambda_r= \mat{
\lambda_1 &  &   \\
 & \lambda_2  &   \\
&  & \ddots &   \\
&  & & \lambda_r   \\
}\ ,
}
$q_J\in\dsZ^J$, and $q_r\in \dsZ^r$ consists of the first $r$ components of $U_R q_J$.
As $q_J$ ranges over $\dsZ^J$, $q_r$ ranges over $\dsZ^r$, resulting in
\eq{
\label{eq:V_SL_alt}
\{ V_{SL} \} = \{ B \Lambda_r  q | q\in\dsZ^r \}\ .
}
Since the SNF \eqnref{eq:SNF_Ma} is a special type of singular value decomposition, we have
\eq{
\{ B x | x\in \dsR^r \} = \col( M_a )\ ,
}
with $\col( M_a )$ the column space of $M_a$.
Therefore, we have $\{ B q | q\in \dsZ^r\} \subset \dsZ^K \cap \col( M_a )$.
On the other hand, since all columns of $U_L$ form a set of bases for $\dsZ^K$, all elements in $\dsZ^K \cap \col( M_a )$ can be expressed as linear combinations of columns of $U_L$ with integer coefficients.
Since the last $K-r$ columns of $U_L$ are not in $\col( M_a )$, all elements in $\dsZ^K \cap \col( M_a )$ can be expressed as linear combinations of the first $r$ columns of $U_L$ with integer coefficients, \ie, $\dsZ^K \cap \col( M_a )  \subset \{ B q | q\in \dsZ^r\}$.
Moreover, since $\{ B q | q\in \dsZ^r\}$ and $\dsZ^K \cap \col( M_a )$ have the same definition of addition and scalar multiplication, we have $\{ B q | q\in \dsZ^r\} = \dsZ^K \cap \col( M_a )$.
Eventually combined with $\mathcal{C}M_a=0$ and $\mathcal{D}M_a=0$, we arrive at
\eq{
\label{eq:V_SL_V_rel}
\{ V_{SL} \}  \subset \{ B q | q\in \dsZ^r\} = \dsZ^K \cap \col( M_a ) \subset \{ V \}\ .
}

\eqnref{eq:V_SL_V_rel} suggests us to derive the DSI set in two steps based on the following expression
\eq{
\label{eq:DSI_set_app_inetermedia_3}
\X=\frac{\{ V \}}{\{ V_{SL} \}}\approx\frac{\{ V \}}{\dsZ^K \cap \col( M_a )}\times \frac{\dsZ^K \cap \col( M_a )}{\{ V_{SL} \}}\ .
}
In the first step, we derive $\frac{\dsZ^K \cap \col( M_a )}{\{ V_{SL} \}}$ from  \eqnref{eq:V_SL_alt} and \eqnref{eq:V_SL_V_rel}, which reads
\eq{
\label{eq:DSI_set_irredSD_app_intermediate}
\frac{\dsZ^K \cap \col( M_a )}{\{ V_{SL} \}} \approx \dsZ_{\lambda_1 \times \lambda_2 \times... \times \lambda_r }\ .
}
So the second step is to derive
\eq{
\frac{\{ V \}}{\dsZ^K \cap \col( M_a )} = \frac{ \dsZ^K\cap \ker \mat{ \mathcal{C} \\ \mathcal{D}}}{\dsZ^K \cap \col( M_a ) }\ .
}

To do so, let us first look at the SNF of $\mat{ \mathcal{C} \\ \mathcal{D}}$
\eq{
\mat{ \mathcal{C} \\ \mathcal{D}} = \widetilde{U}_L\widetilde{\Lambda} \widetilde{U}_R\ .
}
The last $K-\widetilde{r}$ columns of $\widetilde{U}_R^{-1}$ spans $\{ V \}$ with $\widetilde{r}$ the rank of $\mat{ \mathcal{C} \\ \mathcal{D}}$.
We label the matrix formed by the last $K-\widetilde{r}$ columns of $\widetilde{U}_R^{-1}$ as $S$, and label the matrix formed by the last $K-\widetilde{r}$ rows of $\widetilde{U}_R$ as $S^{-1}_L$, where $S^{-1}_L$ is the left inverse of $S$
\eq{
S^{-1}_L S = \mathds{1}_{K-\widetilde{r}}\ .
}
$\{ V \}$ can be rewritten as 
\eq{
\{ V \}= \{ S q | q \in \dsZ^{K-\widetilde{r}} \}\ ,
}
we have
\eqa{
& \{ V \}\cong S^{-1}_L  \{ V \} = \dsZ^{K-\widetilde{r}} \\
&  \col(M_a)\cap \dsZ^K \cong S^{-1}_L  (\col(M_a)\cap \dsZ^K) = \{ S^{-1}_L B q | q\in \dsZ^r\}\ ,
}
resulting in 
\eq{
\frac{\{ V \}}{ \col(M_a)\cap \dsZ^K }\approx \frac{\dsZ^{K-\widetilde{r}}}{\{ S^{-1}_L B q | q\in \dsZ^r\}}\ .
}
Here $\cong$ means being isomorphic.
On the other hand, since $S^{-1}_L a_j \in S^{-1}_L  \{ V \} = \dsZ^{K-\widetilde{r}}$ and the rank of $S^{-1} M_a$ is still $r$, we have
\eq{
\frac{\dsZ^{K-\widetilde{r}}}{\dsZ^{K-\widetilde{r}}\cap \col(S^{-1}_L M_a)}\approx\dsZ^{K-\widetilde{r}-r}\ ,
}
which can be straightforwardly derived by the SNF of $S^{-1}_L M_a$.
So as long as we can verify 
\eq{
\label{eq:transformed_colMaZ}
S^{-1}_L  (\col(M_a)\cap \dsZ^K)=\dsZ^{K-\widetilde{r}}\cap \col(S^{-1}_L M_a)\ ,
}
we have
\eq{
\frac{\{ V \}}{ \col(M_a)\cap \dsZ^K }\approx\dsZ^{K-\widetilde{r}-r}\ ,
}
which, combined with \eqnref{eq:DSI_set_irredSD_app_intermediate} and \eqnref{eq:DSI_set_app_inetermedia_3}, gives \eqnref{eq:DSI_set_IrreSD_app}.

To verify \eqnref{eq:transformed_colMaZ}, let us recall \eqnref{eq:V_SL_V_rel}.
Any element $y$ of $S^{-1}_L  (\col(M_a)\cap \dsZ^K)$ satisfies 
\eq{
y= S^{-1}_L M_a x \text{ and } M_a x \in \dsZ^K\ ,
}
where $x\in \dsR^J$.
Since $S^{-1}_L$ is a $(K-\widetilde{r})\times K$ integer matrix, $M_a x \in \dsZ^K$ infers $S^{-1}_L M_a x\in  \dsZ^{K-\widetilde{r}}$ and thereby  $y\in \dsZ^{K-\widetilde{r}}\cap \col(S^{-1}_L M_a)$, resulting in
\eq{
S^{-1}_L  (\col(M_a)\cap \dsZ^K)\subset\dsZ^{K-\widetilde{r}}\cap \col(S^{-1}_L M_a)\ .
}
On the other hand, for any element $y$ in $\dsZ^{K-\widetilde{r}}\cap \col(S^{-1}_L M_a)$, $y$ has the form 
\eq{
y= S^{-1}_L M_a x \text{ and } y\in \dsZ^{K-\widetilde{r}}\ ,
}
where $x\in \dsR^J$.
Since $S S^{-1}_L V =V $ for any $V\in \{V\}$, we have $S S^{-1}_L M_a =M_a$, resulting in 
\eq{
M_a x = S y \in \dsZ^K\ .
}
Thereby, we have $y\in S^{-1}_L  (\col(M_a)\cap \dsZ^K)$ and thus $\dsZ^{K-\widetilde{r}}\cap \col(S^{-1}_L M_a)\subset S^{-1}_L  (\col(M_a)\cap \dsZ^K)$.
Combined with that fact that $\dsZ^{K-\widetilde{r}}\cap \col(S^{-1}_L M_a)$ and $S^{-1}_L  (\col(M_a)\cap \dsZ^K)$ have the same definition of addition and scalar multiplication, we have 
\eq{
\dsZ^{K-\widetilde{r}}\cap \col(S^{-1}_L M_a)= S^{-1}_L  (\col(M_a)\cap \dsZ^K)\ .
}

\subsection{Hilbert Bases Sets That Span Symmetry Data}
In this part, we provide a general method of finding all Hilbert bases sets that span any given symmetry data $A$, for any given crystalline symmetry group $\G$. 
In \appref{sec:reducible_SD_DSI}, a set of Hilbert bases $\{ a_j\}$ is defined to span $A$ iff $\{ a_j\}$ consists of all distinct columns of an irreducible reduction of $A$.
However, this definition is not convenient for general computation.
Then, we use the following convenient yet equivalent definition for a Hilbert bases set to span $A$.
Namely, a set of Hilbert bases $\{ a_j \}$ spans $A$ iff there exists $c_{jl}\in\dsN$ such that $A_l=\sum_{j} a_j c_{jl}\ \forall l$ and $\sum_{l} c_{jl}\neq 0\ \forall j$.

Now we discuss the method.
Suppose the given symmetry data $A$ has $L$ columns, and $\{ BS\}$ (the set that contains all symmetry contents) for $\G$ in total has $I$ Hilbert bases, labeled as $a_i$ ($i=1,...,I$).
First, find all solutions to 
\eq{
\label{eq:C_equation_gen}
A=\mat{a_1 & ... & a_i & ... & a_I} C
}
for $C\in \dsN^{I\times L}$, and label the solutions as $C^\gamma$ with  $\gamma$ the index labelling the solutions. ($\gamma$ should not be confused with model parameter in \appref{sec:DSI_2+1D_FHOTI}.)
Second, for each solution $C^\gamma$, find all nonzero rows of $C^\gamma$, and then find all the corresponding Hilbert bases, forming a set $\{ a_{j_\gamma}\}$.
Third, all distinct $\{ a_{j_\gamma}\}$ are all Hilbert bases sets that span $A$.

As a demonstration, let us focus on the 1+1D inversion-invariant case.
As shown in \eqnref{eq:1+1D_P_HBs}, we have in total four Hilbert bases $(I=4)$, and thus given any symmetry data $A$, the equation that we should solve is 
\eq{
\label{eq:C_equation_1+1D_P}
A = \mat{ a_1 & a_2 & a_3  & a_4 } C
}
with $C\in \dsN^{4\times L}$.
For the irreducible symmetry data $A$ in \eqnref{eq:1+1D_P_sym_data}, we have only one solution $C$ for \eqnref{eq:C_equation_1+1D_P}, namely
\eq{
C=\left(
\begin{array}{cc}
 1 & 0 \\
 0 & 0 \\
 0 & 0 \\
 0 & 1 \\
\end{array}
\right)\ ,
}
which has two nonzero rows---the first and the fourth.
It means that only one Hilbert bases set $\{a_1, a_4 \}$ spans $A$, coinciding with the conclusion in the main text.
As another example, let us consider the reducible symmetry data in \figref{fig:1+1D_P_reducible_sym_data}, which reads
\eq{
A=\mat{ 1\\ 1\\ 1\\ 1\\}\ .
}
In this case, we have two solutions for \eqnref{eq:C_equation_1+1D_P} as
\eq{
C=\left(
\begin{array}{c}
 0 \\
 1 \\
 1 \\
 0 \\
\end{array}
\right),\left(
\begin{array}{c}
 1 \\
 0 \\
 0 \\
 1 \\
\end{array}
\right)\ .
}
For first solution, the nonzero rows are the second and third, giving us $\{a_2, a_3 \}$; for the second solution, the nonzero rows are the first and fourth, giving us $\{a_1, a_4 \}$.
Thus, the reducible symmetry data in \figref{fig:1+1D_P_reducible_sym_data} is spanned by  $\{a_2, a_3 \}$ or $\{a_1, a_4 \}$, coinciding with \appref{sec:reducible_SD_DSI}.

At last, we emphasize that if two symmetry data are given by the same FGU with different PBZ choices (thus related by the cyclic permutation in \eqnref{eq:cyclic_permutation_P_sym_data}), the method would give the same spanning Hilbert bases sets for them.
It is because the cyclic permutation can only change the order of columns of $C$ in \eqnref{eq:C_equation_gen}, and thus cannot transform a zero row to a nonzero one or vise versa.
It coincides with the fact that $\overline{\{V_{SL}\}}$ in \eqnref{eq:V_SL_set_union_HB_gen} is PBZ-independent.

\clearpage
\newpage

\section{Hilbert Bases for Plane Groups}
\label{app:HB}

In this part, we list the Hilbert bases of $\{BS\}$ for all spinless and spinful plane groups, which are used to provide {\tabPlaneGSpinless}-2.
We label the small irreps of little groups of chosen momenta according to \emph{Bilbao Crystallographic Server}~\cite{Bradlyn2017TQC}.
Given a crystalline symmetry group, \emph{Bilbao Crystallographic Server} sometimes picks more than one momenta in each type (see the definition of type in \appref{sec:sym_data}), but this redundancy can be removed by including extra compatibility relations (or equivalently by considering a larger compatibility matrix $\mathcal{C}$ in \eqnref{eq:comp_rel_gen} and \eqnref{eq:comp_rel_V_gen}).
Therefore, the Hilbert bases and DSIs derived with a larger $\mathcal{C}$ would be equivalent to those derived by picking one momentum in each chosen type.

In each of the following matrices, the first column label the small irreps of little groups of chosen momenta, and all other columns show the Hilbert bases.
Each component of a Hilbert basis labels the number of involved copies of the corresponding small irrep.

\noindent\[\text{spinless p1:}\]

\noindent\[\left(
\begin{array}{c|c}
 \text{V1} & 1 \\
 \text{X1} & 1 \\
 \text{Y1} & 1 \\
 \text{$\Gamma$1} & 1 \\
\end{array}
\right)\]

\noindent\[\text{spinful p1:}\]

\noindent\[\left(
\begin{array}{c|c}
 \overline{\text{V2}} & 1 \\
 \overline{\text{X2}} & 1 \\
 \overline{\text{Y2}} & 1 \\
 \overline{\text{$\Gamma$2}} & 1 \\
\end{array}
\right)\]

\noindent\[\text{spinless p2:}\]

\noindent\[\left(
\begin{array}{c|cccccccccccccccc}
 \text{A1} & 1 & 1 & 1 & 1 & 1 & 1 & 1 & 1 & 0 & 0 & 0 & 0 & 0 & 0 & 0 & 0 \\
 \text{A2} & 0 & 0 & 0 & 0 & 0 & 0 & 0 & 0 & 1 & 1 & 1 & 1 & 1 & 1 & 1 & 1 \\
 \text{B1} & 1 & 1 & 1 & 1 & 0 & 0 & 0 & 0 & 1 & 1 & 1 & 1 & 0 & 0 & 0 & 0 \\
 \text{B2} & 0 & 0 & 0 & 0 & 1 & 1 & 1 & 1 & 0 & 0 & 0 & 0 & 1 & 1 & 1 & 1 \\
 \text{$\Gamma$1} & 1 & 1 & 0 & 0 & 1 & 1 & 0 & 0 & 1 & 1 & 0 & 0 & 1 & 1 & 0 & 0 \\
 \text{$\Gamma$2} & 0 & 0 & 1 & 1 & 0 & 0 & 1 & 1 & 0 & 0 & 1 & 1 & 0 & 0 & 1 & 1 \\
 \text{Y1} & 1 & 0 & 1 & 0 & 1 & 0 & 1 & 0 & 1 & 0 & 1 & 0 & 1 & 0 & 1 & 0 \\
 \text{Y2} & 0 & 1 & 0 & 1 & 0 & 1 & 0 & 1 & 0 & 1 & 0 & 1 & 0 & 1 & 0 & 1 \\
\end{array}
\right)\]

\noindent\[\text{spinful p2:}\]

\noindent\[\left(
\begin{array}{c|cccccccccccccccc}
 \overline{\text{A3}} & 1 & 1 & 1 & 1 & 1 & 1 & 1 & 1 & 0 & 0 & 0 & 0 & 0 & 0 & 0 & 0 \\
 \overline{\text{A4}} & 0 & 0 & 0 & 0 & 0 & 0 & 0 & 0 & 1 & 1 & 1 & 1 & 1 & 1 & 1 & 1 \\
 \overline{\text{B3}} & 1 & 1 & 1 & 1 & 0 & 0 & 0 & 0 & 1 & 1 & 1 & 1 & 0 & 0 & 0 & 0 \\
 \overline{\text{B4}} & 0 & 0 & 0 & 0 & 1 & 1 & 1 & 1 & 0 & 0 & 0 & 0 & 1 & 1 & 1 & 1 \\
 \overline{\text{$\Gamma$3}} & 1 & 1 & 0 & 0 & 1 & 1 & 0 & 0 & 1 & 1 & 0 & 0 & 1 & 1 & 0 & 0 \\
 \overline{\text{$\Gamma$4}} & 0 & 0 & 1 & 1 & 0 & 0 & 1 & 1 & 0 & 0 & 1 & 1 & 0 & 0 & 1 & 1 \\
 \overline{\text{Y3}} & 1 & 0 & 1 & 0 & 1 & 0 & 1 & 0 & 1 & 0 & 1 & 0 & 1 & 0 & 1 & 0 \\
 \overline{\text{Y4}} & 0 & 1 & 0 & 1 & 0 & 1 & 0 & 1 & 0 & 1 & 0 & 1 & 0 & 1 & 0 & 1 \\
\end{array}
\right)\]

\noindent\[\text{spinless pm:}\]

\noindent\[\left(
\begin{array}{c|cccc}
 \text{C1} & 1 & 1 & 0 & 0 \\
 \text{C2} & 0 & 0 & 1 & 1 \\
 \text{$\Gamma$1} & 1 & 0 & 1 & 0 \\
 \text{$\Gamma$2} & 0 & 1 & 0 & 1 \\
 \text{Y1} & 1 & 0 & 1 & 0 \\
 \text{Y2} & 0 & 1 & 0 & 1 \\
 \text{Z1} & 1 & 1 & 0 & 0 \\
 \text{Z2} & 0 & 0 & 1 & 1 \\
\end{array}
\right)\]

\noindent\[\text{spinful pm:}\]

\noindent\[\left(
\begin{array}{c|cccc}
 \overline{\text{C3}} & 1 & 1 & 0 & 0 \\
 \overline{\text{C4}} & 0 & 0 & 1 & 1 \\
 \overline{\text{$\Gamma$3}} & 1 & 0 & 1 & 0 \\
 \overline{\text{$\Gamma$4}} & 0 & 1 & 0 & 1 \\
 \overline{\text{Y3}} & 1 & 0 & 1 & 0 \\
 \overline{\text{Y4}} & 0 & 1 & 0 & 1 \\
 \overline{\text{Z3}} & 1 & 1 & 0 & 0 \\
 \overline{\text{Z4}} & 0 & 0 & 1 & 1 \\
\end{array}
\right)\]

\noindent\[\text{spinless pg:}\]

\noindent\[\left(
\begin{array}{c|c}
 \text{B1} & 1 \\
 \text{B2} & 1 \\
 \text{D1} & 1 \\
 \text{D2} & 1 \\
 \text{$\Gamma$1} & 1 \\
 \text{$\Gamma$2} & 1 \\
 \text{Z1} & 1 \\
 \text{Z2} & 1 \\
\end{array}
\right)\]

\noindent\[\text{spinful pg:}\]

\noindent\[\left(
\begin{array}{c|c}
 \overline{\text{B3}} & 1 \\
 \overline{\text{B4}} & 1 \\
 \overline{\text{D3}} & 1 \\
 \overline{\text{D4}} & 1 \\
 \overline{\text{$\Gamma$3}} & 1 \\
 \overline{\text{$\Gamma$4}} & 1 \\
 \overline{\text{Z3}} & 1 \\
 \overline{\text{Z4}} & 1 \\
\end{array}
\right)\]

\noindent\[\text{spinless cm:}\]

\noindent\[\left(
\begin{array}{c|cc}
 \text{$\Gamma$1} & 1 & 0 \\
 \text{$\Gamma$2} & 0 & 1 \\
 \text{Y1} & 1 & 0 \\
 \text{Y2} & 0 & 1 \\
 \text{V1} & 1 & 1 \\
\end{array}
\right)\]

\noindent\[\text{spinful cm:}\]

\noindent\[\left(
\begin{array}{c|cc}
 \overline{\text{$\Gamma$3}} & 1 & 0 \\
 \overline{\text{$\Gamma$4}} & 0 & 1 \\
 \overline{\text{Y3}} & 1 & 0 \\
 \overline{\text{Y4}} & 0 & 1 \\
 \overline{\text{V2}} & 1 & 1 \\
\end{array}
\right)\]

\noindent\[\text{spinless p2mm:}\]

\noindent\[\left(
\begin{array}{c|cccccccccccccccccccccccc}
 \text{$\Gamma$1} & 1 & 1 & 1 & 1 & 1 & 1 & 1 & 1 & 0 & 0 & 0 & 0 & 0 & 0 & 0 & 0 & 0 & 0 & 0 & 0 & 0 & 0 & 0 & 0 \\
 \text{$\Gamma$2} & 1 & 1 & 1 & 1 & 0 & 0 & 0 & 0 & 1 & 1 & 1 & 1 & 0 & 0 & 0 & 0 & 0 & 0 & 0 & 0 & 0 & 0 & 0 & 0 \\
 \text{$\Gamma$3} & 0 & 0 & 0 & 0 & 0 & 0 & 0 & 0 & 0 & 0 & 0 & 0 & 1 & 1 & 1 & 1 & 1 & 1 & 1 & 1 & 0 & 0 & 0 & 0 \\
 \text{$\Gamma$4} & 0 & 0 & 0 & 0 & 0 & 0 & 0 & 0 & 0 & 0 & 0 & 0 & 1 & 1 & 1 & 1 & 0 & 0 & 0 & 0 & 1 & 1 & 1 & 1 \\
 \text{S1} & 1 & 1 & 0 & 0 & 1 & 0 & 0 & 0 & 1 & 0 & 0 & 0 & 1 & 1 & 0 & 0 & 1 & 0 & 0 & 0 & 1 & 0 & 0 & 0 \\
 \text{S2} & 1 & 1 & 0 & 0 & 0 & 1 & 0 & 0 & 0 & 1 & 0 & 0 & 1 & 1 & 0 & 0 & 0 & 1 & 0 & 0 & 0 & 1 & 0 & 0 \\
 \text{S3} & 0 & 0 & 1 & 1 & 0 & 0 & 1 & 0 & 0 & 0 & 1 & 0 & 0 & 0 & 1 & 1 & 0 & 0 & 1 & 0 & 0 & 0 & 1 & 0 \\
 \text{S4} & 0 & 0 & 1 & 1 & 0 & 0 & 0 & 1 & 0 & 0 & 0 & 1 & 0 & 0 & 1 & 1 & 0 & 0 & 0 & 1 & 0 & 0 & 0 & 1 \\
 \text{X1} & 1 & 0 & 1 & 0 & 1 & 0 & 1 & 0 & 0 & 0 & 0 & 0 & 1 & 0 & 1 & 0 & 0 & 0 & 0 & 0 & 1 & 0 & 1 & 0 \\
 \text{X2} & 1 & 0 & 1 & 0 & 0 & 0 & 0 & 0 & 0 & 1 & 0 & 1 & 1 & 0 & 1 & 0 & 0 & 1 & 0 & 1 & 0 & 0 & 0 & 0 \\
 \text{X3} & 0 & 1 & 0 & 1 & 0 & 0 & 0 & 0 & 1 & 0 & 1 & 0 & 0 & 1 & 0 & 1 & 1 & 0 & 1 & 0 & 0 & 0 & 0 & 0 \\
 \text{X4} & 0 & 1 & 0 & 1 & 0 & 1 & 0 & 1 & 0 & 0 & 0 & 0 & 0 & 1 & 0 & 1 & 0 & 0 & 0 & 0 & 0 & 1 & 0 & 1 \\
 \text{Y1} & 0 & 1 & 1 & 0 & 1 & 0 & 0 & 1 & 0 & 0 & 0 & 0 & 1 & 0 & 0 & 1 & 1 & 0 & 0 & 1 & 0 & 0 & 0 & 0 \\
 \text{Y2} & 0 & 1 & 1 & 0 & 0 & 0 & 0 & 0 & 0 & 1 & 1 & 0 & 1 & 0 & 0 & 1 & 0 & 0 & 0 & 0 & 0 & 1 & 1 & 0 \\
 \text{Y3} & 1 & 0 & 0 & 1 & 0 & 1 & 1 & 0 & 0 & 0 & 0 & 0 & 0 & 1 & 1 & 0 & 0 & 1 & 1 & 0 & 0 & 0 & 0 & 0 \\
 \text{Y4} & 1 & 0 & 0 & 1 & 0 & 0 & 0 & 0 & 1 & 0 & 0 & 1 & 0 & 1 & 1 & 0 & 0 & 0 & 0 & 0 & 1 & 0 & 0 & 1 \\
\end{array}
\right)\]

\noindent\[\text{spinful p2mm:}\]

\noindent\[\left(
\begin{array}{c|c}
 \overline{\text{$\Gamma$5}} & 1 \\
 \overline{\text{S5}} & 1 \\
 \overline{\text{X5}} & 1 \\
 \overline{\text{Y5}} & 1 \\
\end{array}
\right)\]

\noindent\[\text{spinless p2mg:}\]

\noindent\[\left(
\begin{array}{c|cccccc}
 \text{$\Gamma$1} & 1 & 1 & 1 & 0 & 0 & 0 \\
 \text{$\Gamma$2} & 1 & 1 & 0 & 1 & 0 & 0 \\
 \text{$\Gamma$3} & 0 & 0 & 1 & 0 & 1 & 1 \\
 \text{$\Gamma$4} & 0 & 0 & 0 & 1 & 1 & 1 \\
 \text{S1} & 1 & 1 & 1 & 1 & 1 & 1 \\
 \text{X1} & 1 & 1 & 1 & 1 & 1 & 1 \\
 \text{Y1} & 1 & 0 & 1 & 0 & 1 & 0 \\
 \text{Y2} & 1 & 0 & 0 & 1 & 1 & 0 \\
 \text{Y3} & 0 & 1 & 1 & 0 & 0 & 1 \\
 \text{Y4} & 0 & 1 & 0 & 1 & 0 & 1 \\
\end{array}
\right)\]

\noindent\[\text{spinful p2mg:}\]

\noindent\[\left(
\begin{array}{c|cccccc}
 \overline{\text{$\Gamma$5}} & 1 & 1 & 1 & 1 & 1 & 1 \\
 \overline{\text{S2}} & 1 & 1 & 1 & 0 & 0 & 0 \\
 \overline{\text{S3}} & 1 & 0 & 0 & 1 & 1 & 0 \\
 \overline{\text{S4}} & 0 & 1 & 1 & 0 & 0 & 1 \\
 \overline{\text{S5}} & 0 & 0 & 0 & 1 & 1 & 1 \\
 \overline{\text{X2}} & 1 & 1 & 0 & 1 & 0 & 0 \\
 \overline{\text{X3}} & 1 & 0 & 1 & 0 & 1 & 0 \\
 \overline{\text{X4}} & 0 & 1 & 0 & 1 & 0 & 1 \\
 \overline{\text{X5}} & 0 & 0 & 1 & 0 & 1 & 1 \\
 \overline{\text{Y5}} & 1 & 1 & 1 & 1 & 1 & 1 \\
\end{array}
\right)\]

\noindent\[\text{spinless p2gg:}\]

\noindent\[\left(
\begin{array}{c|cccc}
 \text{$\Gamma$1} & 1 & 1 & 0 & 0 \\
 \text{$\Gamma$2} & 1 & 1 & 0 & 0 \\
 \text{$\Gamma$3} & 0 & 0 & 1 & 1 \\
 \text{$\Gamma$4} & 0 & 0 & 1 & 1 \\
 \text{S1} & 1 & 0 & 1 & 0 \\
 \text{S2} & 1 & 0 & 1 & 0 \\
 \text{S3} & 0 & 1 & 0 & 1 \\
 \text{S4} & 0 & 1 & 0 & 1 \\
 \text{X1} & 1 & 1 & 1 & 1 \\
 \text{Y1} & 1 & 1 & 1 & 1 \\
\end{array}
\right)\]

\noindent\[\text{spinful p2gg:}\]

\noindent\[\left(
\begin{array}{c|cccc}
 \overline{\text{$\Gamma$5}} & 1 & 1 & 1 & 1 \\
 \overline{\text{S5}} & 1 & 1 & 1 & 1 \\
 \overline{\text{X2}} & 1 & 1 & 0 & 0 \\
 \overline{\text{X3}} & 0 & 0 & 1 & 1 \\
 \overline{\text{X4}} & 1 & 1 & 0 & 0 \\
 \overline{\text{X5}} & 0 & 0 & 1 & 1 \\
 \overline{\text{Y2}} & 1 & 0 & 1 & 0 \\
 \overline{\text{Y3}} & 0 & 1 & 0 & 1 \\
 \overline{\text{Y4}} & 1 & 0 & 1 & 0 \\
 \overline{\text{Y5}} & 0 & 1 & 0 & 1 \\
\end{array}
\right)\]

\noindent\[\text{spinless c2mm:}\]

\noindent\[\left(
\begin{array}{c|cccccccccccccc}
 \text{$\Gamma$1} & 1 & 1 & 1 & 1 & 1 & 0 & 0 & 0 & 0 & 0 & 0 & 0 & 0 & 0 \\
 \text{$\Gamma$2} & 1 & 1 & 1 & 0 & 0 & 1 & 1 & 0 & 0 & 0 & 0 & 0 & 0 & 0 \\
 \text{$\Gamma$3} & 0 & 0 & 0 & 0 & 0 & 0 & 0 & 1 & 1 & 1 & 1 & 1 & 0 & 0 \\
 \text{$\Gamma$4} & 0 & 0 & 0 & 0 & 0 & 0 & 0 & 1 & 1 & 1 & 0 & 0 & 1 & 1 \\
 \text{Y1} & 0 & 0 & 0 & 1 & 1 & 0 & 0 & 1 & 1 & 1 & 0 & 0 & 0 & 0 \\
 \text{Y2} & 0 & 0 & 0 & 0 & 0 & 1 & 1 & 1 & 1 & 1 & 0 & 0 & 0 & 0 \\
 \text{Y3} & 1 & 1 & 1 & 0 & 0 & 0 & 0 & 0 & 0 & 0 & 1 & 1 & 0 & 0 \\
 \text{Y4} & 1 & 1 & 1 & 0 & 0 & 0 & 0 & 0 & 0 & 0 & 0 & 0 & 1 & 1 \\
 \text{S1} & 2 & 1 & 0 & 1 & 0 & 1 & 0 & 2 & 1 & 0 & 1 & 0 & 1 & 0 \\
 \text{S2} & 0 & 1 & 2 & 0 & 1 & 0 & 1 & 0 & 1 & 2 & 0 & 1 & 0 & 1 \\
\end{array}
\right)\]

\noindent\[\text{spinful c2mm:}\]

\noindent\[\left(
\begin{array}{c|ccc}
 \overline{\text{$\Gamma$5}} & 1 & 1 & 1 \\
 \overline{\text{Y5}} & 1 & 1 & 1 \\
 \overline{\text{S3}} & 2 & 1 & 0 \\
 \overline{\text{S4}} & 0 & 1 & 2 \\
\end{array}
\right)\]

\noindent\[\text{spinless p4:}\]

\noindent\[\left(
\begin{array}{c|cccccccccccccccccccccccccccccccc}
 \text{$\Gamma$1} & 1 & 1 & 1 & 1 & 1 & 1 & 1 & 1 & 0 & 0 & 0 & 0 & 0 & 0 & 0 & 0 & 0 & 0 & 0 & 0 & 0 & 0 & 0 & 0 & 0 & 0 & 0 & 0 & 0 & 0 & 0 & 0 \\
 \text{$\Gamma$2} & 0 & 0 & 0 & 0 & 0 & 0 & 0 & 0 & 1 & 1 & 1 & 1 & 1 & 1 & 1 & 1 & 0 & 0 & 0 & 0 & 0 & 0 & 0 & 0 & 0 & 0 & 0 & 0 & 0 & 0 & 0 & 0 \\
 \text{$\Gamma$3} & 0 & 0 & 0 & 0 & 0 & 0 & 0 & 0 & 0 & 0 & 0 & 0 & 0 & 0 & 0 & 0 & 1 & 1 & 1 & 1 & 1 & 1 & 1 & 1 & 0 & 0 & 0 & 0 & 0 & 0 & 0 & 0 \\
 \text{$\Gamma$4} & 0 & 0 & 0 & 0 & 0 & 0 & 0 & 0 & 0 & 0 & 0 & 0 & 0 & 0 & 0 & 0 & 0 & 0 & 0 & 0 & 0 & 0 & 0 & 0 & 1 & 1 & 1 & 1 & 1 & 1 & 1 & 1 \\
 \text{M1} & 1 & 1 & 0 & 0 & 0 & 0 & 0 & 0 & 1 & 1 & 0 & 0 & 0 & 0 & 0 & 0 & 1 & 1 & 0 & 0 & 0 & 0 & 0 & 0 & 1 & 1 & 0 & 0 & 0 & 0 & 0 & 0 \\
 \text{M2} & 0 & 0 & 1 & 1 & 0 & 0 & 0 & 0 & 0 & 0 & 1 & 1 & 0 & 0 & 0 & 0 & 0 & 0 & 1 & 1 & 0 & 0 & 0 & 0 & 0 & 0 & 1 & 1 & 0 & 0 & 0 & 0 \\
 \text{M3} & 0 & 0 & 0 & 0 & 1 & 1 & 0 & 0 & 0 & 0 & 0 & 0 & 1 & 1 & 0 & 0 & 0 & 0 & 0 & 0 & 1 & 1 & 0 & 0 & 0 & 0 & 0 & 0 & 1 & 1 & 0 & 0 \\
 \text{M4} & 0 & 0 & 0 & 0 & 0 & 0 & 1 & 1 & 0 & 0 & 0 & 0 & 0 & 0 & 1 & 1 & 0 & 0 & 0 & 0 & 0 & 0 & 1 & 1 & 0 & 0 & 0 & 0 & 0 & 0 & 1 & 1 \\
 \text{X1} & 1 & 0 & 1 & 0 & 1 & 0 & 1 & 0 & 1 & 0 & 1 & 0 & 1 & 0 & 1 & 0 & 1 & 0 & 1 & 0 & 1 & 0 & 1 & 0 & 1 & 0 & 1 & 0 & 1 & 0 & 1 & 0 \\
 \text{X2} & 0 & 1 & 0 & 1 & 0 & 1 & 0 & 1 & 0 & 1 & 0 & 1 & 0 & 1 & 0 & 1 & 0 & 1 & 0 & 1 & 0 & 1 & 0 & 1 & 0 & 1 & 0 & 1 & 0 & 1 & 0 & 1 \\
\end{array}
\right)\]

\noindent\[\text{spinful p4:}\]

\noindent\[\left(
\begin{array}{c|cccccccccccccccccccccccccccccccc}
 \overline{\text{$\Gamma$5}} & 1 & 1 & 1 & 1 & 1 & 1 & 1 & 1 & 0 & 0 & 0 & 0 & 0 & 0 & 0 & 0 & 0 & 0 & 0 & 0 & 0 & 0 & 0 & 0 & 0 & 0 & 0 & 0 & 0 & 0 &
0 & 0 \\
 \overline{\text{$\Gamma$6}} & 0 & 0 & 0 & 0 & 0 & 0 & 0 & 0 & 1 & 1 & 1 & 1 & 1 & 1 & 1 & 1 & 0 & 0 & 0 & 0 & 0 & 0 & 0 & 0 & 0 & 0 & 0 & 0 & 0 & 0 &
0 & 0 \\
 \overline{\text{$\Gamma$7}} & 0 & 0 & 0 & 0 & 0 & 0 & 0 & 0 & 0 & 0 & 0 & 0 & 0 & 0 & 0 & 0 & 1 & 1 & 1 & 1 & 1 & 1 & 1 & 1 & 0 & 0 & 0 & 0 & 0 & 0 &
0 & 0 \\
 \overline{\text{$\Gamma$8}} & 0 & 0 & 0 & 0 & 0 & 0 & 0 & 0 & 0 & 0 & 0 & 0 & 0 & 0 & 0 & 0 & 0 & 0 & 0 & 0 & 0 & 0 & 0 & 0 & 1 & 1 & 1 & 1 & 1 & 1 &
1 & 1 \\
 \overline{\text{M5}} & 1 & 1 & 0 & 0 & 0 & 0 & 0 & 0 & 1 & 1 & 0 & 0 & 0 & 0 & 0 & 0 & 1 & 1 & 0 & 0 & 0 & 0 & 0 & 0 & 1 & 1 & 0 & 0 & 0 & 0 & 0
& 0 \\
 \overline{\text{M6}} & 0 & 0 & 1 & 1 & 0 & 0 & 0 & 0 & 0 & 0 & 1 & 1 & 0 & 0 & 0 & 0 & 0 & 0 & 1 & 1 & 0 & 0 & 0 & 0 & 0 & 0 & 1 & 1 & 0 & 0 & 0
& 0 \\
 \overline{\text{M7}} & 0 & 0 & 0 & 0 & 1 & 1 & 0 & 0 & 0 & 0 & 0 & 0 & 1 & 1 & 0 & 0 & 0 & 0 & 0 & 0 & 1 & 1 & 0 & 0 & 0 & 0 & 0 & 0 & 1 & 1 & 0
& 0 \\
 \overline{\text{M8}} & 0 & 0 & 0 & 0 & 0 & 0 & 1 & 1 & 0 & 0 & 0 & 0 & 0 & 0 & 1 & 1 & 0 & 0 & 0 & 0 & 0 & 0 & 1 & 1 & 0 & 0 & 0 & 0 & 0 & 0 & 1
& 1 \\
 \overline{\text{X3}} & 1 & 0 & 1 & 0 & 1 & 0 & 1 & 0 & 1 & 0 & 1 & 0 & 1 & 0 & 1 & 0 & 1 & 0 & 1 & 0 & 1 & 0 & 1 & 0 & 1 & 0 & 1 & 0 & 1 & 0 & 1
& 0 \\
 \overline{\text{X4}} & 0 & 1 & 0 & 1 & 0 & 1 & 0 & 1 & 0 & 1 & 0 & 1 & 0 & 1 & 0 & 1 & 0 & 1 & 0 & 1 & 0 & 1 & 0 & 1 & 0 & 1 & 0 & 1 & 0 & 1 & 0
& 1 \\
\end{array}
\right)\]

\noindent\[\text{spinless p4mm:}\]

\noindent\[\left(
\begin{array}{c|cccccccccccccccccccccccccc}
 \text{$\Gamma$1} & 1 & 1 & 1 & 1 & 1 & 1 & 1 & 0 & 0 & 0 & 0 & 0 & 0 & 0 & 0 & 0 & 0 & 0 & 0 & 0 & 0 & 0 & 0 & 0 & 0 & 0 \\
 \text{$\Gamma$2} & 1 & 0 & 0 & 0 & 0 & 0 & 0 & 1 & 1 & 1 & 1 & 1 & 1 & 0 & 0 & 0 & 0 & 0 & 0 & 0 & 0 & 0 & 0 & 0 & 0 & 0 \\
 \text{$\Gamma$3} & 0 & 0 & 0 & 0 & 0 & 0 & 0 & 1 & 1 & 1 & 1 & 0 & 0 & 1 & 1 & 1 & 0 & 0 & 0 & 0 & 0 & 0 & 0 & 0 & 0 & 0 \\
 \text{$\Gamma$4} & 0 & 1 & 1 & 1 & 1 & 0 & 0 & 0 & 0 & 0 & 0 & 0 & 0 & 1 & 0 & 0 & 1 & 1 & 0 & 0 & 0 & 0 & 0 & 0 & 0 & 0 \\
 \text{$\Gamma$5} & 0 & 0 & 0 & 0 & 0 & 0 & 0 & 0 & 0 & 0 & 0 & 0 & 0 & 0 & 0 & 0 & 0 & 0 & 1 & 1 & 1 & 1 & 1 & 1 & 1 & 1 \\
 \text{M1} & 0 & 1 & 0 & 0 & 0 & 1 & 0 & 1 & 0 & 0 & 0 & 0 & 0 & 0 & 1 & 0 & 0 & 0 & 1 & 1 & 1 & 0 & 0 & 0 & 0 & 0 \\
 \text{M2} & 0 & 0 & 1 & 0 & 0 & 0 & 0 & 0 & 1 & 0 & 0 & 1 & 0 & 0 & 0 & 0 & 1 & 0 & 1 & 0 & 0 & 1 & 1 & 0 & 0 & 0 \\
 \text{M3} & 0 & 0 & 1 & 0 & 0 & 0 & 1 & 0 & 1 & 0 & 0 & 0 & 0 & 0 & 0 & 1 & 0 & 0 & 0 & 0 & 0 & 1 & 1 & 1 & 0 & 0 \\
 \text{M4} & 0 & 1 & 0 & 0 & 0 & 0 & 0 & 1 & 0 & 0 & 0 & 0 & 1 & 0 & 0 & 0 & 0 & 1 & 0 & 1 & 1 & 0 & 0 & 1 & 0 & 0 \\
 \text{M5} & 1 & 0 & 0 & 1 & 1 & 0 & 0 & 0 & 0 & 1 & 1 & 0 & 0 & 1 & 0 & 0 & 0 & 0 & 0 & 0 & 0 & 0 & 0 & 0 & 1 & 1 \\
 \text{X1} & 1 & 0 & 1 & 1 & 0 & 1 & 0 & 1 & 0 & 1 & 0 & 1 & 0 & 0 & 0 & 0 & 0 & 0 & 1 & 1 & 0 & 1 & 0 & 0 & 1 & 0 \\
 \text{X2} & 0 & 0 & 1 & 1 & 0 & 0 & 0 & 1 & 0 & 1 & 0 & 0 & 0 & 1 & 0 & 1 & 0 & 1 & 0 & 1 & 0 & 1 & 0 & 1 & 1 & 0 \\
 \text{X3} & 1 & 1 & 0 & 0 & 1 & 0 & 1 & 0 & 1 & 0 & 1 & 0 & 1 & 0 & 0 & 0 & 0 & 0 & 0 & 0 & 1 & 0 & 1 & 1 & 0 & 1 \\
 \text{X4} & 0 & 1 & 0 & 0 & 1 & 0 & 0 & 0 & 1 & 0 & 1 & 0 & 0 & 1 & 1 & 0 & 1 & 0 & 1 & 0 & 1 & 0 & 1 & 0 & 0 & 1 \\
\end{array}
\right)\]

\noindent\[\text{spinful p4mm:}\]

\noindent\[\left(
\begin{array}{c|cccc}
 \overline{\text{$\Gamma$6}} & 1 & 1 & 0 & 0 \\
 \overline{\text{$\Gamma$7}} & 0 & 0 & 1 & 1 \\
 \overline{\text{M6}} & 1 & 0 & 1 & 0 \\
 \overline{\text{M7}} & 0 & 1 & 0 & 1 \\
 \overline{\text{X5}} & 1 & 1 & 1 & 1 \\
\end{array}
\right)\]

\noindent\[\text{spinless p4gm:}\]

\noindent\[\left(
\begin{array}{c|ccccccccccc}
 \text{$\Gamma$1} & 1 & 1 & 1 & 1 & 0 & 0 & 0 & 0 & 0 & 0 & 0 \\
 \text{$\Gamma$2} & 0 & 0 & 0 & 0 & 1 & 1 & 1 & 1 & 0 & 0 & 0 \\
 \text{$\Gamma$3} & 1 & 0 & 0 & 0 & 1 & 1 & 1 & 0 & 0 & 0 & 0 \\
 \text{$\Gamma$4} & 0 & 1 & 1 & 1 & 0 & 0 & 0 & 1 & 0 & 0 & 0 \\
 \text{$\Gamma$5} & 0 & 0 & 0 & 0 & 0 & 0 & 0 & 0 & 1 & 1 & 1 \\
 \text{M1} & 1 & 1 & 0 & 0 & 1 & 0 & 0 & 0 & 1 & 0 & 0 \\
 \text{M2} & 0 & 0 & 1 & 0 & 0 & 1 & 0 & 1 & 0 & 1 & 0 \\
 \text{M3} & 1 & 0 & 1 & 0 & 0 & 1 & 0 & 0 & 0 & 1 & 0 \\
 \text{M4} & 0 & 1 & 0 & 0 & 1 & 0 & 0 & 1 & 1 & 0 & 0 \\
 \text{M5} & 0 & 0 & 0 & 1 & 0 & 0 & 1 & 0 & 0 & 0 & 1 \\
 \text{X1} & 1 & 1 & 1 & 1 & 1 & 1 & 1 & 1 & 1 & 1 & 1 \\
\end{array}
\right)\]

\noindent\[\text{spinful p4gm:}\]

\noindent\[\left(
\begin{array}{c|cccccccc}
 \overline{\text{$\Gamma$6}} & 1 & 1 & 1 & 1 & 0 & 0 & 0 & 0 \\
 \overline{\text{$\Gamma$7}} & 0 & 0 & 0 & 0 & 1 & 1 & 1 & 1 \\
 \overline{\text{M6}} & 1 & 1 & 0 & 0 & 1 & 1 & 0 & 0 \\
 \overline{\text{M7}} & 0 & 0 & 1 & 1 & 0 & 0 & 1 & 1 \\
 \overline{\text{X2}} & 1 & 0 & 1 & 0 & 1 & 0 & 1 & 0 \\
 \overline{\text{X3}} & 0 & 1 & 0 & 1 & 0 & 1 & 0 & 1 \\
 \overline{\text{X4}} & 1 & 0 & 1 & 0 & 1 & 0 & 1 & 0 \\
 \overline{\text{X5}} & 0 & 1 & 0 & 1 & 0 & 1 & 0 & 1 \\
\end{array}
\right)\]

\noindent\[\text{spinless p3:}\]

\noindent\[\left(
\begin{array}{c|ccccccccccccccccccccccccccc}
 \text{$\Gamma$1} & 1 & 1 & 1 & 1 & 1 & 1 & 1 & 1 & 1 & 0 & 0 & 0 & 0 & 0 & 0 & 0 & 0 & 0 & 0 & 0 & 0 & 0 & 0 & 0 & 0 & 0 & 0 \\
 \text{$\Gamma$2} & 0 & 0 & 0 & 0 & 0 & 0 & 0 & 0 & 0 & 1 & 1 & 1 & 1 & 1 & 1 & 1 & 1 & 1 & 0 & 0 & 0 & 0 & 0 & 0 & 0 & 0 & 0 \\
 \text{$\Gamma$3} & 0 & 0 & 0 & 0 & 0 & 0 & 0 & 0 & 0 & 0 & 0 & 0 & 0 & 0 & 0 & 0 & 0 & 0 & 1 & 1 & 1 & 1 & 1 & 1 & 1 & 1 & 1 \\
 \text{K1} & 1 & 1 & 1 & 0 & 0 & 0 & 0 & 0 & 0 & 1 & 1 & 1 & 0 & 0 & 0 & 0 & 0 & 0 & 1 & 1 & 1 & 0 & 0 & 0 & 0 & 0 & 0 \\
 \text{K2} & 0 & 0 & 0 & 1 & 1 & 1 & 0 & 0 & 0 & 0 & 0 & 0 & 1 & 1 & 1 & 0 & 0 & 0 & 0 & 0 & 0 & 1 & 1 & 1 & 0 & 0 & 0 \\
 \text{K3} & 0 & 0 & 0 & 0 & 0 & 0 & 1 & 1 & 1 & 0 & 0 & 0 & 0 & 0 & 0 & 1 & 1 & 1 & 0 & 0 & 0 & 0 & 0 & 0 & 1 & 1 & 1 \\
 \text{KA1} & 1 & 0 & 0 & 1 & 0 & 0 & 1 & 0 & 0 & 1 & 0 & 0 & 1 & 0 & 0 & 1 & 0 & 0 & 1 & 0 & 0 & 1 & 0 & 0 & 1 & 0 & 0 \\
 \text{KA2} & 0 & 1 & 0 & 0 & 1 & 0 & 0 & 1 & 0 & 0 & 1 & 0 & 0 & 1 & 0 & 0 & 1 & 0 & 0 & 1 & 0 & 0 & 1 & 0 & 0 & 1 & 0 \\
 \text{KA3} & 0 & 0 & 1 & 0 & 0 & 1 & 0 & 0 & 1 & 0 & 0 & 1 & 0 & 0 & 1 & 0 & 0 & 1 & 0 & 0 & 1 & 0 & 0 & 1 & 0 & 0 & 1 \\
 \text{M1} & 1 & 1 & 1 & 1 & 1 & 1 & 1 & 1 & 1 & 1 & 1 & 1 & 1 & 1 & 1 & 1 & 1 & 1 & 1 & 1 & 1 & 1 & 1 & 1 & 1 & 1 & 1 \\
\end{array}
\right)\]

\noindent\[\text{spinful p3:}\]

\noindent\[\left(
\begin{array}{c|ccccccccccccccccccccccccccc}
 \overline{\text{$\Gamma$4}} & 1 & 1 & 1 & 1 & 1 & 1 & 1 & 1 & 1 & 0 & 0 & 0 & 0 & 0 & 0 & 0 & 0 & 0 & 0 & 0 & 0 & 0 & 0 & 0 & 0 & 0 & 0 \\
 \overline{\text{$\Gamma$5}} & 0 & 0 & 0 & 0 & 0 & 0 & 0 & 0 & 0 & 1 & 1 & 1 & 1 & 1 & 1 & 1 & 1 & 1 & 0 & 0 & 0 & 0 & 0 & 0 & 0 & 0 & 0 \\
 \overline{\text{$\Gamma$6}} & 0 & 0 & 0 & 0 & 0 & 0 & 0 & 0 & 0 & 0 & 0 & 0 & 0 & 0 & 0 & 0 & 0 & 0 & 1 & 1 & 1 & 1 & 1 & 1 & 1 & 1 & 1 \\
 \overline{\text{K4}} & 1 & 1 & 1 & 0 & 0 & 0 & 0 & 0 & 0 & 1 & 1 & 1 & 0 & 0 & 0 & 0 & 0 & 0 & 1 & 1 & 1 & 0 & 0 & 0 & 0 & 0 & 0 \\
 \overline{\text{K5}} & 0 & 0 & 0 & 1 & 1 & 1 & 0 & 0 & 0 & 0 & 0 & 0 & 1 & 1 & 1 & 0 & 0 & 0 & 0 & 0 & 0 & 1 & 1 & 1 & 0 & 0 & 0 \\
 \overline{\text{K6}} & 0 & 0 & 0 & 0 & 0 & 0 & 1 & 1 & 1 & 0 & 0 & 0 & 0 & 0 & 0 & 1 & 1 & 1 & 0 & 0 & 0 & 0 & 0 & 0 & 1 & 1 & 1 \\
 \overline{\text{KA4}} & 1 & 0 & 0 & 1 & 0 & 0 & 1 & 0 & 0 & 1 & 0 & 0 & 1 & 0 & 0 & 1 & 0 & 0 & 1 & 0 & 0 & 1 & 0 & 0 & 1 & 0 & 0 \\
 \overline{\text{KA5}} & 0 & 1 & 0 & 0 & 1 & 0 & 0 & 1 & 0 & 0 & 1 & 0 & 0 & 1 & 0 & 0 & 1 & 0 & 0 & 1 & 0 & 0 & 1 & 0 & 0 & 1 & 0 \\
 \overline{\text{KA6}} & 0 & 0 & 1 & 0 & 0 & 1 & 0 & 0 & 1 & 0 & 0 & 1 & 0 & 0 & 1 & 0 & 0 & 1 & 0 & 0 & 1 & 0 & 0 & 1 & 0 & 0 & 1 \\
 \overline{\text{M2}} & 1 & 1 & 1 & 1 & 1 & 1 & 1 & 1 & 1 & 1 & 1 & 1 & 1 & 1 & 1 & 1 & 1 & 1 & 1 & 1 & 1 & 1 & 1 & 1 & 1 & 1 & 1 \\
\end{array}
\right)\]

\noindent\[\text{spinless p3m1:}\]

\noindent\[\left(
\begin{array}{c|cccccccccccc}
 \text{$\Gamma$1} & 1 & 1 & 1 & 0 & 0 & 0 & 0 & 0 & 0 & 0 & 0 & 0 \\
 \text{$\Gamma$2} & 0 & 0 & 0 & 1 & 1 & 1 & 0 & 0 & 0 & 0 & 0 & 0 \\
 \text{$\Gamma$3} & 0 & 0 & 0 & 0 & 0 & 0 & 1 & 1 & 1 & 1 & 1 & 1 \\
 \text{K1} & 1 & 0 & 0 & 1 & 0 & 0 & 2 & 1 & 1 & 0 & 0 & 0 \\
 \text{K2} & 0 & 1 & 0 & 0 & 1 & 0 & 0 & 1 & 0 & 2 & 1 & 0 \\
 \text{K3} & 0 & 0 & 1 & 0 & 0 & 1 & 0 & 0 & 1 & 0 & 1 & 2 \\
 \text{M1} & 1 & 1 & 1 & 0 & 0 & 0 & 1 & 1 & 1 & 1 & 1 & 1 \\
 \text{M2} & 0 & 0 & 0 & 1 & 1 & 1 & 1 & 1 & 1 & 1 & 1 & 1 \\
\end{array}
\right)\]

\noindent\[\text{spinful p3m1:}\]

\noindent\[\left(
\begin{array}{c|cccccccccccc}
 \overline{\text{$\Gamma$4}} & 1 & 1 & 1 & 0 & 0 & 0 & 0 & 0 & 0 & 0 & 0 & 0 \\
 \overline{\text{$\Gamma$5}} & 0 & 0 & 0 & 1 & 1 & 1 & 0 & 0 & 0 & 0 & 0 & 0 \\
 \overline{\text{$\Gamma$6}} & 0 & 0 & 0 & 0 & 0 & 0 & 1 & 1 & 1 & 1 & 1 & 1 \\
 \overline{\text{K4}} & 1 & 0 & 0 & 1 & 0 & 0 & 2 & 1 & 1 & 0 & 0 & 0 \\
 \overline{\text{K5}} & 0 & 1 & 0 & 0 & 1 & 0 & 0 & 1 & 0 & 2 & 1 & 0 \\
 \overline{\text{K6}} & 0 & 0 & 1 & 0 & 0 & 1 & 0 & 0 & 1 & 0 & 1 & 2 \\
 \overline{\text{M3}} & 1 & 1 & 1 & 0 & 0 & 0 & 1 & 1 & 1 & 1 & 1 & 1 \\
 \overline{\text{M4}} & 0 & 0 & 0 & 1 & 1 & 1 & 1 & 1 & 1 & 1 & 1 & 1 \\
\end{array}
\right)\]

\noindent\[\text{spinless p31m:}\]

\noindent\[\left(
\begin{array}{c|ccccccccc}
 \text{$\Gamma$1} & 1 & 1 & 1 & 1 & 0 & 0 & 0 & 0 & 0 \\
 \text{$\Gamma$2} & 1 & 1 & 1 & 0 & 1 & 0 & 0 & 0 & 0 \\
 \text{$\Gamma$3} & 0 & 0 & 0 & 0 & 0 & 1 & 1 & 1 & 1 \\
 \text{K1} & 1 & 0 & 0 & 1 & 0 & 1 & 1 & 0 & 0 \\
 \text{K2} & 1 & 0 & 0 & 0 & 1 & 1 & 1 & 0 & 0 \\
 \text{K3} & 0 & 1 & 1 & 0 & 0 & 0 & 0 & 1 & 1 \\
 \text{KA1} & 0 & 1 & 0 & 1 & 0 & 1 & 0 & 1 & 0 \\
 \text{KA2} & 0 & 1 & 0 & 0 & 1 & 1 & 0 & 1 & 0 \\
 \text{KA3} & 1 & 0 & 1 & 0 & 0 & 0 & 1 & 0 & 1 \\
 \text{M1} & 1 & 1 & 1 & 1 & 0 & 1 & 1 & 1 & 1 \\
 \text{M2} & 1 & 1 & 1 & 0 & 1 & 1 & 1 & 1 & 1 \\
\end{array}
\right)\]

\noindent\[\text{spinful p31m:}\]

\noindent\[\left(
\begin{array}{c|ccccccccc}
 \overline{\text{$\Gamma$4}} & 1 & 1 & 1 & 1 & 0 & 0 & 0 & 0 & 0 \\
 \overline{\text{$\Gamma$5}} & 1 & 1 & 1 & 0 & 1 & 0 & 0 & 0 & 0 \\
 \overline{\text{$\Gamma$6}} & 0 & 0 & 0 & 0 & 0 & 1 & 1 & 1 & 1 \\
 \overline{\text{K4}} & 1 & 0 & 0 & 1 & 0 & 1 & 1 & 0 & 0 \\
 \overline{\text{K5}} & 1 & 0 & 0 & 0 & 1 & 1 & 1 & 0 & 0 \\
 \overline{\text{K6}} & 0 & 1 & 1 & 0 & 0 & 0 & 0 & 1 & 1 \\
 \overline{\text{KA4}} & 0 & 1 & 0 & 0 & 1 & 1 & 0 & 1 & 0 \\
 \overline{\text{KA5}} & 0 & 1 & 0 & 1 & 0 & 1 & 0 & 1 & 0 \\
 \overline{\text{KA6}} & 1 & 0 & 1 & 0 & 0 & 0 & 1 & 0 & 1 \\
 \overline{\text{M3}} & 1 & 1 & 1 & 1 & 0 & 1 & 1 & 1 & 1 \\
 \overline{\text{M4}} & 1 & 1 & 1 & 0 & 1 & 1 & 1 & 1 & 1 \\
\end{array}
\right)\]

\noindent\[\text{spinless p6:}\]

\noindent\[\left(
\begin{array}{c|cccccccccccccccccccccccccccccccccccc}
 \text{$\Gamma$1} & 1 & 1 & 1 & 1 & 1 & 1 & 0 & 0 & 0 & 0 & 0 & 0 & 0 & 0 & 0 & 0 & 0 & 0 & 0 & 0 & 0 & 0 & 0 & 0 & 0 & 0 & 0 & 0 & 0 & 0 & 0 & 0 & 0 &
0 & 0 & 0 \\
 \text{$\Gamma$2} & 0 & 0 & 0 & 0 & 0 & 0 & 1 & 1 & 1 & 1 & 1 & 1 & 0 & 0 & 0 & 0 & 0 & 0 & 0 & 0 & 0 & 0 & 0 & 0 & 0 & 0 & 0 & 0 & 0 & 0 & 0 & 0 & 0 &
0 & 0 & 0 \\
 \text{$\Gamma$3} & 0 & 0 & 0 & 0 & 0 & 0 & 0 & 0 & 0 & 0 & 0 & 0 & 1 & 1 & 1 & 1 & 1 & 1 & 0 & 0 & 0 & 0 & 0 & 0 & 0 & 0 & 0 & 0 & 0 & 0 & 0 & 0 & 0 &
0 & 0 & 0 \\
 \text{$\Gamma$4} & 0 & 0 & 0 & 0 & 0 & 0 & 0 & 0 & 0 & 0 & 0 & 0 & 0 & 0 & 0 & 0 & 0 & 0 & 1 & 1 & 1 & 1 & 1 & 1 & 0 & 0 & 0 & 0 & 0 & 0 & 0 & 0 & 0 &
0 & 0 & 0 \\
 \text{$\Gamma$5} & 0 & 0 & 0 & 0 & 0 & 0 & 0 & 0 & 0 & 0 & 0 & 0 & 0 & 0 & 0 & 0 & 0 & 0 & 0 & 0 & 0 & 0 & 0 & 0 & 1 & 1 & 1 & 1 & 1 & 1 & 0 & 0 & 0 &
0 & 0 & 0 \\
 \text{$\Gamma$6} & 0 & 0 & 0 & 0 & 0 & 0 & 0 & 0 & 0 & 0 & 0 & 0 & 0 & 0 & 0 & 0 & 0 & 0 & 0 & 0 & 0 & 0 & 0 & 0 & 0 & 0 & 0 & 0 & 0 & 0 & 1 & 1 & 1 &
1 & 1 & 1 \\
 \text{K1} & 1 & 1 & 0 & 0 & 0 & 0 & 1 & 1 & 0 & 0 & 0 & 0 & 1 & 1 & 0 & 0 & 0 & 0 & 1 & 1 & 0 & 0 & 0 & 0 & 1 & 1 & 0 & 0 & 0 & 0 & 1 & 1 & 0 &
0 & 0 & 0 \\
 \text{K2} & 0 & 0 & 1 & 1 & 0 & 0 & 0 & 0 & 1 & 1 & 0 & 0 & 0 & 0 & 1 & 1 & 0 & 0 & 0 & 0 & 1 & 1 & 0 & 0 & 0 & 0 & 1 & 1 & 0 & 0 & 0 & 0 & 1 &
1 & 0 & 0 \\
 \text{K3} & 0 & 0 & 0 & 0 & 1 & 1 & 0 & 0 & 0 & 0 & 1 & 1 & 0 & 0 & 0 & 0 & 1 & 1 & 0 & 0 & 0 & 0 & 1 & 1 & 0 & 0 & 0 & 0 & 1 & 1 & 0 & 0 & 0 &
0 & 1 & 1 \\
 \text{M1} & 1 & 0 & 1 & 0 & 1 & 0 & 1 & 0 & 1 & 0 & 1 & 0 & 1 & 0 & 1 & 0 & 1 & 0 & 1 & 0 & 1 & 0 & 1 & 0 & 1 & 0 & 1 & 0 & 1 & 0 & 1 & 0 & 1 &
0 & 1 & 0 \\
 \text{M2} & 0 & 1 & 0 & 1 & 0 & 1 & 0 & 1 & 0 & 1 & 0 & 1 & 0 & 1 & 0 & 1 & 0 & 1 & 0 & 1 & 0 & 1 & 0 & 1 & 0 & 1 & 0 & 1 & 0 & 1 & 0 & 1 & 0 &
1 & 0 & 1 \\
\end{array}
\right)\]

\noindent\[\text{spinful p6:}\]

\noindent\[\left(
\begin{array}{c|cccccccccccccccccccccccccccccccccccc}
 \overline{\text{$\Gamma$7}} & 1 & 1 & 1 & 1 & 1 & 1 & 0 & 0 & 0 & 0 & 0 & 0 & 0 & 0 & 0 & 0 & 0 & 0 & 0 & 0 & 0 & 0 & 0 & 0 & 0 & 0 & 0 & 0 & 0 & 0 &
0 & 0 & 0 & 0 & 0 & 0 \\
 \overline{\text{$\Gamma$8}} & 0 & 0 & 0 & 0 & 0 & 0 & 1 & 1 & 1 & 1 & 1 & 1 & 0 & 0 & 0 & 0 & 0 & 0 & 0 & 0 & 0 & 0 & 0 & 0 & 0 & 0 & 0 & 0 & 0 & 0 &
0 & 0 & 0 & 0 & 0 & 0 \\
 \overline{\text{$\Gamma$9}} & 0 & 0 & 0 & 0 & 0 & 0 & 0 & 0 & 0 & 0 & 0 & 0 & 1 & 1 & 1 & 1 & 1 & 1 & 0 & 0 & 0 & 0 & 0 & 0 & 0 & 0 & 0 & 0 & 0 & 0 &
0 & 0 & 0 & 0 & 0 & 0 \\
 \overline{\text{$\Gamma$10}} & 0 & 0 & 0 & 0 & 0 & 0 & 0 & 0 & 0 & 0 & 0 & 0 & 0 & 0 & 0 & 0 & 0 & 0 & 1 & 1 & 1 & 1 & 1 & 1 & 0 & 0 & 0 & 0 & 0 & 0 &
0 & 0 & 0 & 0 & 0 & 0 \\
 \overline{\text{$\Gamma$11}} & 0 & 0 & 0 & 0 & 0 & 0 & 0 & 0 & 0 & 0 & 0 & 0 & 0 & 0 & 0 & 0 & 0 & 0 & 0 & 0 & 0 & 0 & 0 & 0 & 1 & 1 & 1 & 1 & 1 & 1 &
0 & 0 & 0 & 0 & 0 & 0 \\
 \overline{\text{$\Gamma$12}} & 0 & 0 & 0 & 0 & 0 & 0 & 0 & 0 & 0 & 0 & 0 & 0 & 0 & 0 & 0 & 0 & 0 & 0 & 0 & 0 & 0 & 0 & 0 & 0 & 0 & 0 & 0 & 0 & 0 & 0 &
1 & 1 & 1 & 1 & 1 & 1 \\
 \overline{\text{K4}} & 1 & 1 & 0 & 0 & 0 & 0 & 1 & 1 & 0 & 0 & 0 & 0 & 1 & 1 & 0 & 0 & 0 & 0 & 1 & 1 & 0 & 0 & 0 & 0 & 1 & 1 & 0 & 0 & 0 & 0 & 1
& 1 & 0 & 0 & 0 & 0 \\
 \overline{\text{K5}} & 0 & 0 & 1 & 1 & 0 & 0 & 0 & 0 & 1 & 1 & 0 & 0 & 0 & 0 & 1 & 1 & 0 & 0 & 0 & 0 & 1 & 1 & 0 & 0 & 0 & 0 & 1 & 1 & 0 & 0 & 0
& 0 & 1 & 1 & 0 & 0 \\
 \overline{\text{K6}} & 0 & 0 & 0 & 0 & 1 & 1 & 0 & 0 & 0 & 0 & 1 & 1 & 0 & 0 & 0 & 0 & 1 & 1 & 0 & 0 & 0 & 0 & 1 & 1 & 0 & 0 & 0 & 0 & 1 & 1 & 0
& 0 & 0 & 0 & 1 & 1 \\
 \overline{\text{M3}} & 1 & 0 & 1 & 0 & 1 & 0 & 1 & 0 & 1 & 0 & 1 & 0 & 1 & 0 & 1 & 0 & 1 & 0 & 1 & 0 & 1 & 0 & 1 & 0 & 1 & 0 & 1 & 0 & 1 & 0 & 1
& 0 & 1 & 0 & 1 & 0 \\
 \overline{\text{M4}} & 0 & 1 & 0 & 1 & 0 & 1 & 0 & 1 & 0 & 1 & 0 & 1 & 0 & 1 & 0 & 1 & 0 & 1 & 0 & 1 & 0 & 1 & 0 & 1 & 0 & 1 & 0 & 1 & 0 & 1 & 0
& 1 & 0 & 1 & 0 & 1 \\
\end{array}
\right)\]

\noindent\[\text{spinless p6mm:}\]

\noindent\[\left(
\begin{array}{c|cccccccccccccccccccc}
 \text{$\Gamma$1} & 1 & 1 & 1 & 1 & 1 & 0 & 0 & 0 & 0 & 0 & 0 & 0 & 0 & 0 & 0 & 0 & 0 & 0 & 0 & 0 \\
 \text{$\Gamma$2} & 1 & 1 & 1 & 0 & 0 & 1 & 1 & 0 & 0 & 0 & 0 & 0 & 0 & 0 & 0 & 0 & 0 & 0 & 0 & 0 \\
 \text{$\Gamma$3} & 0 & 0 & 0 & 0 & 0 & 1 & 0 & 1 & 1 & 1 & 1 & 0 & 0 & 0 & 0 & 0 & 0 & 0 & 0 & 0 \\
 \text{$\Gamma$4} & 0 & 0 & 0 & 1 & 0 & 0 & 0 & 1 & 1 & 1 & 0 & 1 & 0 & 0 & 0 & 0 & 0 & 0 & 0 & 0 \\
 \text{$\Gamma$5} & 0 & 0 & 0 & 0 & 0 & 0 & 0 & 0 & 0 & 0 & 0 & 0 & 1 & 1 & 1 & 1 & 0 & 0 & 0 & 0 \\
 \text{$\Gamma$6} & 0 & 0 & 0 & 0 & 0 & 0 & 0 & 0 & 0 & 0 & 0 & 0 & 0 & 0 & 0 & 0 & 1 & 1 & 1 & 1 \\
 \text{K1} & 1 & 0 & 0 & 0 & 1 & 0 & 0 & 1 & 0 & 0 & 1 & 0 & 1 & 1 & 0 & 0 & 1 & 1 & 0 & 0 \\
 \text{K2} & 1 & 0 & 0 & 0 & 0 & 0 & 1 & 1 & 0 & 0 & 0 & 1 & 1 & 1 & 0 & 0 & 1 & 1 & 0 & 0 \\
 \text{K3} & 0 & 1 & 1 & 1 & 0 & 1 & 0 & 0 & 1 & 1 & 0 & 0 & 0 & 0 & 1 & 1 & 0 & 0 & 1 & 1 \\
 \text{M1} & 0 & 1 & 0 & 1 & 1 & 0 & 0 & 1 & 1 & 0 & 0 & 0 & 1 & 0 & 1 & 0 & 1 & 0 & 1 & 0 \\
 \text{M2} & 0 & 1 & 0 & 0 & 0 & 1 & 1 & 1 & 1 & 0 & 0 & 0 & 1 & 0 & 1 & 0 & 1 & 0 & 1 & 0 \\
 \text{M3} & 1 & 0 & 1 & 0 & 0 & 1 & 0 & 0 & 0 & 1 & 1 & 0 & 0 & 1 & 0 & 1 & 0 & 1 & 0 & 1 \\
 \text{M4} & 1 & 0 & 1 & 1 & 0 & 0 & 0 & 0 & 0 & 1 & 0 & 1 & 0 & 1 & 0 & 1 & 0 & 1 & 0 & 1 \\
\end{array}
\right)\]

\noindent\[\text{spinful p6mm:}\]

\noindent\[\left(
\begin{array}{c|cccccc}
 \overline{\text{$\Gamma$7}} & 1 & 1 & 0 & 0 & 0 & 0 \\
 \overline{\text{$\Gamma$8}} & 0 & 0 & 1 & 1 & 0 & 0 \\
 \overline{\text{$\Gamma$9}} & 0 & 0 & 0 & 0 & 1 & 1 \\
 \overline{\text{K4}} & 1 & 0 & 1 & 0 & 1 & 0 \\
 \overline{\text{K5}} & 1 & 0 & 1 & 0 & 1 & 0 \\
 \overline{\text{K6}} & 0 & 1 & 0 & 1 & 0 & 1 \\
 \overline{\text{M5}} & 1 & 1 & 1 & 1 & 1 & 1 \\
\end{array}
\right)\]

\end{document}